\documentclass[a4paper,12pt,times,biblatex,numbered,print,index]{PhDThesisPSnPDF}
\input{Preamble/preamble}

\title{Global Analysis of Nuclear Parton Distribution Functions using Markov Chain Monte Carlo}
\author{Nasim Derakhshanian}

\dept{Department of Theoretical Particle Physics}

\university{The Henryk Niewodniczański Institute of Nuclear Physics\\
Polish Academy of Sciences}
\crest{\includegraphics[width=0.2\textwidth]{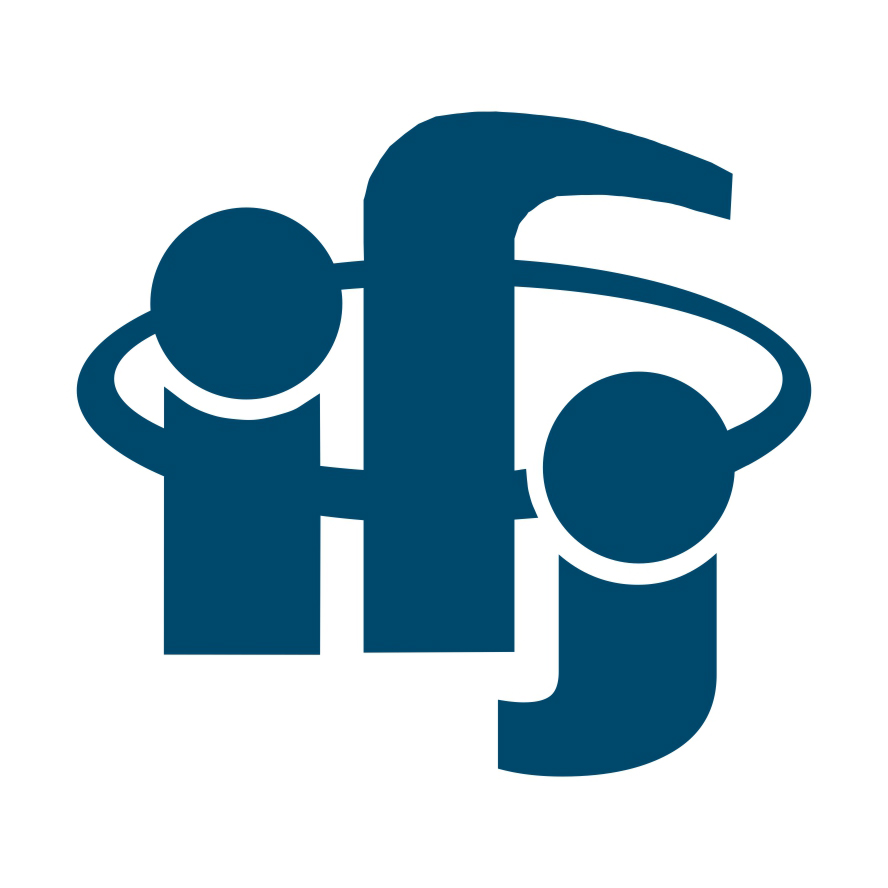}}
\supervisor{Dr. hab A. Kusina}
\degreetitle{Doctor of Philosophy}

\college{IFJ PAN}

\begin{document}

\frontmatter

\maketitle

% ******************************* Thesis Dedidcation ********************************

\begin{dedication} 

To my parents, for making this journey possible.

To my beloved husband, for helping me see it through.

\end{dedication}

% ************************** Thesis Abstract *****************************
% Use `abstract' as an option in the document class to print only the titlepage and the abstract.
\begin{abstract}
Global analyses of nuclear parton distribution functions (nPDFs) have conventionally used the Hessian method for uncertainty estimation. While this approach is computationally efficient, it relies on a local quadratic expansion of the $\chi^2$ function—which essentially assumes that the parameter space follows a Gaussian structure. In the context of nPDFs, where experimental constraints are limited and correlations are strong, this approximation can become inadequate, failing to capture non-Gaussian features such as asymmetric uncertainties, flat directions, and multiple local minima.
In this thesis, a fully Bayesian framework for nPDF determination is developed and implemented using Markov Chain Monte Carlo (MCMC) techniques within the nCTEQ analysis framework. By employing an adaptive Metropolis–Hastings algorithm, the approach allows for direct sampling of the posterior probability distribution, offering a more comprehensive treatment of parameter uncertainties. The resulting analysis reveals a highly non-trivial structure of the parameter space, including pronounced non-Gaussian behavior and the presence of secondary minima, particularly in the valence distributions.
This work features the first global single-nucleus analysis of lead (Pb) PDFs based exclusively on lead data, allowing for a controlled investigation of the intrinsic structure of the fit. This is complemented by a multi-nuclei analysis incorporating data from a range of nuclear targets through a standard analytic $A$-dependence. The comparison between these two setups demonstrates that the inclusion of lighter nuclei significantly constrains quark distributions and change the shape of the extracted lead PDFs, while the gluon distribution remains comparatively stable.
In parallel, a Hessian-based analysis is performed to provide a direct comparison with the Bayesian results. This comparison highlights systematic limitations of the Gaussian approximation, particularly in regions where the likelihood exhibits strong nonlinearity or multi-modality. Overall, the results establish MCMC-based Bayesian inference as a robust and flexible framework for nPDF determination, offering improved reliability in uncertainty quantification and deeper insight into the underlying parameter space.
\end{abstract}

\section*{Streszczenie}
Globalne analizy jądrowych rozkładów partonów (nPDF) tradycyjnie opierają się na metodzie ,,Hessian'' w celu oszacowania niepewności. Choć podejście to jest efektywne obliczeniowo, bazuje ono na lokalnym rozwinięciu kwadratowym funkcji $\chi^2$, co w istocie zakłada gaussowską strukturę przestrzeni parametrów. W kontekście rozkładów nPDF, gdzie ilość danych eksperymentalnych jest ograniczona, a korelacje silne, przybliżenie to może okazać się niewystarczające, nie pozwalając na uchwycenie cech niegaussowskich, takich jak asymetryczne niepewności, płaskie kierunki czy istnienie wielu lokalnych minimów.
W niniejszej rozprawie opracowano i wdrożono w pełni bayesowskie podejście do wyznaczania rozkładów nPDF, wykorzystujące techniki Monte Carlo oparte o łańcuchy Markowa (MCMC), podejście to zostało zaimplementowane w ramach kodu nCTEQ. Dzięki zastosowaniu adaptacyjnego algorytmu Metropolisa-Hastingsa, podejście to umożliwia bezpośrednie próbkowanie rozkładu a posteriori, oferując bardziej kompleksowy opis niepewności parametrów. Przeprowadzona analiza ujawnia wysoce nietrywialną strukturę przestrzeni parametrów, w tym wyraźne zachowania niegaussowskie oraz obecność dodatkowych minimów, szczególnie w rozkładach walencyjnych.
Kluczowym elementem pracy jest pierwsza globalna analiza rozkładów PDF dla pojedynczego jądra atomowego -- ołowiu (Pb), oparta wyłącznie na danych dla ołowiu, co pozwala na kontrolowane badanie wewnętrznej struktury dopasowania. Zostało to uzupełnione analizą wielojądrową, uwzględniającą dane z szeregu różnych tarcz jądrowych poprzez standardową analityczną zależność od liczby masowej~$A$. Porównanie obu tych podejść wykazuje, że włączenie lżejszych jąder znacząco zawęża rozkłady kwarkowe i zmienia kształt wyznaczonych nPDF-ów dla ołowiu, podczas gdy rozkład gluonów pozostaje stosunkowo stabilny.
Równolegle przeprowadzono analizę opartą na metodzie Hessian, aby umożliwić bezpośrednie porównanie z wynikami bayesowskimi. Porównanie to podkreśla systematyczne ograniczenia przybliżenia gaussowskiego, szczególnie w obszarach, w których funkcja wiarygodności wykazuje silną nieliniowość lub wielomodalność. Podsumowując, uzyskane wyniki potwierdzają, że wnioskowanie bayesowskie oparte na metodzie MCMC stanowi solidne i elastyczne narzędzie do wyznaczania rozkładów nPDF, oferując większą wiarygodność w kwantyfikacji niepewności oraz głębszy wgląd w strukturę przestrzeni parametrów.

%\include{Declaration/declaration}
% ************************** Thesis Acknowledgements **************************

\begin{acknowledgements}      

I would like to thank my PhD supervisor, \textbf{Dr. hab. Aleksander Kusina}, for his guidance and support throughout this project. Working with him has been a very valuable experience of my academic journey, and it is something I will truly miss. I appreciate his expertise, patience, and for creating a collaborative environment in which I did not feel alone in this project. 

I am also grateful to my collaborators in the nCTEQ group, especially \textbf{Tomas Je\v{z}o}, \textbf{Peter Risse}, \textbf{Karol Kova\v{r}\'{i}k}, and \textbf{Jan Wissmann}. Being part of such a friendly and professional collaboration was a very valuable experience, and I learned a great deal from working together.

My thanks also go to the members of the Department of Particle Theory (NZ42) for providing a welcoming and friendly working environment. It has been a pleasure to work with them.

This work was supported by the National Science Centre, Poland, under the SONATA BIS grant No. 2019/34/E/ST2/00186. I acknowledge financial support provided by the NAWA-STER Program, project No. PPI/STE/2020/1/00020/U/00001.
\end{acknowledgements}

\section*{Articles}
%\addcontentsline{toc}{chapter}{Articles}
\begin{enumerate}
\item \textbf{Determination of Nuclear PDFs using Markov Chain Monte Carlo Methods}\\
\textbf{N.~Derakhshanian}, P.~Risse, T.~Je\v{z}o, M. Klasen, K.~Kova\v{r}\'{\i}k, A.~Kusina\\
\textit{arXiv:2603.13150 [hep-ph]}, under review in Phys.\ Rev.\ D

\item \textbf{Determination of proton PDF uncertainties with Markov chain Monte Carlo}\\
P.~Risse, \textbf{N.~Derakhshanian}, T.~Je\v{z}o, K.~Kova\v{r}\'{\i}k, A.~Kusina\\
\textit{Phys.\ Rev.\ D \textbf{113} (2026) 054049}\\
\textit{arXiv:2510.16158 [hep-ph]}
\vspace{0.5em}

\item \textbf{Modification of Quark-Gluon Distributions in Nuclei by Correlated Nucleon Pairs}
A.~W.~Denniston, T.~Ježo, A.~Kusina, \textbf{N.~Derakhshanian}, \textit{et al.} (nCTEQ Collaboration)\\
\textit{Phys.\ Rev.\ Lett.\ \textbf{133} (2024) 152502} \\
\textit{arXiv:2312.16293 [hep-ph]}
\end{enumerate}
\vspace{1em}

\section*{Proceedings}

\begin{enumerate}

\item \textbf{Estimating nPDF Uncertainties via Markov Chain Monte Carlo Methods}\\
\textbf{N.~Derakhshanian}, P.~Risse, T.~Je\v{z}o, K.~Kova\v{r}\'{\i}k, A.~Kusina\\
\textit{PoS DIS2024 (2025) 058}\\
\textit{doi:10.22323/1.469.0058}

\item \textbf{Towards a New nCTEQ global nPDF release}\\
P.~Risse, \textbf{N.~Derakhshanian}, P. Duwentäster, \textit{et al.}\\
\textit{Proc.\ DIS2023} \\
\textit{arXiv:2307.07814 [hep-ph]}
\vspace{0.5em}

\item \textbf{Nuclear PDF Determination Using Markov Chain Monte Carlo Methods}\\
\textbf{N.~Derakhshanian}, P.~Risse, T.~Ježo, M.~Klasen, K.~Kovařík, \textit{et al.}\\
\textit{Acta Phys.\ Polon.\ Supp.\ \textbf{16} (2023) 7--A33}\\
\textit{doi:10.5506/APhysPolBSupp.16.7-A33}

\end{enumerate}
% *********************** Adding TOC and List of Figures ***********************

\tableofcontents

\listoffigures

\listoftables

% \printnomenclature[space] space can be set as 2em between symbol and description
%\printnomenclature[3em]

%\printnomenclature

% ******************************** Main Matter *********************************
\mainmatter
\chapter{Introduction}
Understanding the structure of matter at its most fundamental level remains one of the central goals of modern physics. It is now well established that protons and neutrons, collectively termed nucleons, are not elementary particles but rather complex bound states of quarks and gluons, commonly referred to as partons \cite{Gell-Mann:1964ewy, Zweig:1964jf, Feynman:1969ej}. Their dynamics are governed by Quantum Chromodynamics (QCD), the non-abelian gauge theory of the strong interaction \cite{Fritzsch:1973pi, GrossWilczek1973, Politzer:1973fx}. Although the QCD Lagrangian provides a complete description of the strong force at the fundamental level, establishing a connection between its quark and gluon degrees of freedom and the observed properties of hadrons, which are seen in detectors, remains a major theoretical challenge. This difficulty arises from the dual nature of QCD: at short distances, the property of asymptotic freedom \cite{GrossWilczek1973, Politzer:1973fx} enables perturbative calculations, while at long distances the phenomenon of color confinement \cite{wilson1974} makes the dynamics intrinsically complex, preventing us from observing free quarks and gluons.

A significant step toward resolving this challenge came from deep inelastic scattering (DIS) experiments in the late 1960s at SLAC \cite{Bloom:1969kc, Breidenbach:1969kd}, which revealed evidence for point-like constituents inside the proton. These observations led to the development of the Parton model \cite{Feynman:1969wa, Bjorken1975}, providing a simple framework in which the high-energy scattering of leptons from nucleons could be interpreted as incoherent interactions with quasi-free partons. Within this picture, the nonperturbative structure of the nucleon is encoded in parton distribution functions (PDFs), which describe the probability of finding a quark or gluon carrying a fraction $x$ of the longitudinal momentum of a fast-moving nucleon. With the development of QCD and the observation of scaling violations in DIS measurements \cite{Parisi1973, Parisi:2025nob}, it became clear that PDFs are scale-dependent quantities. Their evolution with the probing scale $Q^2$ is governed by the Dokshitzer–Gribov–Lipatov–Altarelli–Parisi (DGLAP) equations \cite{Dokshitzer:1977sg, Gribov:1972ri, ALTARELLI1977298}, derived within perturbative QCD. In contrast, the dependence of PDFs on $x$ reflects the nonperturbative structure of the nucleon which cannot be computed from first principles.\footnote{In the past decade, significant progress has been made toward extracting the $x$-dependence of parton distribution functions directly from first principles using lattice QCD. Nevertheless, the precision of these calculations remains limited, and they are not yet competitive with phenomenological determinations from global analyses \cite{Chu:2025kew}.} Instead, PDFs must be determined phenomenologically through global QCD analyses that compare theoretical predictions with a wide range of experimental data from DIS, Drell–Yan (DY) lepton pair production, and other processes \cite{Forte:2013wc, Ethier:2020way, Ball:2026dyb}.

The predictive power of QCD in high-energy processes, such as DIS, DY production, or other proton–proton collisions at the Large Hadron Collider (LHC), relies on collinear factorization theorems \cite{Collins:1989gx, Collins:1997sr, COLLINS1985104, Bodwin1986}. These theorems establish a rigorous separation between short-distance dynamics, which can be calculated perturbatively as process-dependent hard-scattering cross sections, and the long-distance structure of the nucleon encoded in universal PDFs. Universality ensures that PDFs determined from one class of processes can be applied consistently to others, allowing precise and testable predictions across a wide range of reactions. Within this framework, the cross section for a high-energy hadronic process can be expressed as a convolution of perturbatively calculable coefficient functions with PDFs that describe the nonperturbative distribution of partons inside the incoming hadrons.

Extending the partonic description of hadronic structure from isolated nucleons to nuclei introduces additional complexity. In the early 1980s, DIS measurements on nuclear targets, most notably by the European Muon Collaboration \cite{EuropeanMuon:1983wih}, revealed that the structure functions of bound nucleons differ systematically from those of free nucleons. This observation, known as the EMC effect, marked a turning point in the understanding of nuclear structure at the partonic level. Subsequent experiments confirmed that these deviations persist across a broad range of momentum fractions $x$ and exhibit a characteristic pattern of nuclear modifications \cite{Geesaman:1995yd, Norton2003, Arleo:2025oos}. At small $x$, the parton densities in nuclei are suppressed relative to those in free nucleons due to coherent multiple-scattering effects, a phenomenon known as nuclear shadowing \cite{Armesto:2006ph, KHARZEEV1994361,Kulagin:2004ie}. This suppression is followed by a moderate enhancement at intermediate $x$, commonly referred to as antishadowing \cite{Kulagin:2004ie, Eskola:2009uj, NICOLAEV1975397}. In the region $x \sim 0.3$–$0.7$, the structure functions exhibit the suppression originally observed by the EMC experiment \cite{EuropeanMuon:1983wih}, while at large $x$ the distributions are enhanced by nuclear binding and Fermi motion \cite{Bodek1981}. Together, these observations demonstrate that a nucleus cannot be described as a simple incoherent superposition of independent nucleons, but rather as a correlated many-body system in which the surrounding nuclear medium modifies the underlying structure of quark and gluon distributions.

Within the QCD framework, it is generally assumed that collinear factorization remains applicable to hard scattering processes involving nuclear targets, provided that the probing scale $Q^2$ is sufficiently large to suppress higher-twist contributions associated with the nuclear environment \cite{Qiu:2003cg, Frankfurt:2011cs}. Under this assumption, the short-distance hard-scattering dynamics are identical to those in nucleon scattering, while all nuclear effects are absorbed into modified parton distributions known as nuclear parton distribution functions (nPDFs). These distributions encode how the partonic structure of nucleons is modified in the nuclear environment. As in the case of proton PDFs, nPDFs evolve with the factorization scale $Q^2$ according to the DGLAP equations, while their dependence on the momentum fraction $x$ must be determined phenomenologically through global analyses of experimental data. Over the past two decades, several groups have performed global nPDF extractions using progressively more comprehensive datasets, including DIS, proton–nucleus Drell–Yan production, and more recently vector boson and heavy-flavor production in proton–lead collisions at the LHC. Early global analyses include the HKN framework \cite{Hirai:2007sx}, the DSSZ analysis \cite{deFlorian:2011fp}, the EKPS framework \cite{Eskola:2007my}, and the initial nCTEQ fits \cite{Schienbein:2007fs}, while more recent studies have been performed within the KA framework \cite{Khanpour:2020zyu} and by Helenius and collaborators \cite{Helenius:2021tof}. In parallel, modern global fits such as the nCTEQ analyses \cite{Duwentaster:2022kpv}, the EPPS framework \cite{Eskola:2021nhw}, and the nNNPDF approach \cite{nNNPDF3.0:2022fyi} incorporate increasingly sophisticated theoretical inputs and statistical methodologies in order to determine nuclear modifications of parton distributions and estimate their uncertainties.

Despite this progress, a robust determination of nPDF uncertainties remains challenging. Existing analyses differ in their choice of functional parametrizations, treatment of the nuclear-mass ($A$) dependence, dataset selection, and statistical procedures used for uncertainty estimation. Moreover, nuclear measurements typically provide weaker constraints than those available for proton PDFs, due to their lower precision and more limited kinematic coverage—particularly at very small or large momentum fractions $x$ and for certain flavors. Additional correlations arise from the presence of both protons and neutrons within nuclei. Under the assumption of isospin symmetry, the up-quark distribution in the proton is related to the down-quark distribution in the neutron, and vice versa, coupling the partonic structure of the two nucleon species within the fit. Combined with limited experimental constraints and potential tensions among datasets, these features lead to a highly correlated parameter space in nPDF analyses. Consequently, the $\chi^2$ landscape in nuclear PDF fits is often complex and may deviate significantly from a Gaussian form, exhibiting flat directions in parameter space and, in some cases, multiple local minima. 

Most global nPDF analyses estimate uncertainties using the Hessian method \cite{Stump:2001gu,Pumplin:2001ct}. In this approach, the $\chi^2$ function is expanded quadratically around the best-fit point in the parameter space, implying a multivariate Gaussian approximation for the parameter probability distribution. While computationally efficient, this local approximation cannot capture features such as skewed parameter distributions, strong nonlinear correlations, or the presence of multiple minima in the likelihood surface. In addition, tensions among experimental datasets are often accommodated through a phenomenological tolerance criterion, $\Delta\chi^2>1$, which enlarges the uncertainty bands but weakens the direct probabilistic interpretation of the resulting confidence intervals.

An alternative strategy is the Monte Carlo replica method \cite{ Giele:1998gw, Ball:2009mk, Watt:2012tq, AbdulKhalek:2019mzd}, adopted in the nNNPDF framework. In this approach, pseudo–data replicas are generated by fluctuating the experimental measurements within their uncertainties, and each replica is fitted independently. The resulting ensemble of PDF sets is then interpreted statistically as an estimate of the underlying probability distribution. Although this method avoids the quadratic approximation inherent in the Hessian approach and allows for flexible parametrizations, it does not directly correspond to sampling the posterior distribution of the fit parameters. In particular, when theoretical predictions depend nonlinearly on the parameters or when the likelihood surface deviates from a Gaussian form, the propagation of data fluctuations into parameter space may distort the inferred distribution. As discussed in~\cite{Costantini:2024wby}, under such conditions the replica ensemble may fail to faithfully reproduce the true parameter probability density.

These limitations motivate the adoption of a fully Bayesian framework in which the posterior probability distribution of the parameters is sampled directly. In this approach, uncertainties are defined through probability distributions derived from Bayes’ theorem, providing a statistically consistent interpretation of the results.

Markov Chain Monte Carlo (MCMC) methods \cite{Hogg:2017akh,brooks2011handbook} provide a practical tool for this purpose. MCMC algorithms generate sequences of correlated parameter samples whose stationary distribution converges to the target posterior probability density. By sampling parameters according to Bayes’ theorem, these methods naturally accommodate nonlinear correlations, asymmetric probability distributions, and the presence of multiple minima without relying on Gaussian approximations or tolerance parameters. The resulting ensemble of samples therefore provides direct access to the posterior distribution of model parameters and derived observables. At the same time, this approach introduces additional computational and methodological challenges. Reliable inference therefore requires careful verification of chain convergence and control of autocorrelation effects.

The application of Bayesian sampling methods to parton distribution analyses originated in proton PDF studies. Early work by Giele \textit{et al.}~\cite{Giele:2001mr} demonstrated that MCMC techniques could determine PDFs while simultaneously accounting for experimental systematic uncertainties within a unified statistical framework. The proof-of-principle studies, such as the analysis by Gbedo \textit{et al.}~\cite{Gbedo:2017eyp}, illustrated the feasibility of extracting PDFs from DIS data using MCMC sampling. More recently, several efforts have focused on improving algorithmic efficiency and systematically assessing statistical methodologies for PDF determination. Hunt-Smith \textit{et al.}~\cite{Hunt-Smith:2022ugn,Hunt-Smith:2023ccp} performed detailed comparisons of MCMC, Hessian, and nested sampling approaches in controlled benchmark studies, while Capel \textit{et al.}~\cite{Capel:2024qkm} developed an MCMC-based framework employing forward modeling to predict event-level observables directly. The recently introduced \texttt{Colibri} framework~\cite{Costantini:2025agd,Costantini:2025wxp} further advances Bayesian PDF inference by exploiting orthogonal functional decompositions to improve the statistical properties of PDF parameterizations. Finally, in our recent work~\cite{Risse:2025qlo} we performed a first full proton PDF determination based on MCMC sampling that incorporates both DIS and hadron collider data.

%Building on these developments, and in particular on our proton PDF analysis~\cite{Risse:2025qlo}, the present thesis extends the MCMC framework to nuclear parton distribution functions. We present the first comprehensive global determination of nPDFs based on Markov Chain Monte Carlo sampling using real experimental data. This approach enables a statistically consistent characterization of nuclear PDF uncertainties while allowing the complex structure of the posterior distribution to emerge directly from the data and theoretical model.
To investigate the structure of the nuclear PDF parameter space, two complementary fitting strategies are employed in this thesis. The primary analysis is a dedicated \texttt{Pb-only} fit restricted to experimental data involving lead nuclei. In this setup, the parametrization retains its standard analytic $A$-dependence but is evaluated at fixed $A=208$, enabling a focused exploration of the posterior structure within a single nuclear system. This configuration isolates the intrinsic behavior of the lead fit and allows a detailed study of parameter correlations, flat directions, and possible multiple minima without the influence of cross-nuclear constraints.
In addition, a \texttt{Multi-nuclei} fit is performed by incorporating data from a range of nuclear targets within the same functional framework. The explicit $A$-dependence then links information across different nuclei and provides additional constraints on the lead distributions. Comparing the two setups makes it possible to quantify the impact of these constraints, evaluate the reduction in uncertainties, and assess the stability of the posterior structure under extended data input. In this way, the Bayesian framework developed in this work serves not only for uncertainty quantification but also as a diagnostic tool for studying systematic effects in nuclear PDF modeling.

\subsection*{Aim of the Thesis}
%related works: \cite{Gbedo:2017eyp}
The primary objective of this thesis is to develop and implement a fully Bayesian framework for determining nuclear PDFs using MCMC sampling and to assess its performance relative to the conventional Hessian approach. This work presents the first global nPDF determination based on direct sampling of the posterior distribution, providing a statistically consistent alternative to Gaussian-based uncertainty estimates. A central component of the analysis is the first global extraction of lead PDFs in a single-nucleus setting based exclusively on lead data, enabling a controlled investigation of the underlying parameter space. This is complemented by a \texttt{Multi-nuclei} fit, which incorporates data from different nuclear targets to assess the impact of cross-nuclear constraints on the extracted distributions and their uncertainties.
More specifically, the goals are:
\begin{itemize}
    \item To formulate nuclear PDF fitting within a Bayesian inference framework based on Bayes’ theorem.
    
    \item To implement and validate an adaptive MCMC algorithm suitable for high-dimensional nPDF parameter spaces.
    %\item To perform detailed statistical diagnostics of the Markov chains, including convergence assessment, burn-in removal, autocorrelation analysis, and thinning procedures.
    
    \item To extract nuclear PDFs and quantify their uncertainties directly from the sampled posterior distribution.
    
    \item To compare the Bayesian MCMC results with those obtained using the traditional Hessian method and evaluate systematic differences.
    
    \item To investigate the presence of non-Gaussian features, flat directions, and multiple locally favored minima in nuclear PDF fits.
    
    \item To study the impact of imposed $A$-dependence by comparing a \texttt{Pb-only} analysis with a complementary \texttt{Multi-nuclei} fit.
\end{itemize}

\subsection*{Outline of the Thesis}
This thesis is organized as follows: \textbf{Chapter \ref{chap.2}} reviews the theoretical foundations of QCD relevant to parton distribution functions, including deep inelastic scattering, the parton model, QCD factorization, and DGLAP evolution. \textbf{Chapter \ref{chap.3}} introduces nuclear parton distribution functions, discusses nuclear modifications of PDFs, and describes the general structure of global nPDF analyses, including parameterization strategies and uncertainty estimation methods. \textbf{Chapter \ref{chap.4}} presents the principles of Bayesian inference and formulates the global nuclear PDF fit in a Bayesian framework. \textbf{Chapter \ref{chap.5}} introduces Markov Chain Monte Carlo methods. The theoretical foundations of Markov chains, stationary distributions, ergodicity, and the Metropolis–Hastings algorithm are discussed. \textbf{Chapter \ref{chap.6}} describes the full methodology of the analysis. It details the experimental dataset, theoretical setup, parameterization of nuclear PDFs, construction of the posterior probability, implementation of the MCMC algorithm, and statistical diagnostics of the generated chains. \textbf{Chapter \ref{Chap.7}} presents the results of the \texttt{Pb-only} and \texttt{Multi-nuclei} analyses. Posterior distributions, uncertainty bands, convergence diagnostics, and comparisons with Hessian-based results are discussed. Finally, \textbf{Chapter \ref{chap.8}} presents the conclusions.

%*******************************************************************************
%*********************************** First Chapter *****************************
%*******************************************************************************

\chapter{Theoretical Foundations} \label{chap.2} %Title of the First Chapter
A comprehensive understanding of the internal structure of nucleons and the dynamics of their constituent quarks and gluons is a central objective of modern nuclear and particle physics. Within the framework of Quantum Chromodynamics (QCD), this structure is encoded in parton distribution functions (PDFs), which describe the longitudinal momentum distributions of quarks and gluons inside a fast-moving hadron. Although PDFs are intrinsically nonperturbative objects, they enter physical observables through QCD factorization theorems, enabling a consistent connection between experimentally measurable cross sections and the underlying theory of the strong interaction.

This chapter introduces the theoretical framework necessary for the study of parton distribution functions and their extension to nuclear systems, as developed in later parts of this thesis. Section~\ref{sec:2.1} provides an overview of the Standard Model, with particular emphasis on Quantum Chromodynamics as the non-Abelian gauge theory governing strong interactions. Section~\ref{sec:2.2} discusses deep inelastic scattering (DIS) and the Parton Model, which establish the experimental and conceptual basis for describing hadronic structure in terms of partonic degrees of freedom. Section~\ref{sec:2.3} presents the QCD factorization theorems, which formalize the separation between perturbative short-distance dynamics and nonperturbative long-distance structure, and introduces the DGLAP evolution equations that govern the scale dependence of PDFs. The material presented in this chapter focuses on the aspects of QCD most directly relevant to the determination and interpretation of PDFs in global analyses.
%********************************** %First Section  **************************************
\section{The Standard Model and Quantum Chromodynamics} \label{sec:2.1}
\subsection{Overview of the Standard Model}
The Standard Model of Particle Physics, developed in the 1970s, stands as one of the most remarkable achievements in modern physics. It provides a comprehensive framework for understanding the fundamental particles of the universe and the forces that govern their interactions, with the exception of gravity. The Standard Model is a quantum field theory that unifies three of the four fundamental forces, electromagnetic, weak, and strong interactions, within the realm of quantum mechanics and special relativity.

\begin{figure}[htbp!] 
\centering    
\includegraphics[width=0.8\textwidth]{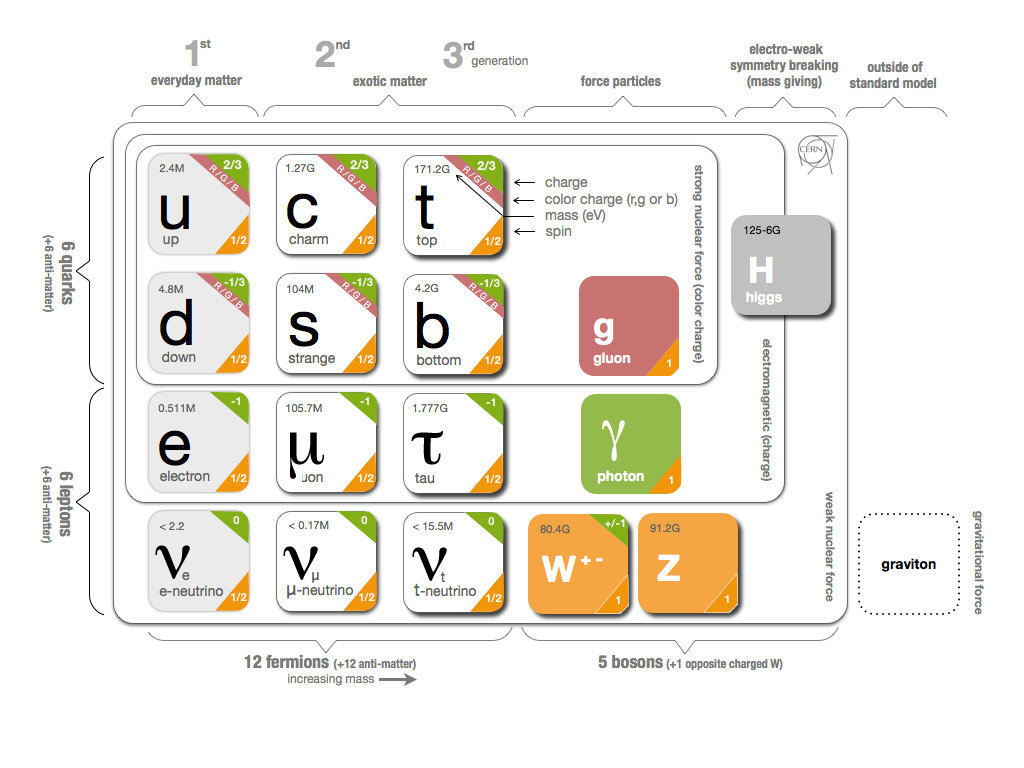}
\caption[Elementary Particles in Standard Model]{Classification of elementary particles according to the Standard Model of particle physics \cite{SMinfographic}.
}
\label{fig:SM}
\end{figure}

Elementary particles in the Standard Model are classified into fermions and bosons, as shown in Fig. \ref{fig:SM}. Fermions are half-integer spin particles and include quarks and leptons, which are the fundamental constituents of matter. Quarks, which are subject to the strong interaction, form hadrons and come in six flavors: up, down, charm, strange, top, and bottom. Leptons consist of electrons, muons, taus, and their associated neutrinos. The Standard Model also incorporates the existence of antiparticles for each fermion. These antiparticles have the same mass as their corresponding particles but possess opposite electric charge and other quantum numbers. For instance, the electron has a positively charged anti-particle called the positron, while the anti-quarks have opposite color charges to quarks. Bosons, on the other hand, are integer-spin particles that mediate the fundamental forces of nature. The photon mediates the electromagnetic force, \wz bosons mediate the weak force, and the gluon mediates the strong force. Additionally, the Higgs boson, discovered in 2012 \cite{HigssATLAS:2012yve}, explains how particles acquire mass through interactions with the Higgs field. 

The Standard Model is a composition of theories developed over the years:
\begin{enumerate}
    \item Quantum Electrodynamics (QED) is the relativistic quantum field theory that describes the interaction of electrically charged particles through the exchange of photons. It accounts for all electromagnetic phenomena involving charged particles and fields. QED is formulated as an abelian gauge theory based on the symmetry group $U(1)$, which reflects the invariance of the wavefunction under local phase transformations. This gauge symmetry governs the interactions between charged fermions and the massless photon, leading to the electromagnetic force. The fact that the photon is massless implies that the electromagnetic interaction has an infinite range. As a result, the static electric and magnetic fields produced by point charges and steady currents fall with the distance as $1/r^2$, consistent with classical electrodynamics.
    \item The electroweak theory unifies the electromagnetic and weak forces into a single framework. It is based on the $SU(2) \times U(1)$ gauge symmetry, where $SU(2)$ describes the weak isospin symmetry and $U(1)$ represents hypercharge symmetry. At high energies, these two symmetries merge into a unified electroweak force. The symmetry is spontaneously broken via the Higgs mechanism, resulting in the emergence of the $W^\pm$, $Z^0$, and photon as mediators of the weak and electromagnetic forces. The weak force, mediated by $W^\pm$ and $Z^0$ bosons, is short-range due to the massive nature of the mediators.
    \item Quantum Chromodynamics is the quantum field theory that describes the fundamental strong force, responsible for binding quarks into protons, neutrons, and other hadrons. It is formulated as a non-abelian gauge theory based on the symmetry group $SU(3)$, where quarks have a type of charge known as ``color.'' Unlike QED, where the photon is electrically neutral and does not self-interact, the QCD force carriers, gluons, carry color charge and can interact with one another. These gluon self-interactions give rise to complex non-linear dynamics that are essential for understanding key features of QCD, such as \emph{asymptotic freedom} and \emph{confinement}. Asymptotic freedom refers to the behavior of the strong coupling constant, which decreases at high energies, allowing quarks to behave as nearly free particles at short distances. At low energies, however, the strong interaction becomes increasingly strong, preventing quarks from being observed in isolation, a phenomenon known as confinement. While confinement has been studied through experiments and lattice QCD calculations \cite{GREENSITE20031}, its underlying mechanism is still not fully understood. Nevertheless, it is widely believed that gluon self-interactions play a central role in the \emph{confinement} mechanism \cite{GrossWilczek1973}. As a result, the strong force appears short-ranged within hadrons, despite being mediated by massless gluons.\footnote{On a larger scale, the residual strong force, also known as the nuclear force, acts between protons and neutrons to bind them within atomic nuclei. This residual force is a remnant of the fundamental strong interaction and operates over a short range, overcoming the electromagnetic repulsion between positively charged protons to hold the nucleus together.}
\end{enumerate}

\subsection{Quantum Chromodynamics}
Quantum Chromodynamics is essential to understanding parton distribution functions. It provides the theoretical framework that describes the strong interactions between quarks and gluons and determines the interpretation of high-energy scattering data. As mentioned, QCD is a quantum field theory characterized by a non-abelian gauge symmetry group, namely the $SU(3)$ group associated with color. The full QCD Lagrangian consists of the classical Yang–Mills part together with gauge-fixing and ghost terms \cite{cteq_handbook}:
\begin{equation}
    \mathcal{L}_{\text{QCD}}= \mathcal{L}_{\text{Clssical}} + \mathcal{L}_{\text{gauge-fixing}} + \mathcal{L}_{\text{ghost}}.
    \label{eq2.1}
\end{equation}
The classical Yang–Mills Lagrangian \cite{Yang:1954ek} describes the propagation of quarks and gluons and their interactions governed by the non-Abelian gauge structure:
\begin{equation}
\mathcal{L}_{\text{Classical}}
= \sum_f \bar{\psi}^f_i \left[(i\gamma^\mu D\mu){ij} - m^f \delta{ij}\right] \psi^f_j
- \frac{1}{4} G^a_{\mu\nu} G^{a\mu\nu}.
\label{eq2.2}
\end{equation}
Here, $f$ labels quark flavors, $i$, $j$ = 1, 2, 3 are fundamental color indices, $\mu$, $\nu$ = 0, ..., 3 are Lorentz indices, and $a$= 1, ..., 8 labels the adjoint color components of the gluon field. The matrix $m^f$ contains the mass of the quark flavor $f$ and $\gamma^\mu$ are the Dirac gamma matrices, describing fermions according to the principles of relativistic quantum mechanics. The quark fields $\psi^f_i$ transform as color triplets, and the covariant derivative, $D_\mu = \partial_\mu - ig_s A_\mu^a T^a$, introduces the gluon field $A_\mu^a$ via the $SU(3)$ generators $T^a$.
The gluon field strength tensor $G^a_{\mu\nu}$
\begin{equation}
    G^a_{\mu\nu} = \partial_\mu A_\nu^a - \partial_\nu A_\mu^a + g_s f^{abc} A_\mu^b A_\nu^c,
\end{equation}
captures the non-Abelian nature of QCD. The structure constants $f^{abc}$ arise from the commutation relations of the $SU(3)$ generators and give rise to the three- and four-gluon self-interaction vertices that distinguish QCD from QED.
To quantize the theory, gauge fixing is required to eliminate redundant degrees of freedom.\footnote{A gluon field $A_\mu^a$ has four components but only two physical polarizations; gauge fixing avoids overcounting equivalent field configurations.} Two commonly used choices are \emph{covariant} gauges and \emph{axial} gauges. In covariant gauges, the gauge-fixing term is expressed as:
\begin{equation}
    \mathcal{L}_{\text{gauge-fixing}} = -\frac{1}{2\xi} (\partial^\mu A_\mu^a)^2,
\end{equation}
where $\xi$ is the gauge parameter. In special cases, such as the \emph{Feynman} gauge \cite{Feynman49} ($\xi=1$), the gluon propagator assumes a particularly simple form. The choice $\xi=0$ corresponds to the \emph{Landau} (or \emph{Lorenz}) gauge \cite{landau1955gauge}, another widely used member of the covariant gauge family, valued for its renormalizability and the simplified structure of the gluon propagator.
In axial or light-cone gauges \cite{Arnowitt62}, one instead imposes the condition  $n \cdot A^a =0$, where $n^\mu$ is a fixed (often light-like) vector. The corresponding gauge-fixing term is
\begin{equation}
    \mathcal{L}_{\text{gauge-fixing}} = -\frac{1}{2} ( n \cdot A^a)^2.
\end{equation}
A key feature of axial gauges is that they do not require ghost fields. This simplifies perturbative calculations by reducing the number of contributing Feynman diagrams. For this reason, axial gauges are sometimes referred to as \emph{physical gauges}: unlike covariant gauges (e.g., the Feynman gauge), they avoid introducing unphysical degrees of freedom associated with gauge redundancy. However, axial gauges are not free of complications—they produce singularities in the gluon propagator when $n \cdot k = 0$, which introduces technical difficulties in loop calculations.
In covariant gauges, the gauge-fixing procedure leads to a nontrivial Jacobian in the path integral, accounted for using the \emph{Faddeev–Popov} method \cite{FADDEEV196729}. This requires the introduction of additional anticommuting scalar fields, the \emph{ghost} fields $c^a$ and $\bar{c}^a$. Although ghosts do not correspond to physical particles and never appear as external states, they play an essential role at the loop level. Their dynamics are governed by the ghost Lagrangian
\begin{equation}
    \mathcal{L}_{\text{ghost}} = \partial^\mu \bar{c}^a (D_\mu c)^a,
\end{equation}
where the covariant derivative generates interactions between ghosts and gluons. Through these interactions, ghost loops cancel contributions from unphysical gluon polarizations introduced by the gauge choice, ensuring both unitarity and gauge invariance of the theory.

The QCD Lagrangian, Eq. (\ref{eq2.1}), can be studied using several complementary theoretical approaches, each suited to different aspects of the strong interaction. One prominent framework is \emph{lattice QCD} \cite{Boyle:2022ncb}, which reformulates the theory on a discretized Euclidean spacetime lattice. By replacing continuous spacetime with a finite grid, lattice QCD avoids many of the analytic difficulties posed by strong coupling in the infrared region and enables first-principles numerical computations of nonperturbative phenomena such as quark confinement, hadron masses, and hadronic matrix elements. The continuum theory is recovered by taking the limit in which the lattice spacing approaches zero. Although remarkably successful at low energies, lattice calculations are computationally intensive due to the large number of degrees of freedom involved and are in some cases limited by the need to formulate the theory in Euclidean spacetime.
Another strategy is provided by \emph{effective field theories} (EFTs) \cite{VanKolck:2002mp}, which simplify the full QCD Lagrangian by focusing on specific energy regimes and using the symmetries of the theory. EFTs are constructed to describe the relevant degrees of freedom in a specific energy regime, capturing the dominant physics without requiring the full dynamics of QCD. 
These approaches are invaluable for capturing the dominant physics in particular regimes without requiring a full treatment of the underlying dynamics.

Among the various approaches, perturbative QCD (pQCD) provides a powerful framework for studying high-energy phenomena in quantum chromodynamics. Its applicability rests on the property of \emph{asymptotic freedom} \cite{Gross2005}, which causes the strong coupling constant $g_s$ to decrease at large momentum transfers, or equivalently, at short distances. As a result, physical observables can be expanded perturbatively in the small parameter $\alpha_s= g_s^2/4 \pi$, allowing reliable calculations for processes involving large energy scales. In this regime, quantities such as scattering cross sections or decay rates are computed using Feynman diagrams with quarks and gluons, expanded order by order in powers of the coupling constant $g_s$. However, perturbative calculations often contain divergent loop integrals, which must be treated carefully to obtain meaningful predictions. These divergences fall into two broad categories: ultraviolet (UV) and infrared (IR). UV divergences arise from contributions at very high momenta (short distances) in loop integrals. Like in other quantum field theories, they reflect the sensitivity of QCD to arbitrarily small distance scales. In pQCD, UV divergences are systematically absorbed into redefined (``renormalized'') parameters of the theory, such as the quark masses and coupling constant—through the process of \emph{renormalization}. This procedure renders physical predictions finite and preserves the consistency and predictivity of the theory. In contrast, IR divergences originate from low-momentum (long-distance) effects. They arise when massless particles, such as gluons, are emitted with vanishing energy (soft divergences) or when two particles become collinear. Such divergences appear both in real-emission diagrams and in virtual loop corrections. Unlike UV divergences, IR divergences cannot be removed by renormalization. Instead, they cancel only when sufficiently inclusive physical observables are considered specifically observables that sum over all unresolved soft and collinear emissions. This cancellation, ensured by the Kinoshita–Lee–Nauenberg (KLN) theorem \cite{Kinoshita:1962ur, LeeNauenberg1964}, guarantees that measurable quantities such as inclusive cross sections remain finite once all degenerate final states are properly accounted for.

\begin{figure}[htbp] 
\centering    
\includegraphics[width=0.7\textwidth]{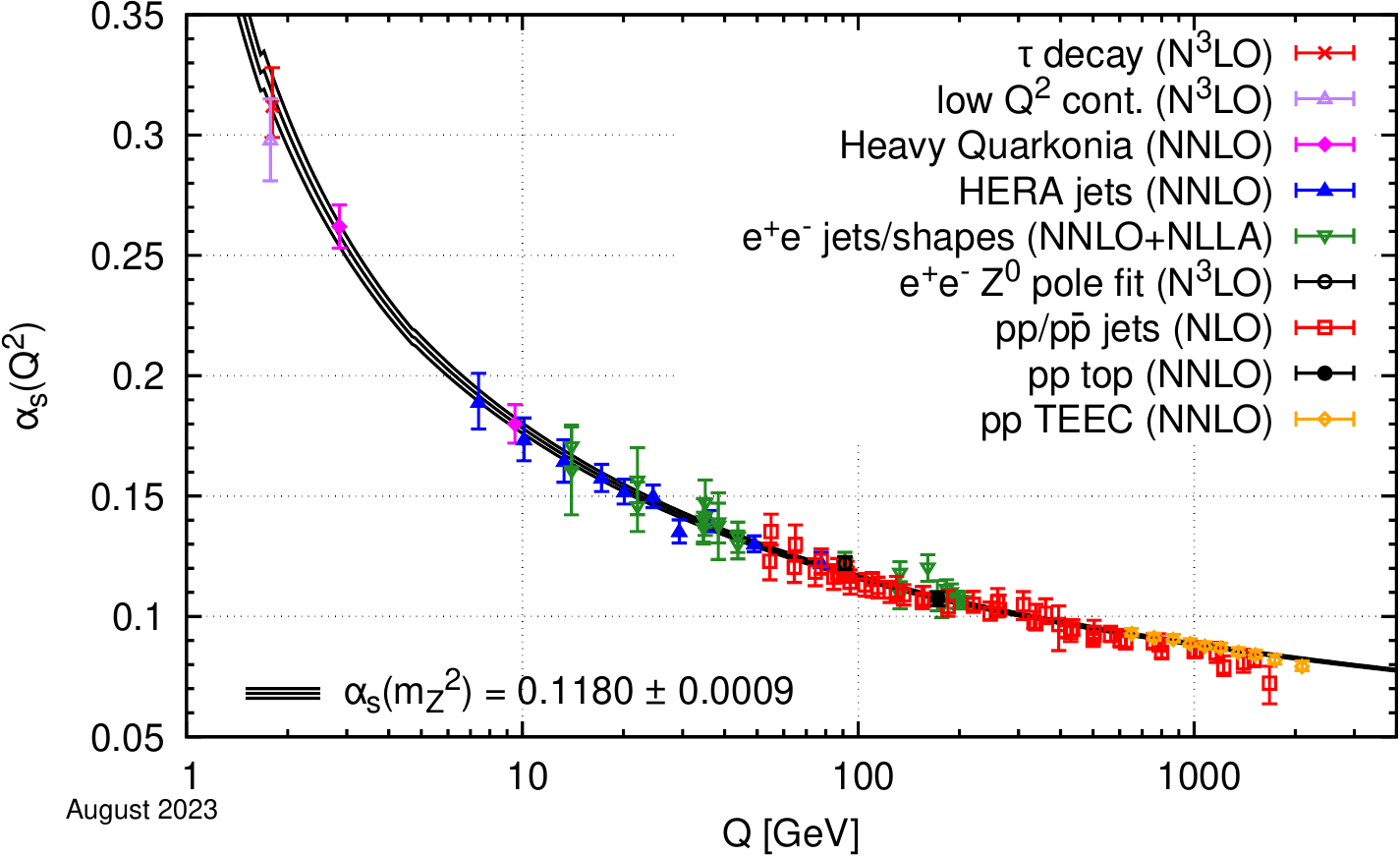}
\caption[QCD running coupling constant]{The running of the QCD coupling constant $\alpha_s(Q^2)$ with energy scale $Q$, illustrating asymptotic freedom at high energies and the approach to the nonperturbative regime at low energies. (from \cite{PDG:2023ofk}).}
\label{fig:alphas}
\end{figure}

The \emph{renormalization} procedure is essential for handling UV divergences in perturbative QCD. It systematically absorbs the infinities that arise in loop diagrams into renormalized (redefined) parameters of the theory, such as the coupling constant and quark masses, thereby yielding finite, physically meaningful predictions. Renormalization introduces a reference energy scale, $\mu_R$, at which these parameters are defined. A key consequence of renormalization is that the coupling constants and quantities such as particle masses become functions of the energy scale. This behavior, known as the \emph{running} of the parameters, is governed by the renormalization group equations (RGE), which determine how the strong coupling or masses evolve with $\mu_R$. For the strong coupling, the RGE is expressed as
\cite{cteq_handbook}
\begin{equation}
    \mu_R \frac{d g_s (\mu_R)}{d \mu_R} = \beta (g_s (\mu_R)),
    \label{eq2.7}
\end{equation}
where the beta function, $\beta(g_s)$, encodes the scale dependence of the coupling and is computed from loop corrections in perturbation theory. For QCD with $N_f$ active quark flavors, the one-loop beta function is
\begin{equation}
    \beta (g_s) = -g_s \frac{g_s^2}{16\pi^2} [11 - \frac{2}{3} N_f].
    \label{eq2.8}
\end{equation}
This result highlights the non-Abelian nature of QCD. The term proportional to 11 originates from gluon self-interactions and dominates over the quark-loop contribution $ \frac{2}{3} N_f$, causing the beta function to be negative. As a consequence, the strong coupling decreases at high energies—a phenomenon known as \emph{asymptotic freedom}.  Solving the RGE at one loop gives the well-known expression for the running of the strong coupling $\alpha_s= g_s^2/4 \pi$
\begin{equation}
    \alpha_s(\mu_R) = \frac{\alpha_s(\mu_{R0})}{1 + 2 b \alpha_s(\mu_{R0}) \ln\left(\frac{\mu_R}{\mu_{R0}}\right)},
    \label{eq2.9}
\end{equation}
where $b= \frac{33-2N_f}{12\pi}$, and $\mu_{R0}$ is the reference energy scale, typically fixed by experimental input. This expression illustrates a fundamental feature of QCD: at high energy scales ($\mu_R \gg \mu_{R0}$), the coupling becomes small, and quarks and gluons behave as though they are nearly free, manifesting asymptotic freedom. At low energy scales ($\mu_R \ll \mu_{R0}$), the coupling grows large, signaling the onset of \emph{confinement}, where quarks are bound inside hadrons. This scale dependence of $\alpha_s$ is strongly supported by experimental measurements over a wide range of energies, as shown in Fig.~\ref{fig:alphas}. This behavior distinguishes QCD from Abelian gauge theories such as QED, in which the coupling increases rather than decreases with energy.
The running of the strong coupling constant $\alpha_s(\mu_R)$ naturally introduces an important nonperturbative energy scale, $\Lambda_{QCD}$. This scale, of the order of a few hundred MeV, marks the transition between the perturbative and nonperturbative regimes of QCD. It arises from the solution of the renormalization group equation and corresponds to the energy at which the coupling becomes large, signaling the breakdown of perturbation theory. Although $\Lambda_{QCD}$ does not appear in the classical QCD Lagrangian, it is determined experimentally and is crucial in characterizing the low-energy behavior of the strong interaction, where confinement and hadronization dominate. The perturbative expression for the running coupling shows that  $\alpha_s(\mu_R)$ increases as the scale $\mu_R$ is lowered. At a certain point, the denominator of Eq.~\eqref{eq2.9} vanishes, causing the coupling to diverge. This divergence is known as the \emph{Landau pole} \cite{Landau:1954nau} and is formally located at $\mu_R = \Lambda_{QCD}$, where the one-loop expression becomes singular. The appearance of the Landau pole indicates that perturbation theory is no longer reliable near this scale. In this strongly coupled region, fully nonperturbative methods, such as lattice QCD, are required to describe the dynamics of quarks and gluons.
%********************************** % Second Section  *************************************
\section{ Deep Inelastic Scattering and the Parton Model}  \label{sec:2.2}

\subsection{Deep Inelastic Scattering}
\begin{figure}[ht!]
    \centering
    \vspace{30pt}
    \begin{fmfgraph*}(200,100)
    % Define labels
    \fmflabel{\(l(k)\)}{il}
    \fmflabel{\(h(p)\)}{ip}
    \fmflabel{\(l'(k')\)}{ol}
    %\fmflabel{\(X\)}{d1}
    \fmflabel{\(X_{1}\)}{oq4}
    \fmflabel{\(X_{2}\)}{oq3}
    \fmflabel{\(...\)}{oq2}
    \fmflabel{\(X_{n}\)}{oq1}

    \fmfleft{ip,il}
    \fmfright{oq1,oq2,oq3,oq4,ol}
    \fmf{fermion}{ip,vp,oq1}
    \fmf{phantom}{ip,vp,oq1}
    \fmf{fermion}{vp,oq1}
    \fmf{fermion}{vp,oq2}
    \fmf{phantom}{ip,vp, oq1}
    \fmf{fermion}{vp,oq3}
    \fmf{fermion}{vp,oq4}
    \fmf{photon,lab.side=right,lab=$V(q)$}{vl,vp}
    \fmf{fermion}{il,vl,ol}
    \fmf{phantom}{il,vl,ol}
    \fmfblob{.15w}{vp}
    \fmfdot{vl}
    \fmffreeze
    \fmfi{plain}{vpath (__ip,__vp) shifted (thick*(0,2))}
    \fmfi{plain}{vpath (__ip,__vp) shifted (thick*(1,-2))}
    \end{fmfgraph*}
    
    % TikZ overlay for the brace and label
    \begin{tikzpicture}[overlay]
        \draw[decorate,decoration={brace,amplitude=10pt},xshift=0.1pt]
        (4.5,3.5) -- (4.5,0.1) node[midway,xshift=18pt] {$X$};
        %\node[anchor=west] at (4,2.2) {$\left.\begin{array}{c} \\ \\ \\ \end{array}\right\} X$};
    \end{tikzpicture}

    \vspace{15pt}
    \caption[Feynman Diagram for DIS process]{Schematic representation of the deep inelastic scattering process. The lepton with momentum $k$ exchanges a virtual boson $V(q)$ with a hadron of momentum $p$. The filled circle denotes the hard partonic subprocess, incorporating the interactions required to produce the inclusive hadronic final state $X = (X_1, \dots, X_n)$ at a given perturbative order. The diagram represents a generic perturbative contribution to the DIS cross section.}
    \label{fig:feynman_diagram}
\end{figure}
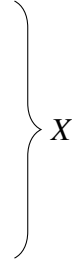

Deep Inelastic Scattering (DIS) is a fundamental process in particle physics used to probe the internal structure of hadrons, leading to the formulation of the Parton Model \cite{Feynman:1969wa, Bjorken1975}. In DIS, a high-energy lepton $\ell$ (such as an electron, muon, or neutrino) scatters off a hadronic target $h$ (e.g., a proton), resulting in a scattered lepton $\ell'$ and an inclusive hadronic final state $X$. The schematic Feynman diagram in Fig.~\ref{fig:feynman_diagram} illustrates this process as the exchange of a virtual gauge boson $V(q)$ (photon, $W^\pm$, or $Z^0$) between the incoming lepton and a partonic constituent of the hadron. The filled circle represents the underlying hard-scattering subprocess, whose detailed structure depends on the perturbative order considered and includes all partonic interactions required to produce the final state $X$.
\begin{fmffile}{dis_diagram}
\end{fmffile}
\paragraph{Kinematics.}
The kinematics of DIS is characterized by several Lorentz-invariant variables. The most relevant independent quantities are
\begin{align*} 
q      &= k - k'                 && \text{(four-momentum transfer)} \\ 
Q^2    &= -q^2 > 0               && \text{(virtuality of the exchanged boson)} \\
x      &= \frac{Q^2}{2 p \cdot q}&& \text{(Bjorken scaling variable)} \\ 
\end{align*}
where $k$ and $k'$ denote the four-momenta of the incoming and outgoing lepton, respectively, and $p$ is the four-momentum of the target hadron with mass $M$. The variable $Q^2$ sets the resolution scale of the probe: larger values of $Q^2$ correspond to shorter-distance sensitivity to the partonic constituents. The dimensionless variable $x$ (Bjorken-$x$) has the interpretation, in the infinite-momentum frame, of the longitudinal momentum fraction carried by the interacting parton and satisfies $0 < x \le 1$.
Additional useful variables include the inelasticity $y = (p \cdot q)/(p \cdot k)$, the energy transfer $\nu = (p \cdot q)/M$, and the invariant mass of the hadronic final state $W^2 = (p + q)^2$, all of which can be expressed in terms of $x$ and $Q^2$. The deep inelastic regime corresponds to $Q^2 \gg M^2$ and $W^2 \gg M^2$, ensuring a hard scattering process and suppressing elastic and resonance contributions, thereby providing access to the short-distance partonic structure of the hadron.
\paragraph{Cross Section.}
In inclusive DIS process, $e(k) + N(p) \rightarrow  e(k') + X(p_X)$, the differential cross section can be written as a contraction of the leptonic tensor, $L^{\mu\nu}$, which describes the electron–photon interaction, and the hadronic tensor, $W_{\mu\nu}$, which encodes  the target’s response to the electromagnetic probe \cite{Blumlein:1996vs,Anselmino:1993tc,robinDIS}. The total center-of-mass energy squared is given by $s=(p + k)^2 $ and $E'$ denotes the energy of the scattered lepton in the LAB frame. The cross section in the one-photon–exchange approximation is
\begin{equation}
    d\sigma = \frac{1}{2s} \frac{e^4}{q^4} L^{\mu\nu} W_{\mu\nu} \frac{d^3 k'}{4\pi^2 E'},
    \label{eq2.12}
\end{equation}
The leptonic tensor $L^{\mu\nu}$ is
\begin{equation}
    L^{\mu\nu}(k,k')= 2[k^\mu k'^\nu + k^\nu k'^\mu - g^{\mu\nu}(k \cdot k' - m_e^2)],
\end{equation}
with $g^{\mu\nu}$ the metric tensor and $m_e$ the lepton mass. The hadronic tensor, $W_{\mu\nu}$, characterizes the response of the proton to the electromagnetic current $J_{\mu}$ in inclusive deep inelastic scattering. It sums over all possible final hadronic states $X$.
\begin{equation}
    W_{\mu\nu}(p, q)=\sum_X \langle X |  J_{\mu} | N(p) \rangle^*  \langle X |  J_{\nu} | N(p)\rangle (2\pi)^3 \delta^{(4)} (p_X -p -q),
\end{equation}
where $p_X$ is the total four-momentum of the state $X$. This tensor is a nonperturbative object that is parameterized in terms of Lorentz-invariant structure functions. For an unpolarized hadron, the most general decomposition of $W_{\mu\nu}$ consistent with Lorentz invariance, gauge invariance, and time-reversal symmetry involves five independent scalar functions $W_1$, $W_2$, $W_3$, $W_4$, and $W_5$, and takes the form
\begin{equation}
    W_{\mu\nu}(x, Q^2) = -g_{\mu\nu}W_1  + \frac{p_\mu p_\nu}{M^2} W_2 -i\epsilon_{\mu\nu\rho\sigma} \frac{p^\rho q^\sigma}{2M^2} W_3 + \frac{q_\mu q_\nu}{M^2} W_4 + \frac{p_\mu q_\nu + p_\nu q_\mu }{M^2} W_5.
\end{equation}
Electromagnetic current conservation ($\partial_\mu J^\mu = 0$) imposes constraints on $W_{\mu\nu}$, reducing the number of independent structures. As a result, the hadronic tensor in inclusive DIS can be expressed in terms of three independent scalar functions.
It is convenient to introduce the structure functions $F_1$, $F_2$, and $F_3$, which are linear combinations of the coefficients $W_i$ and are directly related to measurable cross sections. These functions provide an experimentally relevant parametrization of the hadronic tensor. Their relation to the coefficients $W_i$ is given by
\begin{equation} 
\begin{split}
 &F_1(x, Q^2) = M W_1(x, Q^2),\\
&F_2(x, Q^2) = \frac{p \cdot q}{M} W_2(x, Q^2),\\
&F_3(x, Q^2) = \frac{p \cdot q}{M} W_3(x, Q^2).    
\end{split}
\label{eq2.18}
\end{equation}
In terms of the structure functions, the hadronic tensor takes the form
\begin{equation}
\begin{split}
  W_{\mu\nu}(x, Q^2) & =  \left( -g_{\mu\nu} + \frac{q_\mu q_\nu}{q^2} \right) F_1(x, Q^2) + \frac{ \left( p_\mu - q_\mu \frac{p \cdot q}{q^2} \right)\left( p_\nu - q_\nu \frac{p \cdot q}{q^2} \right)} {p \cdot q} F_2(x, Q^2) \\
& + i\epsilon_{\mu\nu\rho\sigma} \frac{p^\rho q^\sigma}{p \cdot q} F_3(x, Q^2).
\end{split}
\label{eq2.17}
\end{equation}
The function $F_3(x, Q^2)$ is associated with the antisymmetric component of the hadronic tensor and contributes only in the presence of parity-violating interactions, such as charged-current or neutral-current weak exchanges in neutrino or antineutrino scattering. In purely electromagnetic DIS, where parity is conserved, $F_3$ does not appear. Contracting the leptonic tensor with the reduced hadronic tensor yields the double differential cross section for inclusive DIS
\begin{equation}
\begin{split}
    \frac{d^2\sigma}{dxdQ^2} & = \frac{4\pi\alpha^2}{xQ^4} \Big[ xy^2 F_1(x, Q^2) + \left(1-y-\frac{x^2y^2M^2}{Q^2}\right) F_2(x, Q^2) \\
    & \pm y(1-\frac{y}{2}) x F_3(x, Q^2)\Big], 
\end{split}
\end{equation}
where $\alpha=e^2/4 \pi$ is the electromagnetic fine-structure constant. The sign of the $F_3$ term distinguishes neutrino from antineutrino scattering: the upper sign applies to neutrinos and the lower sign to antineutrinos. The structure functions $F_i(x, Q^2)$ are directly accessible through experimental measurements of the DIS cross section. They provide detailed insight into the partonic composition of the target hadron, including both flavor and momentum distributions. Theoretical frameworks such as the parton model and perturbative QCD offer quantitative predictions for these functions, which will be discussed in the following section.

\subsection{The Parton Model}\label{subsec:2.2.2}
The Parton Model, introduced in the late 1960s, provides a direct connection between the structure functions measured in DIS and the underlying parton distribution functions, which characterize how the momentum of a nucleon is distributed among its quarks and gluons. The model was motivated by the surprising results from the DIS experiments at SLAC in 1969 \cite{Bloom:1969kc, Breidenbach:1969kd}, which revealed that protons contain point-like constituents. James Bjorken first proposed the idea of scaling, suggesting that at sufficiently large momentum transfer nucleons behave as collections of nearly free, point-like particles \cite{Bjorken:1969}. Richard Feynman later formalized this picture by introducing the term ``partons'' for these internal constituents \cite{Feynman:1969ej}, which we now identify with quarks and gluons.

A key feature of the Parton Model is its ability to explain the approximate scaling behavior observed in DIS. This phenomenon, known as \emph{Bjorken scaling}, states that at large $Q^2$ and fixed Bjorken $x$, structure functions become nearly independent of $Q^2$:
\begin{equation}
    F_1(x, Q^2) \approx F_1(x), \quad
    F_2(x, Q^2) \approx F_2(x), \quad
    F_3(x, Q^2) \approx F_3(x).
\end{equation}
Such behavior indicates that, at short distances, the partons behave as quasi-free, point-like constituents of the nucleon.
However, increasingly  precise measurements have revealed small but systematic deviations from exact scaling, termed \emph{scaling violations} \cite{Parisi:2025nob, Gurvitz:1995fu}. These deviations arise from the interactions among partons, particularly gluon radiation and quark–gluon splitting, which introduce a weak $Q^2$-dependence into the structure functions. Quantum Chromodynamics provides a quantitative explanation of these effects through the DGLAP evolution equations, which describe how PDFs change with the probing energy scale. In Section~\ref{subsec:2.3.2}, we will explore how QCD predicts these scaling violations and how they lead to the scale evolution of parton distribution functions.

Each of the reduced structure functions carries a distinct physical interpretation. The function $F_1(x, Q^2)$ describes the interaction of transversely polarized virtual photons with the nucleon and is proportional to the probability of absorbing a transversely polarized photon. The structure function $F_2(x, Q^2)$ includes contributions from both longitudinal and transverse photon polarizations and reflects the total probability of scattering off a parton carrying a longitudinal momentum fraction $x$ of the nucleon. The third structure function $F_3(x, Q^2)$ appears in weak interaction DIS and captures the parity-violating effects, allowing one to distinguish between quarks and antiquarks. In the Parton Model, partons are treated as quasi-free particles during the hard scattering, which allows the structure functions to be expressed directly in terms of parton distribution functions. In the massless leading-order approximation, the structure function $F_2(x, Q^2)$ is given by
\begin{equation}
F_2(x) = x \sum_q e_q^2 \left[ q(x) + \bar{q}(x) \right], 
\end{equation}
where $e_q$ is the electric charge of the quark flavor $q$, and $q(x)$ and $\bar{q}(x)$ are the corresponding quark and antiquark PDFs. The structure function $F_1(x, Q^2)$ is related to $F_2(x, Q^2)$ through the \emph{Callan-Gross relation} \cite{Callan1969}:
\begin{equation}
2xF_1(x) = F_2(x),
\end{equation} 
which follows from the assumption that partons are massless spin-$1/2$ fermions that scatter incoherently from the virtual photon. This relation provided crucial early evidence that the charged constituents (quarks) inside the nucleon have spin $1/2$. In QCD, this relation is violated by effects such as gluon radiation and quark–gluon interactions. Even at leading order, quark masses break the Callan–Gross relation. For instance, general-mass schemes such as ACOT (Aivazis–Collins–Olness–Tung) \cite{Aivazis:1993kh, Aivazis:1993pi} include mass-dependent kinematic corrections that generate a nonzero longitudinal structure function, $F_{L}$, even without higher-order QCD corrections. The longitudinal structure function is defined as
\begin{equation}
    F_{L}(x, Q^2)= F_2(x, Q^2)- 2x  F_1(x, Q^2),
\end{equation}
and isolates the contribution from longitudinally polarized virtual photons. In the massless Parton Model and in the absence of QCD corrections, $F_{L}=0$, as implied by the Callan–Gross relation. However, quark mass effects and QCD dynamics both induce a non-zero $F_{L}$, making it a valuable probe of the gluon content of the proton and of the validity of different theoretical schemes across energy scales. The structure function $F_3(x, Q^2)$, also plays a key role, especially in weak interaction processes. In charged-current neutrino scattering, it appears at leading order as
\begin{equation}
F_3(x) = \sum_q \left[ q(x) - \bar{q}(x) \right], 
\end{equation}
providing direct access to quark–antiquark asymmetries in the nucleon and offering important insight into its flavor structure.

Structure functions in DIS not only describe quark distributions inside the nucleon; they also satisfy several fundamental sum rules that impose essential constraints on the parton distribution functions. These sum rules follow from basic conservation laws and provide important consistency checks for the Parton Model and for quantum chromodynamics. The ``momentum sum rule'' ensures that the total momentum of the nucleon must be carried by all of its constituent partons, quarks, antiquarks, and gluons:
\begin{equation}
\sum_{q} \int_{0}^{1} x \left[ q(x) + \bar{q}(x) \right] \, dx + \int_{0}^{1} x \, g(x) \, dx = 1.
\end{equation}
Here $g(x)$ is the gluon PDF, and the sum over quark flavors extends across all active flavors ($u$, $d$, $s$, $c$, ...). In addition to momentum conservation, baryon number and flavor conservation lead to ``number sum rules'' for the valence quark distributions. Defining the valence PDFs as
\begin{equation*}
    u_v(x) = u(x) - \bar{u}(x), \quad
    d_v(x) = d(x) - \bar{d}(x),
\end{equation*}
their integrals must reproduce the known quantum numbers of the nucleon. For the proton, containing two up quarks and one down quark, the sum rules are
\begin{equation}
    \int_{0}^{1} u_v(x) \, dx = 2, \quad \int_{0}^{1} d_v(x) \, dx = 1.
\end{equation}
For the neutron, with one up and two down quarks, the roles of $u_v$ and $d_v$ are reversed:
\begin{equation}
    \int_{0}^{1} u_v(x) \, dx = 1, \quad \int_{0}^{1} d_v(x) \, dx = 2.
\end{equation}
Similarly, since the proton carries zero net strangeness and zero net charm, the corresponding valence components must vanish:
\begin{equation}
    \int_{0}^{1} [s(x) - \bar{s}(x)] \, dx = 0, \quad \int_{0}^{1} [c(x) - \bar{c}(x)] \, dx = 0.
\end{equation}
These relations express the fact that strange and charm quarks in the proton arise entirely from sea-quark fluctuations, not from its valence structure. Together, these sum rules serve as foundational constraints in global PDF analyses, ensuring that extracted PDFs remain consistent with the conserved quantum numbers of QCD.
%********************************** %Third Section  *************************************
\section{QCD Factorization and DGLAP Evolution} \label{sec:2.3}
\subsection{Factorization Theorem}
The QCD \emph{factorization} framework \cite{Collins:1997sr, COLLINS1985104, Bodwin1986, Collins:1987pm} provides the theoretical foundation for connecting perturbative calculations with experimental measurements of high-energy processes. At high momentum transfer $Q^2$ where the strong coupling $\alpha_s(Q^2)$ becomes small due to asymptotic freedom, factorization is expected to hold. In this regime, the factorization theorem enables a systematic separation between short-distance and long-distance contributions. The short-distance dynamics are encoded in perturbatively calculable partonic cross sections, while the long-distance, nonperturbative structure of the hadron is captured by universal Parton Distribution Functions, which describe the momentum and flavor distributions of quarks and gluons.

Within the \emph{collinear factorization} framework \cite{Collins:1989gx}, partons are approximated as moving nearly parallel to the parent hadron’s momentum. Their transverse momenta $P_{\perp}$ are either neglected or integrated out, so the parton distribution functions depend only on the longitudinal momentum fraction $x$.\footnote{Alternative factorization frameworks exist, such as $k_T$-factorization \cite{catani1991high, Hentschinski:2017ayz} and hybrid approaches \cite{vanHameren:2025hyo}, which retain partonic transverse momentum dependence or combine different kinematic regimes. In this work, we restrict ourselves to collinear factorization.} The hadronic cross section then takes the form of a convolution\footnote{The convolution refers to an integral over the parton’s longitudinal momentum fraction. In DIS, it typically appears in the form 
\[
[f \!\otimes\! \hat{\sigma}](x)
= \int_{x}^{1} \frac{dz}{z}\, f(z)\, \hat{\sigma}\!\left(\frac{x}{z}\right),
\]
that represents the contribution of all partonic momentum fractions $z$ compatible with the observed kinematics.} of partonic cross sections with the corresponding PDFs. For the DIS process, the factorized cross section can be written as
\begin{equation}
    \frac{d^2 \sigma}{dx \, dQ^2} 
    = \sum_{i = q, \bar{q}, g} 
     \bigl[ f_i(\mu_F) \, \otimes \hat{\sigma}_{\,i \ell}( \mu_F) \bigr](x,Q^2)
   \;+\; \mathcal{O}\!\Bigl(\tfrac{\Lambda_{\mathrm{QCD}}^2}{Q^2}\Bigr),
\end{equation}
where $x$ is the bjorken scalling variable, $Q^2$ is the virtuality of the exchanged boson, $f_i(\mu_F)$ is the PDF for the parton of flavor $i$ at factorization scale $\mu_F$, and $\hat{\sigma}^{\,i \ell}$ is the perturbatively calculable partonic cross section for scattering of the parton $i$ with the lepton. For hadronic collisions such as Drell--Yan production or inclusive jet production, the factorized cross section takes an analogous form,
\begin{equation}
    \sigma  = \sum_{a,b} \bigl[
    f_{a/H_{1}}(\mu_F)\,\otimes
    f_{b/H_{2}}( \mu_F)\,\otimes
    \sigma_{ab}(\mu_F ) \bigr](Q)
    \;+\;\mathcal{O}\!\Bigl(\tfrac{\Lambda_{\mathrm{QCD}}^{2}}{Q^{2}}\Bigr),
    \label{eq.2.28}
\end{equation}
where $f_{a/H_1}$ and $f_{b/H_2}$ denote the PDFs of partons $a$ and $b$ in the incoming hadrons $H_1$ and $H_2$. The partonic cross section $\hat{\sigma}_{ab}$ is a perturbatively calculable quantity computed at a fixed order in the $\alpha_s$ expansion, such as leading order (LO), next-to-leading order (NLO), or next-to-next-to-leading order (NNLO).
The factorization scale $\mu_F$ separates long- and short-distance dynamics. While its choice is formally arbitrary, it is typically set to the order of the hard scale of the process; for instance, in DIS one usually sets $\mu_F^2 \sim Q^2 = -q^2$, whereas in Drell–Yan production, $\mu_F$ is chosen to be of the order of the invariant mass of the dilepton pair or heavy boson. Although physical observables are independent of $\mu_F$, practical calculations at finite perturbative order retain a residual dependence on the scale due to missing higher-order terms.

An essential property of QCD factorization is that the separation between parton distribution functions and partonic cross sections is not unique. Finite contributions can be redistributed between these components without affecting physical observables. This freedom defines the choice of factorization scheme. In practice, the modified minimal subtraction ($\overline{\mathrm{MS}}$) scheme is the standard convention in most modern applications, although other schemes can also be defined and are related through finite redefinitions~\cite{Collins:2011zzd, Gao:2017yyd, Delorme:2025teo}.

The factorized expression also includes power-suppressed corrections of order  $\mathcal{O}(\Lambda_{\mathrm{QCD}}^2/Q^2)$, which arise from higher-twist contributions in the operator product expansion \cite{Ellis:1982wd, Wilson1969}. In standard collinear factorization, one typically retains only the leading-twist (twist-2) terms, which describe scattering off a single, nearly on-shell parton. Higher-twist effects, associated with complex multi-parton correlations, become relevant at lower momentum scales but are suppressed by powers of $1/Q^2$. Related to these are Target Mass Corrections (TMCs) \cite{Schienbein:2007gr, Ruiz:2023ozv}, which arise from the finite mass $M$ of the nucleon and become non-negligible when $M^2/Q^2$ is significant. Although TMCs lead to power-suppressed contributions similar in scaling to dynamical higher-twist effects, they are purely kinematical in origin and do not represent the multi-parton interactions of genuine higher-twist terms. Instead, they modify the scaling behavior of structure functions, particularly in the large-$x$ region. A common implementation involves replacing the Bjorken variable $x$ with the Nachtmann variable
\begin{equation*}
    \xi = \frac{2x}{1 + \sqrt{1 + \frac{4 x^2 M^2}{Q^2}}}.
\end{equation*}
A complete theoretical treatment, however, is more complicated \cite{Schienbein:2007gr, Ruiz:2023ozv}.

Finally, when heavy quarks, such as charm or bottom, participate in the scattering process, their masses must be treated consistently within the factorization framework. General-mass variable flavor number schemes (GM-VFNS), such as the ACOT scheme \cite{Aivazis:1993kh, Aivazis:1993pi, Collins:1998rz}, FONLL \cite{Cacciari:1998it, Cacciari:2001td}, and the Thorne-Roberts scheme \cite{Thorne:1997ga, Thorne:2006qt}, incorporate heavy-quark mass effects directly into the perturbative coefficients. In these approaches, mass effects enter already at the leading twist and modify both matrix elements and kinematics through explicit dependence on the heavy-quark mass $m_h$. Near the production threshold,\footnote{The threshold region corresponds to $Q^2 \sim m_h^2$, where the available energy is just sufficient to produce a heavy quark-antiquark pair.} modified scaling variables like \ $x \to x (1 + m_h^2/Q^2)$ ensure a smooth transition between the threshold region ($Q^2 \sim  m_h^2$) and the asymptotic massless regime ($Q^2 \gg m_h^2$).

\subsection{DGLAP Evolution Equations}\label{subsec:2.3.2}
As discussed in Sec.~\ref{subsec:2.2.2}, the early Parton Model assumed that DIS structure functions depend only on the Bjorken variable $x$, implying that the internal structure of the proton is independent of the probing scale. However, DIS measurements \cite{Bloom1969, Breidenbach:1969kd, WHITLOW1992475} soon revealed a mild but systematic dependence on the momentum transfer $Q^2$. These \emph{scaling violations} arise from QCD radiation: as the resolving scale increases, partons emit gluons or split into quark-antiquark pairs, thereby redistributing the proton’s momentum among its constituents. The Dokshitzer–Gribov–Lipatov–Altarelli–Parisi (DGLAP) equations \cite{Dokshitzer:1977sg, Gribov:1972ri, ALTARELLI1977298} provide the theoretical framework that quantitatively predicts this $Q^2$-dependence. They govern how parton distribution functions, $f_i(x, \mu_F^2)$, evolve with the factorization scale $\mu_F$ and take the form of the following coupled integro-differential equations:
\begin{equation}
    \frac{\partial f_i(x, \mu_F^2)}{\partial \ln \mu_F^2} =
\frac{\alpha_s(\mu_F^2)}{2\pi} \sum_j \int_x^1 \frac{dz}{z} \, P_{ij}(z, \alpha_s(\mu_F^2)) f_j \left(\frac{x}{z}, \mu_F^2\right).
\label{eq.2.28}
\end{equation}
In DIS, this scale is typically chosen to be $\mu_F = Q$. For simplicity the renormalization scale $\mu_R$ is also often set equal to $\mu_F$ ($\mu_F = \mu_R=Q$) in practical calculations.
Here, $\alpha_s(Q^2)$ is the running strong coupling, which decreases logarithmically at high scales due to asymptotic freedom. The splitting functions $P_{ij}(z)$ encode the probability for a parton $j$ to radiate or split into a parton $i$ carrying a momentum fraction $z$.
The splitting functions are perturbatively calculable order by order in the strong coupling:
\begin{equation}
    P_{ij}(z, \alpha_s)
    = P_{ij}^{(0)}(z)
    + \frac{\alpha_s}{2\pi} P_{ij}^{(1)}(z)
    + \left(\frac{\alpha_s}{2\pi}\right)^2 P_{ij}^{(2)}(z)
    + \dots
    \label{eq.2.31}
\end{equation}
where $P^{(0)}_{ij}(z)$ are the leading-order Altarelli–Parisi kernels \cite{robinDIS}, and higher-order corrections systematically improve the precision of the evolution. At leading order, the splitting functions take the following form
\begin{itemize}
    \item Quark-to-quark splitting:
    \begin{center}
    \begin{minipage}{0.45\textwidth}
    \begin{equation*}
        P_{qq}^{(0)}(z) = C_F \left[\frac{1+z^2}{(1-z)_+} + \frac{3}{2}\,\delta(1-z)\right],
    \end{equation*}
    \end{minipage}
    %\hfill
    \begin{minipage}{0.45\textwidth}
    \centering
    \begin{fmffile}{q_to_qg}
    \begin{fmfgraph*}(100,50)
        \fmflabel{\(q(k)\)}{iq}
        \fmflabel{\(q(zk)\)}{oq}
        \fmflabel{\(g((1-z)k)\)}{og}
        \fmfleft{iq}
        \fmfright{oq,og}
        \fmf{fermion}{iq,v,oq}
        \fmf{gluon}{v,og}
        \fmfdot{v}
    \end{fmfgraph*}
    \end{fmffile}
    \end{minipage}
    \end{center}
     %%%%%%%%%%%%%%%%%%%%%%%%%%%%%%%%%%%%%%%%%%%%%%%%%%%%%%%%%%%%%%%%%%%%5
    \vspace{0.5em}
    \item Quark-to-gluon splitting:
    \begin{center}
    \begin{minipage}{0.45\textwidth}
    \begin{equation*}
        P_{gq}^{(0)}(z) = T_R \left[z^2 + (1-z)^2\right],
    \end{equation*}
    \end{minipage}
    %\hfill
    \begin{minipage}{0.45\textwidth}
    \centering
    \begin{fmffile}{q_to_gq}
    \begin{fmfgraph*}(100,50)
        % Define labels
        \fmflabel{\(q(k)\)}{iq}
        \fmflabel{\(g(zk)\)}{og}
        \fmflabel{\(q((1-z)k)\)}{oq}
        \fmfleft{iq}
        \fmfright{og,oq} 
        % Quark line
        \fmf{fermion}{iq,v,oq} 
        % Gluon emission
        \fmf{gluon,lab.side=right}{v,og}   
        % Vertex
        \fmfdot{v}   
    \end{fmfgraph*}
    \label{fig:q_to_gq}
    \end{fmffile}
    \end{minipage}
    \end{center}
    %%%%%%%%%%%%%%%%%%%%%%%%%%%%%%%%%%%%%%%%%%%%%%%%%%%%%%%%%%%%%%%%%%%%%%%%
    \vspace{0.3em}
    \item Gluon-to-quark splitting:
    \begin{center}
    \begin{minipage}{0.45\textwidth}
    \begin{equation*}
        P_{qg}^{(0)}(z) = C_F \left[\frac{1+(1-z)^2}{z}\right],
    \end{equation*}
    \end{minipage}
    %\hfill
    \begin{minipage}{0.45\textwidth}
    \centering
    \begin{fmffile}{g_to_qqbar}
    \begin{fmfgraph*}(100,50)
        % Define labels
        \fmflabel{\(g(k)\)}{ig}
        \fmflabel{\(q(zk)\)}{oq}
        \fmflabel{\(\bar{q}((1-z)k)\)}{oqb}
        \fmfleft{ig}
        \fmfright{oq,oqb}   
        % Gluon line
        \fmf{gluon}{ig,v}   
        % Quark and antiquark lines
        \fmf{fermion,lab.side=right}{v,oq}
        \fmf{fermion,lab.side=left,tension=1.2}{oqb,v} 
        % Vertex
        \fmfdot{v}   
    \end{fmfgraph*}
    \label{fig:g_to_qqbar}
    \end{fmffile}
    \end{minipage}
    \end{center}
    %%%%%%%%%%%%%%%%%%%%%%%%%%%%%%%%%%%%%%%%%%%%%%%%%%%%%%%%%%%%%%%%%%%%%%%55
    \item Gluon-to-gluon splitting:
    \begin{equation*}
        P_{gg}^{(0)}(z) = 2C_A \left[\frac{z}{(1-z)_+} + \frac{1-z}{z} + z(1-z)\right]
        + \frac{\beta_0}{2}\,\delta(1-z),
    \end{equation*}
    \begin{figure}[h]
    \centering
    \begin{fmffile}{g_to_gg}
    \begin{fmfgraph*}(100,50)
        % Define labels
        \fmflabel{\(g(k)\)}{ig}
        \fmflabel{\(g(zk)\)}{og1}
        \fmflabel{\(g((1-z)k)\)}{og2}
        \fmfleft{ig}
        \fmfright{og1,og2}
        % Gluon line
        \fmf{gluon}{ig,v}
        % Gluon emissions
        \fmf{gluon,lab.side=right}{v,og1}
        \fmf{gluon,lab.side=left}{v,og2}
        % Three-gluon vertex
        \fmfdot{v}
    \end{fmfgraph*}
    %\vspace{15pt}  % Adjust the space here as needed
    \label{fig:g_to_gg}
    \end{fmffile}
    \end{figure}
\end{itemize}
where
\begin{equation}
C_F=\frac{4}{3}, \qquad C_A=3, \qquad T_R=\frac{1}{2}, \qquad
\beta_0 = 11 - \frac{2}{3}N_f ,
\end{equation}
are the SU(3) color factors and the leading coefficient of the QCD beta function (see Eq.~(\ref{eq2.8})), with $N_f$ denoting the number of active quark flavors. Here $\frac{1}{(1-z)_+}$ denotes the standard plus distribution. The diagrams shown above illustrate the corresponding splitting processes: a parent parton with momentum $k$ splits into two daughter partons carrying longitudinal momentum fractions $zk$ and $(1 - z)k$.

The DGLAP splitting functions are known exactly up to next-to-next-to-leading order (NNLO), corresponding to three-loop accuracy in perturbative QCD. These results were obtained by Moch, Vermaseren, and Vogt \cite{Vogt:2004mw, MOCH2014351}, and include both singlet and non-singlet components, available in both Mellin and momentum space.
At the next order, next-to-next-to-next-to-leading order (N\(^3\)LO), the splitting functions are not yet known in full analytic form. However, significant progress has been made through a combination of approximate calculations and theoretical constraints \cite{Cooper-Sarkar:2024crx}. These include results from the large-$n_f$ expansion \cite{Davies:2016jie}, leading-color approximations \cite{Moch:2017uml}, small-$x$ asymptotics derived from high-energy (BFKL) resummation \cite{Bonvini:2018xvt}, and large-$x$ behavior inferred from soft-gluon resummation \cite{Davies:2022ofz}. 
Together, these complementary approaches provide strong constraints on the structure of the N\(^3\)LO splitting functions, enabling increasingly precise phenomenological approximations, although a complete analytic determination remains an open problem.

The DGLAP equations separate parton distribution functions into non-singlet and singlet components, each exhibiting distinct evolution patterns \cite{Block:2010du}. This decomposition simplifies the structure of the evolution equations and reflects the symmetries of the underlying quantum field theory. Non-singlet distributions, such as the valence quark difference $q^{NS}(x, Q^2) = u_v(x, Q^2) - d_v(x, Q^2)$, represent flavor-asymmetric combinations. Because the gluon-to-quark splitting process ($g \rightarrow q \bar{q}$) produces quarks and antiquarks of the same flavor at the same rate, these contributions cancel out in the non-singlet difference. Consequently, non-singlet distributions evolve independently of the gluon distribution. While the constituent quarks carry color charge, the total evolution of the non-singlet density is governed solely by quark-to-quark splittings:
\begin{equation}
    \frac{dq^{NS}(x, Q^2)}{d\ln Q^2}
    = \frac{\alpha_s(Q^2)}{2\pi}
    \int_x^1 \frac{dz}{z}\,
    P_{qq}(z, \alpha_s(Q^2))
    q^{NS}\left(\frac{x}{z}, Q^2\right).
\end{equation}
In contrast, the singlet distribution,
\begin{equation}
    \Sigma(x, Q^2) = \sum_{i=1}^{N_f} \left[ q_i(x, Q^2) + \bar{q}_i(x, Q^2) \right],
\end{equation}
includes the quark and antiquark densities of all active flavors. Unlike the non-singlet case, the singlet distribution couples directly to the gluon distribution $G(x/z, Q^2)$ during evolution. This quark–gluon mixing leads to a coupled system of integro-differential equations:
\begin{equation}
    \frac{d}{d\ln Q^2} 
    \begin{bmatrix} 
        \Sigma(x, Q^2) \\ 
        G(x, Q^2) 
    \end{bmatrix} 
    =
    \frac{\alpha_s(Q^2)}{2\pi} 
    \int_x^1 \frac{dz}{z} 
    \begin{bmatrix} 
        P_{qq}(z) & P_{qg}(z) \\ 
        P_{gq}(z) & P_{gg}(z) 
    \end{bmatrix} 
    \begin{bmatrix} 
        \Sigma(x/z, Q^2) \\ 
        G(x/z, Q^2) 
    \end{bmatrix}.
\end{equation}
At small $x$, the evolution is dominated by the gluon channel. The splitting function $P_{gg}(z)$ behaves as $1/z$ for $z \to 0$, causing the gluon density to rise rapidly with increasing $Q^2$. The quark singlet distribution is likewise enhanced through gluon splitting into quark–antiquark pairs via $P_{qg}(z)$. Efficient numerical implementations of these evolutions, such as those used in QCDNUM \cite{Botje:2010ay}, HOPPET \cite{SALAM2009120}, and APFEL \cite{Bertone:2017gds}, are essential for practical solutions of the DGLAP equations across the wide range of kinematic scales explored at the LHC and future colliders.

The scaling violations observed in DIS arise directly from the DGLAP evolution. As the resolution scale $Q^2$ increases, the rapid growth of the gluon distribution at low $x$ leads to a corresponding rise in the quark singlet distribution. Consequently, the proton structure function  $F_2(Q^2,x)$, which at leading order is primarily sensitive to the quark content, increases with $Q^2$ at small $x$, deviating from the scale-independent behavior predicted by the naive Parton model. This $Q^2$-dependence of $F_2$ provides clear experimental evidence of QCD dynamics and supports the predictive framework of DGLAP evolution.

%****************************** Second Chapter *********************************
%*******************************************************************************
\chapter{Nuclear Parton Distribution Functions}\label{chap.3}
The QCD framework underlying the partonic description of hadronic structure was established in the previous chapter. Within this framework, parton distribution functions (PDFs) encode the nonperturbative longitudinal momentum distributions of quarks and gluons inside a fast-moving nucleon and enter physical observables through collinear factorization. While this framework is well-established for free nucleons, its extension to nuclear systems introduces additional structure arising from the nuclear environment.
Experimental measurements have shown that the partonic distributions in bound nucleons differ systematically from those in free nucleons, indicating that a nucleus cannot be described as a simple superposition of independent constituents. Within the factorization framework, these effects are encoded in nuclear parton distribution functions (nPDFs), which generalize PDFs to nuclear targets. The determination of nPDFs is significantly more challenging than for the free proton. Experimental constraints are more limited in both precision and kinematic coverage, while the need to account for $A$-dependence and isospin effects introduces a more complex parameter space. Consequently, the extracted distributions exhibit larger uncertainties, especially within the more poorly constrained kinematic sectors.

This chapter develops the theoretical and phenomenological framework required for determination of nuclear parton distribution functions. Section~\ref{sec:3.1} briefly summarizes the key properties of proton PDFs relevant for later discussion. Section~\ref{sec:3.2} presents the experimental evidence for nuclear modifications and the characteristic patterns observed in nuclear structure functions. Section~\ref{sec:3.3} introduces the definition of nPDFs and reviews the main approaches used for their determination. Finally, Section~\ref{sec:3.4} outlines the structure of global nPDF analyses, including data selection, parameterization strategies, and methods for uncertainty estimation. This framework provides the basis for the Bayesian nPDF analysis developed in subsequent chapters.

%********************************** %First Section  **************************************
\section{Parton Distribution Functions in the Nucleon}\label{sec:3.1}
\emph{Parton distribution functions}, denoted $f_i(x,Q^2)$, quantify the probability density of finding a parton of type $i$ (quark, antiquark, or gluon) inside a fast-moving nucleon (e.g., a proton), carrying a fraction $x$ of its longitudinal momentum when probed at a resolution scale $Q^2$.\footnote{Note that this probabilistic interpretation is strictly valid only at leading order; beyond leading order, PDFs are scheme-dependent objects defined within QCD factorization and are not necessarily positive definite.} The scale $Q^2$ characterizes the resolving power of the hard interaction: at larger $Q^2$, the probe resolves finer details of the nucleon’s internal structure. Within the QCD factorization framework, PDFs play an essential role by connecting the long-distance, nonperturbative structure of the nucleon to short-distance partonic subprocesses, which are computed perturbatively. In the absence of PDFs, perturbative QCD predictions for hadronic cross sections would be incomplete, as the partonic matrix elements must be weighted by the likelihood of finding the relevant partons in the incoming hadrons.

To clarify the kinematics, consider a proton moving with large momentum along the $z$-axis. In the infinite momentum frame, it is convenient to adopt light-cone coordinates, where the $+$ and $-$ components of a four-vector $v^\mu$ are defined as:
\begin{equation}
    v^{\pm} = \frac{v^0 \pm v^3}{\sqrt{2}}.
\end{equation}
In these coordinates, the momentum of a highly boosted proton takes the form
\begin{equation}
    P = (P^+, P^-, P_{\perp}) = \left( P^+, \frac{m_p^2}{2P^+}, 0 \right),
\end{equation}
where $P^+$ is the dominant component (since the proton is highly boosted along $+z$), $m_p$ is the proton mass, and we have chosen the frame such that the transverse momentum vanishes ($P_{\perp}=0$). A parton inside the proton carries momentum $k$, and its fraction of the proton’s $+$-momentum is defined as:
\begin{equation}
    x=\frac{k^+}{P^+}, \quad  0 < x < 1.
\end{equation}
This longitudinal momentum fraction is the fundamental variable of PDFs and, in the DIS process, is equivalent to the Bjorken $x$ variable. Thus, $x$ provides a direct measure of how the proton’s momentum is shared among its quark and gluon constituents.

The PDFs can be defined rigorously through matrix elements of quark and gluon field operators, which encode the nonperturbative correlations among partons inside the proton \cite{Kovarik:2019xvh}. For a quark of flavor $i$, the collinear PDF is given by the gauge-invariant light-cone correlator
\begin{equation}
    f_{i/N}(x, Q^2) = \frac{1}{4\pi} \int d y^- e^{-i x P^+ y^-} 
    \left\langle P \middle| \bar{\psi}_i(0, y^-, 0_{\perp}) \gamma^+ \mathcal{W}(y^-, 0) \psi_i(0) \middle| P \right\rangle.
\end{equation}
Here, $\psi_i$ and $\bar \psi_i$ are the quark field operators, and the Dirac matrix $\gamma^+$ projects the component aligned with the large proton momentum. The separation $y^-$ between the fields lies along the light-cone direction conjugate to $P^+$, and the Fourier factor $e^{-i x P^+ y^-}$ extracts the longitudinal momentum fraction $x$. The Wilson line $ \mathcal{W}(y^-, 0)$ ensures the gauge invariance by connecting the two spacetime points; the resulting operator is therefore a bilocal gauge-invariant operator.\footnote{A bilocal operator consists of field operators evaluated at two distinct spacetime points connected by a Wilson line, ensuring gauge invariance under local gauge transformations.} Similarly, the gluon PDF is defined in terms of the field-strength tensor as
\begin{equation}
    f_{g/N}(x, Q^2) = \frac{1}{2\pi x P^+} \int d y^- e^{-i x P^+ y^-} 
    \left\langle P \middle| F^{+ \mu}(0, y^-, 0_{\perp}) \mathcal{W}(y^-, 0) F^{+}_{\ \mu}(0) \middle| P \right\rangle,
\end{equation}
where $F^{\mu\nu}$ is the gluon field-strength tensor. As in the quark case, the light-cone separation and Fourier transform project out the longitudinal momentum fraction $x$, and the Wilson line ensures gauge invariance of the nonlocal operator.

Since these matrix elements are taken between identical proton states, they are referred to as forward matrix elements. The profound consequence of the QCD Factorization Theorem is the \emph{universality} of these functions: PDFs are process-independent and do not depend on the specific hard scattering process in which they are measured (e.g., DIS, Drell-Yan, jet production). Factorization allows the cross-section for any hard scattering process to be expressed as a convolution of these universal PDFs with process-specific, perturbatively calculable hard-scattering coefficients. Although the PDFs themselves cannot be computed perturbatively, their evolution with the resolution scale $Q^2$  is governed by the DGLAP equations (see Sec. \ref{subsec:2.3.2}), which capture the effects of parton radiation and splitting as the probe’s resolution changes. At low resolution scales, the proton is dominated by its valence quark content. As $Q^2$ grows, , gluon radiation becomes increasingly important, populating the small-$x$ region and causing gluons to carry an ever larger fraction of the proton’s momentum at high energies.

As PDFs characterize the probability densities of partons inside a nucleon, they must satisfy several sum rules that reflect fundamental QCD conservation laws, most notably baryon (quark) number conservation and momentum conservation. These sum rules are imposed at the initial scale $Q_0$, and the DGLAP evolution equations ensure that they remain satisfied at all higher scales. Consequently, sum rules provide essential internal consistency conditions in global QCD analyses and help constrain the behavior of PDFs across energy scales.\footnote{Sum rules can be factorization-scheme dependent. In schemes other than $\overline{\text{MS}}$, higher-order corrections may modify the explicit form of the quark number and momentum integrals, leading to scheme-dependent expressions for the individual parton contributions \cite{Collins:2011zzd, Delorme:2025teo}. However, the underlying conservation laws remain unchanged, and physical observables are unaffected by these scheme choices.}
\begin{itemize}
    \item \textbf{Quark Number Sum Rule:} The total number of valence quarks in a nucleon is fixed by its quantum numbers. Valence quarks are defined as the excess of quarks over antiquarks of a given flavor, and their number is obtained by integrating the difference between quark and antiquark PDFs:
    \begin{equation}
    \int_0^1 \left[ q_i(x, Q^2) - \bar{q}_i(x, Q^2) \right]\, dx = N_i^{\text{valence}}.
    \end{equation}
    For the proton, this leads to the well-known relations:
    \begin{equation*}
    \int_0^1 [u(x,Q^2) - \bar{u}(x,Q^2)]\, dx = 2, \quad
    \int_0^1 [d(x,Q^2) - \bar{d}(x,Q^2)]\, dx = 1,
    \end{equation*}
    \begin{equation*}
    \int_0^1 \left[q_i(x,Q^2) - \bar{q}_i(x,Q^2)\right] \, dx = 0, 
    \qquad i = s, c, b, t,
    \end{equation*}
    which encode its valence structure. Because quark number is a strictly conserved quantity, these relations hold at any scale $Q^2$; the DGLAP evolution equations preserve them exactly.
    \item \textbf{Momentum Sum Rule:} In QCD, all partons must together carry the total momentum of the nucleon. This leads to the momentum sum rule:
    \begin{equation}
    \int_0^1  \left[ \sum_q x q(x, Q^2) + \sum_{\bar{q}} x \bar{q}(x, Q^2) + x g(x, Q^2) \right] \, dx = 1.
    \end{equation}
    Although the sum rule itself holds at any scale, the momentum distribution among partons evolves with $Q^2$. As $Q^2$ increases, gluon radiation and parton splitting, captured by the DGLAP equations, shift momentum from valence quarks toward gluons and sea quarks.
\end{itemize}
Since PDFs encode the non-perturbative structure of hadrons and describe the long-distance QCD dynamics, they cannot be computed using perturbation theory. \emph{Lattice QCD} \cite{LIN2018107, Alexandrou:2015rja} provides a first-principles, nonperturbative framework to evaluate QCD observables by discretizing spacetime and numerically computing the path integral. Historically, lattice computations were limited to a few first Mellin moments of PDFs, which do not contain sufficient information to reconstruct the full $x$-dependence. However recent developments, particularly the introduction of quasi-PDFs \cite{Ji:2013dva, Lin:2014zya} and pseudo-PDFs \cite{Radyushkin:2017cyf, Joo:2020spy}, represent a major conceptual step forward. These approaches compute spatial correlation functions of boosted hadrons in Euclidean space, which are accessible to lattice simulations. Although these correlators differ from the light-cone PDFs appearing in high-energy scattering, they can be systematically matched to them using perturbative factorization formulas. This matching corrects for differences between the lattice-observable quantities and the light-cone PDFs relevant for phenomenology. As a result, these methods provide a practical path toward determining the full $x$-dependence of PDFs from first principles \cite{LIN2025104177}. Despite these advances, lattice calculations are not yet sufficient for use in high-precision phenomenological applications. Consequently, the extraction of PDFs for phenomenology still relies on global QCD analyses \cite{Kovarik:2019xvh, Ethier:2020way}. In these analyses, PDFs are parameterized at an initial scale $Q^2_0$, evolved to higher scales, and fitted to a wide range of high-energy experimental data, such as DIS, Drell–Yan, jet production, and electroweak boson production in hadronic collisions. The resulting distributions, $ f_i(x, Q^2)$, provide a universal, scale-dependent description of partonic structure that links experimental measurements with perturbative QCD predictions.
%
%********************************** %Second Section  **************************************
\section{Nuclear Effects}\label{sec:3.2}
\begin{figure}[htbp!] 
\centering    
\includegraphics[width=0.7\textwidth]{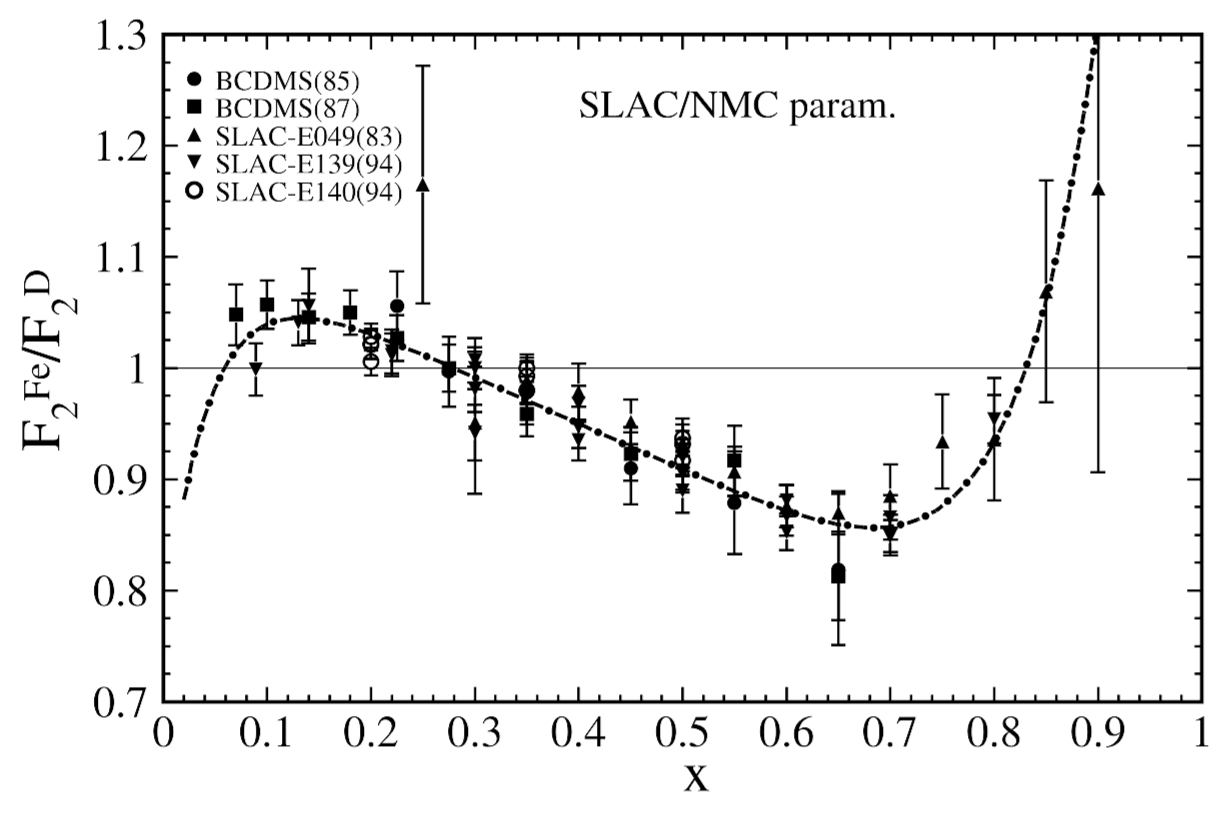}
\caption[Nuclear correction ratio]{Nuclear correction ratio $F_{2}^{\mathrm{Fe}} / F_{2}^{\mathrm{D}}$, the figure taken from \cite{Schienbein:2009kk}.}
\label{fig:nuclearcorr}
\end{figure}

DIS measurements on nuclear targets have revealed that nucleons bound inside nuclei do not behave like isolated, free nucleons. A naive expectation was that the structure function of a nucleus nearly equals the sum of the structure functions of its constituent protons and neutrons. However, experimental data show clear deviations from this picture. The first compelling evidence came from the European Muon Collaboration (EMC) in the early 1980s, which observed unexpected modifications in the DIS structure function of iron compared to deuterium \cite{EuropeanMuon:1983wih}. These deviations, known as the EMC effect, reveal that the nuclear environment induces significant and non-trivial changes to the partonic structure of nucleons. Since then, a broad range of experiments has confirmed and refined our understanding of these nuclear modifications. A convenient observable for characterizing these effects is the nuclear modification ratio, defined as
\begin{equation}
    R_A (x) \equiv  \frac{F_2^A(x)}{F_2^D(x)},
\end{equation}
where $F_2^A(x)$ is the per‐nucleon structure function of a nucleus with mass number $A$ and $F_2^D(x)$ is the corresponding quantity measured in deuterium, typically treated as a substitute for a system of nearly free nucleons. If nuclear binding had no influence on partonic dynamics, one would expect $R_A (x)=1$ for all Bjorken $x$.\footnote{This expectation holds exactly only for isoscalar nuclei, where the number of protons equals the number of neutrons. For non-isoscalar nuclei, even without nuclear modifications, differences in the up- and down-quark content of the target relative to deuterium lead to deviations from unity.}
A similar quantity is defined at the level of parton distributions through the nuclear modification factor
\begin{equation}
    R_i^A(x,Q^2) \equiv \frac{f_i^{p/A}(x,Q^2)}{f_i^p(x,Q^2)},
\end{equation}
where $f_i^{p/A}$ denotes the PDF of parton flavor $i$ in a proton bound inside a nucleus with mass number $A$, and $f_i^p$ is the corresponding free-proton PDF.\footnote{Nuclear modification ratios can also be defined at the level of measured cross sections, e.g.\ as ratios of lepton–nucleus to lepton–deuterium DIS cross sections.}
The observed values of $R_A (x)$, however, differ significantly from this naive expectation. As shown in Fig.~\ref{fig:nuclearcorr}, DIS measurements from SLAC \cite{Gomez1994} and BCDMS \cite{hepdata.30200} (with isoscalar corrections applied for iron) reveal a characteristic pattern of nuclear modifications across Bjorken $x$, demonstrating that the nuclear environment changes the partonic structure of nucleons in a nontrivial way.

\begin{enumerate}
    \item \textbf{Shadowing} ($x \lesssim 0.1$):
    At very low $x$, the nuclear structure function is suppressed relative to that of free nucleons, yielding $R_A (x) < 1$. This suppression, known as \emph{shadowing} \cite{Armesto:2006ph, KHARZEEV1994361,Kulagin:2004ie}, is believed to arise from coherent multiple scattering and the overlap of partons from different nucleons. In this regime, gluon recombination and coherence effects reduce the effective parton densities, especially for gluons and sea quarks. Shadowing grows with nuclear mass and, in the context of leading-twist QCD, is generally predicted to become only logarithmically less pronounced at higher $Q^2$, as DGLAP evolution resolves the partonic structure more finely and gradually dilutes the effective coherence and recombination effects.
    \item \textbf{Anti-Shadowing} ($0.1  \lesssim x \lesssim 0.3$):
    In the intermediate region, we observe a slight enhancement, with \(R_A(x)>1\). This \emph{anti-shadowing} behavior \cite{Kulagin:2004ie, Eskola:2009uj, NICOLAEV1975397} is often interpreted in connection with global conservation laws, particularly the momentum sum rule: the suppression at low \(x\) and in the EMC region typically requires compensating strength at other values of \(x\). Although the underlying mechanism remains under study, models show that it may result from constructive interference between different exchange processes or from short-range dynamical correlations among nucleons \cite{Brodsky:2009bg, Frankfurt:2011cs, Kulagin:2004ie}.
    \item \textbf{The EMC Effect} ($0.3  \lesssim x \lesssim 0.7$):
    In the medium-\(x\) region, the structure function shows a notable suppression, typically around 10–15\% for heavy nuclei, such that \(R_A(x)<1\). This phenomenon—the \emph{EMC effect} \cite{Geesaman:1995yd}, first observed by the European Muon Collaboration \cite{EuropeanMuon:1983wih}, demonstrates that the quark structure of nucleons is modified in the nuclear environment. Its origin is still not fully resolved, but several mechanisms have been proposed:
    \begin{itemize}
        \item \textbf{Nucleon binding and off-shell effects:}  
    Bound nucleons are off their mass shell due to nuclear binding, which may change their internal structure and thus their parton distributions \cite{AKULINICHEV1985485}.
        \item \textbf{Short-range correlations (SRCs):}  
        Much evidence points to SRCs, strong and short distance interactions between nucleon pairs, as a major contributor \cite{Weinstein:2010rt, Hen:2012fm, Hen:2013oha}. In SRC pairs (predominantly neutron–proton), nucleons carry momenta well above the Fermi momentum and are strongly off-shell. Their internal quark structure can be significantly modified in this dense local environment, redistributing the parton momentum density toward larger Bjorken-$x$ and suppressing the intermediate-$x$ region, consistent with the observed EMC suppression.
    \end{itemize}
    The EMC effect becomes stronger with increasing mass number \(A\) and shows very weak \(Q^2\)-dependence, indicating that it is driven by nonperturbative nuclear dynamics.
    \item \textbf{Fermi Motion} ($ x \gtrsim 0.7$):
    At large \(x\), \(R_A(x)\) rises above unity. This enhancement originates from the intrinsic \emph{Fermi motion} of nucleons bound inside a nucleus \cite{Bodek1981}. Because nucleons possess significant momentum even in the nuclear ground state, a DIS probe can interact with a nucleon moving in the direction of the virtual photon, making its partons appear to carry a larger momentum fraction \(x\). This kinematic broadening shifts the nuclear structure function toward higher \(x\) and can even produce contributions for \(x>1\).\footnote{In nuclear DIS, the region $x>1$ corresponds to scattering from nucleons belonging to the high-momentum tail of the nuclear wave function. These momenta arise from short-range correlations or other multi-nucleon configurations, leading to partons that carry momentum fractions beyond the kinematic limit of a single free nucleon at rest ($x=1$), which is significantly larger than the average momentum per nucleon. This regime is sometimes described in terms of ``superfast'' quarks; see \cite{Freese:2015ebu} for a detailed discussion.} Such high-\(x\) tails have been observed directly in Jefferson Lab measurements \cite{Seely:2009gt, Fomin:2010ei} up to \(x\sim1.4\), and are especially pronounced for heavier nuclei with larger average Fermi momentum.
\end{enumerate}

%********************************** %Third Section  **************************************
\section{Nuclear Parton Distribution Functions}\label{sec:3.3}
Nuclear parton distribution functions describe the momentum distributions of quarks and gluons inside nuclei, extending the concept of PDFs from free nucleons to nucleons bound within a nucleus. The nuclear environment modifies the partonic structure in nontrivial ways, such that a nucleus cannot be treated as a simple incoherent sum of free protons and neutrons. Binding effects, nucleon off-shellness, short-range correlations, and coherent multi-nucleon interactions all modify the parton momentum distributions relative to those in isolated nucleons (as illustrated in Fig.~\ref{fig:nuclearcorr}). These nuclear modifications are not incorporated in a universal manner across all theoretical treatments. In some approaches, the nuclear PDFs themselves, denoted $f_i^A(x_A, Q^2)$, are defined as fundamental nonperturbative distributions in which shadowing, antishadowing, the EMC effect, and Fermi motion are embedded directly in the initial nuclear structure. Other approaches introduce nuclear effects through multiplicative modification factors applied to free-proton PDFs, effectively treating the nuclear environment as a deviation from the proton baseline.

Various theoretical and phenomenological approaches have been developed to describe the behavior of nuclear modifications. Generally, they may be categorized into ``data-driven global fits'' and ``microscopic (physics-based) models,'' though modern analyses often combine elements of both:\footnote{A third category involves first-principles calculations from Lattice QCD. However, current results are limited to very light nuclei, such as deuterium and $^3$He, and are typically restricted to low moments (e.g.\ the first moment) of the distributions; see, e.g., \cite{Winter:2017bfs, Constantinou:2020hdm, Detmold:2020snb}.}
\begin{itemize}
    \item \textbf{Data‐Driven Global Fits}: Analogous to the determination of free-proton PDFs, nuclear PDFs can be extracted from experimental data through flexible functional forms constrained in global QCD analyses (see \cite{Klasen:2023uqj}). In this approach, nuclear effects are encoded directly in the fitted functions, either as nuclear-dependent PDFs or as nuclear modification factors applied to the proton baseline. These functions are treated as universal, process-independent quantities determined from data rather than from explicit nuclear dynamics. Modern nPDF sets such as EPPS21 \cite{Eskola:2021nhw}, nCTEQ15HQ \cite{Duwentaster:2021ioo}, and nNNPDF3.0 \cite{nNNPDF3.0:2022fyi} adopt this strategy, offering quantitative uncertainty estimates and broad phenomenological coverage. However, these fits do not attempt to interpret the underlying nuclear mechanisms that generate the observed modifications.
    
    \item \textbf{Microscopic and physics-based models}: To explain the physical origin of nuclear modifications, microscopic models incorporate explicit nuclear physics effects into the calculation of nuclear structure functions. Many such models focus on a single mechanism relevant in a specific kinematic region. For example, models based on short-range correlations (SRC) \cite{Subedi:2008zz, nCTEQ:2023cpo} address the high-$x$ region by describing the large relative momenta of nucleon pairs inside nuclei. While these approaches provide valuable insight, their applicability is typically limited to the domain where the dominant mechanism is expected to operate. By contrast, comprehensive frameworks seek to combine several nuclear mechanisms into a unified description valid across a broad range of $x$ and $Q^2$. A common example is the convolution approach, in which the nucleus is treated as a system of bound nucleons with a momentum distribution described by a spectral function. The nuclear structure function is obtained by folding the bound-nucleon PDF with this distribution, naturally generating Fermi motion and binding effects \cite{Bodek1981}. Additional mechanisms, such as the modification of the bound nucleon’s internal structure due to off-shellness, are included through dedicated functions that model the PDF dependence on the nucleon virtuality \cite{Kulagin1994, Alekhin:2022uwc}. The Kulagin–Petti (KP) model \cite{Kulagin:2016zkk, Ru:2016wfx} represents such a multi-component framework. It integrates several distinct mechanisms: (1) Fermi motion and nuclear binding; (2) off-shell corrections to the nucleon PDFs; (3) contributions from meson-exchange currents (MEC), which enhance the low-$x$ pion-induced component \cite{Ericson:1983um}; and (4) coherent interactions of the exchanged boson with multiple nucleons, generating nuclear shadowing \cite{FRANKFURT1981215}. By combining these effects within a consistent theoretical structure, the KP model successfully reproduces the observed $x$, $Q^2$, and mass-number dependence of nuclear structure function ratios, as well as related observables in nuclear Drell–Yan \cite{Kulagin:2015lkm} and electroweak-boson production in proton–lead collisions \cite{Ru:2016wfx}.
\end{itemize}

In high-energy nuclear interactions, it is commonly assumed that the collinear factorization framework can be extended to hard scattering processes involving nuclei. Within this framework, nuclear PDFs play the same conceptual role as free-nucleon PDFs: they encode all long-distance, nonperturbative physics, while the short-distance partonic subprocesses remain calculable in perturbative QCD. Under this assumption, the cross section for proton–nucleus or nucleus–nucleus collisions can be written in direct analogy with the free-proton case as
\begin{equation}
\sigma = \sum_{a,b} \bigl[
f_a^A(\mu_{F}) \, \otimes \, f_b^B(\mu_{F}) \, \otimes \, \hat{\sigma}_{ab}(\mu_{F})
\bigr] (Q)+
\mathcal{O}\Bigl(\tfrac{\Lambda_{\mathrm{QCD}}^{2}}{Q^{2}}\Bigr),
\label{eq:2.28_nuclear}
\end{equation}
where $f_a^A$ and $f_b^B$ are the nPDFs for partons $a$ and $b$ in nuclei $A$ and $B$, respectively, and $\hat{\sigma}_{ab}$ denotes the perturbatively calculable partonic hard-scattering cross section, which is assumed to be independent of the nuclear environment. The term $\mathcal{O}(\Lambda_{\mathrm{QCD}}^{2}/Q^{2})$ represents higher-twist contributions, which are suppressed at sufficiently large values of $Q$.

However, it is important to emphasize that collinear factorization for nuclear processes has not been fully proven within QCD. While factorization theorems have been formally established for several inclusive processes involving free protons, such as DIS and Drell–Yan production, their extension to nuclear systems introduces additional theoretical challenges \cite{Collins:2011zzd}. In a nuclear environment, effects such as multiple parton scattering, coherent initial-state interactions, and nuclear-enhanced higher-twist contributions (including final-state partonic energy loss) can, in principle, violate the assumptions underlying standard factorization proofs \cite{Accardi:2004gp, Albacete:2014fwa}. Despite these complications, numerous theoretical studies have examined the validity of collinear factorization in proton–nucleus ($pA$) collisions, particularly for inclusive observables at high momentum transfer $Q^2$. These analyses indicate that leading-twist collinear factorization remains a robust and reliable approximation in specific kinematic regions (most notably at moderate to large Bjorken-$x$ and sufficiently high $Q^2$) where coherence effects, power-suppressed corrections, and small-$x$ dynamics are expected to be suppressed \cite{Qiu:2004qk, Qiu:2006da, Frankfurt:2011cs}. In practice, the phenomenological success of collinear factorization provides strong empirical evidence for its use in the nuclear environment. Factorized calculations accurately describe a broad class of observables in both $pA$ and $AA$ collisions, particularly for color-neutral final states such as Drell–Yan processes and electroweak boson production. This empirical consistency is further supported by the success of global nPDF analyses—such as EPPS21 \cite{Eskola:2021nhw}, nCTEQ15 \cite{Kovarik:2015cma}, and nNNPDF3.0 \cite{nNNPDF3.0:2022fyi}—which reproduce a broad range of nuclear data across multiple processes within the collinear framework. In these analyses, nuclear modifications such as shadowing, antishadowing, the EMC effect, and Fermi motion are incorporated into universal, process-independent nPDFs. Under the assumption that collinear factorization holds, the scale evolution of nPDFs is governed by the standard DGLAP equations:
\begin{equation}
    \frac{\partial f_i^A(x, Q^2)}{\partial \ln Q^2} = \frac{\alpha_s(Q^2)}{2\pi} 
    \sum_j \int_x^1 \frac{dz}{z} P_{ij}(z) f_j^A(x/z, Q^2),
\end{equation}
where $P_{ij}$ are the universal QCD splitting functions. Thus, while the initial conditions of the nuclear PDFs encode nonperturbative nuclear effects, their evolution with $Q^2$ proceeds identically to that of free-nucleon PDFs.

Nuclear PDFs can be understood from two complementary viewpoints. At the ``nucleus'' level, the distribution $f_i^{A,\text{nucleus}}(x_A, Q^2)$ represents the probability of finding a parton of flavor $i$ carrying a fraction $x_A \in (0,1]$ of the total nuclear momentum $P_A$. The variable 
\begin{equation*}
    x_A = \frac{Q^2}{2P_A \cdot q}
\end{equation*}
is the physical Bjorken‑$x$ measured in DIS on a nuclear target. For practical purposes, especially in global nPDF analyses comparing different nuclei, it is often convenient to work with a per-nucleon momentum fraction, $x_N = A\,x_A$, which corresponds to the fraction of momentum carried by the parton relative to an average nucleon. This rescaled variable spans $x_N \in (0,A]$, although in global QCD analyses one typically restricts to $x_N \leq 1$.\footnote{For the physical interpretation of the region $x_N > 1$, see \cite{Freese:2015ebu}. In principle, this region should also be accounted for in the DGLAP evolution.} In this representation, nuclear PDFs are interpreted on a per-nucleon basis, which facilitates their parameterization, evolution, and comparison across different nuclear systems within the standard QCD factorization framework. For a nucleus composed of $Z$ protons and $N=A-Z$ neutrons and assuming isospin symmetry, the per-nucleon nuclear PDF can be decomposed into bound-proton and bound-neutron contributions as
\begin{equation}
    f_i^{(A,Z)}(x_N, Q^2) \equiv f^A_i=\frac{Z}{A} f_i^{p/A}(x_N, Q^2) + \frac{A-Z}{A} f_i^{n/A}(x_N, Q^2),
    \label{eq.3.12}
\end{equation}
where $f_i^{p/A}$ and $f_i^{n/A}$ are the PDFs of protons and neutrons bound inside the nucleus. This decomposition provides a consistent bridge between nuclear PDFs and the nucleonic structure of the nucleus, with the bound neutron PDF typically obtained by applying isospin symmetry to the bound proton distributions $(u \leftrightarrow d)$.

Sum rules provide further constraints on nuclear PDFs, ensuring that fundamental conservation laws remain satisfied even after accounting for nuclear modifications. The momentum sum rule, applied to per-nucleon nPDFs, takes the following form:
\begin{equation}
    \int_0^1 dx \left[ \sum_q x f_q^A(x, Q^2) + \sum_{\bar{q}} x f_{\bar{q}}^A(x, Q^2) + x f_g^A(x, Q^2) \right] = 1.
\end{equation}
In addition, quark-number sum rules impose constraints on the total number of valence quarks in the nucleus. For a nucleus with $Z$ protons and $N$ neutrons, the valence up- and down-quark distributions satisfy
\begin{equation}
    \int_0^1 dx \, u_v^A(x, Q^2) = \frac{2Z+N}{A}, \qquad \int_0^1 dx \, d_v^A(x, Q^2) = \frac{Z+2N}{A}.
\end{equation}
Here $u_v^A(x, Q^2)$ and $d_v^A(x, Q^2)$ denote the per-nucleon valence quark distributions inside the nucleus. These constraints ensure that the net number of valence quarks correctly matches the isospin composition of the nuclear system.

%%%%%%%%%%%%%%%%%%%%%%%%%%%%% Fourth section %%%%%%%%%%%%%%%%%%%%%%%%%%%%%%%%%%%%%%
\section{Global Analysis of nuclear PDFs}\label{sec:3.4}
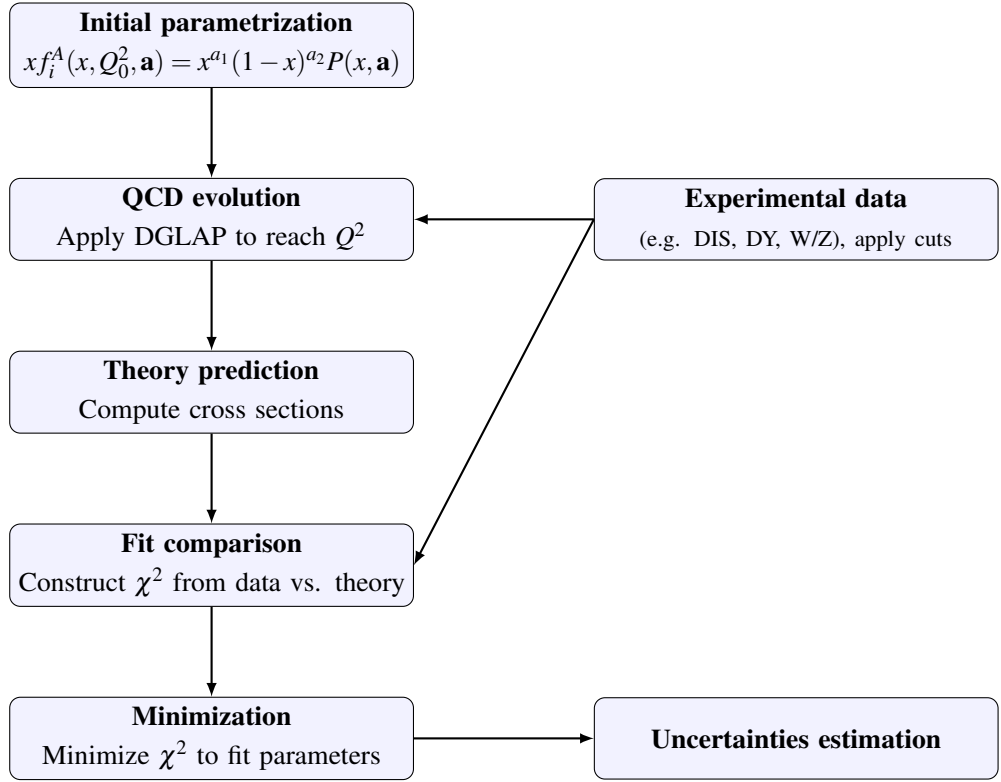
\begin{figure}[H]
\centering
\begin{tikzpicture}[scale=0.85, transform shape, 
  node distance=1.4cm and 2.8cm, auto,
  block/.style = {rectangle, draw=black, fill=blue!5, 
                  text width=6cm, align=center,
                  rounded corners, minimum height=3em},
  arrow/.style = {-latex, thick}
  ]
% --- Nodes (vertical flow) ---
\node[block] (input) {
  \textbf{ Initial parametrization}\\
  $xf_i^A(x,Q_0^2,\mathbf{a}) = x^{a_1}(1 - x)^{a_2} P(x,\mathbf{a} )$
};
\node[block, below=of input] (dglap) {
  \textbf{QCD evolution}\\
  Apply DGLAP to reach $Q^2$
};

\node[block, below=of dglap] (theory) {
  \textbf{Theory prediction}\\
  Compute cross sections
};

\node[block, below=of theory] (chi2) {
  \textbf{Fit comparison}\\
  Construct $\chi^2$ from data vs. theory
};

\node[block, below=of chi2] (minim) {
  \textbf{Minimization}\\
  Minimize $\chi^2$ to fit parameters
};

% --- Right-side nodes ---
\node[block, right=of dglap] (data) {
  \textbf{Experimental data}\\
  \footnotesize (e.g. DIS, DY, W/Z), apply cuts
};
\draw[arrow] (data.west) -- (dglap.east);
\draw[arrow] (data.west) -- (chi2.east);

\node[block, right=of minim] (uncert) {
  \textbf{Uncertainties estimation}
};
\draw[arrow] (minim.east) -- (uncert.west);

% --- Main vertical arrows ---
\draw[arrow] (input) -- (dglap);
\draw[arrow] (dglap) -- (theory);
\draw[arrow] (theory) -- (chi2);
\draw[arrow] (chi2) -- (minim);

\end{tikzpicture}
\caption{Flowchart representation of a generic global nuclear PDF analysis.}
\label{fig:globalqcd}
\end{figure}

Global analyses of nuclear PDFs aim to determine the momentum distributions of quarks and gluons within bound nucleons across a wide range of nuclei. This is achieved by confronting theoretical predictions, constructed within the framework of QCD collinear factorization and evolved using the DGLAP equations, with experimental data from high-energy processes. The fits are additionally constrained by QCD sum rules, ensuring physical consistency.
A typical global nPDF analysis proceeds through a well-defined sequence of steps, summarized schematically in Fig.~\ref{fig:globalqcd}. The procedure begins with the selection of experimental data, which defines the kinematic coverage and constraining power of the analysis. Given the selected data, an initial parametrization for the parton distributions is specified at a starting scale $Q_0^2$, usually chosen to be of order $1~\mathrm{GeV}^2$. This parametrization, expressed in terms of a functional form with free parameters, provides the nonperturbative input for each parton flavor. The distributions are then evolved to the relevant kinematic scales using the DGLAP equations.
Theoretical predictions for the experimental observables are subsequently computed using the evolved PDFs, and a global $\chi^2$ function is constructed to quantify the agreement between theory and data. Minimization procedures are applied to determine the best-fit parameter set, and dedicated uncertainty-estimation methods are employed to assess the reliability and precision of the extracted nPDFs.
This methodology has been implemented in several global analyses, including EPPS21 \cite{Eskola:2021nhw}, nCTEQ15HQ \cite{Duwentaster:2022kpv}, nNNPDF3.0 \cite{nNNPDF3.0:2022fyi}, TUJU21 \cite{Helenius:2021tof}, and KSASG20 \cite{Khanpour:2020zyu}. While these analyses have the same underlying structure, they differ in their choices of parametrization, data selection, minimization algorithms, and uncertainty quantification frameworks.
%%%%%%%%%%%%%%%%%
\subsection{Experimental Input and Data Selection}
The selection and treatment of experimental data are critical elements of any global nPDF analysis. The datasets span a wide kinematic range and include a diverse set of processes: fixed-target deep inelastic scattering on nuclear targets, proton–nucleus Drell–Yan processes, electroweak boson production in $pA$ collisions at the LHC, as well as measurements involving heavy-flavor and quarkonium production. These processes provide complementary sensitivity to different parton flavors and probe distinct regions of the parton momentum fraction $x$, thereby enabling constraints on both quark and gluon distributions across the relevant phase space. To ensure consistency with the theoretical framework of collinear factorization and DGLAP evolution, standard kinematic cuts are applied. Measurements taken at low $Q^2$ or low hadronic invariant mass $W^2$, where higher-twist contributions and other nonperturbative effects become non-negligible, are typically excluded from the fits. In addition, careful treatment is required for experimental normalization uncertainties and correlated systematic errors in order to maintain statistical consistency and ensure a robust extraction of the parton distributions. A summary of key observables used in global nPDF analyses, together with their dominant flavor sensitivities and characteristic $x$-coverage, is provided in Table~\ref{tab:nPDF_observables}.
\begin{table}[H]
\centering
\caption[Summary of key observables used in global nPDF  fits]{Summary of key observables used in global nPDF  fits, including their dominant flavor sensitivity and typical kinematic coverage in momentum fraction $x$.}
\begin{tabular}{|p{5.5cm}|p{6.5cm}|p{3.2cm}|}
\hline
\textbf{Observable / Process} & \textbf{Dominant Sensitivity} & \textbf{Typical $x$-range Probed} \\
\hline
Charged-lepton DIS \newline (\(e+A \rightarrow e+X\)) & Valence quarks at large \(x\); sea quarks at intermediate \(x\) & \(x \gtrsim 0.01\) \\
\hline
Neutrino DIS (\(\nu+A\)) & Flavor separation in sea quarks; strange quark sensitivity & \(x \gtrsim 0.01\) \\
\hline
Drell–Yan (\(p+A \rightarrow \mu^+\mu^-+X\)) & Sea quarks & \(0.01 \lesssim x \lesssim 0.2\) \\
\hline
\wz boson production (\(p+A\)) & Sea and valence quarks at intermediate \(x\); some gluon sensitivity & \(0.01 \lesssim x \lesssim 0.2\) \\
\hline
Inclusive jets/dijets (\(p+A\)) & Gluons at mid-to-high \(x\) & \(x \gtrsim 0.05\) \\
\hline
Open heavy flavor (e.g., D/B mesons) & Gluons at small to intermediate \(x\) & \(10^{-5} \lesssim x \lesssim 0.1\) \\
\hline
Quarkonium (e.g., \(J/\psi, \Upsilon\)) & Gluons at small \(x\) & \(x \lesssim 10^{-2}\) \\
%\hline
%Photoproduction / UPCs & Gluons at very small \(x\) & \(x \lesssim 10^{-3}\) \\
\hline
\end{tabular}
\label{tab:nPDF_observables}
\end{table}
%%%%%%%%%%%%%%%%%%%%%%%%%%%

\subsection{Parameterization} \label{sec:3.4.2}
In global analyses of nuclear PDFs, the partonic structure of nuclei is described by extending the standard proton PDF formalism to incorporate nuclear modifications. Under the assumption of isospin symmetry between protons and neutrons, the nuclear PDFs are expressed as a linear combination of the bound-proton and bound-neutron PDFs, as given in Eq. (\ref{eq.3.12}). This decomposition provides the starting point for most global nPDF analyses. In practice, fitting frameworks typically parametrize the PDFs of bound protons at an initial scale $Q_0$, which is usually chosen to be of order $1 \mathrm{GeV}$. The corresponding bound-neutron PDFs are obtained by applying isospin symmetry transformations to the proton input. However, different analyses introduce distinct choices regarding the functional form of the initial parametrization, the manner in which nuclear dependence is implemented, the optimization strategy used in the global fit, and the method adopted for estimating uncertainties. These methodological differences lead to variations among the available nPDF sets, even when they are constrained by similar data.
\begin{itemize}
    \item \textbf{nCTEQ}: The nCTEQ15 global analysis \cite{Kovarik:2015cma}, together with its following extensions (e.g., nCTEQ15WZ \cite{Kusina:2020lyz}, nCTEQ15HQ \cite{Duwentaster:2022kpv}), parameterizes the bound proton PDFs at the initial scale $Q_0=1.3 \mathrm{GeV}$ using functional forms inspired by the CTEQ proton PDF framework.\footnote{Similar functional forms are also employed in the TUJU21 \cite{Helenius:2021tof} and KSASG20 analyses, which likewise adopt CTEQ-like initial parametrizations with nucleus-dependent coefficients.} For each parton flavor $i$, the input distribution is written as
    \begin{equation}
        xf_{i}^{p/A}(x,Q_{0})=c_{0} \, x^{c_{1}^i}(1-x)^{c_{2}^i} \, e^{c_{3}^i x}(1+e^{c_{4}^i} \,x)^{c_{5}^i},
    \end{equation}
    where the coefficients $c_j$ ($j=0, \dots,5$) are free fit parameters and the flavor index $i$ spans the combinations \(u_v,\ d_v,\ g,\ \bar u + \bar d,\ s + \bar s,\ s - \bar s.\). Nuclear dependence is introduced by allowing each coefficient to vary smoothly with the mass number $A$
    \begin{equation}
        c_{j} \to c_{j}(A) \equiv p_{j} + a_{j}(1-A^{-b_{j}}),
    \end{equation}
    where $p_j$ corresponds to the parameters defining the free proton PDFs. In the limit $A=1$, all nuclear modifications vanish, and the distributions reduce to the underlying proton baseline (e.g., CTEQ6 \cite{Pumplin:2002vw}). More recently, the nCTEQ framework has incorporated SRC effects \cite{nCTEQ:2023cpo}. In this formulation, the bound-proton PDF is expressed as a linear combination of a free-proton component and an SRC-induced high-momentum component:
    \begin{equation}
        f_i^{p/A}(x,Q^2)=(1 - C_p^A)\, f_i^{\text{free-proton}}(x,Q^2)\;+\; C_p^A\, f_i^{\text{SRC}}(x,Q^2),
    \end{equation}
    where \(f_i^{\text{free-proton}}\) denotes the baseline proton PDF and \(f_i^{\text{SRC}}\) represents the high-momentum tail associated with nucleons in SRC pairs. The mixing parameter \(C_p^A\) is related to the number of SRC pairs. This hybrid approach improves the description of the EMC region and the high-\(x\) behavior observed in nuclear DIS data, and provides additional physical insight into the extracted nPDFs. 
    \item \textbf{EPPS}: In the EPPS21 analysis \cite{Eskola:2021nhw}, nuclear effects are introduced through multiplicative modification factors applied to a chosen baseline proton PDF
    \begin{equation}
    f_{i}^{p/A}(x, Q^2) = R_{i}^{A}(x, Q^2) \cdot f_{i}^{p}(x, Q^2),
    \end{equation}
    where $f_{i}^{p/A}$ denotes the bound proton PDF and $R_{i}^{A}(x, Q^2)$ represents the nuclear modification factor for flavor \(i\). At the initial scale \(Q_0^2\), the modification factor is parametrized piecewise in three regions of Bjorken \(x\):
    \begin{equation}
    R_i^{p/A}(x, Q_0^2) = 
    \begin{cases}
    a_{0i} + a_{1i} \, \frac{x - x_{ai}}{x_{ai} - a_{2i}} \, e^{-h (x - x_{ai})^2}, & x \leq x_{ai} \\[8pt]
    b_{0i} \, x^{b_{1i}} (1 - x)^{b_{2i}} e^{b_{3i} x}, & x_{ai} < x \leq x_{ei} \\[8pt]
    c_{0i} + c_{1i}(c_{2i} - x)(1 - x)^{-\beta_i}, & x_{ei} < x \leq 1
    \end{cases}
    \end{equation}
    where $x_{ai}$ and $x_{ei}$ are the transition points separating the shadowing, EMC, and Fermi-motion regions, while \(a_i\), \(b_i\), \(c_i\), and \(\beta_i\) are flavor-dependent fit parameters. The nuclear mass-number dependence is modeled through a scaling relation:
    \begin{equation}
    R_i^{p/A}(x, Q_0^2) = 1 + \left( R_i^{p/A_{\text{ref}}}(x, Q_0^2) - 1 \right) \left( \frac{A}{A_{\text{ref}}} \right)^{\gamma_i}, \quad \gamma_i > 0,\quad A_{\text{ref}} = 12,
    \end{equation}
    where $A_{\text{ref}}$ is the reference nucleus and $\gamma_i$ controls the strength of the \(A\)-scaling.
    \item \textbf{nNNPDF}: In the nNNPDF framework, including the recent nNNPDF3.0 global analysis \cite{nNNPDF3.0:2022fyi}, nuclear PDFs are determined using a model-independent approach based on artificial neural networks. Instead of selecting a fixed functional form for the \(x\)-dependence, the PDFs are expressed as flexible functions learned directly from data. At the initial scale \(Q_0^2\), the bound-proton PDFs are parametrized as
    \begin{equation}
    x f_i^{p/A}(x, Q_0^2) = B_i \, x^{\alpha_i} (1 - x)^{\beta_i} \, \mathrm{NN}_i(x, A), \quad i = \Sigma, T_3, T_8, V, V_3, g,
    \end{equation}
    where \(\mathrm{NN}_i(x,A)\) represents the output of a neural network that takes both the momentum fraction $x$ and the atomic mass number $A$ as input features. The terms $x^{\alpha_i} $ and $(1 - x)^{\beta_i}$ are ``preprocessing'' factors that control the asymptotic behavior at small and large $x$, respectively. The PDFs are expressed in the evolution basis consisting of the quark singlet $\Sigma$, the gluon $g$, and non-singlet combinations $V, V_3, T_3$ and $T_8$. In this approach, the \(A\)-dependence is not imposed through explicit scaling assumptions (e.g., power laws in \(A\)); rather, it is learned implicitly from the global dataset, with nucleon-level structure incorporated through convolution with proton PDFs. Consequently, the fit is performed simultaneously across multiple nuclei, making the resulting nPDFs minimally biased by theoretical assumptions about functional forms or nuclear systematics.
\end{itemize}
%%%%%%%%%%%%%%%%%%%%%%%%%%%%%%%%%%%%%%%%%%
\subsection{Minimization and Parameter Optimization} \label{sec:3.4.3}
Once the initial parametrizations of nuclear PDFs are specified at the starting scale $Q_0$ and evolved to the relevant energy scales using the DGLAP equations, global analyses must determine the values of the free parameters that best describe the experimental data. This is accomplished by minimizing a figure-of-merit function—typically the chi-squared ($\chi^2$) statistic—which quantifies the agreement between theoretical predictions and measurements. A general form of the $\chi^2$ function used in global nPDF analyses is
\begin{equation} 
\chi^2(\mathbf{a}) = \sum_{i,j} (D_i - T_i(\mathbf{a})) \, (C^{-1})_{ij} \, (D_j - T_j(\mathbf{a})),
\label{eq:3.22} 
\end{equation}
where $\mathbf{a}=\{a_1, a_2, \dots, a_n\}$ denotes the parameters defining the PDFs at the input scale, $D_i$ are the experimental data points, $T_i(\mathbf{a})$ are the corresponding theoretical predictions, and $C$ is the total covariance matrix incorporating both statistical and systematic uncertainties from the data. The covariance matrix typically takes the form
\begin{equation} 
C_{T,ij}(\mathbf{a}) = \sigma_i^2 \delta_{ij} + \sum_\alpha \bar{\sigma}_{i\alpha} \bar{\sigma}_{j\alpha},
\label{eq.7} 
\end{equation}
where $\sigma_i$ denotes the uncorrelated (statistical) uncertainty of the $i$-th data point and $\bar{\sigma}_{i\alpha}$ represents the contribution from correlated systematic uncertainty source $\alpha$. Normalization uncertainties constitute an additional correlated contribution and will be treated separately in Chap. \ref{chap.6}. After constructing the $\chi^2$ function, the next step is the \emph{minimization} procedure, in which $\chi^2$ is minimized with respect to the parameter vector $\mathbf{a}$ to obtain the best-fit nPDFs. Different global analyses adopt different optimization strategies. 

In the \textbf{nCTEQ} framework, the global fit is performed by adjusting the parameters of the chosen nPDF functional forms to minimize the total $\chi^2$ constructed from the full experimental covariance matrix. The numerical minimization is carried out using a combination of algorithms, primarily the Nelder–Mead \texttt{SIMPLEX} \cite{Nelder:1965zz} and \texttt{BFGS} quasi-Newton \cite{HEAD1985264, Shanno1970ConditioningOQ} methods, which are iteratively alternated to ensure robust convergence in the highly non-linear parameter space characteristic of global nuclear PDF fits.

The \textbf{EPPS} framework follows a structurally similar $\chi^2$-based fitting strategy. However, it does not fit bound-proton PDFs directly. Instead, it parametrizes nuclear modification factors $R_i^{A}(x, Q^2)$ applied to a chosen free-proton baseline PDF set. This separation between baseline and nuclear effects leads to different parameterization choices and constraints, but the underlying optimization logic, deterministic $\chi^2$ minimization followed by Hessian uncertainty estimation, remains broadly aligned with the nCTEQ methodology. In contrast, the \textbf{nNNPDF} approach is different from traditional deterministic minimization. Rather than identifying a single global minimum, it constructs an ensemble of Monte Carlo replicas of the experimental dataset by sampling uncertainties directly \cite{Forte:2002fg}. Each replica is fitted independently using neural-network parametrizations, with minimization carried out by stochastic optimization algorithms, typically variants of stochastic gradient descent, applied to a loss function analogous to a $\chi^2$. The resulting ensemble of independently trained networks encodes the probabilistic distribution of nPDFs implied by the data.

\subsection{Uncertainty estimation}
After determining the best-fit parameters (or, in replica approaches, an ensemble of fitted networks), the next step is to quantify the uncertainties associated with the extracted nPDFs. In standard approaches, these uncertainties reflect only the propagation of experimental statistical and systematic errors under the assumed parametrization and theoretical framework. Two widely used strategies for estimating PDF uncertainties are the \emph{Hessian} \cite{Pumplin:2001ct,Pumplin_2002} and the \emph{Monte Carlo (MC) replica} \cite{Giele:1998gw, Watt:2012tq} methods. The Hessian method, used by nCTEQ and EPPS, approximates the $\chi^2$ function near its minimum by a multivariate quadratic form, enabling the construction of error eigenvector sets that represent statistically independent directions in parameter space. In contrast, the MC replica method, used by nNNPDF, generates many fluctuated versions of the experimental dataset and performs a separate fit for each, thereby sampling the full probability distribution of PDFs without imposing fixed functional forms. Both strategies have advantages: the Hessian method is computationally efficient for parametrizations with relatively few parameters, while the MC replica method offers a flexible and non-parametric approach, particularly well suited to neural-network PDF representations.
%%%%%%%%%%%%%%%%%%%%%%%%
\subsubsection{Hessian Method}
The Hessian approach to estimating parameter uncertainties in a fit begins by approximating the $\chi^2$ function in the vicinity of its global minimum by a multivariate quadratic form. Let $\mathbf{a}_0$ denote the vector of best-fit parameters, at which $ \chi^2_0 \equiv \chi^2(\mathbf{a}_0) = \min\limits_{a}\,\chi^2(\mathbf{a})$. Near the minimum, provided the likelihood surface is smooth and the deviations are sufficiently small, the $\chi^2$ function can be approximated by a quadratic Taylor expansion:
\begin{equation} 
\chi^2(\bm{a}) \approx \chi^2_0 + \sum_{i=1}^n \sum_{j=1}^n H_{ij}(a_i - a_{0,i})(a_j - a_{0,j}), 
\label{eq:hessian-expansion} 
\end{equation} 
where $n$ is the total number of independent fit parameters and the coefficients $H_{ij}$ form the Hessian matrix $\mathbf{H}$, defined by the second-order partial derivatives of the $\chi^2$ function, evaluated at the minimum $\bm{a}_0$
\begin{equation} 
H_{ij} = \left. \frac{1}{2} \frac{\partial^2 \chi^2}{\partial a_i \partial a_j} \right|_{\bm{a} = \mathbf{a}_0}. 
\end{equation} 
The Hessian matrix encodes the curvature of the $\chi^2$ hypersurface around the minimum and captures both the sensitivity of each parameter and their mutual correlations.\footnote{In practice, the Hessian matrix is typically evaluated numerically. Fitting frameworks, such as the standard package \texttt{MINUIT} \cite{James:Minuit}, provide automated procedures (e.g., the HESSE algorithm) to compute the required second-order partial derivatives through finite-difference techniques.} To make practical use of the Hessian, one diagonalizes it:
\begin{equation}
\bm{H} = \bm{V} \bm{\Lambda} \bm{V}^T
\end{equation}
where $\bm{V}$ is the orthogonal matrix whose columns are the eigenvectors ${\bm{v}^{(k)}}$ and $\bm{\Lambda}$ is the diagonal matrix of eigenvalues ${\lambda_k}$. The eigenvectors identify $n$ independent directions in parameter space, and the eigenvalues quantify the curvature along those directions. Specifically, directions with large eigenvalues are tightly constrained by data, while small eigenvalues correspond to weakly constrained or nearly flat directions. By transforming the coordinate system into the Hessian eigenvector basis, the parameter displacements can be expressed in terms of the uncorrelated error variables $z_k$
\begin{equation}
    z_k=\sum_{i} V_{ik} \, \frac{a_i -a_{0,i}}{\sqrt{\lambda_k}}.
\end{equation}
In this transformed coordinate system, the quadratic expansion of the $\chi^2$ function simplifies to its canonical form:
\begin{equation}
\chi^2(\bm{a}) \approx \chi^2_0 + \sum_{k=1}^{n} z_k^2.
\end{equation}
This canonical form establishes a direct correspondence between movements in the eigenvector directions and increases in $\chi^2$. The set of error PDFs is generated by defining displacements $t_k$ along each of these independent directions, which are governed by a tolerance criterion
\begin{equation}
t_k = \pm T_k,
\end{equation}
where $T_k$ is the tolerance factor associated with the $k$-th eigenvector direction. The tolerance values are chosen such that the total increase in $\chi^2$ from the best-fit point $\mathbf{a}_0$ satisfies:
\begin{equation}
\Delta \chi^2 = \sum_{k=1}^{n} t_k^2.
\end{equation}
The quantity $\Delta \chi^2$ defines the permitted rise in $\chi^2$ above its minimum $\chi^2_0$, corresponding to a chosen confidence level. For perfectly consistent data with purely Gaussian uncertainties, $\Delta \chi^2 = 1$ yields a 68\% confidence interval for a single parameter, as follows from Wilks’ theorem~\cite{Wilks:1938dza}. In realistic global PDF fits, however, datasets often exhibit systematic tensions, correlated uncertainties, and theoretical inconsistencies that violate these ideal assumptions. As a result, larger and sometimes direction-dependent tolerances are required to obtain uncertainty estimates that faithfully reflect the constraints imposed by the data. There is no unique prescription for defining the tolerance, and different groups adopt phenomenologically motivated choices tailored to their datasets and fitting strategies. The CT18 analysis \cite{Hou:2019efy} adopts a global tolerance approach, in which a \emph{fixed} $\Delta \chi^2$ is calibrated to correspond to a chosen confidence level (e.g., $\Delta \chi^2 \approx 35$ for 68\% C.L. and $\Delta \chi^2 = 100$ for 90\% C.L.), thereby accommodating tensions among diverse experimental datasets. Following a similar philosophy, the nCTEQ15 analysis \cite{Kovarik:2015cma} also employs a fixed global tolerance, typically $\Delta \chi^2 = 35$.
In contrast, the MMHT \cite{Harland-Lang:2019pla}, MSTW \cite{Martin:2009iq}, and MSHT \cite{Bailey:2020ooq} analyses use a dynamic tolerance procedure, in which each eigenvector direction in parameter space is assigned its own tolerance $T_k$. These tolerances are determined by requiring that all datasets remain within their specified confidence limits along each direction.

Using the eigenvector basis, PDF error sets are constructed by displacing the parameters from the minimum along eigendirection $k$ by an amount:
\begin{equation}
\Delta \mathbf{a}^{(k)} = \pm T_k \sqrt{\frac{\Delta \chi^2}{\lambda_k}} \bm{v}^{(k)},
\end{equation}
yielding two error sets $S_k^{\pm}$ for each direction $k$. For $n$ fitted parameters, the full Hessian error ensemble consists of $2n$ error sets plus the central (best-fit) value. 
To propagate these uncertainties to a physical observable $\mathcal{O}$, one evaluates the observable using each of the $S_k^\pm$ sets. The symmetric Hessian uncertainty is given by
\begin{equation}
\Delta \mathcal{O} = \frac{1}{2} \sqrt{ \sum_{k=1}^{n} \left[ \mathcal{O}(S_k^+) - \mathcal{O}(S_k^-) \right]^2 }.
\end{equation}
This estimate is appropriate when deviations from the central prediction are approximately Gaussian and symmetric. If the underlying $\chi^2$ surface is asymmetric or the observable depends nonlinearly on the parameters, an asymmetric prescription is preferred:
\begin{equation}
\Delta \mathcal{O}^{\pm} = \sqrt{ \sum_{k=1}^{n} \left[ \max \left(0, \pm \left( \mathcal{O}(S_k^\pm) - \mathcal{O}_0 \right) \right) \right]^2 },
\end{equation}
where $\mathcal{O}_0 = \mathcal{O}(\mathbf{a}_0)$ is the central prediction. This approach captures situations in which positive and negative displacements have different impacts on the observable.

While the Hessian method is computationally efficient and has been widely used in global PDFs and nPDFs analyses, it also has several limitations that may impact the reliability of the resulting uncertainty estimates:
\begin{itemize}
    \item Assumption of Gaussianity: The method relies on a quadratic expansion of the $\chi^2$ function around its minimum, effectively assuming that the parameter uncertainties follow a multivariate Gaussian distribution. In practice, this assumption can fail, especially in poorly constrained directions of parameter space, leading to underestimated or inaccurate uncertainty bands.
    \item Dependence on Tolerance Choice: The uncertainty estimates are sensitive to the chosen tolerance \(\Delta\chi^{2}\), which is not fixed by first principles and must be specified empirically. A tolerance that is too small fails to account for dataset tensions or systematic inconsistencies, while a tolerance that is too large yields overly conservative uncertainties.
    \item Sensitivity to Local Minima: The Hessian approach assumes that the expansion is performed around the global minimum of the \(\chi^{2}\) surface. In high-dimensional parameter spaces, however, multiple local minima may exist. If the optimization converges to a local rather than the global minimum, the resulting uncertainties reflect only the local curvature of $\chi^2$ and may fail to capture the full structure of the parameter space, thereby misrepresenting the true uncertainties.
\end{itemize}
\subsubsection{Monte Carlo Method}
An alternative to the Hessian approach for estimating PDF uncertainties is the Monte Carlo replica method, which is particularly well suited to flexible, non-parametric fitting strategies. Unlike the Hessian method, whose accuracy relies on the assumption that the $\chi^2$ surface is approximately quadratic near its minimum, the MC method makes no assumptions about the linearity or Gaussian behavior in parameter space. Instead, it samples directly from the full probability distribution implied by the data, allowing it to naturally capture non-linear correlations, multi-modal structures, and non-Gaussian uncertainty profiles. The main idea is to generate a large ensemble of statistically equivalent pseudo-datasets, each obtained by fluctuating the experimental measurements within their uncertainties. Each replica is then treated as an independent dataset and refitted, yielding an ensemble of PDF replicas that approximates the underlying probability distribution.

Formally, for a given measurement $D_i$ with statistical uncertainty $\sigma_i^{\text{stat}}$ and correlated systematic uncertainties $\sigma_{i\alpha}^{\text{sys}}$, the $r$-th replica is generated as
\begin{equation}
    D_i(r) = D_i + r_i^{\text{stat}} \, \sigma_i^{\text{stat}} + \sum_{\alpha} r_{i\alpha}^{\text{sys}} \, \sigma_{i\alpha}^{\text{sys}},
\end{equation}
where the random variables $r_i^{\text{stat}}$ and $r_{i\alpha}^{\text{sys}}$ are drawn from a standard normal distribution. This construction ensures that both uncorrelated and correlated experimental uncertainties are correctly propagated, preserving the full covariance structure of the original data. After generating $N_{\text{rep}}$ such replicas, typically in the range of 100–1000, each dataset is independently fitted, producing a corresponding set of PDF replicas ${f_i^{(r)}(x, Q^2)}$. The resulting ensemble represents an empirical sampling of the probability distribution for the PDFs. For any observable $\mathcal{O}$ dependent on the PDFs, the MC estimates of the central value and associated uncertainty are then given by
\begin{equation}
    \langle \mathcal{O} \rangle = \frac{1}{N_{\text{rep}}} \sum_{r=1}^{N_{\text{rep}}} \mathcal{O}\big(f_i^{(r)}\big),
\end{equation}
\begin{equation}
    \Delta \mathcal{O} = \sqrt{ \frac{1}{N_{\text{rep}}} \sum_{r=1}^{N_{\text{rep}}} \left( \mathcal{O}\big(f_i^{(r)}\big) - \langle \mathcal{O} \rangle \right)^2 }.
\end{equation}

This method offers several strengths. It automatically incorporates parameter correlations and fully accounts for nonlinear dependencies of observables on the PDFs. It is particularly robust in extrapolation regions, such as very small-$x$ or large-$x$, where data constraints are weak and theoretical uncertainties are substantial. Moreover, the MC framework easily accommodates highly flexible parametrizations, including neural networks, thereby avoiding biases associated with specific functional forms and enabling the fit to capture complex features in the data. Despite these advantages, the MC replica approach also faces both conceptual and practical challenges.

Recent analyses \cite{Costantini:2024wby} have shown that the ensemble of replicas does not necessarily correspond to a statistically consistent posterior distribution, especially when theoretical predictions depend nonlinearly on the fit parameters. A detailed study in this paper demonstrates that the replica ensemble reproduces the correct posterior only in the special case of linear models with Gaussian errors. In realistic PDF and nPDF fits, the Monte Carlo procedure may therefore deviate from the true parameter distribution, potentially leading to a systematic underestimation of uncertainties, particularly in extrapolation regions. These findings indicate that, while flexible, the MC method does not in general provide a mathematically controlled sampling of the underlying probability distribution. Furthermore, the statistical interpretation of MC ensembles is only valid under restrictive assumptions, and therefore the resulting uncertainty bands cannot automatically be regarded as confidence intervals in a strict probabilistic sense. On a practical level, the MC methodology is also significantly more computationally demanding than the Hessian approach. Each replica requires an independent fit, often hundreds to thousands of them, significantly increasing both the cost of the analysis and the computational resources needed for further phenomenological applications. This makes the method powerful but resource-intensive, especially when using flexible parametrizations such as neural networks. In such approaches, the training procedure typically requires splitting the available data into training and validation subsets, which can be particularly limiting in the context of nPDF fits where the dataset is relatively sparse, effectively reducing the amount of data directly constraining the fit.
%!TEX root = ../thesis.tex
%*******************************************************************************
%****************************** Third Chapter *********************************
%*******************************************************************************

\chapter{Bayesian Inference } \label{chap.4}
In the previous chapter, global analyses of nuclear parton distribution functions were formulated within a deterministic framework based on $\chi^2$ minimization, in which a single best-fit parameter set is obtained and uncertainties are estimated using methods such as the Hessian approach or Monte Carlo replicas. While these techniques efficiently propagate experimental uncertainties, they rely on approximations, most notably the quadratic expansion of the $\chi^2$ function in the Hessian method, and do not provide a complete probabilistic description of the parameter space. In particular, the Hessian approach assumes near-Gaussian behavior around the minimum, while replica methods infer parameter uncertainties indirectly through data fluctuations.
Bayesian inference provides a conceptually different and statistically consistent framework for parameter estimation. In this approach, model parameters are treated as random variables, and their probability distributions represent the degree of belief conditioned on available information. Bayes’ theorem combines prior knowledge with the likelihood derived from experimental data to construct a posterior probability distribution that encodes uncertainties, correlations, and possible degeneracies in the parameter space. Unlike traditional $\chi^2$-based methods, this framework does not rely on quadratic approximations or tolerance criteria and instead provides a direct probabilistic interpretation of the inferred parameters.

This chapter introduces the principles of Bayesian inference and their application to global nPDF analyses. Section~\ref{sec:4.1} presents the foundations of Bayesian inference and the interpretation of probability distributions in the context of parameter estimation. Section~\ref{sec:4.2} reviews the statistical description of experimental uncertainties and the role of probability distributions and covariance matrices. Section~\ref{sec:4.3} discusses the construction of likelihood functions for both uncorrelated and correlated datasets. Section~\ref{sec:4.4} describes parameter estimation within the Bayesian framework, including the extraction of summary statistics and credible intervals. Finally, Section~\ref{sec:4.5} formulates the global nuclear PDF fit in a Bayesian setting, in which the parameters defining the nPDFs are determined directly from their posterior distribution. The material presented in this chapter provides the statistical foundation for the Markov Chain Monte Carlo methods introduced in the following chapter.
%%%%%%%%%%%%%%%%%%%%%%%%%%%%%%%%%%%%%%%%%%%%%%%%%%%%%%%%%%%%%%%
\section{Principles of Bayesian Inference}\label{sec:4.1}
\emph{Bayesian inference} \cite{gelman2013bayesian, Gregory_2005} provides a probabilistic framework for data analysis in which both observed quantities and unknown parameters are described by probability distributions. In contrast to the frequentist perspective, where probability is interpreted strictly as a long-run frequency, Bayesian probability quantifies uncertainty about any unknown quantity. This includes not only inherently random outcomes but also fixed yet unobserved parameters and future predictions. Within this framework, unknown parameters are treated as uncertain and assigned probability distributions that represent our state of knowledge prior to observing the data. \emph{Bayes’ theorem} then provides the mechanism for updating this prior information with empirical evidence through the likelihood function, yielding a \emph{posterior distribution} that encodes all current information about the parameters, including their uncertainties and correlations. This approach yields not a single point estimate, but a full probabilistic characterization of the parameter space. Conceptually, Bayesian data analysis can be viewed as an iterative process with three main steps. First, one constructs a probability model that includes all relevant components—data, parameters, and hidden variables—and reflects prior knowledge as well as the structure of the measurement process. Second, this model is updated in light of the observed data to obtain the posterior distribution, which expresses the current state of knowledge about the unknown quantities. Third, the model is evaluated by assessing how well it explains the data, the plausibility and scientific meaning of its conclusions, and the sensitivity of those conclusions to modeling assumptions. Based on this evaluation, the model may be refined or extended, and the cycle repeated to improve the analysis.

\subsection*{Bayes' Theorem}
Bayesian inference is fundamentally based on \emph{Bayes' theorem}, which provides the mathematical framework for updating prior knowledge about unknown parameters using observed data. Let $\bm{\theta}$ denote the vector of unknown model parameters and $\bm{D}$ the observed data. In the Bayesian framework, $\bm{\theta}$ is treated as a random variable whose uncertainty is described by probability distributions. Bayes’ theorem expresses how the prior distribution is revised once data are observed:
\begin{equation}
    P( \bm{\theta}| \mathbf{D}) =  \frac{ P( \bm{\theta}) \cdot P(\mathbf{D} | \bm{\theta})}{P(\mathbf{D})}.
    \label{Bayes' formula}
\end{equation}
Here, $P( \bm{\theta})$ is the \emph{prior distribution}, representing knowledge or assumptions about the parameters before observing the data. $P(\mathbf{D} | \bm{\theta})$ is the \emph{likelihood}, describing how probable the observed data are for a given choice of parameters. $P( \bm{\theta}| \mathbf{D})$ is the \emph{posterior distribution}, which encodes the updated state of knowledge after incorporating the information provided by the data. $P(\mathbf{D})$ is the \emph{marginal likelihood} or \emph{evidence}, defined as
\begin{equation}
    P(\mathbf{D})=\int P(\boldsymbol{\bm{\bm{\theta}}}) \cdot P(\mathbf{D} \mid \boldsymbol{\bm{\theta}}) \,d\boldsymbol{\bm{\theta}}.
\end{equation}
The evidence ensures that the posterior is properly normalized and also plays a key role in Bayesian model comparison through Bayes factors. In practice, evaluating the marginal likelihood is often challenging. For models with many parameters or non-conjugate priors,\footnote{A prior is called conjugate if the resulting posterior distribution belongs to the same family as the prior. Conjugate priors simplify computation but are not always available or appropriate for complex models.} the evidence integral is analytically intractable\footnote{An integral is analytically intractable when no closed-form expression exists and numerical approximation is required.} and computationally expensive to approximate. For this reason, many Bayesian analyses work with the \emph{unnormalized posterior},
\begin{equation}
    P( \bm{\theta}|\mathbf{D}) \propto P( \bm{\theta}) \cdot P(\mathbf{D} | \bm{\theta}),
\end{equation}
which is sufficient for inference tasks such as locating the maximum a posteriori (MAP) estimate.  The MAP estimate identifies the parameter values that make the posterior distribution largest—that is, the values that are most plausible after combining prior knowledge with the information contained in the data. Since the evidence does not depend on $\bm{\theta}$, it does not affect the location of the posterior maximum. The proportional form is also crucial for computational techniques such as Markov Chain Monte Carlo (MCMC). These methods explore the posterior distribution by comparing relative posterior values, rather than requiring the normalization constant. As a result, they enable efficient Bayesian inference even for complex models where the evidence term is impossible to compute directly.

The prior distribution, $P(\bm{\theta})$, encapsulates information available before the data are observed. Priors may incorporate results from previous experiments, theoretical constraints, or expert judgment. They may also be chosen to be weakly informative, designed to minimize their influence on the posterior. The choice of prior is an important part of Bayesian modeling and should be considered carefully, especially in situations with limited data.

The likelihood $P(\mathbf{D} | \bm{\theta} )$ quantifies how compatible the observed data are with a given choice of parameters. It specifies, for each possible value of $\bm{\theta}$, the probability that the model assigns to observing the dataset $\bm{D}$. If the data points $\{D_1, D_2, \dots, D_n \}$ are assumed to be conditionally independent given $\bm{\theta}$, the likelihood factorizes:
\begin{equation}
    P( \mathbf{D} | \bm{\theta}) = \prod_{i=1}^n P(D_i |\bm{\theta} ).
\end{equation}
For example, if each observation $D_i$ is modeled as a Gaussian with known variance $\sigma^2_i$ and mean prediction $\mu_i(\bm{\theta})$, then
\begin{equation}
    P(D_i | \bm{\theta}) = \frac{1}{\sqrt{2\pi \sigma_i^2}} \exp\left(-\frac{(D_i - \mu_i(\bm{\theta}))^2}{2\sigma_i^2} \right).
\end{equation}
Gaussian likelihoods are widely used when modeling measurement errors or continuous-valued data, particularly when the uncertainties are well approximated by normal (Gaussian) fluctuations around a predicted mean. More general likelihood formulations—particularly for correlated data—will be discussed in Sec.~\ref{sec:likelihood}.

In Bayesian inference, the posterior distribution captures all available information, including the prior knowledge and the data evidence, and provides a full probabilistic characterization of uncertainty in the model parameters. Several types of summaries can be extracted from the posterior, depending on the inferential objective:
\begin{itemize}
    \item Point estimation: Representative parameter values can be obtained from the posterior using summary statistics such as the mean, median, or mode. A common choice is the posterior mean,
    \begin{equation*}
        \mathbb{E}[\bm{\theta}] = \int \bm{\theta} \, P(\bm{\theta} \mid \mathbf{D}) \, d\bm{\theta}.
    \end{equation*}
   These summaries provide single-valued estimates while still being grounded in the full posterior distribution.
    \item Interval estimation: Bayesian credible intervals quantify uncertainty by identifying ranges of parameter values with a specified proportion of the posterior probability. A 95\% credible interval \([\bm{\theta}_L, \bm{\theta}_U]\) is defined such that
    \begin{equation*}
        \int_{\bm{\theta}_L}^{\bm{\theta}_U} P(\bm{\theta} \mid \mathbf{D}) \, d\bm{\theta} = 0.95,
    \end{equation*}
    meaning that, conditional on the model and data, the parameter lies in this interval with probability 95\%. Unlike frequentist confidence intervals, credible intervals enable a direct probabilistic interpretation.
    \item Prediction: Bayesian inference naturally extends to predicting new or unobserved data. The posterior predictive distribution integrates over parameter uncertainty:
    \begin{equation*}
        P(\tilde y \mid \mathbf{D}) = \int P(\tilde y \mid \bm{\theta})\,P(\bm{\theta} \mid \mathbf{D})\,d\bm{\theta},
    \end{equation*}
    which yields a principled prediction for a future observation \(\tilde y\). This accounts not only for the inherent randomness in the data-generating process but also for uncertainty in the parameter estimates themselves.
\end{itemize}
%%%%%%%%%%%%%%%%%%%%%%%%%%%%%%%%%%%%%%%%%%%%%%%%%%%%%%%%%%%%%%%%%%%%%5
\section{Probability Distributions and Statistical Foundations}\label{sec:4.2}
A rigorous treatment of probability distributions is essential for Bayesian modeling and statistical inference. This section introduces the fundamental building blocks used to describe uncertainty: probability mass functions for discrete outcomes, probability density functions for continuous variables, and the associated concepts of joint, marginal, and conditional distributions. We then discuss independence, which allows simplification of complex probabilistic structures through factorization, and conclude with two foundational limit theorems—the Law of Large Numbers and the Central Limit Theorem—that underpin many analytical approximations and justify the behavior of Monte Carlo methods used later in this chapter.

\subsection*{Probability Mass and Density Functions}
Whether probability is assigned to specific outcomes or to intervals depends on the nature of the random variable. Two fundamental constructs are used for this purpose: the \emph{probability mass function} (PMF) for discrete variables and the \emph{probability density function} (PDF) for continuous ones.
\begin{itemize}
    \item \textbf{Probability Mass Function (PMF):} For a discrete random variable $X$, the PMF $p_X(x)$ specifies the probability that $X$ takes the value $x$:
    \begin{equation}
        P(X = x) = p_X(x).\footnote{In this notation, $X$ is a random variable and $x$ is a possible outcome. $P$ denotes the probability of an events, and $p_X(x)$ is the function assigning probabilities to discrete outcomes.}
    \end{equation}
    The PMF satisfies $p_X(x) \geq 0$ for all $x$, and $\displaystyle \sum_{x} p_{X}(x) = 1$. It therefore provides an exact probability for $X$.
    \item \textbf{Probability Density Function (PDF):}  
    For a continuous random variable \(X\), probabilities are assigned to intervals rather than individual points. Because continuous variables take values on an uncountable set, the probability of any exact value is zero:
    \begin{equation}
       P(X = x) = 0.
    \end{equation}
    Instead, probabilities are computed using the PDF \(f_X(x)\) via
    \begin{equation}
      P(a \leq X \leq b) = \int_a^b f_X(x)\,dx,
    \end{equation}
    with \(f_X(x) \ge 0\) and normalization
    \begin{equation}
        \int_{-\infty}^{\infty} f_X(x)\,dx = 1.
    \end{equation}
    The PDF therefore describes how probability is distributed over the continuous range of outcomes.
\end{itemize}

\subsection*{Joint, Marginal, and Conditional Distributions; Independence}
Once a random variable is identified as discrete or continuous, these concepts naturally extend to collections of variables through their \emph{joint}, \emph{marginal}, and \emph{conditional} distributions. These constructions describe how multiple variables relate to one another and form the foundation for expressing dependence and independence.

\begin{itemize}
    \item \textbf{Joint Distribution:} The joint probability distribution of two (or more) random variables describes the probability of observing their outcomes simultaneously.
    
    For discrete variables, the joint PMF is:
    \begin{equation}
        P(X = x, Y = y) = p_{X,Y}(x, y).
    \end{equation}

    For continuous variables, the joint PDF $f_{X,Y}(x, y)$ characterizes the density of probability over \(\mathbb{R}^2\). The probability that the pair \((X, Y)\) lies in a region \(A \subset \mathbb{R}^2\) is given by
    \begin{equation}
        P\left((X, Y) \in A\right) = \iint_A f_{X,Y}(x, y) \, dx \, dy.
    \end{equation}

    \item \textbf{Marginal Distribution:} The marginal distribution describes the behavior of a single variable regardless of the values taken by the others. It is obtained by summing or integrating the joint distribution over the unwanted variable.

    For discrete variables:
    \begin{equation}
        p_X(x) = \sum_y p_{X,Y}(x, y).
    \end{equation}

    For continuous variables:
    \begin{equation}
        f_X(x) = \int_{-\infty}^{\infty} f_{X,Y}(x, y) \, dy.
    \end{equation}

    \item \textbf{Conditional Distribution:} A conditional distribution quantifies the probability of one variable given that another has been observed.

    For discrete variables:
    \begin{equation}
        P(X = x \mid Y = y) = \frac{p_{X,Y}(x, y)}{p_Y(y)} \quad \text{if } p_Y(y) > 0.
    \end{equation}

    For continuous variables:
    \begin{equation}
        f_{X \mid Y}(x \mid y) = \frac{f_{X,Y}(x, y)}{f_Y(y)} \quad \text{if } f_Y(y) > 0.
    \end{equation}
\end{itemize}

\subsection*{Independence and Factorization}
With joint, marginal, and conditional distributions defined, we can formally express the notion of independence between random variables. Intuitively, two variables are independent if knowing the value of one provides no information about the other.
\paragraph{Definition.}
Two random variables \( X \) and \( Y \) are \emph{independent} if their joint distribution factorizes into the product of their marginals:
\begin{equation}
    P(X = x, Y = y) = P(X = x) \cdot P(Y = y) \quad \text{for all } x, y.
\end{equation}
Equivalently, independence holds if the conditional distribution of one variable reduces to its marginal:
\begin{equation}
    P(X = x \mid Y = y) = P(X = x) \quad \text{for all } x, y \,\,\text{ with } \, P(Y = y) > 0.
\end{equation}
The same definitions apply in the continuous case using probability densities:
\begin{equation}
    f_{X,Y}(x, y) = f_X(x) \cdot f_Y(y) \quad \Leftrightarrow \quad f_{X \mid Y}(x \mid y) = f_X(x),
\end{equation}
for all \( x, y \) with \( f_Y(y) > 0 \).

Independence assumptions play a central role in probabilistic modeling. A common example arises in statistical inference, where individual measurements \( D_1, \dots, D_n \) are assumed to be conditionally independent given a parameter \( \theta \). Under this assumption, the likelihood takes the factorized form of
\begin{equation}
    P(\mathbf{D} \mid \theta) = \prod_{i=1}^n P(D_i \mid \theta),
\end{equation}
a decomposition that greatly simplifies analytical calculations and enables efficient sampling-based methods such as Markov Chain Monte Carlo.

\subsection*{Limit Theorems}
Understanding the behavior of sample-based quantities as the number of observations grows is essential for both analytical calculations and simulation-driven methods in Bayesian inference. Two fundamental theorems—the Law of Large Numbers and the Central Limit Theorem (see Sec.~1.6 of \cite{givens2012computational})—provide the theoretical basis for approximating expectations, characterizing estimator variability, and justifying the use of Monte Carlo methods throughout this chapter.

\paragraph{Law of Large Numbers (LLN).}  
Let \( X_1, X_2, \dots, X_n \) be independent and identically distributed (i.i.d.) random variables with finite mean \( \mu = \mathbb{E}[X_i] \). Their sample mean,
\begin{equation}
    \overline{X}_n = \frac{1}{n} \sum_{i=1}^n X_i,
\end{equation}
converges almost surely\footnote{An event occurs \emph{almost surely} (a.s.) if it happens with probability 1. A sequence of random variables \( X_n \) converges almost surely to \( X \) if \( P\left( \lim_{n \to \infty} X_n = X \right) = 1 \). This is stronger than convergence in probability, which only requires that the chance of large deviations goes to zero.} to the true mean \( \mu \) as the sample size increases:
\begin{equation}
    \overline{X}_n \xrightarrow{\;a.s.\;} \mu \quad \text{as } \, \, n \to \infty.
\end{equation}
The LLN ensures that empirical averages approximate population means in the long run. In Bayesian computation, especially in posterior expectation calculations and simulation-based methods such as MCMC, the LLN justifies the replacement of theoretical expectations by sample averages obtained from sufficiently long Monte Carlo runs \cite{brooks2011handbook}.

\paragraph{Central Limit Theorem (CLT).}  
While the LLN guarantees convergence of the sample mean, the CLT characterizes the distribution of its fluctuations for large but finite \( n \). Under the i.i.d. assumption and finite variance \( \sigma^2 = \mathrm{Var}(X_i) \), the Central Limit Theorem states:
\begin{equation}
    \frac{\sum_{i=1}^{n} (X_i - \mu)}{\sigma \sqrt{n}}
    \;\xrightarrow{\;d\;} \; \mathcal{N}(0, 1) \quad \text{as } \, n \to \infty,\footnote{Convergence in distribution (denoted \( \xrightarrow{d} \)) means that the cumulative distribution functions of a sequence of random variables converge pointwise to the cumulative distribution function of a limiting random variable at all continuity points. Unlike almost sure or probabilistic convergence, it does not imply convergence of individual sample, only that the overall distributional shape approaches a limiting form.}
\end{equation}
or equivalently, the distribution of the sample mean approaches a normal distribution:
\begin{equation}
    \overline{X}_n \;\approx\; \mathcal{N}\!\left(\mu, \tfrac{\sigma^2}{n} \right)
    \quad \text{for large } n.
\end{equation}
In both frequentist and Bayesian contexts, the CLT justifies approximating the sampling distribution of an estimator by a Gaussian centered at the maximum likelihood (or maximum a posteriori) estimate. It also explains why normal-based credible intervals provide good approximations when the posterior is sufficiently concentrated and approximately symmetric (see Chapters 4 and 5 of \cite{gelman2013bayesian}).

%%%%%%%%%%%%%%%%%%%%%%%%%%%%%%%%%%%%%%%%%%%%%%%%%%%%%%%%%%%%%%%%%%%%%%%
\section{Likelihood Function} \label{sec:likelihood}\label{sec:4.3}
In statistical modeling, the likelihood function quantifies how well a model explains the observed data given a particular choice of parameters. It encodes the assumptions about the data-generating process and provides the foundation for both frequentist (e.g., maximum likelihood estimation) and Bayesian inference (via Bayes’ theorem). A standard simplifying assumption is that measurement errors are independent across observations. Under this assumption, the likelihood factorizes into a product over individual data points, leading to a widely used formulation. However, independence is often violated in practice. Many experimental datasets, especially in high-energy and nuclear physics, contain non-negligible correlations arising from shared systematic uncertainties, calibration effects, detector responses, or normalization shifts. In such cases, a factorized likelihood is no longer appropriate, and a more general formulation is required. This section develops the likelihood function in both scenarios. We begin by deriving the likelihood under independent Gaussian errors and then extend the formulation to correlated data using the multivariate normal distribution and its associated covariance matrix. Finally, we introduce the log-likelihood, which is algebraically simpler, numerically more stable, and forms the basis for practical parameter estimation.

\subsection{Likelihood Function for Uncorrelated Data}
Let \( \{D_k\}_{k=1}^N \) denote a set of observed data points and \( \{T_k(\bm{\theta})\}_{k=1}^N \) the corresponding model predictions, where \( \bm{\theta} \) is the vector of model parameters. Assume that each measurement \( D_k \) is affected by an independent Gaussian error with known \( \sigma_k \) standard deviation. Under this independence assumption, the likelihood of the full dataset factorizes into a product of single-point Gaussian likelihoods:
\begin{equation}
P(\mathbf{D} \mid \bm{\theta}) = \prod_{k=1}^{N} \frac{1}{\sqrt{2\pi \sigma_k^2}} \exp\left( -\frac{(T_k(\bm{\theta}) - D_k)^2}{2\sigma_k^2} \right).
\end{equation}
It is convenient to express this in terms of the uncorrelated chi-squared statistic,
\begin{equation}
\chi^2_{\mathrm{u}}(\bm{\theta}) := \sum_{k=1}^{N} \frac{(T_k(\bm{\theta}) - D_k)^2}{\sigma_k^2},
\end{equation}
so that the likelihood takes the compact form
\begin{equation}
P(\mathbf{D} \mid \theta) \propto \exp\left( -\frac{1}{2} \chi^2_{\mathrm{u}}(\theta) \right),
\end{equation}
with the full normalization factor given by
\begin{equation}
\prod_{k=1}^{N} \frac{1}{\sqrt{2\pi \sigma_k^2}}.
\end{equation}
This formulation is appropriate when measurement errors are uncorrelated across data points. In many realistic datasets, however, experimental systematics induce correlations among measurements, requiring a more general multivariate likelihood, which we develop in the next subsection.

\subsection{Likelihood Function for Correlated Data}\label{sec:4.3.2}
To construct a likelihood function that incorporates correlations among measurements, we begin with probability density of an $N$-dimensional standard normal random vector \( \mathbf{Z} \sim \mathcal{N}(\mathbf{0}, \mathbf{I}) \), where \( \mathbf{I} \) denotes the \( N \times N \) identity matrix. In this distribution, each component \( Z_i \) is independently distributed as \( \mathcal{N}(0, 1) \), and all off-diagonal covariances vanish i.e., \( \mathrm{Cov}(Z_i, Z_j) = 0 \) for \( i \ne j \). The corresponding density is
\begin{equation}
f_{\mathbf{Z}}(\mathbf{z}) = \frac{1}{(2\pi)^{N/2}} \exp\left( -\frac{1}{2} \mathbf{z}^\top \mathbf{z} \right).
\end{equation}
To model correlated data, we introduce an affine transformation\footnote{An affine transformation consists a linear transformation followed by a translation. In vector form, it has the structure \( \mathbf{y} = \mathbf{A} \mathbf{x} + \mathbf{b} \), where \( \mathbf{A} \) is a matrix and \( \mathbf{b} \) a constant vector. Affine transformations preserve straight lines and parallelism but not necessarily distances or angles.} 
\[
\mathbf{D} = \bm{\mu} + \mathbf{A} \mathbf{Z},
\]
where \( \bm{\mu} \in \mathbb{R}^N \) is the mean vector (typically representing model predictions) and \( \mathbf{A} \in \mathbb{R}^{N \times N} \) is an invertible matrix that introduces the desired pattern of correlations. The covariance matrix of \( \mathbf{D} \) is then:
\[
\mathbf{C} = \mathrm{Cov}[\mathbf{D}] = \mathbf{A} \mathbf{A}^\top.
\]
Affine transformations preserve Gaussian structure. Therefore,
\[
\mathbf{D} \sim \mathcal{N}(\boldsymbol{\mu}, \mathbf{C}),
\]
with mean and covariance given by
\begin{equation}
\begin{aligned}
\mathbb{E}[\mathbf{D}] &= \mathbb{E}[\boldsymbol{\mu} + \mathbf{A} \mathbf{Z}] = \boldsymbol{\mu} + \mathbf{A} \cdot \mathbb{E}[\mathbf{Z}] = \boldsymbol{\mu}, \\[6pt]
\mathrm{Cov}[\mathbf{D}] &= \mathrm{Cov}[\mathbf{A} \mathbf{Z}] = \mathbf{A} \cdot \mathrm{Cov}[\mathbf{Z}] \cdot \mathbf{A}^\top = \mathbf{A} \mathbf{I} \mathbf{A}^\top = \mathbf{A} \mathbf{A}^\top = \mathbf{C}.
\end{aligned}
\end{equation}
To derive the probability density of $\mathbf{D}$, we apply the multivariate change-of-variables formula. The inverse transformation of \( \mathbf{D} = \boldsymbol{\mu} + \mathbf{A} \mathbf{Z} \) is
\[
\mathbf{z} = \mathbf{A}^{-1}(\mathbf{d} - \boldsymbol{\mu}),
\]
where \( \mathbf{d} \) is a realization of \( \mathbf{D} \). The change-of-variables rule gives
\begin{equation}
f_{\mathbf{D}}(\mathbf{d}) = f_{\mathbf{Z}}(\mathbf{A}^{-1}(\mathbf{d} - \boldsymbol{\mu})) \cdot \left| \det(\mathbf{A}^{-1}) \right|.
\end{equation}
Substituting the standard multivariate normal density for \( f_{\mathbf{Z}} \) yields
\begin{equation}
f_{\mathbf{D}}(\mathbf{d}) = \frac{1}{(2\pi)^{N/2}} \exp\left( -\frac{1}{2} (\mathbf{d} - \boldsymbol{\mu})^\top (\mathbf{A}^{-1})^\top \mathbf{A}^{-1} (\mathbf{d} - \boldsymbol{\mu}) \right) \cdot \frac{1}{|\det(\mathbf{A})|}.
\end{equation}
To simplify, we use:
\begin{equation}
(\mathbf{A}^{-1})^\top \mathbf{A}^{-1} = (\mathbf{A} \mathbf{A}^\top)^{-1} = \mathbf{C}^{-1}, \quad \text{and} \quad |\mathbf{C}| = |\mathbf{A}|^2 \;\Rightarrow\; |\mathbf{A}| = |\mathbf{C}|^{1/2}.
\end{equation}
Substituting these identities gives the familiar multivariate normal density:
\begin{equation}
f_{\mathbf{D}}(\mathbf{d}) = \frac{1}{(2\pi)^{N/2} |\mathbf{C}|^{1/2}} \exp\left( -\frac{1}{2} (\mathbf{d} - \boldsymbol{\mu})^\top \mathbf{C}^{-1} (\mathbf{d} - \boldsymbol{\mu}) \right).
\label{eq:multivariate_gaussian_density}
\end{equation}

\subsubsection*{Multivariate Gaussian Likelihood Function}
In modeling applications, the mean of a multivariate normal distribution is typically not fixed but depends on a set of theoretical predictions. Let \( \mathbf{T}(\bm{\theta}) \in \mathbb{R}^N \) denote the vector of model predictions for a parameter vector \( \bm{\theta} \). These predictions represent the expected values of the data under the model. To represent the observed data vector \(\mathbf{D} \in \mathbb{R}^N \), we assume a multivariate normal distribution centered at \(\mathbf{T}(\bm{\theta}) \) with covariance matrix \(\mathbf{C} \)
\[
\mathbf{D} \sim \mathcal{N}(\mathbf{T}(\bm{\theta}), \mathbf{C}).
\]
This formulation captures the idea that deviations of the observed data from the theoretical predictions arise from measurement uncertainties encoded in \(\mathbf{C} \), including both variances and correlations. Starting from the general multivariate normal density in Eq.~(\ref{eq:multivariate_gaussian_density}), replacing the mean vector \( \boldsymbol{\mu} \) with \( \mathbf{T}(\bm{\theta}) \) and evaluating at the observed data yields the correlated-data likelihood:
\begin{equation}
P(\mathbf{D} \mid \bm{\theta}) = \frac{1}{(2\pi)^{N/2} |\mathbf{C}|^{1/2}} \exp\left( -\frac{1}{2} (\mathbf{D} - \mathbf{T}(\bm{\theta}))^\top \mathbf{C}^{-1} (\mathbf{D} - \mathbf{T}(\bm{\theta})) \right).
\label{eq:4.35}
\end{equation}
Here, the residuals \( \mathbf{D} - \mathbf{T}(\bm{\theta}) \) quantify the deviation between prediction and data, while the normalization ensures that the density integrates to one. This formulation incorporates the full covariance structure and is the basis for both maximum-likelihood and Bayesian treatments of correlated data.
The quadratic form in the exponent motivates the definition of the generalized chi-squared statistic:
\begin{equation}
\chi^2(\bm{\theta}) := (\mathbf{D} - \mathbf{T}(\bm{\theta}))^\top \mathbf{C}^{-1} (\mathbf{D} - \mathbf{T}(\bm{\theta})),
\label{eq:4.35}
\end{equation}
which is the Mahalanobis distance\footnote{The Mahalanobis distance \cite{Mahalanobis1936} generalizes the Euclidean distance by rescaling and rotating the space according to the covariance matrix. It measures how many standard deviations the vector $\mathbf{D}$ lies from the mean $\mathbf{T}(\bm{\theta})$, accounting for correlations among components.} between the data and model prediction. Using this definition, the likelihood takes the compact form:
\begin{equation}
P(\mathbf{D} \mid \bm{\theta}) = \frac{1}{(2\pi)^{N/2} |\mathbf{C}|^{1/2}} \exp\left( -\frac{1}{2} \chi^2(\bm{\theta}) \right).
\label{eq:4.36}
\end{equation}
When the covariance matrix \( \mathbf{C} \) is diagonal—i.e., the measurements are uncorrelated—this reduces to the familiar uncorrelated chi-squared expression:
\begin{equation}
\chi^2_u(\bm{\theta}) = \sum_{k=1}^N \frac{(D_k - T_k(\bm{\theta}))^2}{\sigma_k^2}.
\end{equation}
In this special case, each residual contributes independently to the total discrepancy, weighted only by the inverse of its corresponding variance.

\subsubsection*{Log-Likelihood Function}
In practical statistical inference, whether for numerical optimization, maximum-likelihood estimation, or Bayesian sampling, it is often preferable to work with the \emph{log-likelihood} function. The log-likelihood, denoted \( \ell(\bm{\theta}) \), is simply the natural logarithm of the likelihood:
\begin{equation}
\ell(\bm{\theta}) := \log P(\mathbf{D} \mid \bm{\theta}).
\end{equation}
Applying the logarithm to the multivariate Gaussian likelihood in Eq.~(\ref{eq:generalized_likelihood}) gives
\begin{equation}
\ell(\bm{\theta}) = -\frac{N}{2} \log(2\pi) - \frac{1}{2} \log |\mathbf{C}| - \frac{1}{2} \chi^2(\bm{\theta}),
\end{equation}
where \( \chi^2(\bm{\theta}) \) is the generalized chi-squared statistic defined in Eq.~(\ref{eq:4.35}). Working with the log-likelihood offers several advantages. First, taking logarithms converts products into sums, simplifying algebraic expressions and improving numerical stability, especially important in high-dimensional problems where the likelihood values can be extremely small and subject to numerical underflow. Second, when the covariance matrix \( \mathbf{C} \) does not depend on the parameters \( \bm{\theta} \), the terms $-\frac{N}{2} \log(2\pi) - \frac{1}{2} \log |\mathbf{C}|$ become constants with respect to \( \bm{\theta} \). In such cases, maximizing the log-likelihood is exactly equivalent to minimizing the generalized chi-squared statistic,
\[
\operatorname*{arg\,max}_{\bm{\theta}}\,\ell(\bm{\theta})
\;\;\Longleftrightarrow\;\;
\operatorname*{arg\,min}_{\bm{\theta}}\,\chi^{2}(\bm{\theta}),
\]
which is the basis for the extensive application of chi-squared minimization under Gaussian error assumptions.
%%%%%%%%%%%%%%%%%%%%%%%%%%%%%%%%%%%%%%%%%%%%%%%%%%%%%%%%%%%%%%%%%%%%%%%%
\section{Parameter Estimation and Inference}
Estimating model parameters and quantifying their associated uncertainties are fundamental in statistical inference. Both frequentist and Bayesian approaches rely on \emph{estimators}, functions of the observed data that provide approximations to unknown parameters. This section formalizes the concept of an estimator, reviews the framework of maximum likelihood estimation and its key theoretical properties, and introduces essential summary statistics and interval-based methods used to express uncertainty.

\paragraph{Estimators}
Let \( \bm{\theta} \) denote a parameter (or vector of parameters) in a statistical model, and let \( \mathbf{D} = \{D_1, \dots, D_n\} \) denote the observed data, regarded as a realization of an underlying random sample. An \emph{estimate} \( \hat{\bm{\theta}} \) is the numerical value obtained from the observed data, while the corresponding \emph{estimator} is the statistic \( \hat{\bm{\theta}}(\mathbf{D}) \) viewed as a function of the random data-generating process. Because it depends on random observations, the estimator is itself a random variable. The quality of an estimator is typically assessed using the following criteria:
\begin{itemize}
    \item Bias: The difference between the estimator’s expected value and the true parameter. An estimator is \emph{unbiased} if \( \mathbb{E}[\hat{\bm{\theta}}] = \bm{\theta} \).
    \item Consistency: An estimator is \emph{consistent} if it converges in probability to the true value as the sample size grows. Intuitively, a consistent estimator becomes increasingly accurate with more data.
    \item Efficiency: Among unbiased estimators, an efficient estimator has the smallest variance. Efficiency reflects how well an estimator uses the available information.
    \item Variance: Variance captures the degree of fluctuation of \( \hat{\bm{\theta}} \) across repeated samples. Lower variance indicates greater stability and reliability.
\end{itemize}

\subsection*{Maximum Likelihood Estimation}\label{sec:4.4}
Among the various approaches to parameter estimation, the \emph{maximum likelihood estimator} (MLE) \cite{casella2002statistical, wasserman2013all} is one of the most widely used because of its strong theoretical guarantees and broad applicability. The main idea is to select the parameter value under which the observed data are most probable according to the specified statistical model.

Given independent observations \( D_1, \dots, D_n \), the likelihood function of the model  \( P(\mathbf{D} \mid \bm{\theta}) \), the MLE is defined as:
\begin{equation}
\hat{\bm{\theta}}_{\mathrm{MLE}} = \arg\max_{\bm{\theta}}  \ell(\bm{\theta}),
\end{equation}
where \( \ell(\bm{\theta}) = \log P(\mathbf{D} \mid \bm{\theta}) \) is the log-likelihood. Under standard \emph{regularity conditions},\footnote{These include assumptions such as differentiability of the likelihood, parameter identifiability, and the existence of a unique maximizer. Formal statements may be found in \cite{casella2002statistical, wasserman2013all}.} the MLE satisfies several important large-sample properties:
\begin{enumerate}
    \item Consistency: The MLE converges in probability\footnote{A sequence of estimators $\hat{\theta}$ \emph{converges in probability} ($\xrightarrow{p}$) to \( \theta_0 \) if, for every $\varepsilon$ \[P\bigl(\,|\hat{\theta}_{n} - \theta_{0}| > \varepsilon\,\bigr) \;\longrightarrow\; 0 \quad \text{as } n \to \infty.\] This means that the estimator becomes arbitrarily close to the true value with high probability as the sample size increases.} to the true parameter value \( \bm{\theta}_0 \) as the sample size grows:
    \begin{equation}
        \hat{\bm{\theta}}_{\mathrm{MLE}} \xrightarrow{p} \bm{\theta}_0 \quad \text{as } \, n \to \infty.
    \end{equation}
    Thus, with enough data, the estimator will be arbitrarily close to the truth with high probability.
    \item Asymptotic Normality: For large samples, the distribution of the MLE is approximately Gaussian:
    \begin{equation}
    \sqrt{n}\,(\hat{\bm{\theta}}_{\mathrm{MLE}} - \bm{\theta}_0) \xrightarrow{d} \mathcal{N}\left(0,\, I^{-1}(\bm{\theta}_0)\right),
    \end{equation}
    where \( I(\bm{\theta}_0) \) is the Fisher information matrix \cite{Wald1949},
    \[
    I(\bm{\theta}) = -\mathbb{E} \left[ \nabla^2_{\bm{\theta}}\, \ell(\bm{\theta}) \right].
    \]
    This result provides an approximate sampling distribution for the estimator and forms the basis for large-sample confidence intervals and hypothesis tests.
    \item Asymptotic Efficiency: Among all consistent and asymptotically unbiased estimators, the MLE achieves the minimum possible asymptotic variance. In this sense, it is optimal in large samples.
\end{enumerate}
These properties justify the extensive use of the MLE as a point estimator. In practice, the asymptotic normality result is particularly valuable because it enables uncertainty quantification through Gaussian approximations, even in complex models where exact distributions are analytically intractable.

\subsection*{Statistical Measures}
Reliable parameter estimation requires clear definitions of the statistical quantities used to summarize data and probability distributions. These quantities fall into two categories:
\begin{itemize}
\item \emph{Population parameters}, which characterize features of an underlying probability distribution (typically unknown);
\item \emph{Sample statistics}, which are computed from observed data and serve as estimators of the corresponding population values.
\end{itemize}

\paragraph{Mean, Variance, and Covariance.}
Let \( X \) be a random variable with a probability distribution \( p(x) \). The \emph{population mean} or expected value is defined as 
\begin{equation}
    \mathbb{E}[X]
    \;=\;
    \int x\,p(x)\,dx
    \quad \text{(or } \sum_{x} x\,p(x) \text{ if } X \text{ is discrete)}.
\end{equation}
It represents the long-run average outcome of repeated realizations of \( X \).
The \emph{population variance} measures the spread of the distribution around its mean:
\begin{equation}
    \mathrm{Var}(X)
    \;=\;
    \mathbb{E}\bigl[(X - \mathbb{E}[X])^2\bigr].
\end{equation}
For two random variables \( X \) and \( Y \), the \emph{covariance} quantifies their joint variability:
\begin{equation}
    \mathrm{Cov}(X, Y)
    \;=\;
    \mathbb{E}\bigl[(X - \mathbb{E}[X])(Y - \mathbb{E}[Y])\bigr].
\end{equation}
A positive covariance indicates that the variables tend to increase together, while a negative covariance indicates an inverse relationship.

\paragraph{Sample Statistics.}
In practice, only a finite dataset is available, and empirical quantities approximate population parameters. Given observations \( X_1, \dots, X_n \), the key sample statistics are:
\begin{equation}
\begin{array}{lll}
\text{Sample mean:}       & \displaystyle \bar{X} = \frac{1}{n} \sum_{i=1}^{n} X_i, \\[10pt]
\text{Sample variance:}   & \displaystyle \hat{\mathrm{Var}}(X) = \frac{1}{n - 1} \sum_{i=1}^{n} (X_i - \bar{X})^2, \\[10pt]
\text{Sample covariance:} & \displaystyle \hat{\mathrm{Cov}}(X, Y) = \frac{1}{n - 1} \sum_{i=1}^{n} (X_i - \bar{X})(Y_i - \bar{Y}).
\end{array}
\end{equation}
These empirical measures capture the central tendency (mean), dispersion (variance), and linear dependence (covariance) of observed data and play a central role in both frequentist and Bayesian inference.

\subsection*{Confidence Intervals vs.\ Credible (Bayesian) Intervals}  
Statistical inference quantifies parameter uncertainty through either frequentist \emph{confidence intervals} (CIs) or Bayesian \emph{credible intervals} (CrIs). Although they often numerically coincide in large-sample settings, their conceptual interpretations are fundamentally different.

\paragraph{Frequentist Confidence Intervals.} In the frequentist framework, the parameter vector  $\bm{\theta}$ is treated as a fixed but unknown quantity, while the data $\bm{D}$ are considered a random realization  from an underlying probabilistic mechanism. A $(1-\alpha)\times100\%$ \emph{confidence interval} characterizes the long-run performance of the estimation procedure rather than the uncertainty about $\bm{\theta}$ itself.  Formally, if one were to repeat the data-generating process indefinitely and compute a new interval for each dataset, then $(1-\alpha)\times100\%$ of those intervals would contain the true parameter value $\bm{\theta}$. Thus, the interval is random whereas the parameter is not. Under standard regularity conditions and for sufficiently large sample sizes, the maximum-likelihood estimator $\hat{{\bm{\theta}}}$ is approximately normally distributed. In this case, a classical large-sample confidence interval may be written as
\begin{equation}
\text{CI} = \hat{\bm{\theta}} \pm z_{1-\alpha/2} \times \mathrm{SE}(\hat{\bm{\theta}}),
\end{equation}
where $z_{1-\alpha/2}$ is the $(1-\alpha/2)$-quantile of the standard normal distribution, and $\mathrm{SE}(\hat{\bm{\theta}})$ denotes the standard error of the estimator (e.g., $z_{0.975} \approx 1.96$ for $\alpha = 0.05$). The latter is often approximated using the inverse of the observed Fisher information $I(\hat{\bm{\theta}})$
\begin{equation}
\mathrm{SE}(\hat{\bm{\theta}}) \approx \sqrt{ \frac{I^{-1}(\hat{\bm{\theta}})}{n}},
\end{equation}
with $n$ the sample size. These approximations are widely used in global QCD analyses, where the Hessian of the likelihood (or $\chi^2$) function is the primary mechanism for propagating uncertainties.

\paragraph{Bayesian Credible Intervals.}
In contrast, the Bayesian framework treats $\bm{\theta}$ as a random variable whose uncertainty is fully described by the posterior distribution $P(\bm{\theta}\mid \mathbf{D})$, obtained by updating prior information through Bayes’ theorem. A $(1-\alpha)\times100\%$ \emph{credible interval} is any interval $[\bm{\theta}_{L},\bm{\theta}_{U}]$ satisfying
\begin{equation}
\int_{\bm{\theta}_L}^{\bm{\theta}_U} P(\bm{\theta} \mid D) \, d\bm{\theta} = 1 - \alpha.
\end{equation}
This interval has a direct probabilistic interpretation: conditional on the data and the prior, the probability that $\bm{\theta}$ lies within $[\bm{\theta}_{L},\bm{\theta}_{U}]$ is $1 - \alpha$.
Two forms of credible intervals are commonly used:
\begin{itemize}
    \item \textbf{Quantile-based} intervals, defined by the $\alpha/2$ and $1-\alpha/2$ posterior quantiles. These are straightforward to compute from MCMC samples and correspond directly to the percentile method used in practical analyses (see Sec.~\ref{sec:6.3.5}).
    
    \item \textbf{Highest Posterior Density (HPD)} intervals, which contain the region of parameter space with the highest posterior probability density subject to covering a total probability mass $1 - \alpha$. In practical implementations based on MCMC samples, such regions are often approximated by selecting parameter configurations that provide the best agreement with the data (see Sec.~\ref{sec:6.3.5}).\footnote{HPD regions can capture non-Gaussian, correlated, and even multi-modal structures of the posterior distribution, providing a more faithful representation of the underlying parameter space than quadratic (Hessian) approximations. Their construction, however, relies on adequate sampling and becomes increasingly challenging in high-dimensional spaces.}
\end{itemize}
Unlike confidence intervals, credible intervals quantify uncertainty about the parameter itself rather than the long-run frequency properties of an interval-generating procedure. However, as noted in \cite{gelman2013bayesian}, under broad regularity conditions, ``asymptotically a $100(1-\alpha)\%$ central posterior interval for $\bm{\theta}$ has the property that in repeated samples of $D$, $100(1-\alpha)\%$ of the intervals include the true value $\bm{\theta}_0$.''
This asymptotic agreement arises because, for large datasets, the likelihood becomes sharply peaked and dominates the influence of the prior, yielding an approximately Gaussian posterior centered at the maximum likelihood estimator. Consequently, Bayesian credible intervals and frequentist confidence intervals often agree closely in practice, particularly when weakly informative or noninformative priors are employed.
%%%%%%%%%%%%%%%%%%%%%%%%%%%%%%%%%%%%%%%%%%%%%%%%%%%%%%%%%%%%%%%%%%%%%%%%%%%%%%%%%
\section{Bayesian Formulation of a Global Nuclear PDF Fit}\label{sec:4.5}
In a Bayesian formulation of a global nuclear Parton Distribution Function analysis, the determination of parton distributions is treated as a full probabilistic inference problem. Instead of finding a single best-fit parameter point accompanied by Hessian eigenvector sets, the analysis aims to characterize the complete posterior distribution of the nPDF parameters given the data. This reformulation fundamentally changes the structure of a global PDF analysis: the primary objective shifts from locating a minimum of the global $\chi^2$ function to sampling the underlying probability density in parameter space; uncertainties are represented not through a local quadratic expansion of $\chi^2$ but through ensembles of posterior samples; and prior information, whether uniform or theory-motivated, can be incorporated in a statistically consistent way. Because the mapping from the parameter vector $\bm{\theta}$ to the theoretical predictions $\mathbf{T}(\bm{\theta})$ is highly nonlinear due to flexible parametrizations and QCD evolution, the posterior typically develops non-Gaussian features and nontrivial parameter correlations. These structures are naturally propagated to the PDFs and to all derived observables through the posterior sample ensemble.

Let $\bm{\theta}$ denote the vector of parameters defining the nPDFs at the input scale $Q_0$. For any set of parameters, the PDFs are evolved via the DGLAP equations and convolved with perturbative partonic cross sections to obtain theoretical predictions $\mathbf{T}(\bm{\theta})$. The experimental data are collected in the vector $\mathbf{D}$ with an associated covariance matrix $\mathbf{C}$. Assuming non-informative priors and Gaussian uncertainties, the posterior distribution follows from Bayes’ theorem and is proportional to the multivariate normal likelihood,
\begin{equation}
    P(\bm{\theta}\mid\mathbf{D}) \propto P(\mathbf{D} \mid \bm{\theta}) = \frac{1}{(2\pi)^{N/2} |\mathbf{C}|^{1/2}} \exp\left( -\frac{1}{2} \chi^2(\bm{\theta}) \right),
    \label{eq4.50}
\end{equation}
with the generalized chi-squared
\[
\chi^2(\bm{\theta}) = (\mathbf{D} - \mathbf{T}(\bm{\theta}))^\top \mathbf{C}^{-1} (\mathbf{D} - \mathbf{T}(\bm{\theta})).
\]
Since the resulting posterior cannot be computed analytically in realistic global analyses, numerical sampling through Markov Chain Monte Carlo methods will be employed. MCMC algorithms generate representative draws from the posterior, enabling the computation of marginal parameter distributions, credible intervals, and correlation patterns without relying on Gaussian or linearity assumptions.

From a methodological perspective, the Bayesian approach offers a more robust statistical framework than the traditional Hessian method. Rather than depending on the curvature of $\chi^2$ at a single minimum, Bayesian inference derives uncertainties from the full shape of the likelihood, allowing it to accommodate asymmetric distributions, long tails, or mild multimodality. Such features arise naturally in nPDF fits due to limited data constraints and the nonlinear dependence of observables on the fitted parameters, particularly through flexible parametrizations. It also provides a principled way to integrate prior information or theoretical constraints. However, these advantages come with challenges: Bayesian analyses are computationally more demanding, often requiring large numbers of likelihood evaluations; they may exhibit sensitivity to prior choices when data are sparse; and ensuring convergence and decorrelation of MCMC samples demands careful diagnostics. In practice, the Bayesian and Hessian approaches should be viewed as complementary. The Hessian method remains efficient and effective when the likelihood is approximately quadratic, whereas the Bayesian framework provides a more flexible and comprehensive treatment of uncertainties when the parameter space exhibits complex structure. A detailed discussion of the MCMC methodology employed in this work is presented in the next chapter.
%!TEX root = ../thesis.tex
%*******************************************************************************
%****************************** Fourth Chapter *********************************
%*******************************************************************************
\chapter{Markov Chain Monte Carlo (MCMC) Methods} \label{chap.5}
The Bayesian framework introduced in the previous chapter defines the posterior probability distribution of the model parameters but does not, by itself, provide a practical means of evaluating or sampling it in high-dimensional spaces. In the context of global nuclear PDF analyses, the parameter space is typically large and highly correlated, making direct numerical integration or grid-based approaches infeasible. Consequently, an efficient computational strategy is required to explore the posterior distribution beyond local approximations around a single best-fit point.
Markov Chain Monte Carlo (MCMC) methods provide such a framework. These methods generate sequences of correlated samples whose stationary distribution converges to the target posterior, enabling a direct numerical representation of parameter uncertainties, correlations, and non-Gaussian features. Unlike traditional approaches based on local curvature, MCMC techniques allow for a global exploration of the parameter space and are therefore well suited for problems in which the likelihood surface exhibits complex structure. At the same time, their practical implementation requires careful control of convergence, autocorrelation, and sampling efficiency, particularly in high-dimensional applications such as global nPDF fits.

This chapter develops the computational and theoretical foundations of MCMC methods used in this work. Section~\ref{sec:5.1} introduces Monte Carlo sampling and motivates the transition to Markov-chain–based techniques. Section~\ref{sec:5.2} presents the fundamental properties of Markov chains, including detailed balance, stationarity, and ergodicity. Section~\ref{sec:5.3} describes commonly used MCMC algorithms, with particular emphasis on the Metropolis–Hastings algorithm employed in this analysis. The material presented in this chapter provides the methodological basis for the MCMC implementation and analysis described in the next Chapter.
\section{From Monte Carlo Sampling to MCMC}\label{sec:5.1}
\emph{Monte Carlo} (MC) methods \cite{dunn2022exploring, robert2013monte} provide a general framework for approximating integrals that are analytically intractable by converting them to problems of stochastic sampling. Consider the expectation $I = \mathbb{E}_\pi \left[ h(\theta) \right]$, where $h(\theta)$ is a function of interest and $\pi$ is the target probability distribution. If independent samples $\theta_1, \theta_2, \dots, \theta_n$ can be drawn from $\pi$, the expectation can be estimated by the empirical average
\begin{equation}
\hat{I}_n = \frac{1}{n} \sum_{i=1}^{n} h(\theta_i).
\end{equation}
By the law of large numbers, $\hat{I}_n$ converges to the true value $I$ as $n \to \infty$, and under mild conditions the central limit theorem ensures that the estimator converges with an uncertainty that decreases as $\mathcal{O}(n^{-1/2})$ \cite{robert2013monte}. Thanks to their conceptual simplicity and favorable convergence properties, standard Monte Carlo techniques form the basis of a wide range of numerical tools used in physics, statistics, and engineering. Their practical implementation relies on high-quality pseudo-random number generators, which produce the independent random variates or multi-dimensional random vectors required for sampling.\footnote{In practice, these samples are generated using pseudo-random number generators (PRNGs), where “pseudo” indicates that the sequence is produced deterministically from an initial seed but is designed to mimic independent random draws. PRNGs typically generate uniform variates on $(0,1)$, which are then transformed into samples from more general distributions, such as Gaussian or multivariate distributions, using methods like inverse transform sampling, rejection sampling, or Box–Muller transformations. The statistical quality of the PRNG is essential for ensuring the reliability of Monte Carlo estimators \cite{robert2013monte, gentle2003random}.}

In many applications, however, direct sampling from $\pi$ is not feasible because the distribution is known only up to a normalizing constant or exhibits a complex functional form. Classical approaches such as importance sampling \cite{Doucet2000} and rejection sampling \cite{martino2011generalization} address this limitation by introducing an auxiliary proposal distribution $q(\theta)$ from which sampling is straightforward. In importance sampling, expectations are computed using weighted samples, where the weights account for the mismatch between $q$ and $\pi$. In rejection sampling, candidate points drawn from $q$ are accepted with probability proportional to $\pi/q$.\footnote{Unlike MCMC, where samples are correlated through the Markov chain, rejection sampling produces independent samples from the target distribution.} These methods are effective when the proposal distribution closely approximates the target and remain well suited for low- to moderate-dimensional problems. However, their efficiency declines rapidly in high-dimensional settings or when the target distribution exhibits strong correlations, heavy tails, or multiple modes. In such cases, importance weights can become highly variable, and acceptance probabilities in rejection sampling can become prohibitively small. These challenges are particularly relevant in Bayesian inference, where posterior distributions often arise from complex likelihood functions combined with informative priors, making the construction of efficient proposal distributions difficult.

\emph{Markov Chain Monte Carlo} (MCMC) methods \cite{gilks1995markov, brooks2011handbook, neal1993mcmc} overcome these limitations by adopting a different strategy. Rather than generating independent samples, MCMC constructs a Markov chain whose transition kernel is designed such that its stationary distribution coincides with the target distribution $\pi(\theta)$. Under appropriate conditions, most notably detailed balance and ergodicity (discussed in Section~\ref{sec:5.2}), the chain converges to the desired distribution, and after an initial thermalization phase, the generated samples can be used to estimate expectations with respect to $\pi$. Although successive samples are correlated, the resulting sequence provides a consistent estimator of posterior expectations once autocorrelations are properly accounted for. A key advantage of MCMC methods is that they do not require knowledge of the normalization constant of the target distribution, which is typically intractable in high-dimensional problems. Furthermore, MCMC scales more effectively than classical sampling methods when exploring complex parameter spaces, allowing for efficient sampling of distributions with nonlinear correlations, non-Gaussian features, or extended degeneracies.
%%%%%%%%%%%%%%%%%%%%%%%%%%%%%%%%%%%%%%%%%%%%%%%%%%%%%%%%%%%%%%%%%%%%%%%%%%%%%%%%%%%%%%%%
\section{ Fundamentals of Markov Chains}\label{sec:5.2}
This section introduces the theory of Markov chains \cite{roberts2004general, norris1998markov, gamerman1997markov}, which form the mathematical basis of MCMC methods. A Markov chain is a particular type of stochastic process characterized by the Markov property: the future evolution of the system depends only on its current state and not on the sequence of events that occurred before it. This ``memoryless'' property underlies the iterative sampling structure used in MCMC algorithms. We begin by introducing the definition and basic properties of Markov chains, then discuss stationary distributions and their relationship to target distributions in Bayesian inference. Finally, we describe the concepts of ergodicity and convergence, which ensure that a Markov chain will eventually generate samples distributed according to its stationary distribution.

\subsection{ Definition and properties}
A \emph{Markov chain} is a sequence of random variables ($\bm{X}_0, \bm{X}_1, \dots$) taking values in a state space $S$ and satisfying the Markov property. This property states that the future evolution of the chain depends only on its present state, not on the path previously taken. Formally, for all $n \ge 0$
\begin{equation}
    P(\mathbf{X}_{n+1} = \mathbf{x}_{n+1} \mid \mathbf{X}_n = \mathbf{x}_n, \mathbf{X}_{n-1} = \mathbf{x}_{n-1}, \ldots, \mathbf{X}_0 = \mathbf{x}_0) = P(\mathbf{X}_{n+1} = \mathbf{x}_{n+1} \mid \mathbf{X}_n = \mathbf{x}_n).
\end{equation}
A Markov chain is fully characterized by its transition probabilities
\begin{equation}
P_{ij} = P(X_{n+1} = j \mid X_n = i).
\end{equation}
which describe the probability of moving from state $i$ to state $j$ in one step. These probabilities are collected into the transition matrix
\begin{equation}
\bm{P} =
\begin{bmatrix}
P_{11} & P_{12} & \cdots & P_{1s} \\
P_{21} & P_{22} & \cdots & P_{2s} \\
\vdots & \vdots & \ddots & \vdots \\
P_{s1} & P_{s2} & \cdots & P_{ss}
\end{bmatrix},
\end{equation}
where $S= \{1, 2, \dots, s \}$ is the (finite) state space. To represent valid transition probabilities, the entries of $\bm{P}$ must satisfy:
\begin{enumerate}
    \item Non-negativity: All probabilities are non-negative,
    \begin{equation}
    P_{ij} \geq 0 \quad \forall \, i, j.
    \end{equation}
    \item Row sums to one: The total probability of transitioning from any given state $i$ to any other state must equal one,
    \begin{equation}
    \sum_{j} P_{ij} = 1 \quad \forall \, i.
    \end{equation}
\end{enumerate}
These conditions ensure that each row of $\bm{P}$ forms a proper probability distribution over the next-state possibilities. A Markov chain is time-homogeneous if the transition probabilities do not depend on the step index $n$. Under this assumption, the probability of reaching state $j$ from state $i$ in exactly $n$ steps is given by the $(i,j)$-entry of the matrix power
\begin{equation}
    \bm{P}^{(n)} = \bm{P}^n, \quad P_{ij}^{(n)} = P(X_n=j \mid X_0=i).
\end{equation}
The behavior of a Markov chain is strongly shaped by structural properties of its state space. Several classifications are fundamental:
\begin{itemize}
    \item Accessible states: State $j$ is accessible from state $i$ if $P_{ij}^{(n)} > 0$ for some finite $n$, meaning the chain can reach $j$ starting from $i$.
    \item Communication: States $i$ and $j$ communicate if each is accessible from the other. This partitions the state space into communication classes.
    \item Irreducibility: A chain is irreducible if every state communicates with every other state. The chain can then reach any state from any starting point.
    \item Periodicity: A state $i$ has period $d$ if returns to $i$ occur only in multiples of  $d$. A state is aperiodic if $d=1$. A chain is aperiodic if all states are aperiodic.
    \item Recurrent and Transient States: A state is recurrent if the chain returns to it with probability 1 (possibly infinitely often). It is transient if return is not guaranteed.
    \item Absorbing States: A state $i$ is absorbing if, once entered, the process cannot leave it. Formally, $P_{ii}=1$ and $P_{ij}=0$ for all $j \ne i$.
\end{itemize}
These classifications form the foundation for analyzing Markov chains. In the MCMC context, irreducibility and aperiodicity are particularly important, as they ensure that the chain can explore the entire parameter space and eventually converge to a unique stationary distribution corresponding to the target posterior.
\subsection{Stationary Distributions}
A key concept in analyzing the long-term behavior of a Markov chain is the \emph{stationary distribution} (or invariant distribution). For a finite state space $S= \{1, 2, \dots, s \}$, a probability vector $\bm{\pi} = (\pi_1, \pi_2, \ldots, \pi_s)$ is called stationary if it satisfies
\begin{equation}
\bm{\pi} \bm{P} = \bm{\pi},
\end{equation}
meaning that if the chain is initialized according to $\bm{\pi}$, its distribution remains unchanged under the transition matrix $\bm{P}$. More explicitly, for each state $j$,
\begin{equation}
\pi_j = \sum_{i \in S} \pi_i \, P_{ij},
\end{equation}
which expresses a global balance condition, the total probability flowing into the state $j$ equals the total probability flowing out. A stronger form of balance, often easier to verify, is the \emph{detailed balance} condition. A distribution $\bm{\pi}$ satisfies detailed balance with respect to $\bm{P}$ if
\begin{equation}
\pi_i P_{ij} = \pi_j P_{ji} \quad \forall \, i, j,
\label{eq:5.10}
\end{equation}
i.e., the probability flux from state $i$ to state $j$ matches exactly the flux from $j$ to $i$. Detailed balance is sufficient but not necessary for a stationary distribution. If a chain satisfies detailed balance, then global balance automatically holds, ensuring that $\bm{\pi}$ is stationary. 

In the context of continuous state spaces, the transition mechanism is described not by a matrix but by a transition kernel $T(\bm{X}, \bm{X}')$, where it represents the probability density of moving from $\bm{X}$ to $\bm{X}'$. The stationary condition generalizes to
\begin{equation}
\pi(\bm{X}') = \int \pi(\bm{X}) \, T(\bm{X}, \bm{X}') \, d\bm{X},
\end{equation}
which is the continuous analogue of the global balance equation. The detailed balance condition likewise becomes
\begin{equation}
    \pi(\bm{X}) \, T(\bm{X}, \bm{X}') = \pi(\bm{X}') \, T(\bm{X}', \bm{X}).
\end{equation}
Although a Markov chain may possess a stationary distribution without satisfying detailed balance, the detailed balance condition plays a central role in the design of MCMC algorithms. There are two reasons: First, detailed balance provides a straightforward route to establishing stationarity; verifying a local condition involving pairs of states is generally far easier than solving a high-dimensional global balance equation. Second, many commonly used MCMC algorithms—including the Metropolis–Hastings method—are explicitly constructed to satisfy detailed balance by design, ensuring that the target distribution is stationary. This significantly simplifies algorithm construction while guaranteeing the correct invariant distribution \cite{Chib1995}.

\subsection{Ergodicity and convergence}\label{sec:5.2.3}
For a Markov chain to be useful in applications such as MCMC, it is essential to understand the conditions under which the chain converges to a stationary distribution. Two structural properties play a central role: irreducibility and aperiodicity. Together, these properties define \emph{ergodicity}. A Markov chain is called ergodic if it is both:
\begin{itemize}
    \item Irreducible: every state can be reached from any other state in a finite number of steps. This ensures that the chain can explore the entire state space and does not become confined to a subset.
    \item Aperiodic: the chain does not get trapped in cycles; that is, returns to a state can occur at irregular intervals rather than only at multiples of some integer greater than one.
\end{itemize}
These conditions prevent the chain from decomposing into isolated communication classes or becoming trapped in periodic oscillations, both of which would inhibit full exploration of the state space and prevent convergence to a unique stationary distribution.

\paragraph{The Fundamental Theorem of Markov Chains,} often referred to as the \emph{Ergodic Theorem}, establishes the precise conditions under which a Markov chain converges to a unique stationary distribution \cite{meyn2009markov, LukeTierney}. For a finite state-space Markov chain that is irreducible and aperiodic, the theorem guarantees the existence of a unique stationary distribution $\bm{\pi}$ satisfying
\begin{equation}
\bm{\pi} \bm{P} = \bm{\pi}, \quad \quad \sum_{x \in S} \pi(x) = 1.
\end{equation}
Furthermore, the distribution of the chain after $n$ steps converges to $\bm{\pi}$ regardless of the initial state $\bm{X}_0 $. Formally, for any subset $A \subseteq S$,
\begin{equation}
\lim_{n \to \infty} P^n(X_0, A) = \pi(A),
\end{equation}
where $P^n(X_0, A)$ denotes the probability that the chain is in set $A$ after $n$ transitions. This convergence is typically quantified using the \emph{total variation} (TV) distance, defined for two measures $\mu$ and $\nu$ on a discrete state space by
\begin{equation}
\|\mu - \nu\|_{\mathrm{TV}} = \frac{1}{2} \sum_{x \in S} |\mu(x) - \nu(x)|.
\end{equation}
In terms of TV distance, the convergence result is
\begin{equation}
    \lim_{n \to \infty} \| P^n(X_0, \cdot) - \pi(\cdot) \|_{\mathrm{TV}} = 0.
\end{equation}
In many cases, including most chains used in practical MCMC algorithms, this convergence occurs at a geometric (exponential) rate: there exist constants $\rho<1$ and $C>0$ such that
\begin{equation}
     \| P^n(X_0, \cdot) - \pi(\cdot) \|_{\mathrm{TV}} \le C \rho^n.
    \label{eq:5.16}
\end{equation}
Geometric ergodicity is particularly desirable because it ensures rapid convergence to stationarity and facilitates the use of standard MCMC diagnostics.

\subsubsection*{Continuous-State Generalization}
For Markov chains on a continuous state space $S \subset \mathbb{R}^d$, the transition matrix is replaced by a transition kernel $T(x, A)$ or, when a density exists, $T(x, x')$. The stationary distribution is now described by a density $\pi(x)$ (or more generally, a probability measure), which must satisfy the stationarity equation
\begin{equation}
    \pi(A) = \int_{S} \pi(x)\, T(x, A)\, dx \quad \text{for all measurable } A \subseteq S.
\end{equation}
If a density $T(x, x')$ exists, this can be written as
\[
\pi(A) = \int_{S} \pi(x)\, T(x, x')\, dx.
\]
Under conditions analogous to irreducibility and aperiodicity, commonly expressed as \emph{Harris recurrence} \cite{LukeTierney} and aperiodicity in the continuous setting—the continuous-state Markov chain admits a unique stationary distribution $\pi$, and the distribution of $X_n$ converges to $\pi$ in total variation:
\begin{equation}
    \lim_{n \to \infty} \, \| \mathcal{L}(X_n) - \pi \|_{\mathrm{TV}} = 0,
\end{equation}
where $\mathcal{L}(X_n)$ denotes the law (distribution) of $X_n$. Many MCMC kernels, including the Metropolis–Hastings algorithm, are constructed to satisfy the detailed balance condition
\[
\pi(x) \, T(x, x') = \pi(x') \, T(x', x),
\]
which ensures that $\pi$ is stationary. As in the discrete case, geometric ergodicity may hold under additional regularity conditions. When present, it guarantees that the distribution of $X_n$ approaches $\pi$ at an exponential rate. This exponential convergence provides stronger theoretical guarantees for MCMC convergence and ensures that the central limit theorem holds, thereby supporting the validity of asymptotic variance estimates for Monte Carlo averages.
%%%%%%%%%%%%%%%%%%%%%%%%%%%%%%%%%%%%%%%%%%%%%%%%%%%%%%%%%%%%%%%%%%%%%%%%%%%%%%%%%%
\section{MCMC algorithms} \label{sec:5.3}
Markov Chain Monte Carlo methods form a broad class of algorithms designed to sample from complex probability distributions by constructing a Markov chain whose stationary distribution is the desired target distribution $\pi(\mathbf{x})$. MCMC is central to Bayesian inference, statistical physics, and numerous applications where direct sampling from high-dimensional or analytically intractable distributions is not feasible. Unlike classical Monte Carlo methods, which rely on independent and identically distributed (i.i.d.) samples, MCMC produces dependent samples by sequentially updating the state of a Markov chain. Crucially, the dependence between samples decays as the lag between them increases, meaning correlations are smaller at larger lags. This property allows the chain to approximate independent sampling from $\pi$ once convergence is achieved.

Formally, let $\{\mathbf{X}_t\}_{t=0}^{\infty}$ denote a Markov chain generated by an MCMC algorithm, where each state $\mathbf{X}_t \in \mathbb{R}^n$ represents a draw from the $n$-dimensional target distribution.\footnote{Here, $\mathbf{X}_t = (\mathbf{X}_t^{(1)}, \mathbf{X}_t^{(2)}, \dots, \mathbf{X}_t^{(n)}) $ denotes a complete sample from the $n$-dimensional posterior or target distribution $\pi(\mathbf{X})$.} A fundamental property of such a chain is that its dependence on earlier states diminishes over time. In particular, for sufficiently large lags $k$, the chain exhibits asymptotic independence:
\begin{equation*}
    \lim_{k \to \infty} P(\mathbf{X}_{t+k} \mid \mathbf{X}_t) = P(\mathbf{X}_{t+k}),
\end{equation*}
indicating that $\mathbf{X}_t$ and $\mathbf{X}_{t+k}$ become approximately independent as $k$ increases. This asymptotic independence is a necessary condition for the practical application of MCMC, as it ensures that the influence of the initial starting point $\mathbf{X}_0$ gradually disappears. As the chain evolves, its distribution converges toward the invariant distribution $\pi$:
\begin{equation*}
    \lim_{t \to \infty} P(\mathbf{X}_{t} \mid \mathbf{X}_0) = \pi(\mathbf{x}).
\end{equation*}
After a sufficiently long burn-in period, during which the chain ``forgets'' its initial state, the subsequent draws can be regarded as approximate samples from $\pi$, subject only to the manageable autocorrelation effects inherent in the sequence. All MCMC algorithms rely on two foundational principles:
\begin{itemize}
\item \textbf{Stationarity:} The transition mechanism must ensure that the stationary distribution of the Markov chain is exactly the target distribution $\pi$. This is often enforced by designing transitions to satisfy global balance or the more convenient detailed balance condition.
\item \textbf{Efficient Exploration:} The chain must explore the state space effectively, avoiding slow mixing, unstable behavior, or confinement to restricted regions. Appropriate proposal mechanisms, acceptance rules, and tunable algorithmic parameters are essential for good performance, particularly in high-dimensional problems. 
\end{itemize}
Different MCMC algorithms implement these principles through distinct strategies for proposing and accepting new states. In the following, we review several widely used MCMC methods:

\subsubsection{Metropolis-Hastings Algorithm}
The Metropolis-Hastings (MH) algorithm \cite{metropolis1953equation, Hastings:1970aa, robert2004metropolis} is one of the most fundamental and widely used MCMC methods. It constructs a Markov chain by proposing a new state from a flexible, user-specified proposal distribution. The critical step is the acceptance rule, which decides whether to move to the new state or remain in the current one. This acceptance probability is carefully constructed to satisfy the detailed balance condition with respect to the target distribution, $\pi(\bm{x})$. This mechanism ensures that the target distribution $\pi$ is the unique stationary distribution of the resulting chain, guaranteeing eventual convergence.
A defining strength of MH is its generality: the proposal kernel can be symmetric (like in the original Metropolis algorithm) or asymmetric. This flexibility makes MH the backbone of many practical MCMC implementations, particularly in complex analyses like global nPDF fitting where the target posterior density lacks a closed-form normalization constant. We will discuss the Metropolis-Hastings algorithm in detail in the following section.

\subsubsection{Gibbs Sampling}
Gibbs sampling \cite{geman1984stochastic} is a specialized and often highly efficient form of the Metropolis–Hastings algorithm. It is designed for multivariate problems where the full conditional distributions are obtainable. Instead of proposing a move for the entire parameter vector, Gibbs sampling updates each component sequentially, drawing directly from its full conditional distribution while holding all other components fixed. For a parameter vector $\mathbf{X} = (\mathbf{X}^{(1)}, \dots, \mathbf{X}^{(n)}) $, an iteration cycles through sampling:
\[
X^{(i)} \sim \pi\!\left( X^{(i)} \mid X^{(-i)} \right), \quad i = 1, \dots, n,
\]
where $X^{(-i)}$ denotes all components except $X^{(i)}$. Because each update is drawn from an exact conditional distribution, the acceptance probability for every step is one (100\% accepted). This can make Gibbs sampling extremely efficient, especially in structured models like hierarchical Bayesian models. However, its use is restricted to cases where these conditional distributions are available in closed form or are easy to sample from, which often makes it impractical for complex, tightly coupled systems found in many QCD and PDF-fitting applications.

\subsubsection{Hamiltonian Monte Carlo (HMC)}
Hamiltonian Monte Carlo \cite{Duane:1987de}, also known as Hybrid Monte Carlo, is an advanced MCMC technique designed to improve sampling efficiency in high-dimensional or strongly correlated parameter spaces. HMC introduces auxiliary momentum variables and simulates Hamiltonian dynamics. This simulation uses the gradient of the log-target density to generate proposals that follow approximate trajectories of constant energy. This allows HMC to propose distant states in parameter space while maintaining a high probability of acceptance. While HMC is computationally more demanding per iteration, as it requires calculating the gradient of the log-target density, it virtually eliminates the slow random-walk behavior characteristic of simpler MH chains. As a result, HMC often achieves superior mixing and much faster exploration of complex posterior landscapes. These advantages make it a highly popular choice in modern large-scale Bayesian inference, often used along with automatic differentiation tools. 

\subsubsection{Slice Sampling}
Slice sampling \cite{neal2003slice} provides an adaptive and tune-free MCMC strategy based on uniformly sampling from the region under the graph of the target density, $\pi$. The method introduces an auxiliary variable, $u$, to define a ``slice'' of the target distribution: $S = \{\, x \mid \pi(x) \ge u \,\}$. The algorithm alternates between drawing a new value for $u$ and then sampling $x$ uniformly from the slice $S$. This procedure is appealing because it adaptively determines appropriate step sizes for the chain, eliminating the need to manually tune proposal scales. Slice sampling is robust and performs well for many unimodal or moderately complex posterior distributions. However, its efficiency can degrade significantly in high-dimensional settings, as the task of identifying and sampling uniformly from the resulting slice $S$ often becomes computationally expensive, requiring many evaluations of the target density.

\subsection{The Metropolis–Hastings Algorithm} \label{sec:5.3.1}
The Metropolis-Hastings algorithm constructs a Markov chain with stationary distribution $\pi(\mathbf{x})$ by generating candidate (proposed) states from a proposal distribution and applying an acceptance rule to ensure correct sampling from the target distribution. At iteration $t$, given the current state $\bm{X}$, the algorithm draws a candidate state $\bm{X}^*$ from a proposal kernel $q(\mathbf{X}^* \mid \mathbf{X})$, and then either accepts or rejects this candidate according to an acceptance probability chosen to enforce detailed balance. This mechanism ensures that $\pi$ is invariant for the chain and, under suitable conditions, that samples from the chain asymptotically follow $\pi$. For any two distinct states $\bm{X}$ and $\bm{X}^*$,  the transition probability of the MH chain is
\begin{equation}
    P(\mathbf{X} \to \mathbf{X}^*) \;=\;
    q(\mathbf{X}^* \mid \mathbf{X}) \, \alpha(\mathbf{X}, \mathbf{X}^*),
\end{equation}
where the acceptance probability is defined as
\begin{equation}
\alpha \bigl(\mathbf{X}, \mathbf{X}^{*}\bigr)
\;=\;
\min \,\Biggl (\,1,\;
\frac{\pi(\mathbf{X}^{*})\,q\bigl(\mathbf{X}\mid \mathbf{X}^{*}\bigr)}{\pi(\mathbf{X})\,q\bigl(\mathbf{X}^{*}\mid \mathbf{X}\bigr)}
\Biggr).
\label{eq:5.17}
\end{equation}
If the proposal is rejected, the chain remains at the current state, giving the diagonal transition probability, $P(\mathbf{X} \to \mathbf{X})$. This is the complement of the probability of moving to any other state
\begin{equation}
    P(\mathbf{X} \to \mathbf{X}) \;=\; 1-\int q(\mathbf{X}^* \mid \mathbf{X}) \, \alpha(\mathbf{X}, \mathbf{X}^*) \, d\mathbf{X}^*.
\end{equation}
The MH acceptance rule is specifically constructed to satisfy the detailed balance condition with respect to the target distribution:
\begin{equation}
    \pi(\mathbf{X})\,P(\mathbf{X} \to \mathbf{X}^*)
    \;=\;
    \pi(\mathbf{X}^*)\,P(\mathbf{X}^* \to \mathbf{X}).
\end{equation}
Substituting the transition probabilities yields
\begin{equation}
    \pi(\mathbf{X}) q(\mathbf{X}^* \mid \mathbf{X}) \min\left(1, \frac{\pi(\mathbf{X}^*) q( \mathbf{X} \mid \mathbf{X}^*)}{\pi(\mathbf{X}) q(\mathbf{X}^* \mid \mathbf{X})}\right) = 
\pi(\mathbf{X}^*) q(\mathbf{X} \mid \mathbf{X}^*) \min\left(1, \frac{\pi(\mathbf{X}) q(\mathbf{X}^* \mid \mathbf{X})}{\pi(\mathbf{X}^*) q( \mathbf{X} \mid \mathbf{X}^*)}\right).
\end{equation}
Multiplying $\pi(\mathbf{X}) q(\mathbf{X}^* \mid \mathbf{X})$ into the minimum produces
\[
\min\left(\pi(\mathbf{X}) q(\mathbf{X}^* \mid \mathbf{X}), \pi(\mathbf{X}^*) q(\mathbf{X} \mid \mathbf{X}^*)\right),
\]
since the ratio inside the minimum cancels those terms. By symmetry, the right-hand side of the detailed balance equation reduces to the same expression. Thus, both sides are equal, confirming that detailed balance holds. Because detailed balance implies stationarity, the MH algorithm guarantees that $\pi$ is the invariant distribution of the Markov chain it constructs.

To ensure that the chain converges to $\pi$ from any initial state, it must also be ergodic, meaning both irreducible and aperiodic \cite{robert2013monte}:
\begin{itemize}
    \item Irreducibility: The MH chain is irreducible if the proposal distribution has full support, meaning that every state $\mathbf{X}^*$ is accessible from every other state $\mathbf{X}$ in a single step with a non-zero probability:
    \[
    q(\mathbf{X}^* \mid \mathbf{X}) >0 \quad\text{for all} \, \mathbf{X}, \mathbf{X}^*.
    \]
    Proposals such as multivariate Gaussians centered at the current state satisfy this condition, ensuring that any region of the state space can eventually be reached.
    \item Aperiodicity: The algorithm is inherently aperiodic because the possibility of proposal rejection produces self-transitions:
    \[
    \mathbf{X}_{t+1}=\mathbf{X}_t.
    \]
    This mechanism introduces a non-zero probability of remaining in the same state, preventing deterministic cycling and guaranteeing aperiodicity.\footnote{Aperiodicity in the MH algorithm is guaranteed by the rule (the transition kernel) that governs the chain, not by the outcome of any particular run. The rule is constructed so that a self-transition (remaining in the current state due to a rejected proposal) is always a theoretical possibility with a non-zero probability. Even if the chain, by chance, accepts every single proposal and appears to be constantly moving, the potential for a self-transition is mathematically built into its definition, which is sufficient to ensure the chain cannot fall into a fixed recurring cycle.}
\end{itemize}
When both conditions hold, the MH chain converges to the target distribution in total variation distance. This theoretical guarantee underlies the practical use of MH sampling in Bayesian inference: after a sufficient burn-in period, the states of the chain approximate draws from the posterior distribution.

A detailed discussion of practical considerations involved in implementing MCMC algorithms, such as initialization strategies, burn-in determination, convergence diagnostics, autocorrelation analysis, thinning, and error estimation, is presented in the next chapter. These steps are essential to ensuring that the samples obtained from the MH algorithm form a reliable and representative approximation of the target distribution in high-dimensional Bayesian analyses.

\chapter{Methodology} \label{chap.6}
This chapter presents the methodological framework used to determine the posterior probability distribution of nuclear parton distribution function (nPDF) parameters from experimental data within the Bayesian approach introduced in the previous chapters. Building on the theoretical foundations of QCD, the formulation of global nPDF analyses, and the Markov Chain Monte Carlo (MCMC) methods discussed earlier, this chapter provides a complete and systematic workflow for performing a global nPDF fit. The focus is on the construction of the likelihood, the implementation of MCMC sampling, and the statistical procedures used to extract parameter estimates and uncertainties from the resulting posterior distribution.

The chapter is organized as follows. Section~\ref{sec:6.1} describes the experimental datasets included in the analysis together with the theoretical framework used to compute the corresponding observables. In addition, the parametrization of nuclear PDFs at the input scale is introduced, establishing the model parameters that are estimated in the Bayesian fit. Section~\ref{sec:6.2} presents the
construction of the posterior probability distribution and the sampling strategy used to
explore the high-dimensional parameter space. In particular, the adaptive Metropolis-Hastings algorithm employed in this work is described, together with the definition of the likelihood function, the choice of prior distributions, and the practical aspects of chain initialization and evolving multiple Markov chains. Section~\ref{sec:6.3} focuses on the statistical analysis of the generated chains. Convergence diagnostics are performed using trace plots and cumulative mean stabilization, followed by a detailed study of autocorrelations within the Markov chains. The integrated autocorrelation time is estimated using the $\Gamma$-method and used to determine an appropriate thinning strategy for reducing correlations between successive samples. The chapter then introduces the statistical procedures used to characterize the posterior sample and estimate parameter uncertainties. Finally, Section~\ref{sec:6.4} provides a brief overview of the full analysis procedure, summarizing the steps from the definition of the likelihood to the extraction of posterior distributions and
uncertainties.

%%%%%%%%%%%%%%%%%%%%%%%%%%%%%%%%%%%%%%%%%%%%%%%%%%%%%%%%%%%%%%%%%%%%%%%%%%%
\section{Nuclear PDF Analysis Setup}\label{sec:6.1}
This section describes the experimental input and theoretical framework used in the nuclear PDF analysis. For each dataset, the corresponding perturbative QCD calculation is presented, highlighting its sensitivity to specific partonic flavors and kinematic regions. This formulation directly defines the likelihood function used in the MCMC sampling.
The analysis includes electroweak boson production in proton–lead ($p$Pb) collisions, neutrino deep inelastic scattering on lead, heavy-flavor and quarkonium production, as well as fixed-target DIS and Drell–Yan nuclear ratios. Theoretical predictions are computed using nPDFs parameterized at an initial scale and evolved to the relevant kinematics. Experimental observables incorporate full statistical, systematic, and normalization uncertainties. Their comparison with theory determines the likelihood and constrains the nPDF parameters.

\subsection{Experimental Data and Theoretical Framework}
The selection of experimental input and the definition of the corresponding analysis strategy are central to a reliable determination of nuclear PDFs. In this work, two complementary global analyses are performed in order to balance physics reach with computational feasibility. The first is a \texttt{Pb-only} analysis, which incorporates exclusively experimental data on lead nuclei and is the primary focus of this thesis. The second is a \texttt{Multi-nuclei} analysis, which extends the dataset to include measurements performed on a wide range of nuclear targets. Performing both analyses allows us to test assumptions related to the mass-number ($A$) dependence of nuclear PDFs and to assess the stability of the results when data from different nuclei are included.

\subsubsection{Pb-only Analysis}
The \texttt{Pb-only} analysis is restricted to experimental data involving lead nuclei ($A = 208$). This setup eliminates the need to model the $A$-dependence of nuclear effects, significantly simplifying the theoretical framework and reducing computational cost. It therefore enables the generation of large MCMC samples, allowing the posterior distribution to be resolved with high statistical precision and facilitating a detailed study of parameter correlations, convergence, and non-Gaussian features.
The main limitation of this approach is the lack of direct constraints on the $A$-dependence of nPDFs. Consequently, the \texttt{Pb-only} analysis cannot determine nuclear scaling independently but instead provides a controlled environment for testing the assumptions underlying $A$-dependent global fits and the subsequent \texttt{Multi-nuclei} analysis.
The dataset includes electroweak vector boson production in proton–lead collisions, neutrino DIS on lead, and heavy-flavor production in proton–lead collisions. These processes probe complementary kinematic regions and partonic flavors, providing sufficient constraints for a robust determination of Pb nPDFs at $A=208$.

\paragraph{Electroweak Vector Boson Production:} The production of electroweak vector bosons ($W^\pm$ and $Z$) in proton-lead ($p$Pb) collisions provides a particularly clean probe of parton-level dynamics in nuclei. In this analysis, we include measurements of $W^\pm$ and $Z$ production collected by LHC experiments during Runs I and II. The datasets and observables are summarized in Table~\ref{tab:WZdata}.
Theoretical predictions for the $W^\pm$ and $Z$ cross sections are computed at next-to-leading order (NLO) in perturbative QCD using the \texttt{MCFM} framework~\cite{Campbell:2015qma}, which employs Monte Carlo integration techniques to handle the relevant final-state kinematics. To achieve the computational efficiency required for repeated $\chi^2$ evaluations in a global fit, these calculations are interfaced with the \texttt{APPLgrid} library~\cite{Carli:2010rw}. In this approach, the hard-scattering perturbative coefficients are precomputed using \texttt{MCFM} and stored in lookup grids. This allows the theoretical cross sections to be obtained via a fast convolution of the stored grids with the nuclear PDFs, avoiding the need to repeat the time-consuming Monte Carlo integration for each evaluation of the $\chi^2$.
At leading order, electroweak boson production is dominated by quark-antiquark annihilation subprocesses,
\[
q + \bar{q}' \to W^{\pm}, \qquad q + \bar{q} \to Z^0,
\]
which makes these observables directly sensitive to the flavor composition of quark and antiquark distributions in the nucleus. At NLO, gluon–quark scattering channels also contribute,
\[
g + q \to W + \bar{q}',
\]
contribute, introducing an indirect sensitivity to the gluon nuclear PDF through QCD corrections.
Since electroweak bosons are color singlets and do not participate in strong interactions, their production is largely unaffected by final-state nuclear effects. Consequently, $W^\pm$ and $Z$ observables provide a theoretically well-controlled probe of initial-state nuclear modifications. They play a central role in constraining the flavor structure of nuclear PDFs, particularly the light-quark and antiquark distributions, over a broad range of momentum fractions. The analysis is restricted to rapidity distributions in perturbatively safe kinematic regions, and no additional threshold resummation or specialized kinematic cuts are required. Further discussion of the constraints from electroweak boson production in $p$Pb collisions on nPDFs can be found in Ref.~\cite{Kusina:2020lyz}.
\begin{table}[t]
    \centering
    \caption[$W^\pm$ and $Z$ production data from LHC experiments]{$W^\pm$ and $Z$ production data from LHC experiments. $N_{\text{data}}$ denotes the number of data points after/before applying kinematic cuts.}
    %%%%%%%%%%%%%%%%%%%%%%%%%%%%%%%%%%%%%%%%%%%%%%%%
% LHC W/Z data (Filtered)
%%%%%%%%%%%%%%%%%%%%%%%%%%%%%%%%%%%%%%%%%%%%%%%%
\begingroup
\small
\renewcommand{\arraystretch}{0.95}  % default is 1.0
\begin{tabular}{|l l c c c c|}
  \hline
  \multicolumn{6}{|c|}{LHC $p$Pb $W^{\pm}/Z$ data $d\sigma/dy$} \\
  \hline
  Exp. & Observable & $\sqrt{s}$ [TeV] & ID & Ref. & $N_{\text{data}}$ \\
  \hline
  \multirow{3}{*}{ATLAS}
    & $Z$     & 5.02 & 6215 & \cite{Aad:2015gta}        & 14/14 \\
    & $W^{+}$ & 5.02 & 6213 & \cite{AtlasWpPb}          & 10/10 \\
    & $W^{-}$ & 5.02 & 6211 & \cite{AtlasWpPb}          & 10/10 \\
  \hline
  \multirow{5}{*}{CMS}
    & $Z$     & 5.02 & 6235 & \cite{Khachatryan:2015pzs}& 12/12 \\
    & $W^{+}$ & 5.02 & 6233 & \cite{Khachatryan:2015hha}& 10/10 \\
    & $W^{-}$ & 5.02 & 6231 & \cite{Khachatryan:2015hha}& 10/10 \\
    & $W^{+}$ & 8.16 & 6234 & \cite{CMS:2019leu}        & 24/24 \\
    & $W^{-}$ & 8.16 & 6232 & \cite{CMS:2019leu}        & 24/24 \\
  \hline
  LHCb
    & $Z$     & 5.02 & 6275 & \cite{Aaij:2014pvu}       & 2/2 \\
  \hline
  \multirow{2}{*}{ALICE}
    & $W^{+}$ & 5.02 & 6253 & \cite{ALICE:2016rzo}      & 2/2 \\
    & $W^{-}$ & 5.02 & 6251 & \cite{ALICE:2016rzo}      & 2/2 \\
  \hline
  \textbf{Total} & & & & & \textbf{120/120} \\
  \hline
\end{tabular}
\endgroup

\iffalse
\begingroup
\footnotesize
\renewcommand{\arraystretch}{0.95}
\begin{tabular}{|@{}l@{} l@{} c@{} c@{} c@{} c@{}|}
  \hline
  \multicolumn{6}{|c|}{LHC $p$Pb $W^{\pm}/Z$ data $d\sigma/dy$} \\
  \hline
  Exp. & Observable & $\sqrt{s}$ [TeV] & ID & Ref. & $N_{\text{data}}$ \\
  \hline
  \multirow{3}{*}{ATLAS}
    & $Z$     & 5.02 & 6215 & \cite{Aad:2015gta}        & 14/14 \\
    & $W^{+}$ & 5.02 & 6213 & \cite{AtlasWpPb}          & 10/10 \\
    & $W^{-}$ & 5.02 & 6211 & \cite{AtlasWpPb}          & 10/10 \\
  \hline
  \multirow{5}{*}{CMS}
    & $Z$     & 5.02 & 6235 & \cite{Khachatryan:2015pzs}& 12/12 \\
    & $W^{+}$ & 5.02 & 6233 & \cite{Khachatryan:2015hha}& 10/10 \\
    & $W^{-}$ & 5.02 & 6231 & \cite{Khachatryan:2015hha}& 10/10 \\
    & $W^{+}$ & 8.16 & 6234 & \cite{CMS:2019leu}        & 24/24 \\
    & $W^{-}$ & 8.16 & 6232 & \cite{CMS:2019leu}        & 24/24 \\
  \hline
  LHCb
    & $Z$     & 5.02 & 6275 & \cite{Aaij:2014pvu}       & 2/2 \\
  \hline
  \multirow{2}{*}{ALICE}
    & $W^{+}$ & 5.02 & 6253 & \cite{ALICE:2016rzo}      & 2/2 \\
    & $W^{-}$ & 5.02 & 6251 & \cite{ALICE:2016rzo}      & 2/2 \\
  \hline
  \textbf{Total} & & & & & \textbf{120/120} \\
  \hline
\end{tabular}
\endgroup
\fi 
    \label{tab:WZdata}
\end{table}

\paragraph{Neutrino DIS:} Neutrino DIS provides unique and essential constraints on the flavor structure of nuclear PDFs that are not accessible through charged-lepton DIS alone. In this analysis, we include inclusive charged-current (CC) DIS measurements from the CHORUS collaboration on a lead target, as summarized in Table~\ref{tab:nudata}.
Unlike neutral-current (NC) DIS, neutrino-induced CC interactions probe specific quark flavors through the exchange of a $W$ boson, thereby introducing additional sensitivity to the flavor decomposition of the nuclear PDFs. In particular, neutrino DIS plays a crucial role in constraining the strange quark distribution in nuclei, which remains poorly determined when relying solely on electromagnetic probes. While the inclusive neutrino DIS cross section is dominated by charged-current scattering off light quarks, charm production via
\[
\nu_\mu + s \to \mu^- + c,
\]
provides direct and unique sensitivity to the strange-quark distribution. Consequently, neutrino DIS measurements offer an important complement to electroweak boson production and heavy-flavor data included in the global fit, especially in the low-to-intermediate Bjorken-$x$ region. Historically, tensions have been reported between neutrino DIS and charged-lepton DIS measurements, particularly at low and intermediate values of $x$~\cite{Schienbein:2009kk, Kovarik:2010uv, Paukkunen:2013grz}. This issue was investigated in detail in the nCTEQ15$\nu$ analysis~\cite{Muzakka:2022wey}, which demonstrated that a consistent global description can be achieved once correlated experimental uncertainties are properly treated and appropriate data selection criteria are applied. Motivated by these findings, we include the CHORUS CC DIS data in the present \texttt{Pb-only} analysis. These measurements provide a large number of statistically precise data points on a lead target and are found to be compatible with the other experimental inputs considered here.

Theoretical predictions for neutrino-induced DIS cross sections are computed at NLO in perturbative QCD using the simplified ACOT-$\chi$ scheme (SACOT-$\chi$)~\cite{Kramer:2000hn,Risse:2025smp} at NLO to compute the corresponding structure functions. This scheme is based on the ACOT (Aivazis–Collins–Olness–Tung) framework \cite{Aivazis:1993kh,Aivazis:1993pi} and belongs to the class of general-mass variable flavor number schemes (GM-VFNS)~\cite{Thorne:2010pa}, which consistently account for heavy-quark mass effects across the full kinematic range of DIS. Within GM-VFNS, the number of active parton flavors evolves dynamically with the energy scale $Q$. Heavy quarks are treated as massless partons above their mass threshold, while their full mass dependence is retained in the hard-scattering coefficient functions governing flavor-creation processes. The SACOT-$\chi$ prescription introduces a modified scaling variable, $\chi=x(1+\frac{m_h^2}{Q^2})$, where $m_h$ denotes the heavy-quark mass. This rescaling enforces the correct kinematic threshold behavior when $Q^2 \sim m_h^2$ and ensures a smooth transition to the massless limit at high $Q^2$, thereby avoiding unphysical contributions from heavy-quark PDFs near threshold. The scheme is therefore both theoretically robust and numerically stable, making it particularly well suited for describing charm production in neutrino DIS.

The calculations are performed using the \texttt{APFEL++} framework~\cite{Bertone:2017gds, Risse:2023rxd}, a flexible and efficient tool for computing DIS structure functions and cross sections within collinear factorization. \texttt{APFEL++} implements QCD evolution at fixed order and supports general-mass schemes such as SACOT-$\chi$. In addition, its grid-based design allows for fast evaluation of theory predictions during the global fit, making it well suited for iterative likelihood evaluations within the MCMC framework.
To ensure that the theoretical predictions remain within the perturbative regime of QCD and to suppress potential higher-twist and target-mass effects, standard DIS kinematic cuts are applied:
\begin{equation*}
Q^2 > 4.0~\mathrm{GeV}^2, 
\qquad \text{and} \qquad 
W^2 > 12.25~\mathrm{GeV}^2.
\end{equation*}
\begin{table}[t]
    \centering
    \caption[Neutrino DIS data from the CHORUS experiment]{Neutrino DIS data from the CHORUS experiment. $N_{\text{data}}$ denotes the number of data points after/before applying kinematic cuts.}
    %%%%%%%%%%%%%%%%%%%%%%%%%%%%%%%%%%%%%%%%%%%%%%%%
% DIS Neutrino data
%%%%%%%%%%%%%%%%%%%%%%%%%%%%%%%%%%%%%%%%%%%%%%%%%
\begingroup
\small
\renewcommand{\arraystretch}{0.95}  % default is 1.0
\begin{tabular}{|l l c c c |}
  \hline
  \multicolumn{5}{|c|}{Neutrino DIS data} \\
  \hline \hline
  Exp. & Observable & ID & Ref. & $N_{\text{data}}$ \\
  \hline
  \multirow{2}{*}{Chorus}
    & $\nu$       & 5946 & \multirow{2}{*}{\cite{Onengut:2005kv}} & 412/607 \\
    & $\bar{\nu}$ & 5947 &                                        & 412/607 \\
  \hline \hline
  \textbf{Total} &       &      &                                  & \textbf{824/1214} \\
  \hline
\end{tabular}
\endgroup
    \label{tab:nudata}
\end{table}

\paragraph{Heavy-Flavor and Quarkonium Production:} Open heavy-flavor and quarkonium production in proton-lead ($p$Pb) collisions provide powerful constraints on the nuclear gluon distribution, since the dominant partonic channel for both processes is gluon-gluon fusion over a wide kinematic range. Open heavy-flavor meson production, such as $D^0$ and $B^+$ mesons, is dominated at leading order by
\[
g + g \to Q + \bar{Q},
\]
where $Q$ denotes a heavy quark (charm or bottom). The produced heavy quark and antiquark subsequently hadronize into heavy-light mesons, e.g.\ $D^0 = c\bar{u}$ and $B^+ = b\bar{u}$. Quarkonium production, such as $J/\psi$ and $\Upsilon$ states, also originates predominantly from gluon-gluon fusion,
\[
g + g \to Q\bar{Q}[n] \rightarrow \text{quarkonium},
\]
where the heavy-quark pair is produced in an intermediate quantum state $[n]$ before undergoing a nonperturbative transition into a bound quarkonium state. The $J/\psi$ meson corresponds to a $c\bar{c}$ bound state (charmonium), while the $\Upsilon$ family corresponds to $b\bar{b}$ bound states (bottomonium). The datasets included in this analysis are summarized in Table~\ref{tab:HQdata} and comprise measurements of open heavy-flavor mesons and quarkonium states from LHC experiments.
\begin{table}[!h]
    \centering
    \caption[Heavy flavor and quarkonium production data from the LHC]{Heavy flavor and quarkonium production data from the LHC. $N_{\text{data}}$ denotes the number of data points after/before applying kinematic cuts.}
    %%%%%%%%%%%%%%%%%%%%%%%%%%%%%%%%%%%%%%%%%%%%%%%%
% LHC heavy quark production in proton-lead collsions
%%%%%%%%%%%%%%%%%%%%%%%%%%%%%%%%%%%%%%%%%%%%%%%%
\begingroup
\small
\renewcommand{\arraystretch}{0.95}  % default is 1.0
\begin{tabular}{|l l c c c c|}
    \hline
    \multicolumn{6}{|c|}{LHC heavy flavour production in $p$Pb $d^2\sigma/dydp_t$}\\
    \hline
    \hline
    Exp. & Observable & $\sqrt{s}$ [TeV] & \;ID & Ref. & $N_{\text{data}}$ \\
    \hline 
    \hline
    %%%%%%%%%%%%%%% ALICE
    \multirow{10}{*}[-0.0em]{ALICE} 
    & $\text{prompt}~D^0$ & \;\;\;5.02 & 3101 & \cite{ALICE:2014xjz} & 6/10 \\
    & $\text{incl.}~J/\Psi$ & \;\;\;5.02 & 3103 & \cite{ALICE:2013snh} & 0/12 \\
    & $\text{incl.}~J/\Psi$ & \;\;\;5.02 & 3104 & \cite{ALICE:2015sru} & 10/25 \\
    & $\text{incl.}~J/\Psi$ & 8.16 & 3112 & \cite{ALICE:2018mml} & 9/24 \\
    & $\text{incl.}~\Upsilon(1S)$ & 8.16 & 3114 & \cite{ALICE:2019qie} & 3/10 \\
    & $\text{prompt}~D^0$ & \;\;\;5.02 & 3122 & \cite{ALICE:2019fhe} & 13/21 \\
    & $\text{prompt}~D^0$ & \;\;\;5.02 & 3123 & \cite{ALICE:2016yta} & 6/11 \\
    & $\text{incl.}~\Psi(2S)$ & \;\;\;8.16 & 3126 & \cite{ALICE:2020vjy} & 3/10 \\
    & $\text{incl.}~\Psi(2S)$ & \;\;\;5.02 & 3127 & \cite{ALICE:2014cgk} & 2/8 \\
    & $\text{incl.}~\Upsilon(1S)$ & \;\;\;8.16 & 3110 & \cite{ALICE:2014ict} & 0/4 \\
    \cline{1-6}
    %%%%%%%%%%%%%%% ATLAS
    \multirow{7}{*}[-0.0em]{ATLAS} 
    & $\text{incl.}~\Upsilon(1S)$  & \;\;\;5.02 & 3109 & \cite{ATLAS:2017prf} & 6/8 \\
    & $\text{non-pr.}~J/\Psi$ & \;\;\;5.02 & 3116 & \cite{ATLAS:2017prf} & 8/8 \\
    & $\text{prompt}~J/\Psi$ & \;\;\;5.02 & 3117 & \cite{ATLAS:2017prf} & 8/8 \\
    & $\text{prompt}~J/\Psi$ & \;\;\;5.02 & 3118 & \cite{ATLAS:2015mpz} & 10/10 \\
    & $\text{non-pr.}~J/\Psi$ & \;\;\;5.02 & 3119 & \cite{ATLAS:2015mpz} & 10/10 \\
    & $\text{prompt}~\Psi(2S)$ & \;\;\;5.02 & 3124 & \cite{ATLAS:2017prf} & 8/8 \\
    & $\text{non-pr.}~\Psi(2S)$ & \;\;\;5.02 & 3125 & \cite{ATLAS:2017prf} & 8/8 \\
    \cline{1-6}
    %%%%%%%%%%%%%%% CMS
    \multirow{3}{*}[-0.0em]{CMS\;} 
    & $\text{prompt}~\Psi(2S)$ & \;\;\;5.02 & 3115 &\cite{CMS:2018gbb} & 17/17 \\
    & $\text{prompt}~J/\Psi$ & \;\;\;5.02 & 3120 &\cite{CMS:2017exb} & 51/53 \\
    & $\text{non-pr.}~J/\Psi$ & \;\;\;5.02 & 3121 &\cite{CMS:2017exb} & 51/53 \\
    \cline{1-6}
    %%%%%%%%%%%%%%% LHCb
    \multirow{5}{*}[-0.0em]{LHCb} 
    & $\text{prompt}~D^0$ & \;\;\;5.02 & 3102 & \cite{LHCb:2017yua} & 53/92 \\
    & $\text{prompt}~J/\Psi$ & \;\;\;8.16 & 3105 & \cite{LHCb:2017ygo} & 88/140 \\
    & $\text{non-pr.}~J/\Psi$ & \;\;\;8.16 & 3106 & \cite{LHCb:2017ygo} & 88/140 \\
    & $\text{non-pr.}~J/\Psi$ & \;\;\;5.02 & 3107 & \cite{LHCb:2013gmv} & 25/40 \\
    & $\text{prompt}~J/\Psi$ & \;\;\;5.02 & 3108 & \cite{LHCb:2013gmv} & 25/40 \\
    & $\text{incl.}~\Upsilon(1S)$ & 5.02 & 3111 & \cite{LHCb:2014rku} & 0/2 \\
    & $\text{incl.}~\Upsilon(1S)$ & 8.16 & 3113 & \cite{LHCb:2018psc} & 36/46 \\
    \hline 
    \hline 
    \textbf{Total} &  &  & & & \hspace{-0.35cm}\textbf{544/717} \\
    \hline
\end{tabular}
\endgroup
    \label{tab:HQdata}
\end{table}

Incorporating these observables into a global nPDF fit is nontrivial, particularly for quarkonium production, where no universally accepted first-principles framework provides both precise and computationally efficient predictions across all relevant kinematics. To incorporate these diverse processes into the global fit efficiently, we therefore adopt a data-driven approach rather than relying exclusively on NLO perturbative QCD calculations. Following the strategy developed in the nCTEQ15HQ analysis~\cite{Duwentaster:2022kpv}, the single-inclusive production cross section is modeled assuming gluon-gluon dominance and an effective $2\to2$ kinematic description of the hard scattering. Within this framework, the production cross section is expressed as a convolution of the initial-state gluon PDFs with an effective scattering matrix element. This matrix element is parameterized using a \emph{Crystal Ball} functional form \cite{Kom:2011bd, Kusina:2017gkz}, whose parameters control the overall normalization and the shape of the transverse-momentum distribution. These parameters are determined by fits to corresponding proton-proton baseline data for each final state. In contrast to the treatment of neutrino DIS discussed earlier, where standard nCTEQ structure function calculations employ the SACOT-$\chi$ implementation of the GM-VFNS, the Crystal Ball parameterization for hadronic heavy-flavor and quarkonium production allows complex short-distance dynamics, hadronization effects, and possible higher-order contributions to be absorbed into a small set of effective parameters. This data-driven formulation offers significant computational advantages for repeated likelihood evaluations within the MCMC framework, making it particularly well suited for large-scale sampling. Its validity has been carefully assessed in the nCTEQ15HQ analysis through detailed comparisons with established NLO calculations. For open heavy-flavor meson production, the effective approach was benchmarked against NLO predictions in the GM-VFNS, demonstrating good agreement in both shape and normalization. For quarkonium production, comparisons with NLO calculations based on Non-Relativistic QCD (NRQCD)\footnote{Nonrelativistic QCD \cite{Bodwin:1994jh} is an effective field theory that describes the production and decay of heavy quarkonium by separating short-distance perturbative dynamics from long-distance nonperturbative matrix elements associated with quarkonium formation. It provides a systematic expansion in the heavy-quark velocity inside the bound state.} showed consistency within theoretical uncertainties.
In our analysis, the heavy-flavor and quarkonium datasets are restricted to kinematic regions where collinear factorization and the data-driven modeling approach are expected to be reliable. We follow the kinematic selections adopted in the nCTEQ15HQ analysis,
\begin{equation*}
  P_T > 3.0~\mathrm{GeV}, 
  \qquad \text{and} \qquad
  -4.0<y<4.0.
\end{equation*}
The lower transverse-momentum cut suppresses contributions from the low-$p_T$ region, where nonperturbative and higher-twist effects become increasingly important and the simple Crystal Ball parametrization becomes unreliable. The rapidity restriction ensures compatibility with the phase-space coverage of the included LHC measurements. These selections are imposed both in the proton-proton baseline determination and in the nPDF fit.

\subsubsection{Multi-nuclei Analysis}
To explore the $A$-dependence of nuclear parton distribution functions, we perform a second global fit that incorporates experimental data from a broad range of nuclear targets beyond lead. While this \texttt{Multi-nuclei} analysis is statistically less precise than the \texttt{Pb-only} fit due to reduced MCMC sampling, it provides an essential cross-check of the conclusions drawn from the \texttt{Pb-only} study and offers insight into the systematics of nuclear modifications across different nuclei.
The \texttt{Multi-nuclei} fit includes all datasets used in the \texttt{Pb-only} analysis, electroweak vector boson production in $p$Pb collisions, heavy-flavor production, and neutrino DIS on lead, and extends the dataset by incorporating fixed-target DIS and Drell-Yan measurements from a wide range of experiments and nuclear targets. These additional data provide sensitivity to both valence and sea-quark distributions and allow nuclear effects to be probed across a broader range of Bjorken-$x$. In particular, the following datasets are included:
\begin{table}
    \centering
    \caption[NC DIS data including ratios to deuterium]{NC DIS data including ratios to deuterium, $F_2^A/F_2^D$ and $\sigma_{\text{DIS}}^{A}/\sigma_{\text{DIS}}^{D}$, as well as ratios between two nuclei, $F_{2}^{A}/F_{2}^{A'}$. The quantity $N_{\text{data}}$ denotes the number of data points after/before applying kinematic cuts.}
    \begingroup
\small
\renewcommand{\arraystretch}{0.95}
\begin{tabular}{|l l c c c|}
\hline
\multicolumn{5}{|c|}{\textbf{NC DIS data}} \\
\hline
Ratio & Experiment & ID &Ref. & $N_{\text{data}}$ \\
\hline
\multicolumn{5}{c}{$\mathbf{F_2^A/F_2^D}$ and $\bm{\sigma_{\text{DIS}}^{A}/\sigma_{\text{DIS}}^{D}}$} \\
\hline
${}^3\mathrm{He}/\mathrm{D}$ & Hermes & 5156 & \cite{HERMES:2002llm} & 17/182 \\
\hline
\multirow{2}{*}{${}^4\mathrm{He}/\mathrm{D}$} & NMC-95,re & 5124 & \cite{Amaudruz:1995tq} & 12/18 \\
 & SLAC-E139 & 5141 & \cite{Gomez1994} & 3/18 \\
\hline
$\mathrm{Li}/\mathrm{D}$ & NMC-95 & 5115 & \cite{Arneodo:1995cs} & 11/24 \\
\hline
$\mathrm{Be}/\mathrm{D}$ & SLAC-E139 & 5138 & \cite{Gomez1994} & 3/17 \\
\hline
\multirow{6}{*}{$\mathrm{C}/\mathrm{D}$} & FNAL-E665-95 & 5125 & \cite{Adams:1995is} & 3/11 \\
 & SLAC-E139 & 5139 & \cite{Gomez1994} & 2/7 \\
 & EMC-88 & 5107 & \cite{Ashman:1988bf} & 9/9 \\
 & EMC-90 & 5110 & \cite{Arneodo:1989sy} & 0/9 \\
 & NMC-95 & 5113 & \cite{Arneodo:1995cs} & 12/24 \\
 & NMC-95,re & 5114 & \cite{Amaudruz:1995tq} & 12/18 \\
\hline
\multirow{2}{*}{$\mathrm{N}/\mathrm{D}$} & Hermes & 5157 & \cite{HERMES:2002llm} & 19/175 \\
 & BCDMS-85 & 5103 & \cite{Bari:1985ga} & 9/9 \\
\hline
\multirow{2}{*}{$\mathrm{Al}/\mathrm{D}$} & SLAC-E049 & 5134 & \cite{Bodek:1983ec} & 0/18 \\
 & SLAC-E139 & 5136 & \cite{Gomez1994} & 3/17 \\
\hline
\multirow{4}{*}{$\mathrm{Ca}/\mathrm{D}$} & NMC-95,re & 5121 & \cite{Amaudruz:1995tq} & 12/18 \\
 & FNAL-E665-95 & 5126 & \cite{Adams:1995is} & 3/11 \\
 & SLAC-E139 & 5140 & \cite{Gomez1994} & 2/7 \\
 & EMC-90 & 5109 & \cite{Arneodo:1989sy} & 0/9 \\
\hline
\multirow{5}{*}{$\mathrm{Fe}/\mathrm{D}$} & SLAC-E049 & 5131 & \cite{Bodek:1983qn} & 2/14 \\
 & SLAC-E139 & 5132 & \cite{Gomez1994} & 6/23 \\
 & SLAC-E140 & 5133 & \cite{Dasu:1993vk} & 0/10 \\
 & BCDMS-87 & 5101 & \cite{Benvenuti:1987az} & 10/10 \\
 & BCDMS-85 & 5102 & \cite{Bari:1985ga} & 6/6 \\
\hline
\multirow{3}{*}{$\mathrm{Cu}/\mathrm{D}$} & EMC-93 & 5104 & \cite{Ashman:1992kv} & 9/10 \\
 & EMC-93 (chariot) & 5105 & \cite{Ashman:1992kv} & 9/9 \\
 & EMC-88 & 5106 & \cite{Ashman:1988bf} & 9/9 \\
\hline
$\mathrm{Kr}/\mathrm{D}$ & Hermes & 5158 & \cite{HERMES:2002llm} & 12/167 \\
\hline
$\mathrm{Ag}/\mathrm{D}$ & SLAC-E139 & 5135 & \cite{Gomez1994} & 2/7 \\
\hline
$\mathrm{Sn}/\mathrm{D}$ & EMC-88 & 5108 & \cite{Ashman:1988bf} & 8/8 \\
\hline
$\mathrm{Xe}/\mathrm{D}$ & FNAL-E665-92 & 5127 & \cite{Adams:1992nf} & 2/10 \\
\hline
$\mathrm{Au}/\mathrm{D}$ & SLAC-E139 & 5137 & \cite{Gomez1994} & 3/18 \\
\hline
$\mathrm{Pb}/\mathrm{D}$ & FNAL-E665-95 & 5129 & \cite{Adams:1995is} & 3/11 \\
\hline
\multicolumn{5}{c}{$\mathbf{F_2^A/F_2^{A'}}$ } \\
\hline
$\mathrm{C}/\mathrm{Li}$ & NMC-95,re & 5123 & \cite{Amaudruz:1995tq} & 7/25 \\
\hline
$\mathrm{Ca}/\mathrm{Li}$ & NMC-95,re & 5122 & \cite{Amaudruz:1995tq} & 7/25 \\
\hline
$\mathrm{Be}/\mathrm{C}$ & NMC-96 & 5112 & \cite{Arneodo:1996rv} & 14/15 \\
\hline
$\mathrm{Al}/\mathrm{C}$ & NMC-96 & 5111 & \cite{Arneodo:1996rv} & 14/15 \\
\hline
\multirow{2}{*}{$\mathrm{Ca}/\mathrm{C}$} & NMC-95,re & 5120 & \cite{Amaudruz:1995tq} & 7/25 \\
 & NMC-96 & 5119 & \cite{Arneodo:1996rv} & 14/15 \\
\hline
$\mathrm{Fe}/\mathrm{C}$ & NMC-96 & 5143 & \cite{Arneodo:1996rv} & 14/15 \\
\hline
$\mathrm{Sn}/\mathrm{C}$ & NMC-96 & 5159 & \cite{Arneodo:1996ru} & 111/146 \\
\hline
$\mathrm{Pb}/\mathrm{C}$ & NMC-96 & 5116 & \cite{Arneodo:1996rv} & 14/15 \\
\hline
\hline
\textbf{Total} &  &  &  & \textbf{415/1209} \\
\hline
\end{tabular}
\endgroup

    \label{tab:NC_DIS}
\end{table}
\begin{itemize}
    \item \textbf{Neutral-Current (NC) DIS:} These data originate from charged-lepton scattering experiments such as SLAC, NMC, EMC, and BCDMS, and cover a wide variety of nuclear targets, including deuterium, helium, lithium, beryllium, carbon, calcium, iron, and heavier nuclei. The NC DIS datasets summarized in Table~\ref{tab:NC_DIS} include both ratios to deuterium, $F_2^A/F_2^D$ and $\sigma_{\mathrm{DIS}}^{A}/\sigma_{\mathrm{DIS}}^{D}$, as well as ratios between two nuclear targets, $F_2^{A}/F_2^{A'}$. Theoretical predictions for the NC DIS observables are computed within the collinear factorization framework at NLO in perturbative QCD. Consistent with the neutrino DIS analysis described earlier, heavy-quark mass effects are treated via the SACOT-$\chi$ GM-VFNS, and structure functions are evaluated using the \texttt{APFEL++} library. To ensure the applicability of perturbative QCD and to suppress regions strongly affected by higher-twist contributions, standard DIS kinematic cuts of
    \[
    Q^2 > 4.0~\mathrm{GeV}^2, \qquad W^2 > 12.25~\mathrm{GeV}^2
    \]
    are imposed. The NC DIS data play a central role in constraining quark distributions through the nuclear structure function $F_2^A(x,Q^2)$. They are particularly effective in determining valence-quark distributions and provide precise constraints in the EMC region at large $x$.

    \item \textbf{Drell--Yan (DY) production:} The DY data are taken from fixed-target proton-nucleus experiments, specifically FNAL-E772-90 and FNAL-E886-99, which measured dimuon production via quark-antiquark annihilation,
    \[
    q + \bar{q} \to \gamma^* \to \mu^+ \mu^-.
    \]
    These measurements, summarized in Table~\ref{tab:DY}, are reported as ratios of DY cross sections for heavier nuclear targets (e.g., C, Ca, Fe, W) relative to lighter reference nuclei (D or Be). Theoretical predictions for DY production are computed at NLO in perturbative QCD within the standard collinear factorization framework, and the calculations are performed directly using an internal code without precomputed interpolation grids. The DY process is particularly sensitive to nuclear sea-quark distributions, especially $\bar{u}$ and $\bar{d}$, at moderate to large values of $x$. As such, these data provide complementary constraints to DIS measurements and play an important role in disentangling nuclear effects in the valence- and sea-quark sectors.
\end{itemize}
\begin{table}[!h]
        \centering
        \caption[Drell–Yan data as cross-section ratios of heavy to light nuclei]{Drell–Yan data as cross-section ratios of heavy to light nuclei; $N_{\text{data}}$ denotes the number of data points after/before applying kinematic cuts.}
        %%%%%%%%%%%%%%%%%%%%%%%%%%%%%%%%%%%%%%%%%%%%%%%%
% DY data sigmaA/sigmaA'
%%%%%%%%%%%%%%%%%%%%%%%%%%%%%%%%%%%%%%%%%%%%%%%%
\begingroup
\small
\renewcommand{\arraystretch}{0.95}  % default is 1.0
\begin{tabular}{|l l c c c|}
    \hline
    \multicolumn{5}{|c|}{Fixed target DY data $\mathbf{\sigma_{DY}^{pA}/\sigma_{DY}^{pA'}}$}\\
    \hline
    \hline
    Nucleus & Experiment &  ID & Ref. &  $N_{\text{data}}$ \\
    \hline
    C/D     & FNAL-E772-90 & 5203 & \cite{Alde:1990im}  & 9/9 \\
    \hline 
    Ca/D & FNAL-E772-90 & 5204 & \cite{Alde:1990im}  & 9/9 \\
    \hline 
    Fe/D & FNAL-E772-90 & 5205 & \cite{Alde:1990im}  & 9/9 \\
    \hline 
    W/D & FNAL-E772-90 & 5206 & \cite{Alde:1990im}  & 9/9 \\
    \hline 
    Fe/Be & FNAL-E886-99 & 5201 & \cite{Vasilev:1999fa}  & 28/28 \\
    \hline 
    W/Be & FNAL-E886-99 & 5202 & \cite{Vasilev:1999fa}  & 28/28 \\
    \hline 
    \hline 
    \textbf{Total} &  &  &   & \textbf{92/92} \\
    \hline 
\end{tabular}
\endgroup
%%%%%%%%%%%%%%%%%%%%%%%%%%%%%%%%%%%%%%%%%%%%%%%%
        \label{tab:DY}
    \end{table}
While including several nuclear targets significantly broadens the kinematic coverage of the analysis, it also introduces substantial computational challenges. In the \texttt{Multi-nuclei} fit, theoretical predictions must be evaluated separately for each nucleus, as the parton distributions undergo independent QCD evolution and are subject to nuclear modifications that depend explicitly on both the mass number $A$ and the atomic number $Z$. Although the high-speed interpolation grids for each nucleus are constructed only once and reused throughout the fit, the evaluation of theory predictions for multiple nuclei requires separate initialization and consideration of nucleus-specific inputs at each likelihood evaluation. In practice, each evaluation of the global $\chi^2$ function involves looping over a larger number of datasets and nuclear targets, initializing the corresponding nuclear PDFs, and performing the required convolutions with partonic cross sections. This leads to a significantly higher computational cost per likelihood evaluation compared to the \texttt{Pb-only} case. As a consequence, the effective sampling rate of the MCMC algorithm is reduced for a fixed timeframe, resulting in a smaller total number of samples in the \texttt{Multi-nuclei} fit relative to the \texttt{Pb-only} analysis. Nevertheless, the available statistics remain sufficient for a stable determination of the nPDF parameters and their associated uncertainties. For this reason, the \texttt{Multi-nuclei} fit is treated as a complementary and exploratory study, while the primary results of this thesis are based on the high-statistics \texttt{Pb-only} analysis.

\subsubsection{Treatment of Normalization Uncertainties}
Experimental datasets used in global PDF analyses often carry an overall normalization uncertainty, typically associated with luminosity measurements or global efficiency factors. While this uncertainty can be interpreted as an overall rescaling of the data, in practice the experimental measurements are kept fixed and the uncertainty is incorporated through the theoretical predictions. This approach preserves the integrity of the published data and ensures a statistically consistent treatment within the fit.
A naive implementation of normalization uncertainties, such as directly rescaling the data points, can lead to the so-called \emph{d’Agostini bias}~\cite{DAGOSTINI1994306}, in which the fitted theory is systematically biased toward lower values. To avoid this effect, normalization uncertainties are treated at the level of the theoretical predictions.

A useful way to motivate this treatment is through the introduction of a multiplicative nuisance parameter $N$, leading to an extended $\chi^2$ function
\[
\chi^2(\theta, N) = \sum_i \frac{(D_i - N\,T_i(\theta))^2}{\sigma_i^2}
+ \frac{(N - 1)^2}{\sigma_{\text{norm}}^2},
\]
where $\{D_i\}$ denotes the experimental measurements, $T_i(\theta)$ are the theoretical predictions evaluated at parameter point $\theta$, and $\sigma_i$ are the uncorrelated statistical uncertainties. The parameter $\sigma_{\text{norm}}$ represents the normalization uncertainty, which is incorporated as a Gaussian constraint on $N$ centered at unity.
While this formulation is statistically sound, it poses practical challenges for sampling-based methods such as Markov Chain Monte Carlo. At each point in the parameter space $\theta$, the nuisance parameter $N$ would need to be either optimized or marginalized numerically, which becomes computationally inefficient when many datasets with independent normalization uncertainties are included. To avoid the explicit introduction of normalization nuisance parameters in the fit, we employ the so-called \emph{$T$-method}~\cite{Muzakka:2022wey}. While conceptually related to the iterative $t_0$-method commonly used in NNPDF analyses~\cite{Ball:2009qv}, the $T$-method provides an analytic marginalization of normalization uncertainties under the assumption of Gaussian errors, making it particularly well suited for an MCMC-based fitting framework. This approach allows normalization uncertainties to be consistently propagated into the likelihood without increasing the dimensionality of the parameter space. The analytic marginalization results in an additional contribution to the covariance matrix,
\[
C_{\text{norm},ij} = \sigma_{\text{norm}}^2\, T_i(\theta)\, T_j(\theta),
\]
which reflects the fully correlated nature of normalization uncertainties across all data points within a given dataset.
The total covariance matrix used in the analysis is therefore given by
\begin{align}
C_{T,ij}(\theta) =
\sigma_i^2\,\delta_{ij}
+ \sum_\alpha \bar{\sigma}_{i\alpha}\,\bar{\sigma}_{j\alpha}
+ \sigma_{\text{norm}}^2\, T_i(\theta)\, T_j(\theta),
\label{eq:6.1}
\end{align}
where the first term accounts for uncorrelated statistical uncertainties, the second term represents correlated systematic uncertainties from source $\alpha$, and the final term encodes the normalization uncertainty obtained through the $T$-method.
With this definition, the $\chi^2$ function entering the likelihood becomes
\[
\chi^2(\theta) =
\sum_{i,j}
\bigl[D_i - T_i(\theta)\bigr]\,
\bigl(C_T^{-1}(\theta)\bigr)_{ij}\,
\bigl[D_j - T_j(\theta)\bigr].
\]
This formulation consistently incorporates statistical, systematic, and normalization uncertainties into the posterior distributions sampled by the MCMC algorithm. Although the dependence of the covariance matrix on the theory predictions increases the computational cost of the fit, since the covariance matrix must be recomputed at each step in parameter space, this approach is essential for avoiding normalization-induced biases and for achieving a statistically robust determination of nuclear PDFs that faithfully propagates correlated uncertainties and captures non-Gaussian features in the final error estimates.

\subsection{Nuclear PDF Parameterization} \label{sec:6.1.2}
This study employs the nCTEQ framework \cite{Kovarik:2015cma} to parameterize nuclear parton distribution functions. Under the assumption of isospin symmetry, the PDF of a nucleus with atomic number $Z$ and mass number $A$, denoted $f_i^{(A,Z)}(x,Q)$, is written as a linear combination of the bound proton ($f^{p/A}$) and neutron PDFs ($f^{n/A}$),
\begin{equation}
f_{i}^{(A,Z)}(x,Q)= \frac{Z}{A} f_{i}^{p/A}(x,Q) + \frac{A-Z}{A} f_{i}^{n/A}(x,Q),
\label{eq:6.2}
\end{equation}
where the bound neutron distributions are obtained from the bound proton PDFs by isospin symmetry. The bound proton PDFs are parameterized at the initial scale $Q_0 = 1.3~\mathrm{GeV}$,\footnote{The choice $Q_0 = 1.3~\mathrm{GeV}$ corresponds to the charm-quark mass scale, above which charm is treated as an active parton in the PDF evolution.} using the CJ15 proton PDFs~\cite{Accardi:2016qay} as the baseline reference. For most prton flavors, the functional form at the initial scale is taken to be
\begin{equation}
\begin{split}
xf_i^{p/A}(x,Q_0) &= c_0^ix^{c^i_1}(1-x)^{c^i_2}\left(1+c^i_3\sqrt{x}+c^i_4x\right), \\
i &= {u_v,\bar{u}+\bar{d},\bar{d}-\bar{u},g,s+\bar{s}},
\end{split}
\label{eq:6.3}
\end{equation}
which provides sufficient flexibility to describe the distributions over the full $x$ range. An exception is made for the down-valence distribution, for which additional freedom is introduced through a coupling to the up-valence PDF,
\begin{equation}
\begin{split}
    xd_v^{p/A}(x,Q_0) &= c^{d_v}_0 \Big[
      x^{c^{d_v}_1} (1-x)^{c^{d_v}_2} (1+c^{d_v}_3\sqrt{x}+c^{d_v}_4x) \\
      &\qquad\qquad+ c^{d_v}_5x^{c^{d_v}_6} \,xu_v^{p/A}(x,Q_0)
      \Big],
\end{split}  
\label{eq:6.4}
\end{equation}
allowing for a more flexible description of the relative behavior of the valence quark distributions at intermediate and large $x$. To account for a possible light-sea flavor asymmetry, the ratio of down to up antiquark distributions is parameterized independently as
\begin{equation}
\begin{split}
  \frac{\bar{d}}{\bar{u}}(x,Q_0) &= c_0^{i}x^{c_1^{i}}(1-x)^{c_2^{i}} + 1 + c_3^{i}x(1-x)^{c_4^{i}},\\
  i&=\bar{d}/\bar{u}.
\end{split}
\end{equation}
In the present analysis, no asymmetry is allowed in the strange sector, such that $s(x) = \bar{s}(x)$ at the input scale. Nuclear modifications are introduced through an explicit dependence of the PDF parameters on the nuclear mass number $A$. Each coefficient $c^i_j$ is written as
\begin{equation}
c^i_j(A) = p_j + a_j\ln A + b_j\ln^2A,
\label{eq:6.6}
\end{equation}
which guarantees consistency with the free-proton limit at $A = 1$, where the coefficients reduce to the baseline values $p_j$. In practice, the proton baseline parameters $p_j$ are taken directly from the CJ15 analysis~\cite{Accardi:2016qay}, as summarized in Table ~\ref{tab:proton-baseline}, and are held fixed throughout the fit. All nuclear effects are therefore encoded in the coefficients $a_j$ (and, in general, $b_j$). This parameterization is applied consistently in both the \texttt{Pb-only} and \texttt{Multi-nuclei} analyses. In the \texttt{Pb-only} fit, where data from a single nuclear target are considered, the dependence on $\ln A$ effectively reduces to a constant evaluated at $A=208$. In principle, one could instead fit nucleus-specific parameters directly. However, expressing the nuclear modifications through Eq.~\eqref{eq:6.6} ensures that the fitted coefficients $a_j$ maintain a clear interpretation as nuclear-modification parameters and remain directly comparable between the \texttt{Pb-only} and \texttt{Multi-nuclei} fits. To fully specify the PDF parameterization at the input scale, number and momentum sum rules are imposed to constrain the normalization coefficients $c_0$. The valence number sum rules are used to fix the normalizations of the $u_v$ and $d_v$ distributions for each nucleus. The momentum sum rule is enforced to ensure that the total momentum carried by all partons equals that of the parent nucleon. As a consequence, the normalization of the light sea, $\bar{u}+\bar{d}$, is fixed. Since the momentum sum rule provides only a single constraint, the remaining normalization coefficients of the gluon and $s+\bar{s}$, must be treated separately, either by fitting them to data or by imposing additional physical assumptions.
\begin{table}[t]
    \centering
    \caption[Fixed proton baseline parameters taken from CJ15]{Fixed proton baseline parameters \( p_j \) at $Q_0=1.3$ GeV, taken from CJ15 \cite{Accardi:2016qay}.}
    \begin{tabular}{c|cccccc}
\textsc{Parameter}	& $u_v$
		& $d_v$
		& $\bar d+\bar u$
		& $\bar d/\bar u$
		& $g$				
            & $s+\bar s$\\ \hline
$p_0$		& --
		& --
		& --
		& 35712
		& 0.3887
            & 0.50002\\
$p_1$		& 0.61537
		& 1.1595
		& -0.20775
		& 4.0249
		& 0.60307
		& -0.20775\\
$p_2$		& 3.5433
		& 6.5514
		& 8.3286
		& 20.154
		& 6.4812
		& 8.3286\\
$p_3$		& 0
		& -3.503
		& 0
		& 17.0
		& -3.3064
		& 0\\
$p_4$		& 3.4609
		& 4.6787
		& 14.606
		& 51.156
		& 3.1721
		& 14.606\\ 
$p_5$       & 0
            & 0.0036
            & 0
            & 0
            & 0
            & 0\\ 
$p_6$       & --
            & 2.0
            & --
            & --
            & --
            & --\\
\hline
\end{tabular}
    \label{tab:proton-baseline}
\end{table}
\begin{table}[t]
    \centering
    \caption{Fitted parameters $a_j$ used in the nPDF analysis.}
    \begin{tabular}{c|cccccc}
\textsc{Parameter}	& $u_v$
		& $d_v$
		& $\bar d+\bar u$
		& $\bar d/\bar u$
		& $g$				
        & $s+\bar s$\\ \hline
$a_0$	& \texttt{sum rule}
		& \texttt{sum rule}
		& \texttt{sum rule}
		& 0
		& 0
        & 0\\
$a_1$	& \texttt{fit}
		& \texttt{fit}
		& \texttt{fit}
		& 0
		& \texttt{fit}
		& 0\\
$a_2$	& \texttt{fit}
		& \texttt{fit}
		& \texttt{fit}
		& 0
		& \texttt{fit}
		& 0\\
$a_3$	& \texttt{fit}
		& \texttt{fit}
		& 0
		& 0
		& 0
		& 0\\
$a_4$	& 0
		& 0
		& 0
		& 0
		& 0
		& 0\\ 
$a_5$   & 0
        & 0
        & 0
        & 0
        & 0
        & 0\\ 
$a_6$   & --
        & 0
        & --
        & --
        & --
        & --\\
\hline
\end{tabular}
    \label{tab:fit-parameters}
\end{table} 
For the purposes of this study, the fit is restricted to a total of ten nuclear-modification parameters $a_j$, as listed in Table~\ref{tab:fit-parameters}. This choice represents a balance between computational efficiency and sufficient flexibility to capture the dominant nuclear effects. Specifically, three parameters control the nuclear modification of the valence distributions, while two parameters are assigned to the $\bar{u}+\bar{d}$ and gluon PDFs. To further reduce complexity and stabilize the fit, all quadratic coefficients $b_j$ are set to zero, corresponding to a linear dependence of the nuclear modifications on $\ln A$. All nuclear-modification parameters not included in the fit are fixed to zero. The fitted parameters $a_j$ are assigned initial starting values at the beginning of the MCMC sampling and are subsequently allowed to vary freely during the fit in response to the experimental data, thereby generating the nuclear modifications required by the global analysis.

%%%%%%%%%%%%%%%%%%%%%%%%%%%%%%%%%%%%%%%%%%%%%%%%%%%%%%%%%%%%%%%%%%%%%%%%%%%
\section{MCMC Implementation} \label{sec:6.2}
\subsection{Algorithm}\label{sec:6.2.1}
The Metropolis-Hastings (MH) algorithm is a fundamental and widely used Markov Chain Monte Carlo method for generating a sequence of samples from a target probability distribution $\pi$, as discussed in Sec.~\ref{sec:5.3.1}.  In this framework, the Markov chain evolves in the parameter space of the model. At each iteration $t$, the state of the chain is represented by a parameter vector $\bm{\theta}_{t} =[ \theta_{t}^1$, $\theta_{t}^2$,..., $\theta_{t}^d]^T\in \mathbb{R}^d$, where $d$ denotes the dimensionality of the nPDF parameter space. Each state $\bm{\theta}_{t}$ therefore corresponds to a complete configuration of the model parameters, in this case a full set of nPDF parameters. At iteration $t$, the MH algorithm proposes a new candidate state $\bm{\theta}^*_{t+1}$ by drawing from a proposal distribution $q_t \equiv q(\bm{\theta}^*_{t+1}\mid\bm{\theta}_{t} )$, which is conditioned on the current state.
A common and convenient choice is a multivariate Gaussian (normal) proposal
\[
\bm{\theta}^*_{t+1} \sim q(\bm{\theta}^*_{t+1}\mid\bm{\theta}_{t} )= \mathcal{N}_d(\bm{\theta}_{t}, \mathbf{C}_0),
\]
where $\mathbf{C}_0$ is a fixed positive-definite covariance matrix. The diagonal entries of $\mathbf{C}_0$ control the marginal step sizes (variances) along each parameter direction, while the off-diagonal entries give the correlations between parameters and thus determine the orientation of proposed moves in parameter space. The proposed candidate $\bm{\theta}^*_{t+1}$ is accepted with probability
\begin{equation}
\alpha \bigl(\bm{\theta}_{t}, \bm{\theta}^{*}_{t+1}\bigr)
\;=\;
\min\!\Biggl\{\,1,\;
\frac{\pi(\bm{\theta}^{*}_{t+1})\,q\bigl(\bm{\theta}_{t}\mid \bm{\theta}^{*}_{t+1}\bigr)}{\pi(\bm{\theta}_{t})\,q\bigl(\bm{\theta}^{*}_{t+1}\mid \bm{\theta}_{t}\bigr)}
\Biggr\},
\label{eq:6.7}
\end{equation}
where $\pi(.)$ denotes the target distribution, which in a Bayesian setting corresponds to the posterior probability density. If the proposal is accepted, the chain is updated to $\bm{\theta}_{t+1}= \bm{\theta}^{*}_{t+1}$; otherwise, the chain remains at its current state, $\bm{\theta}_{t+1}= \bm{\theta}_{t}$. This acceptance–rejection mechanism fully defines the transition kernel $P(\bm{\theta}_t, \cdot)$, which governs the probability distribution of the next state $\bm{\theta}_{t+1}$ given the current state $\bm{\theta}_{t}$.\footnote{
The kernel $P(\bm{\theta}_t, d\bm{\theta})$ has two contributions:: 
(i) a continuous component $q(\bm{\theta}^{*}\mid \bm{\theta}_t)\,\alpha(\bm{\theta}_t,\bm{\theta}^{*})\,d\bm{\theta}$, corresponding to transitions to $\bm{\theta}^{*}\neq \bm{\theta}_t$; and 
(ii) a point mass at $\bm{\theta}_t$ with probability 
$1 - \int_{\mathbb{R}^d} q(\bm{\theta}^{*}\mid \bm{\theta}_t)\,\alpha(\bm{\theta}_t,\bm{\theta}^{*})\,d\bm{\theta}^{*}$,
corresponding to rejection.} This process is repeated for a large number of iterations to construct a Markov chain whose stationary distribution is $\pi(\bm{\theta})$.
Unlike traditional Monte Carlo methods, in which rejected proposals are discarded, rejected moves in the MH algorithm lead to repeated entries of the current state in the chain. These repetitions are an essential feature of the algorithm and ensure that the detailed balance condition is satisfied. Detailed balance, in turn, guarantees convergence of the Markov chain to the target distribution under mild regularity conditions~\cite{robert2013monte}.

\subsubsection{The Adaptive Metropolis-Hastings proposal }
Although widely applicable, the standard MH algorithm can become inefficient in high-dimensional parameter spaces or when the target distribution exhibits strong correlations among parameters. The main challenge is to select an optimal proposal distribution, which determines both the typical step size and the orientation of proposed moves in parameter space. Since the shape of the target distribution is generally unknown, selecting an optimal proposal is nontrivial. A poorly chosen proposal can significantly impact the convergence and efficiency of the Markov chain, requiring a large number of likelihood evaluations to obtain statistically meaningful samples.\footnote{A poorly chosen proposal distribution may be too narrow, leading to high acceptance rates but strong autocorrelations and slow exploration of parameter space, or too broad, resulting in low acceptance rates and a Markov chain that becomes effectively stuck.}
The \emph{Adaptive Metropolis-Hastings} (AMH) algorithm, introduced by Haario et al. \cite{haario2001adaptive}, addresses these limitations by dynamically adapting the covariance matrix of the proposal distribution during the sampling process. The core idea is to use the information accumulated from the history of the Markov chain to tune the proposal distribution as the simulation progresses. By learning the scale and correlation structure of the target distribution, the algorithm can propose more efficient moves, thereby improving sampling efficiency already at relatively early stages of the chain. The AMH algorithm can be broken down into two phases:
\begin{enumerate}
    \item Initialization and Pre-adaptation
    
    The AMH algorithm begins with an initial ``pre-adaptation'' phase, during which it operates as a standard Metropolis–Hastings sampler. An initial parameter vector $\bm{\theta}_0$ is chosen, and a fixed proposal covariance matrix $\mathbf{C}_0$ is specified. A common choice for $\mathbf{C}_0$ is a diagonal matrix with small, constant variances. This fixed proposal is used for the first $N_{\text{pre}}$ iterations, where $N_{\text{pre}}$ is a user-defined parameter. The purpose of this phase is to allow the chain to explore the general area of the target distribution and to accumulate an initial set of samples before adaptive updates are applied.

    \item Adaptive phase
    
    After the initial $N_{\text{pre}}$ iterations, the algorithm transitions to the adaptive phase. For each subsequent iteration $t>N_{\text{pre}}$, the covariance matrix of the proposal distribution, $\mathbf{C}_t$, is updated using the full history of the Markov chain. To avoid the prohibitive computational cost and memory requirements associated with storing all past samples, the covariance matrix is updated recursively, relying only on information from the current iteration. At iteration $t$, the updated covariance matrix $\mathbf{C}_t$ is computed as:
    \begin{equation}
    \mathbf{C}_t = \frac{1}{t-1} \left( \mathbf{S}_{t}- t \,\bm{\mu}_{t} \bm{\mu}_{t}^T \right),
    \label{eq:6.8}
    \end{equation}
    where $\mathbf{C}_t$ provides an empirical estimate of the covariance structure of the target distribution and explicitly captures the correlations between the model parameters ($ \theta_{t}^1, \theta_{t}^2, \dots, \theta_{t}^d$) inferred from the accumulated Markov chain samples. Here $\bm{\mu}_{t}$ denotes the running mean of the sampled parameter vectors and $\mathbf{S}_{t}$ is the accumulated cross-product matrix. These quantities are updated recursively according to\footnote{These recursive relations are based on the original definitions: \[ S_{t} = \sum_{i=0}^{t-1} \theta_i \theta_i^{T}, \quad \mu_{t} = \frac{1}{t} \sum_{i=0}^{t-1} \theta_i,\] which correspond to the uncentered second moment matrix and the sample mean computed from all samples up to iteration $t-1$.}
    \begin{itemize}
        \item Cross-product matrix: $\mathbf{S}_{t} = \mathbf{S}_{t-1}+  \bm{\theta}_t \bm{\theta}_t^T$,
        \item Mean vector: $\bm{\mu}_{t} = \bm{\mu}_{t-1} + \frac{1}{t} (\bm{\theta}_t-\bm{\mu}_{t-1} )$.
    \end{itemize}
    This recursive updating scheme allows the proposal covariance matrix to continuously adapt to the evolving estimate of the target distribution’s correlation structure, while maintaining computational efficiency. As a result, the AMH algorithm achieves more effective exploration of parameter space compared to a fixed-proposal MH sampler, particularly in high-dimensional and strongly correlated settings.
\end{enumerate}
With the adaptive covariance matrix $\mathbf{C}_t$ defined, the proposal distribution for the next iteration can be constructed. The AMH algorithm employs a mixture random-walk proposal, which is a crucial design choice that preserves the correct ergodic properties of the algorithm. Since the adaptive scheme introduces a dependence on the full history of the chain, the resulting process is formally non-Markovian. The use of a mixture proposal ensures that the resulting stochastic process remains ergodic and converges to the desired target distribution. Specifically, the candidate state $\bm{\theta}^*_{t+1}$ is generated from a weighted combination of an adaptive and a fixed proposal,
\begin{equation}
\bm{\theta}^*_{t+1} \sim q(\bm{\theta}^*_{t+1}|\bm{\theta}_{t}) = (1-\beta) \, \mathcal{N}(\bm{\theta}_{t}, s_d \, \mathbf{C}_{t}) + \beta \, \mathcal{N}(\bm{\theta}_{t}, \mathbf{C}_0),
\end{equation}
where the first term represents the adaptive component and accounts for the majority of proposed moves. The scaling factor $s_d$ controls the overall step size in parameter space and therefore plays a central role in balancing efficient exploration against a reasonable acceptance rate. Extensive empirical and theoretical studies have been devoted to the optimal tuning of this scaling factor. In particular, Gelman \emph{et al.}~\cite{Gelman1996} showed that for random-walk Metropolis algorithms targeting multivariate Gaussian distributions, an asymptotically optimal choice is $s_d=\frac{(2.4)^2}{d}$, where $d$ is the dimensionality of the parameter space. This choice is designed to achieve an acceptance rate of approximately 23.4\% \cite{Gelman1997}, which has been found to provide an optimal balance between rapid convergence and efficient exploration of the target distribution. The second term in the mixture proposal corresponds to a fixed Gaussian component with covariance $ \mathbf{C}_0$ and is included to ensure the theoretical validity of the adaptive algorithm. The small mixing probability $\beta \in (0, 1)$ guarantees a non-zero probability of proposing moves independent of the adaptive covariance. This ensures irreducibility of the Markov chain and is a necessary condition for convergence to a unique stationary distribution. Given the mixture proposal, a candidate $\bm{\theta}^*_{t+1}$ is drawn and accepted or rejected according to the standard Metropolis-Hasting rule, Eq. (\ref{eq:6.7}). The accept–reject mechanism therefore remains unchanged relative to the standard MH algorithm, with all adaptations entering solely through the proposal distribution.

One crucial aspect of adaptive algorithms is ensuring their convergence to the correct target distribution, despite their formally non-Markovian nature. In adaptive schemes, the proposal kernel depends on the entire history of the chain through the covariance matrix $\mathbf{C}_{t}$, which complicates the standard proofs of ergodicity for Markov chains. Nevertheless, the Adaptive Metropolis–Hastings algorithm is known to remain asymptotically valid provided that two key theoretical conditions, established in Ref.~\cite{Roberts2006}, are satisfied:
\begin{itemize}
    \item Diminishing Adaptation: This condition requires that the influence of the adaptive updates on the chain’s dynamics decreases with time, ensuring that the algorithm eventually stabilizes. Formally, diminishing adaptation is satisfied if the total variation distance between successive transition kernels, $P_{t}$ and $P_{t+1}$, vanishes asymptotically,
    \[
    \sup_{\bm{\theta}\in \mathbb{R}^d}\left\| P_{t+1}(\bm{\theta},\cdot) - P_t(\bm{\theta},\cdot)\right\|
    \xrightarrow{p} 0 \quad (t\to\infty),
    \]
    where $\left\| \cdot \right\|$ denotes the total variation norm.\footnote{The supremum ( $\sup$) is taken over all starting points $\bm{\theta} \in \mathbb{R}^d$. This ensures that the difference between successive transition kernels vanishes uniformly throughout the entire parameter space, rather than only for specific regions or initial conditions.} In the AMH algorithm, this condition is satisfied because the covariance matrix $\mathbf{C}_{t}$ is updated with an effective step size of order $\mathcal{O}(1/t)$ (see Eq. (\ref{eq:6.8})).  As a result, changes to the proposal kernel become progressively smaller at large $t$, and the chain asymptotically ``forgets'' the adaptation, thereby recovering ergodic behavior.
    \item Bounded Proposal Component: The second condition, closely related to the requirement of simultaneous uniform ergodicity, is the presence of a non-degenerate fixed component in the proposal distribution. In the AMH algorithm, this is ensured by the non-zero mixing parameter $\beta$ in the proposal kernel. The fixed Gaussian component prevents the adaptive proposal from collapsing or becoming poorly conditioned and guarantees that the chain has global access to the full parameter space.  Even if the empirically estimated covariance matrix $\mathbf{C}_{t}$ is poorly defined or numerically unstable, this bounded component ensures that the chain can continue to propose reasonable moves and maintain irreducibility.
\end{itemize}

\paragraph{Reset to Mean.}
In practical applications, a significant challenge for adaptive algorithms is the potential for the chain to become overly influenced by its initial states, particularly when the starting point lies far from the high-probability region of the target distribution. During such non-equilibrium phases, the empirical covariance matrix of the chain is not yet a reliable estimator of the true posterior structure, and relying heavily on these early samples can result in a poorly adapted proposal that slows down convergence and mixing. To reduce this effect, we employ a \emph{Reset to Mean} (RTM) mechanism within the AMH proposal. The RTM mechanism periodically resets the self-adapted proposal distribution to a fixed, non-adaptive covariance matrix $\mathbf{C}_{0}$. This procedure reduces the influence of the initial starting point and early, potentially misleading correlations that do not accurately reflect the structure of the posterior distribution. By occasionally ``forgetting'' its accumulated history, the chain is given an opportunity to re-explore parameter space and escape regions where it may have become inefficiently trapped. A key aspect of the RTM mechanism is the construction of an appropriate fixed covariance matrix. In this analysis, $\mathbf{C}_{0}$ is defined by taking the absolute values of the diagonal elements of the current adaptive covariance matrix, scaling them by a small factor of 0.05, and using these values as the diagonal entries of a diagonal matrix, thereby discarding all off-diagonal correlation terms. When an RTM is triggered, the pre-adaptation counters and run-control parameters are reset to their initial values, returning the algorithm to its non-adaptive phase for the subsequent $N_{\text{pre}}$ iterations.

\paragraph{Multiplicity.}
The output of the AMH algorithm is stored in a \texttt{CSV} file that records the full history of the Markov chain. For each state of the chain, the file contains:
\begin{itemize}
    \item Parameter values: the numerical values of the nPDF parameters at that iteration;
    \item Chi-squared ($\chi^2$) value: the goodness-of-fit corresponding to that parameter point;
    \item Multiplicity: an integer specifying how many consecutive iterations the chain remains at the same parameter point due to rejected proposals.
\end{itemize}
\emph{Multiplicity} provides a compact representation of the Markov chain by encoding the probability density without explicitly storing repeated copies of identical states. In practical terms, a state with multiplicity $m$ is statistically equivalent to having been sampled $m$ times. For example, a parameter point with multiplicity $10$ contributes to posterior estimates in exactly the same way as ten identical samples at that location in parameter space.
Beyond its role as a bookkeeping device, multiplicity also carries diagnostic information. Large multiplicity values indicate regions where proposed moves are frequently rejected, which may signal inefficient local exploration or inadequate tuning of the proposal distribution. However, such repetitions are not pathological; rather, they are an essential feature of the Metropolis–Hastings algorithm and ensure that the stationary distribution is sampled correctly.
Crucially, multiplicity must be taken into account when computing posterior expectations, distributions, and summary statistics. States with higher multiplicity contribute proportionally more to posterior averages, reflecting the longer Monte Carlo time spent in those regions. Proper treatment of multiplicity therefore guarantees an unbiased reconstruction of the posterior distribution and preserves the correct probabilistic weighting of all sampled parameter configurations.

\paragraph{Validity and Challenges.}
The adaptive MH algorithm updates the $\chi^2$-based covariance matrix at each proposal stage using an efficient recursive scheme, which avoids the need for repeated recomputation of full empirical covariances and substantially reduces the associated computational cost. This feature makes the algorithm particularly attractive for applications involving high-dimensional parameter spaces. In contrast to more advanced MCMC techniques, such as Hybrid or Hamiltonian Monte Carlo (HMC)~\cite{Duane:1987de}, which require multiple evaluations of the $\chi^2$ function per iteration to integrate Hamiltonian dynamics, the AMH algorithm performs only a single $\chi^2$ evaluation per proposal, resulting in a significantly lower per-step computational overhead. Despite these advantages, the AMH algorithm can encounter difficulties when the posterior distribution exhibits multi-modal structure. In such cases, the absence of an explicit mechanism for proposing long-distance moves between distant modes may cause the Markov chain to become confined to a local region of high probability, with the adaptive procedure reinforcing the local covariance structure and further suppressing transitions between separated minima.\footnote{Previous global nPDF studies suggest that such features are not expected in the present analysis. In particular, earlier nCTEQ fits~\cite{Kovarik:2015cma} indicate that the relevant $\chi^2$ surface is generally dominated by a single, well-defined global minimum and is therefore well suited to adaptive random-walk sampling.} Nevertheless, it is important to emphasize that the AMH algorithm does not inherently prevent the identification of secondary minima. When multiple independent chains are initialized from different regions of parameter space, or when sufficiently long chains are generated, the MCMC framework can still reveal the presence of additional local minima. Evidence for such structures may appear in characteristic convergence patterns, clustering of sampled points, or large fluctuations in parameter values as the algorithm attempts to transition between modes. In this sense, although AMH is not specifically designed for efficient multimodal sampling, it remains capable of diagnosing secondary minima should they exist, provided that appropriate convergence and consistency checks are carried out.

\subsection{Posterior Construction for Nuclear PDF Sampling} \label{sec:6.2.2}
In this study, we perform a full Bayesian inference of nuclear PDFs by sampling from their posterior distribution using the Adaptive Metropolis-Hastings algorithm. This approach is grounded in Bayes’ theorem, reviewed in Sec.~\ref{sec:4.1}, which relates the posterior distribution to the prior information and the likelihood function. Since the AMH algorithm requires only the ratio of posterior probabilities between successive states of the Markov chain, the normalization constant appearing in Bayes’ theorem (Eq.~(\ref{Bayes' formula})) cancels out and does not need to be evaluated explicitly. Consequently, the posterior distribution of the nPDF parameters $\bm{\theta}$\footnote{The PDFs are specified through functional forms in $x$ and $Q^2$, whose coefficients are parameterized as $c_j(A) = p_j + a_j\ln A + b_j\ln^2A$. In principle, all three parameters, $p_j$, $a_j$, and $b_j$, can be determined from data through the fit. However, in global nuclear PDF analyses, $p_j$ is kept fixed, as explained in Sec.~\ref{sec:6.1.2}.} given the experimental data $\mathbf{D}$ and the covariance matrix $\mathbf{C}$ is sampled from a distribution proportional to the product of the likelihood and the prior,
\begin{equation}
    P( \bm{\theta}|\mathbf{D}, \mathbf{C}) \propto P( \bm{\theta}) \cdot P(\mathbf{D} | \bm{\theta}, \mathbf{C}).
\end{equation}
In practice, this implies that the target distribution of the AMH algorithm is the unnormalized posterior, and the quantity that drives the MCMC dynamics is the log-likelihood function. As discussed in Sec.~\ref{sec:4.3.2}, when experimental measurements exhibit correlated uncertainties, the likelihood assumes a quadratic form. Under the assumption of Gaussian uncertainties, the log-likelihood can be written in terms of a generalized chi-squared statistic defined in Eq. (\ref{eq:4.35}) as
\begin{equation}
    \log P(\mathbf{D} \mid \mathbf{\bm{\theta}}, \mathbf{C}) = -\frac{1}{2} \chi^2(\mathbf{\bm{\theta}}, \mathbf{D}, \mathbf{C}).
\end{equation}
Here, $\mathbf{T}(\bm{\theta})$ denotes the vector of the theoretical predictions corresponding to the experimental data points, evaluated as a function of the parameter vector $\bm{\theta}$.
In practical implementations of the AMH algorithm, calculations are performed in the logarithmic domain rather than directly using probability values. This is essential because both the likelihood  $P(\mathbf{D} \mid \mathbf{\bm{\theta}}, \mathbf{C})$ and the prior $P( \bm{\theta})$ can become extremely small in high-dimensional parameter spaces, leading to numerical underflow. Working in log-space transforms products into sums and significantly improves numerical stability. Accordingly, the MH acceptance probability (Eq.~(\ref{eq:6.7})) is evaluated using log-probabilities. Recall that the Metropolis-Hastings algorithm accepts a proposed state $\bm{\theta}^*$ with probability
\begin{equation}
\alpha \bigl(\bm{\theta}, \bm{\theta}^*\bigr)
\;=\;
\min\!\Biggl\{\,1,\;
\frac{\pi(\bm{\theta}^*)\,q\bigl(\bm{\theta}\mid \bm{\theta}^*\bigr)}{\pi(\bm{\theta})\,q\bigl(\bm{\theta}^*\mid \bm{\theta}\bigr)}
\Biggr\}.
\end{equation}
where $\pi(\bm{\theta}) \propto P(\mathbf{D} | \bm{\theta}, \mathbf{C})\cdot P( \bm{\theta})$ denotes the unnormalized posterior distribution and $q(\cdot|\cdot)$ is the proposal distribution. For symmetric proposals satisfying $q\bigl(\bm{\theta}\mid \bm{\theta}^*\bigr)= q\bigl(\bm{\theta}^*\mid \bm{\theta}\bigr)$, the acceptance probability simplifies to a ratio of posterior probabilities,
\[
\alpha \bigl(\bm{\theta}, \bm{\theta}^*\bigr)
\;=\;
\min\!\Biggl\{\,1,\;
\frac{P(\mathbf{D} | \bm{\theta}^*, \mathbf{C})\cdot P( \bm{\theta}^*)}{P(\mathbf{D} | \bm{\theta}, \mathbf{C})\cdot P( \bm{\theta})}
\Biggr\}.
\]
Taking the logarithm of the numerator and denominator yields the log-posterior difference:
\begin{equation}
    \Delta = \left[ \log P( \mathbf{D}\mid \bm{\theta}^*, \mathbf{C}) + \log P(\bm{\theta}^*) \right] - \left[ \log P(\mathbf{D}\mid \bm{\theta}, \mathbf{C}) + \log P(\bm{\theta}) \right],
\end{equation}
which is used to compute the acceptance probability in log-space as
\[
\alpha(\bm{\theta}, \bm{\bm{\theta}}^*) = \min\{1, \exp(\Delta)\}.
\]
In the implementation used in this work, the Metropolis–Hastings acceptance step is performed entirely in the logarithmic domain. The code computes the log-posterior difference $\Delta$, referred to internally as the acceptance ratio. If $\Delta> 0$, the proposed state is accepted unconditionally. Otherwise, a random number $t$ is drawn from a uniform distribution on $(0,1)$, and the proposal is accepted if $\Delta > \log(t)$. This procedure is mathematically equivalent to the standard acceptance rule, but offers improved numerical stability and robustness for high-dimensional posterior distributions.

\subsubsection{Prior Distributions}
In the Bayesian framework, the prior distribution $P(\bm{\theta})$ encodes information about the model parameters before incorporating experimental data. In the present analysis, most fit parameters are assigned flat priors over sufficiently broad domains. Within the region explored by the Markov chains, these priors are effectively non-informative and cancel in the Metropolis–Hastings acceptance ratio. As a result, they do not influence the shape of the posterior distribution, and the parameters are primarily constrained by the experimental data through the likelihood function.\footnote{Additional constraints are implicitly imposed through parameters that are not fitted. These include, for example, the enforcement of momentum and valence quark number sum rules, as well as fixed baseline quantities entering the PDF parameterization.} 

\begin{figure}[h]
    \centering
    \includegraphics[width=0.9\textwidth]{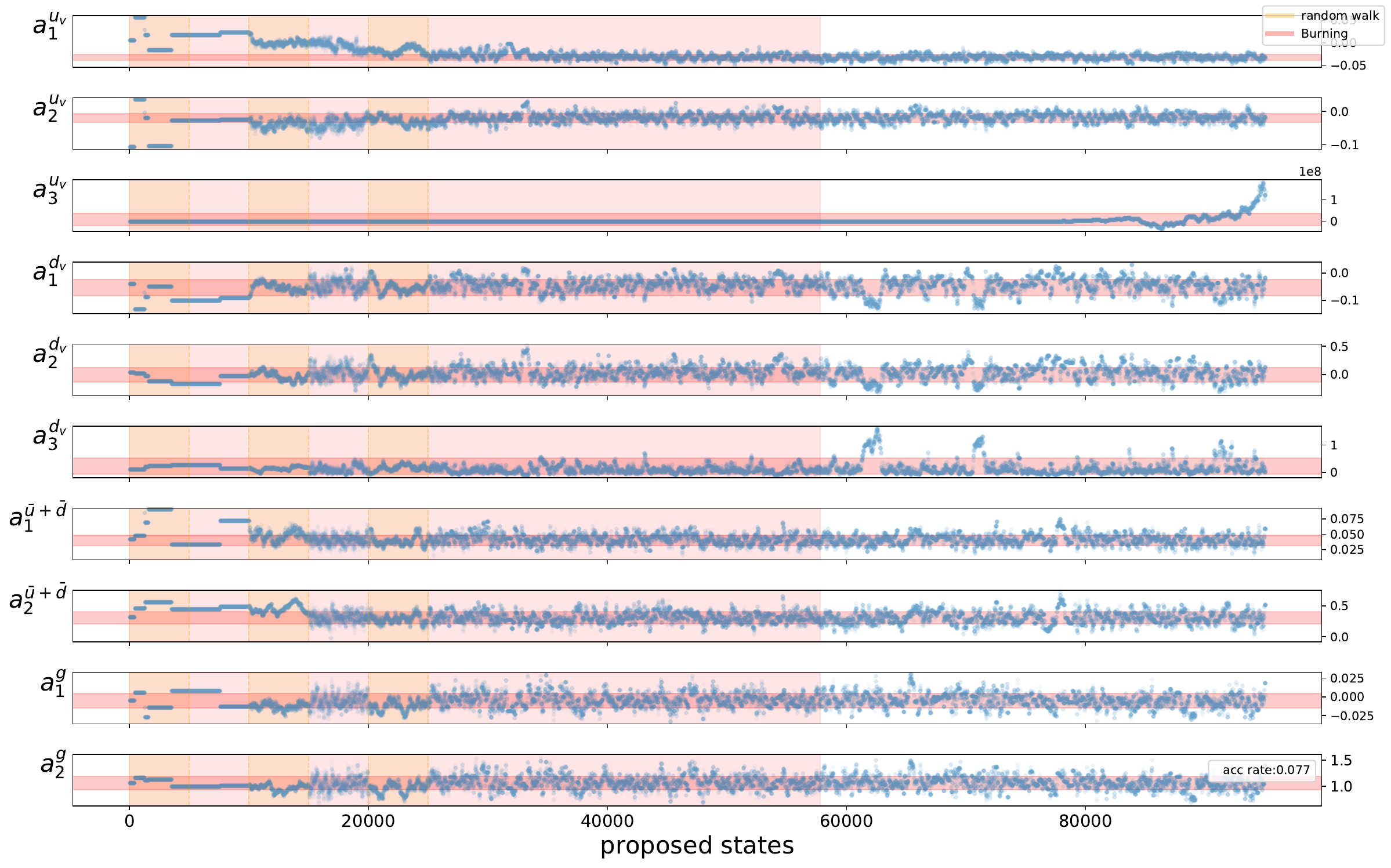}
    \caption[MCMC trace plot illustrating the drift of the $a_3^{u_v}$ parameter in the \texttt{Pb-only} analysis]{MCMC trace plot for the \texttt{Pb-only} analysis, showing the evolution of the fitted nPDF parameters as a function of the number of MCMC steps. The parameter $a_3^{u_v}$ exhibits large drifts in the absence of an explicit prior, indicating that it is only weakly constrained by the likelihood and that the posterior is effectively flat along this direction.
    }
    \label{fig:chain_runaway_uv_a3}
\end{figure}

\begin{figure}[h]
    \centering
    \includegraphics[width=0.65\textwidth]{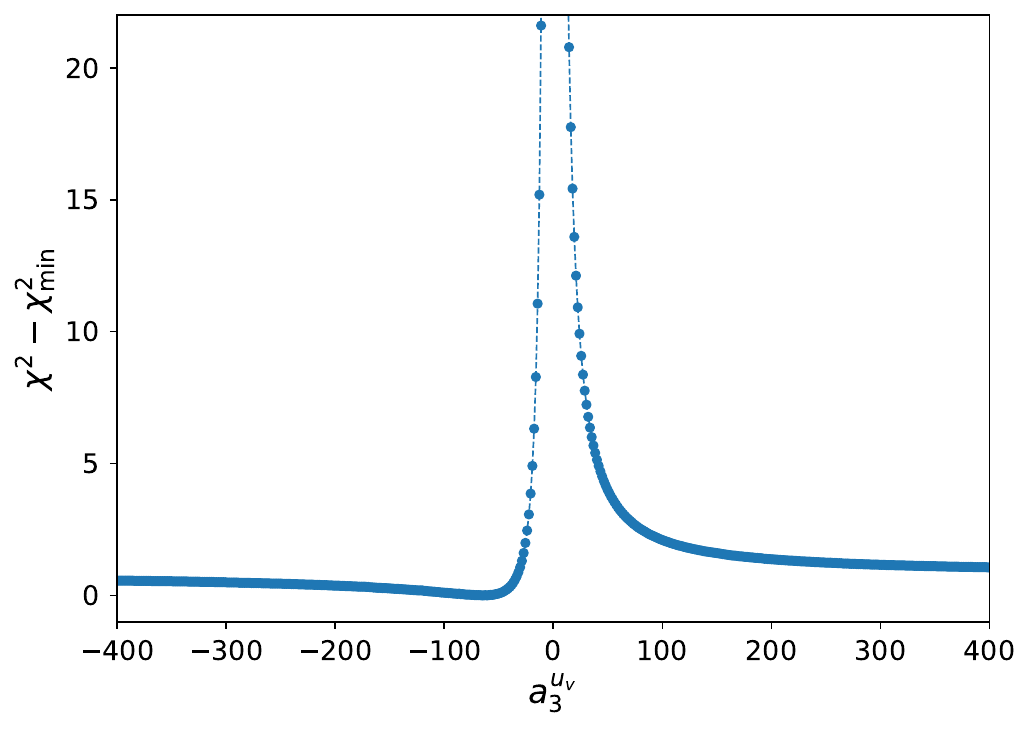}
    \caption[Scan of the \(\chi^2\) function along the $a_3^{u_v}$ parameter in the \texttt{Pb-only} analysis]{Scan of the \(\chi^2\) function along the parameter $a_3^{u_v}$ in the \texttt{Pb-only} analysis. The parameter $a_3^{u_v}$ is varied while all other parameters are fixed at their values corresponding to the MCMC global minimum of the full fit. The profile exhibits a nearly constant $\chi^2$ over a broad interval, indicating weak sensitivity of the likelihood to this parameter, together with a sharp, localized increase around $a_3^{u_v}=0$ that strongly disfavors this region.
    }
    \label{fig:chi2_scan_uv_a3}
\end{figure}

An exception arises in the \texttt{Pb-only} analysis, where the dataset is restricted to measurements involving lead nuclei. In this case, the parameter $a_3^{u_v}$, which controls the polynomial modulation of the up-valence quark distribution in Eq.~(\ref{eq:6.3}), exhibits unstable behavior during MCMC sampling when left unconstrained. As illustrated in Fig.~\ref{fig:chain_runaway_uv_a3}, the parameter undergoes large drifts across the parameter space, indicating that it is only weakly constrained by the likelihood.
To investigate this behavior, a dedicated scan of the $\chi^2$ function along the $a_3^{u_v}$ direction was performed, as shown in Fig.~\ref{fig:chi2_scan_uv_a3}. In this scan, $a_3^{u_v}$ is varied while all other parameters are fixed at their values corresponding to the MCMC global minimum. The resulting profile is characterized by a nearly constant $\chi^2$ over a broad interval, demonstrating that the likelihood is largely insensitive to this parameter. In addition, a sharp, localized increase in $\chi^2$ is observed around $a_3^{u_v} =0$, forming a narrow barrier that strongly disfavors this specific region while leaving the surrounding parameter space effectively unconstrained. This combination of a flat direction and a localized exclusion region explains the observed sampling behavior.


A common approach in such situations is to fix poorly constrained parameters to a specific value. While this removes flat directions in parameter space, it also reduces the flexibility of the fit and suppresses potential non-Gaussian correlations. In contrast, the MCMC framework naturally accommodates such structures by sampling the full posterior distribution. For this reason, instead of fixing $a_3^{u_v}$, we introduce a bounded uniform prior,
\footnote{In this work, we distinguish between \emph{flat} and \emph{uniform} priors. A flat prior refers to an improper prior with constant density over an unbounded domain, while a uniform prior denotes a proper distribution with finite support. The latter is used to enforce physically reasonable parameter ranges and ensure numerical stability.}
\[
a_3^{u_v} \sim \mathcal{U}(-300,\;300),
\]
which regularizes the flat direction and prevents the sampler from exploring unphysical regions of parameter space while preserving the ability of the data to determine the posterior where meaningful constraints are present. The $\chi^2$ scan confirms that the fit quality remains effectively unchanged across most of this interval, indicating that the role of the prior is to stabilize the sampling rather than to impose a phenomenological constraint.
From a sampling perspective, the presence of the localized high-$\chi^2$ region around $a_3^{u_v}=0$ does not significantly affect the efficiency of the MCMC algorithm. Because this region occupies only a small volume of the parameter space, the Markov chain can efficiently traverse it and explore the extended, nearly degenerate likelihood regions on either side. Although proposals into this disfavored region lead to reduced acceptance probabilities, this effect remains localized and does not degrade the overall mixing or effective sample size of the chain.

\subsection{Initialization of MCMC chains}
The setup of the MCMC algorithm requires careful specification of the initial conditions in order to ensure efficient sampling from the posterior distribution.
As part of the analysis strategy adopted in this work, multiple independent Markov chains are run in parallel, with each chain initialized in the vicinity of the minimum found in a Hessian fit. This approach provides a robust basis for convergence assessment through comparisons across chains and simultaneously increases the total number of posterior samples available for statistical analysis. The initialization procedure begins with a preliminary deterministic fit to locate the Hessian minimum, which serves as a physically meaningful reference point in parameter space. Starting the MCMC chains from completely random points in the full parameter domain was found to be highly inefficient, as such initializations typically lie in regions of extremely low posterior probability and lead to prolonged burn-in periods with poor acceptance rates. Initializing the chains near the Hessian minimum significantly reduces this issue and accelerates convergence. Specifically, each MCMC chain is initialized by perturbing the Hessian best-fit parameters according to
\[
\theta_i^{(0)} = \theta_i^{\min} \cdot (1 + \delta_i), \quad \delta_i \in [-0.2,\ 0.2]
\]
where $\theta_i^{\min}$ is the value of the $i$-th parameter at the Hessian minimum and $\delta_i$ is a uniformly distributed random perturbation. This initialization ensures that the chains begin in a physically meaningful and high-probability region of the parameter space, while the random perturbation introduces sufficient diversity among the initial states to enable comprehensive exploration of the posterior distribution. To maintain statistical independence between chains, each chain is assigned a unique random seed. This guarantees that the stochastic proposal sequences and initializations differ across chains, which is necessary for meaningful convergence diagnostics and for identifying potential multi-modal features in the posterior.
%%%%%%%%%%%%%%%%%%%%%%%%%%%%%%%%%%%%%%%%%%%%%%%%%%%%%%%%%%%%%%%%%%%%%%
\section{MCMC Analysis}\label{sec:6.3}
Once the posterior distribution is defined\footnote{In this context, the posterior distribution refers to the target analytical density function, defined as the product of the likelihood and priors (see Eq.(\ref{eq4.50})). The analysis that follows focuses on the stochastic samples generated from this function.} and sampled via the Adaptive MH algorithm, we analyze the generated chains to ensure the validity of our Bayesian inference. This section describes the diagnostic procedures used to evaluate sample quality and extract robust statistical estimates. We begin by examining the convergence behavior of the Markov chains using a set of diagnostic tools designed to verify whether the chains have reached their stationary distribution. Once convergence is established, we determine the appropriate thermalization period (burn-in) and remove initial samples that may be biased by the starting conditions. We then study the autocorrelation structure of the post-burn-in samples to quantify statistical dependence and estimate the effective number of independent samples. Based on this analysis, a thinning procedure is applied to reduce redundancy and retain a representative subset of approximately uncorrelated samples. The resulting ensemble is subsequently used to estimate posterior distributions and provide robust uncertainty quantification for the nuclear PDFs.

\subsection{Convergence Assessment}
A fundamental requirement in any MCMC analysis is to verify that the generated chains have converged to their stationary distribution $\pi(\theta)$, which in a Bayesian context corresponds to the target posterior. Convergence ensures that the sampled parameter values are representative of the posterior distribution rather than being influenced by the initial conditions or transient behavior during the early stages of the simulation. If convergence is not achieved, statistical estimates derived from the chains, such as posterior means, variances, or credible intervals, may be biased and unreliable.

As discussed in Sec.~\ref{sec:5.2.3}, under mild regularity conditions such as irreducibility and aperiodicity, a Markov chain is guaranteed to converge in distribution to a unique stationary distribution. In practice, however, convergence can only be assessed using finite-length chains, and the true posterior distribution is not known analytically. Consequently, convergence diagnostics are inherently heuristic: they do not provide rigorous guarantees but instead offer practical indicators that the chains have likely reached stationarity and are sampling efficiently from the posterior~\cite{brooks2011handbook,Hogg:2017akh}. In this work, convergence is assessed using complementary diagnostic tools, including trace plots and cumulative mean plots. While these diagnostics do not establish convergence in a strict mathematical sense, they provide valuable qualitative and quantitative insight into the stability of the chains and the reliability of the sampled posterior.

\subsubsection{Trace plots}
\begin{figure}[htbp!]
    \centering
    \includegraphics[width=0.8\textwidth]{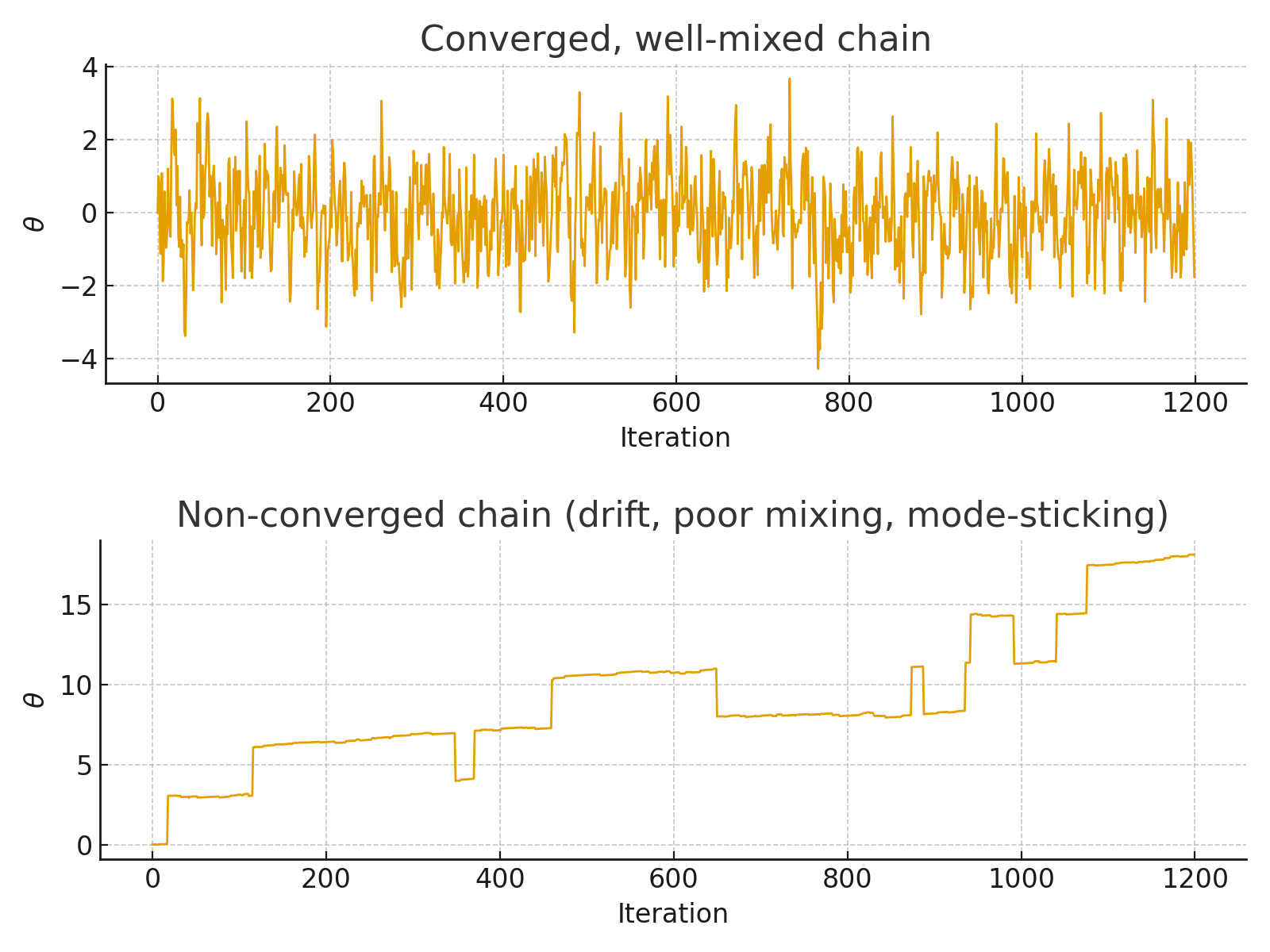}
    \caption[Convergence Assessment using trace plots]{Pedagogical example of trace plots: (top) a well-mixed, converged chain fluctuating around a stable mean; (bottom) a non-converged chain showing strong drift, poor mixing, and mode-sticking events.}
    \label{fig:trace_plot}
\end{figure}

Trace plots provide a time-series visualization of the sampled values of a given parameter across MCMC iterations, $\{\theta_i^{(0)}, \theta_i^{(1)},\ \theta_i^{(2)},\ \ldots,\ \theta_i^{(T)}\}$. For a well-converged Markov chain, the trace typically exhibits rapid, random-looking fluctuations around a stable central value, with no long-term trends or systematic drifts. Such behavior indicates that the chain has sufficiently explored the target distribution and is no longer strongly influenced by its initial conditions. In contrast, problematic trace plots may reveal a variety of convergence issues. These include slow mixing, where the chain remains confined to limited regions of parameter space for extended periods; systematic trends or drifts, signaling that the chain has not yet reached stationarity; and mode-sticking behavior, in which the chain becomes trapped in a local mode of a multimodal posterior. These issues can bias parameter estimates and lead to a substantial underestimation of the associated uncertainties. For this reason, visual inspection of trace plots remains a standard and powerful diagnostic tool in MCMC analyses~\cite{roy2019}. To illustrate the qualitative differences between well-behaved and problematic sampling behavior, Fig.~\ref{fig:trace_plot} shows pedagogical examples of synthetic trace plots that are not derived from the present analysis. The upper panel displays a well-mixed and converged chain, characterized by stable fluctuations around a constant mean, indicative of effective exploration of the parameter space and loss of dependence on the starting point. The lower panel shows a non-converged chain exhibiting several common issues simultaneously: a pronounced drift over iterations, poor mixing manifested as long periods of near-constant values, and occasional mode-sticking behavior in which the chain jumps between restricted subregions of the parameter space.
In this analysis, trace plots are generated for all fitted nPDF parameters and examined across multiple independently initialized chains. A detailed presentation and discussion of these diagnostic plots, together with the corresponding MCMC results, are provided in Chapter~\ref{Chap.7}.

\subsubsection{Cumulative Mean Plots}
\begin{figure}[!h]
    \centering
    \includegraphics[width=0.8\textwidth]{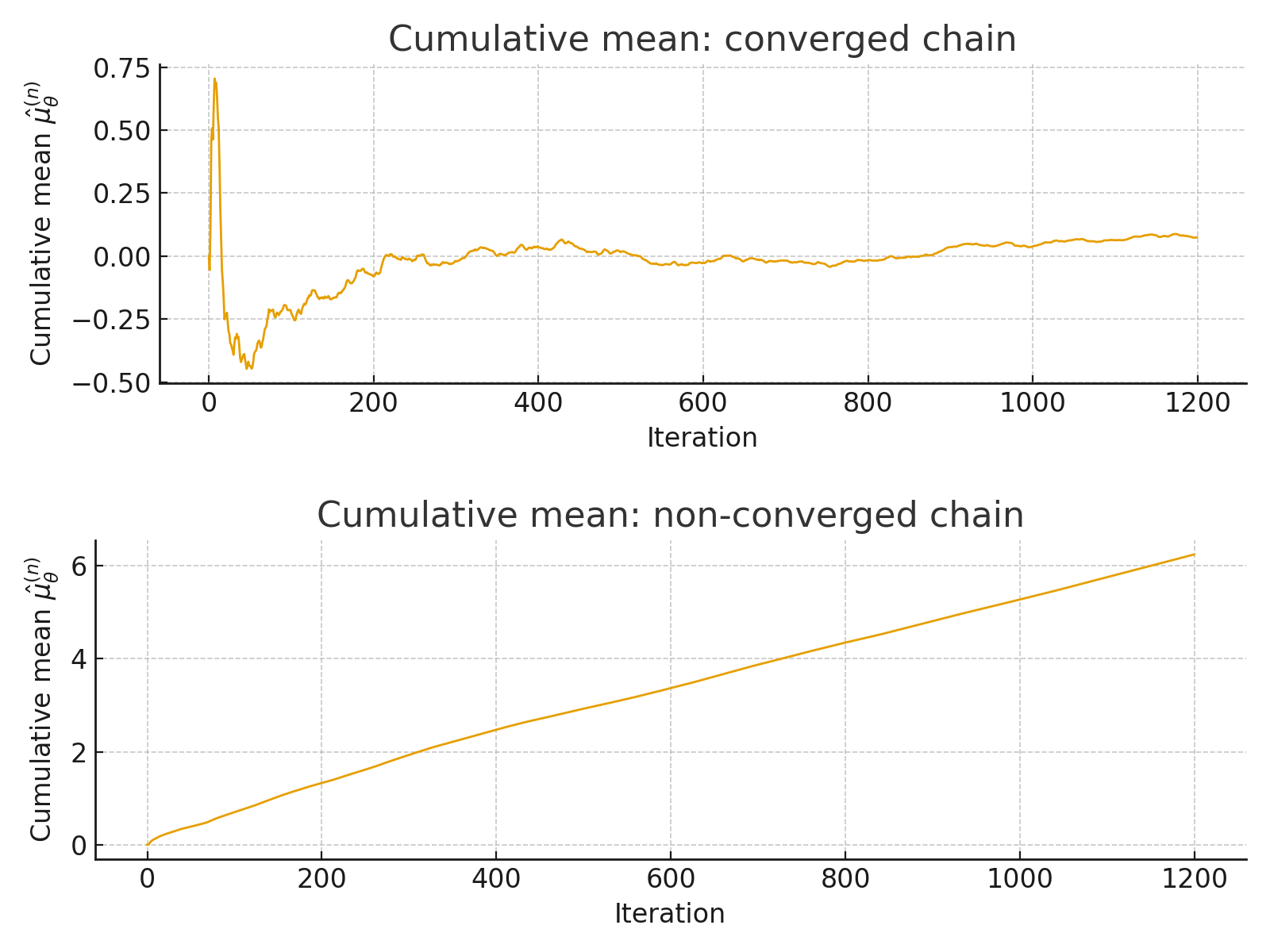}
    \caption[Convergence Assessment using cumulative mean evolution]{Pedagogical example of illustrative cumulative mean evolution: (top) a converged chain whose running average stabilizes rapidly; (bottom) a non-converged chain exhibits persistent drift and instability in the cumulative mean.}
    \label{fig:cummean_plot}
\end{figure}

\emph{Cumulative mean} plots provide a complementary diagnostic by tracking the running average of a parameter as a function of the number of MCMC iterations. If the chain has reached stationarity, this running average is expected to stabilize around the true posterior expectation value $\mathbb{E}_{\pi}[\theta]$ as the number of samples increases, with fluctuations diminishing as more information is accumulated. A flat cumulative mean trajectory therefore indicates that the influence of the initial state has largely disappeared and that the chain is producing consistent estimates of the posterior mean. However, such stabilization should be regarded as a necessary but not sufficient condition for convergence. While instability or persistent drift in the cumulative mean is a clear indication that stationarity has not yet been achieved, a stable cumulative mean alone does not guarantee global convergence. In particular, a chain that remains confined to a local mode of a multi-modal posterior may exhibit an apparently stable running average while failing to explore the full target distribution.
To illustrate these behaviors, Fig.~\ref{fig:cummean_plot} shows pedagogical examples of cumulative mean plots constructed from synthetic data and not derived from the present analysis. The upper panel displays a converged chain whose cumulative mean rapidly stabilizes around a constant value. In contrast, the lower panel shows a non-converged chain with a drifting and unstable cumulative mean, reflecting the same convergence issues observed in its corresponding trace plot.
In this analysis, the cumulative mean plots are computed and examined for all fitted parameters across multiple independent chains. Consistent stabilization of the cumulative means across chains, together with the evidence provided by trace plots, increases confidence that the MCMC samples are representative of the stationary posterior distribution. Detailed results and representative examples are presented in Chapter~\ref{Chap.7}.
\subsubsection{Multi-Diagnostic Approach}
Visual diagnostics, in particular trace plots and cumulative mean plots, provide the primary tools for assessing convergence in this analysis. These methods offer direct insight into the stability, mixing, and stationarity of the Markov chains and therefore serve as the main criteria for determining whether the sampling procedure has reached a reliable regime. Since such diagnostics are inherently qualitative, they are complemented by simple numerical indicators that characterize the efficiency of the sampling process.
In particular, we monitor the ``acceptance rate'' of the Metropolis-Hastings algorithm and the corresponding ``multiplicity,'' which quantifies the number of consecutive iterations for which the chain remains at the same state due to rejected proposals (see Sec.~\ref{sec:6.2.1}). These two quantities are closely related: a low acceptance rate leads to large multiplicities, reflecting reduced mobility of the chain and increased autocorrelation among samples. Conversely, excessively high acceptance rates typically indicate that proposal steps are too small, resulting in slow exploration of the parameter space. 

In this analysis, the acceptance rates are found to be approximately $0.09$ for the \texttt{Pb-only} chains and $0.196$ for the \texttt{Multi-nuclei} chains. These values indicate a reasonably balanced proposal scale, allowing for sufficient exploration of the parameter space while maintaining an acceptable level of statistical efficiency. The corresponding multiplicity behavior is consistent with this picture and does not exhibit signs of poor sampling.
Overall, these numerical indicators provide a useful consistency check on the behavior observed in the visual diagnostics but do not replace them. The final assessment of convergence is therefore based primarily on the stability and agreement of the trace plots and cumulative mean plots across independent chains.

\subsection{Thermalization}
A crucial practical step in MCMC analysis is the removal of the initial segment of each Markov chain, commonly referred to as the \emph{thermalization} or the \emph{burn-in} period. During this early phase, the chain remains influenced by its initial state $\bm{\theta}_0$ and has not yet reached the stationary distribution $\pi(\bm{\theta})$. As a result, samples generated in this regime are generally biased and cannot be regarded as representative draws from the posterior distribution. The appropriate length of the burn-in period, denoted by $N_{\text{burn}}$, is not fixed a priori and is typically determined heuristically. Its value depends on several factors, including the choice of initial conditions, the efficiency of the proposal distribution, and the overall complexity of the target posterior. Initializing the chains near regions of high posterior probability—as done in this work by perturbing the Hessian minimum—can substantially reduce the number of iterations required for thermalization. Nevertheless, even with such informed initialization, a finite burn-in phase remains necessary. In practice, the selection of $N_{\text{burn}}$ is determined by a careful examination of the convergence diagnostics discussed in the previous section. Once the trace plots, cumulative mean plots, and complementary numerical indicators demonstrate stable and stationary behavior, the initial non-equilibrium portion of each chain is discarded. The remaining post-burn-in samples are then treated as representative draws from the target posterior distribution and are used in all subsequent statistical analyses.

\subsection{Autocorrelation Function}
In MCMC simulations, the sequence of generated parameter states represents a stationary stochastic process. A notable characteristic of this process is the presence of \emph{autocorrelation}, i.e., temporal dependence between successive samples. This dependence arises naturally from the Markov property, whereby each state  $\bm{\theta}_{i+1}$ is conditionally dependent on the preceding state $\bm{\theta}_{i}$ As a consequence, successive samples are statistically correlated, which fundamentally reduces the statistical efficiency of MCMC sampling relative to independent and identically distributed (i.i.d.) sampling.

The primary objective of an MCMC simulation is to estimate the expectation value of a scalar observable $X=f(\bm{\theta})$, where $f:\mathbb{R}^d \to \mathbb{R}$ is a function defined on the parameter space. Given a sequence of $N$ samples $\{ \bm{\theta}_i \}_{i=1}^N$ drawn from the stationary distribution, the expectation value $\mu = \mathbb{E}[f(\bm{\theta})]$ is estimated using the sample mean:
\begin{equation}
    \bar{X}_N = \frac{1}{N} \sum_{i=1}^{N} f(\bm{\theta}_i).
    \label{eq:6.15}
\end{equation}
In the idealized case of i.i.d.\ samples, the variance of this estimator is given by
\[
\operatorname{Var}(\bar{X}_N) = \frac{\sigma^2}{N},
\]
where $\sigma^2= \operatorname{Var}(f(\bm{\theta}))$ denotes the true population variance of the observable.\footnote{In statistical inference, it's important to distinguish between population parameters, such as the true mean $\mu = \mathbb{E}[X]$ and variance $\sigma^{2} = \operatorname{Var}(X)$, which are fixed but unknown, and sample-based estimators, such as the sample mean $\bar{X}_N$ and sample variance $S^{2}$, which are computed from finite data.
While the population variance $\sigma^{2}$ appears in theoretical expressions, 
the sample variance $S^{2}$ is used in practice to estimate uncertainty, for example in the Monte Carlo standard error defined in Eq. (\ref{eq:6.17}).} Since $\sigma^2$ is generally unknown, it is estimated in practice using the sample variance,
\begin{equation}
    S^2 =  \frac{1}{N-1} \sum_{i=1}^{N} (f(\bm{\theta}_i) - \bar{X}_N)^2,
    \label{eq:6.16}
\end{equation}
leading to the Monte Carlo standard error (MCSE),
\begin{equation}
    \sigma_{\text{MC}} = \frac{S}{\sqrt{N}}.
    \label{eq:6.17}
\end{equation}

In MCMC applications, however, the samples are generally correlated, and the above expression for the MCSE underestimates the true uncertainty. To properly account for this dependence, we introduce the autocovariance function $\Gamma(k)$, which measures the covariance between observable values separated by a lag $k$
\begin{equation}
    \Gamma(k) = \operatorname{Cov}(f(\bm{\theta}_i), f(\bm{\theta}_{i+k})) = \mathbb{E}[(f(\bm{\theta}_i) - \mu)(f(\bm{\theta}_{i+k}) - \mu)].
\end{equation}
The marginal variance corresponds to zero lag, $\Gamma(0) =\sigma^2$. The normalized autocovariance defines the autocorrelation function (ACF),
\begin{equation}
    \rho(k) = \frac{\Gamma(k)}{\Gamma(0)}=\frac{\Gamma(k)}{\sigma^2}.
\end{equation}
For a stationary Markov chain, the ACF satisfies $\rho(0) = 1$ and $\rho(k) = \rho(-k)$, and for an ergodic chain it decays to zero as  $k \to \infty$. The variance of the sample mean $\operatorname{Var}(\bar{X}_N)$ for correlated samples can be derived from the general expression for the variance of a sum of dependent random variables. Writing $X_i= f(\bm{\theta_i})$, one obtains
\begin{align}
    \operatorname{Var}(\bar{X}_N) &= \operatorname{Var}\left(\frac{1}{N}\sum_{i=1}^N X_i\right) \\
    &= \frac{1}{N^2} \operatorname{Var}\left(\sum_{i=1}^N X_i\right) \\
    &= \frac{1}{N^2} \sum_{i=1}^N \sum_{j=1}^N \operatorname{Cov}(X_i, X_j).
    \label{eq:variance_double_sum}
\end{align}
For a stationary chain, the covariance depends only on the lag $k=|i-j|$, allowing the double sum to be reorganized by grouping terms according to their lag $k \in \{0,1,\dots,N-1\}$. Specifically:
\begin{itemize}
    \item For lag $k=0$: there are $N$ terms $\Gamma(0)$, coming from all $i=j$.
    \item For lag $k=1$: there are $2(N-1)$ terms, corresponding to all $i$ and $j$ with $|i-j|=1$.
    \item For a general lag $k \in [1, N-1]$: there are $2(N-k)$ terms.
\end{itemize}
This yields
\[
\sum_{i=1}^{N} \sum_{j=1}^{N} \Gamma(|i-j|) = N \Gamma(0) + 2 \sum_{k=1}^{N-1} (N-k)\,\Gamma(k),
\]
and hence
\[
\operatorname{Var}(\bar{X}_N) 
= \frac{1}{N^2}\left[ N\Gamma(0) + 2 \sum_{k=1}^{N-1} \left(N-k\right)\Gamma(k) \right].
\]
In the asymptotic limit $N \to \infty$, and assuming that the autocovariance decays sufficiently rapidly, $\frac{N-k}{N} \to 1$ for fixed $k$, allowing the upper limit of the sum to be extended to infinity. One then obtains
\[
    \operatorname{Var}(\bar{X}_N) \approx \frac{1}{N} \left[ \Gamma(0) + 2 \sum_{k=1}^{\infty} \Gamma(k) \right].
\]
Expressing the autocovariances in terms of the autocorrelation function $\Gamma(k)= \rho(k) \sigma^2$, the variance of the sample mean becomes
\begin{equation}
    \operatorname{Var}(\bar{X}_N) \approx \frac{\sigma^2}{N} \left[ 1 + 2 \sum_{k=1}^{\infty} \rho(k) \right].
    \label{eq:6.22}
\end{equation}
Equivalently, using the symmetry property $\rho(k)=\rho(-k)$, this expression can be written as
\begin{equation}
    \operatorname{Var}(\bar{X}_N) \approx \frac{\sigma^2}{N} \sum_{k=-\infty}^{\infty} \rho(k).
    \label{eq:6.23}
\end{equation}
where the infinite sum over autocorrelations quantifies the statistical inefficiency induced by correlations in the Markov chain. Relative to the ideal i.i.d.\ case, the variance of the estimator is inflated by a factor proportional to the total integrated autocorrelation across all time lags.

\subsubsection{Integrated Autocorrelation Time}
While the autocorrelation function $\rho(k)$ quantifies the statistical dependence between samples separated by a lag $k$, the overall impact of these correlations on the efficiency of an MCMC simulation is conveniently summarized by a single quantity: the \emph{integrated autocorrelation time} (IAT), denoted $\tau_{\text{int}}$. Starting from the asymptotic expression for the variance of the sample mean given in Eq.~(\ref{eq:6.22}),
\[
     \operatorname{Var}(\bar{X}_N)\approx \frac{\sigma^2}{N} \left(  1 + 2\sum_{k=1}^{\infty} \rho(k) \right),
\]
and recalling that $\rho(0)=1$ by definition and that $\rho(k)=\rho(-k)$, we define the integrated autocorrelation time as
\begin{equation}
    \tau_{\text{int}} = \sum_{k=-\infty}^{\infty} \rho(k) = \frac{1}{2} + \sum_{k=1}^{\infty} \rho(k).
    \label{eq:6.25}
\end{equation}
With this definition, the variance of the sample mean can be written compactly as
\begin{equation}
    \operatorname{Var}(\bar{X}_N) \approx \frac{\sigma^{2}}{N} \cdot 2\tau_{\text{int}}
\end{equation}
The integrated autocorrelation time $\tau_{\text{int}}$ quantifies the degree of statistical inefficiency induced by correlations in the Markov chain. Intuitively, it represents the average number of consecutive MCMC samples required to obtain one effectively independent observation. Larger values of $\tau_{\text{int}}$ indicate stronger persistence of correlations and, consequently, reduced sampling efficiency. On this basis, the \emph{effective sample size} $N_{\text{eff}}$ is defined as the number of independent samples that would yield the same statistical precision as the correlated MCMC sample,
\begin{equation}
    N_{\text{eff}}=\frac{N}{2\tau_{\text{int}}}.
\end{equation}
Accordingly, the standard error of the sample mean must be corrected to account for autocorrelation,
\begin{equation}
    \sigma_{\text{MCMC}} \approx \sqrt{\frac{2\tau_{\text{int}}}{N}} \cdot \sigma = \frac{\sigma }{\sqrt{N_{\text{eff}}}}.
\end{equation}
This expression provides the appropriate generalization of the Monte Carlo standard error for correlated samples generated by a Markov chain. In practice, estimating $\tau_{\text{int}}$ is essential for assessing the reliability and efficiency of MCMC simulations. Chains with small $\tau_{\text{int}}$ mix rapidly and yield a large number of effective samples, whereas chains with large $\tau_{\text{int}}$ require substantially longer runs to achieve the same level of statistical precision. For a more detailed discussion of autocorrelation analysis in MCMC, see Refs.~\cite{Wolff:2003sm,geyer1992,givens2012computational}.

\subsubsection{$\Gamma$-method}\label{gamma-method}
The $\Gamma$-method \cite{Wolff:2003sm} provides a practical and statistically robust framework for estimating the integrated autocorrelation time $\tau_{\text{int}}$ from time series generated by MCMC simulations. As defined in Eq.~\eqref{eq:6.25}, $\tau_{\text{int}}$ is given by an infinite sum over the normalized autocorrelation function $\rho(k)$. In practice, however, this sum cannot be evaluated exactly from finite data. The $\Gamma$-method resolves this issue through an \emph{automatic windowing procedure} that truncates the sum in a controlled and statistically justified manner. Rather than summing autocorrelations up to an arbitrarily chosen maximum lag, the $\Gamma$-method determines an optimal cutoff window, denoted $W_{\text{opt}}$, such that contributions from $\rho(k)$ beyond this point are negligible compared to the statistical noise. For a given window size $W$, the truncated estimator of the integrated autocorrelation time is defined as
\begin{equation}
    \widehat{\tau}_{\text{int}}(W) = \frac{1}{2} + \sum_{k=1}^{W} \rho(k).
\end{equation}
The window size $W$ is increased iteratively until a suitable stopping criterion is met, with the aim of capturing the dominant correlation structure while excluding noisy, non-informative long-lag contributions that would otherwise destabilize the estimate. To guide the choice of $W_{\text{opt}}$, a decay factor $g_W$ is introduced to monitor the expected residual correlation at lag $W$,
\begin{equation}
    g_W = \exp\left(-\frac{W}{\tau_{\text{exp}}}\right).
\end{equation}
where $\tau_{\text{exp}}$ denotes a characteristic exponential decay time, which is approximated self-consistently from the observed autocorrelation behavior and the current estimate of $\tau_{\text{int}}$. The window size is increased until $g_W$ becomes sufficiently small—indicating that the remaining correlations are negligible compared to statistical fluctuations—or until a predefined maximum window size is reached. The resulting value is taken as the optimal cutoff $W_{\text{opt}}$. Once $W_{\text{opt}}$ is determined, the integrated autocorrelation time is accumulated recursively according to
\begin{align*}
    \tau[0] &= \frac{1}{2}, \\
    \tau[W] &= \tau[W - 1] + \rho[W], \quad \text{for } W = 1, \dots, W_{\text{opt}}.
\end{align*}
The statistical uncertainty associated with the estimated autocorrelation time is approximated by
\begin{equation}
    \tau_{\text{err}}(W) = 2 \cdot \tau[W] \cdot \sqrt{ \frac{W + 0.5 - \tau[W]}{N} },
\end{equation}
where $N$ denotes the total number of MCMC samples (after thermalization). This expression accounts for both the finite length of the time series and the residual noise in the estimated autocorrelation function. Using the optimal window $W_{\text{opt}}$, the variance of the sample mean corrected for autocorrelation effects is given by
\begin{equation}
    \operatorname{Var}(\bar{X}_N) \approx \frac{\sigma^2}{N} \cdot 2\tau[W_{\text{opt}}],
\end{equation}
and the corresponding standard error becomes
\begin{equation}
    \sigma_{\text{MCMC}} \approx \sqrt{\frac{2\tau[W_{\text{opt}}]}{N}} \cdot \sigma.
\end{equation}
The strength of the $\Gamma$-method lies in its ability to balance two competing sources of error: systematic bias arising from truncating the autocorrelation sum too early, and increased statistical noise resulting from including poorly determined long-lag correlations. By adaptively selecting the window $W_{\text{opt}}$, the method provides a reliable and statistically well-controlled estimate of $\tau_{\text{int}}$, the effective sample size, and the associated Monte Carlo uncertainties. For further implementation details and performance studies, we refer to Ref.~\cite{Wolff:2003sm}.

\subsection{Thinning Method}\label{sec:6.3.4}
As discussed previously, the $\Gamma$-method provides a statistically rigorous framework for estimating the integrated autocorrelation time $\tau_{\text{int}}$ and for quantifying the impact of autocorrelations on uncertainty estimates. In the present analysis, the MCMC chains generated for the nuclear PDF fits exhibit a substantial degree of autocorrelation, reflecting the high dimensionality of the parameter space and the complexity of the posterior distribution. While the $\Gamma$-method allows these correlations to be treated correctly, the explicit evaluation of $\tau_{\text{int}}$ for each parameter and for each observable of interest can become computationally demanding in large-scale analyses. Moreover, the full MCMC chains contain a significant amount of redundant information due to strong correlations between successive samples. Retaining all samples would therefore be inefficient both from a computational and a data-management perspective, particularly given the goal of constructing and publicly releasing the final nPDF sets. For these reasons, we adopt the \emph{thinning} method \cite{link2012thinning,riabiz2022optimal} as a complementary strategy. Thinning is used primarily to simplify error estimation by producing approximately uncorrelated samples and to reduce the total number of nPDF replicas that need to be stored and distributed, while preserving the essential statistical properties of the posterior distribution.

Thinning involves retaining only every $\eta$-th sample from the original chain and discarding the rest, where $\eta$ denotes the thinning rate. Starting from a chain of length $N$,
\[ 
\{\bm{\theta}_{1}, \bm{\theta}_{2}, \dots, \bm{\theta}_{N}\} \;\longrightarrow\; \{\bm{\theta}_{1}, \bm{\theta}_{1+\eta}, \bm{\theta}_{1+2\eta}, \dots\, \bm{\theta}_{1+(\left\lfloor \tfrac{N}{\eta} \right\rfloor-1)\eta} \}.
\]
By construction, this procedure increases the separation between retained samples and substantially suppresses autocorrelation, so that the resulting subsampled chain more closely approximates an independent and identically distributed sequence. In the idealized limit of completely uncorrelated samples, the integrated autocorrelation time reaches its theoretical minimum,
\[
(\tau_{\text{int}})^{\text{ideal}} = \tfrac{1}{2},
\]
which corresponds to a scenario where the autocorrelation function $\rho(k)$ vanishes for all lags $k \geq 1$ in Eq. (\ref{eq:6.25}). To approach this ideal scenario in practice, we estimate $\tau_{\text{int}}$ for each parameter using the $\Gamma$-method described in Sec.~\ref{gamma-method}. Based on these estimates, the thinning rate $\eta$ is determined through an iterative procedure. Starting from a small initial value of $\eta$, we recompute $\tau_{\text{int}}$ for the thinned chain. If the resulting autocorrelation time satisfies $\tau_{\text{int}} \approx 0.5$ within statistical uncertainty, the procedure is terminated and the retained samples are treated as effectively independent. Otherwise, the thinning interval is increased incrementally, and the procedure is repeated until the desired reduction in autocorrelation is achieved. This iterative approach ensures that the retained samples provide a faithful approximation to independent draws from the target posterior distribution.

The application of the thinning in this analysis is motivated by two main considerations:
\begin{enumerate}
    \item Simplification of Error Analysis: By reducing autocorrelation to negligible levels, the thinned chain can be treated as effectively uncorrelated. This permits the use of the standard Monte Carlo error estimate (Eq. (\ref{eq:6.17})) without the need for observable-specific autocorrelation corrections.
    \item Data reduction and practical output: A key objective of this work is to produce a compact yet statistically representative ensemble of nuclear PDFs for public release in the form of LHAPDF grids~\cite{Buckley:2014ana}. Retaining the full correlated MCMC chains would be computationally inefficient and largely redundant. Thinning allows a substantial reduction in the number of stored replicas while preserving the essential features of the posterior distribution, resulting in nPDF sets that are both practical for phenomenological applications and statistically robust.
\end{enumerate}
Since the thinning procedure yields samples that are approximately uncorrelated and independent, the outputs from multiple independently initialized chains can be merged into a single ensemble without introducing bias. This combined chain increases the effective sample size and creates a robust, unified Monte Carlo representation of the posterior distribution.

\subsection{Parameter Estimation and Uncertainty}\label{sec:6.3.5}
Parameter estimation and uncertainty quantification in this work are performed using the ensemble of nPDF replicas obtained from the merged, thinned MCMC chains described in the previous section. After convergence, thermalization, thinning, and the combination of independent chains, the resulting ensemble provides a statistically robust Monte Carlo representation of the posterior distribution. Let $\{ \bm{\theta}^{(i)} \}_{i=1}^{N_{\rm rep}}$ denote the set of MCMC-sampled parameter vectors, where each $\bm{\theta}^{(i)}$ represents a complete configuration of the fitted nPDF parameters and $N_{\rm rep}$ is the number of independent samples retained after burn-in and thinning. From each parameter set, the corresponding nuclear parton distribution functions are computed as
\[
f_i(x, Q^2) = f(x, Q^2; \bm{\theta}^{(i)}),
\]
yielding an ensemble of $N_{\rm rep}$ nPDF replicas $\{ f_i(x, Q^2) \}_{i=1}^{N_{\rm rep}}$. In principle, this combined ensemble of approximately independent samples could be used directly to estimate uncertainties through standard Monte Carlo estimators. However, for practical purposes and for clarity in interpreting the results, we summarize the posterior information using a set of well-established uncertainty estimation methods. These methods provide complementary perspectives on the structure of the posterior distribution and are commonly employed in global PDF analyses to construct confidence and credible intervals. In our analysis, we employ several approaches to estimate uncertainties, each based on a different statistical interpretation and offering complementary insights. Uncertainty quantification can be performed at two distinct but closely related levels:
\begin{itemize}
  \item On the sampled parameter values $\{\bm{\theta}^{(i)}\}$ directly, which provides insights into the posterior of the fit parameters;
  \item On the derived nPDFs $\{f_i(x, Q^2)\}$ and on observables evaluated from them (e.g., cross sections), which enables the propagation of parameter uncertainties to physically meaningful distributions and phenomenological predictions.
\end{itemize}
The methods described below focus primarily on uncertainty estimation at the level of the nPDF replicas and derived observables, as these constitute the central phenomenological output of the analysis.
\begin{enumerate}
    \item \textbf{Monte Carlo Standard Error (MCSE):}
    In this approach, the central value of an observable \(f(x,Q^2)\) is defined as the ensemble mean over the set of replicas,
    \[
    \mu(x,Q^2) = \frac{1}{N_{\rm rep}} \sum_{i=1}^{N_{\rm rep}} f_i(x,Q^2),
    \]
    where \(N_{\rm rep}\) denotes the total number of replicas. The associated uncertainty is quantified by the sample standard deviation,
    \[
    S(x,Q^2) =
    \sqrt{\frac{1}{N_{\rm rep}-1}\sum_{i=1}^{N_{\rm rep}}\left[f_i(x,Q^2) - \mu(x,Q^2)\right]^2}.
    \]
    Assuming approximate normality of the posterior distribution, the symmetric \(68.3\%\) confidence interval is then constructed as
    \[
    \text{Band:}\quad[\, \mu(x,Q^2) - S(x,Q^2),\; \mu(x,Q^2) + S(x,Q^2) \,].
    \]
    This method is efficient and straightforward when the posterior is close to Gaussian. However, it may underestimate uncertainties if the posterior distribution is skewed or exhibits nontrivial structure, such as long tails or multiple modes.\footnote{Strictly speaking, this construction does not correspond to the Monte Carlo standard error in the statistical sense. The true MCSE quantifies the uncertainty on the estimated mean and is given by \(S(x,Q^2)/\sqrt{N_{\rm rep}}\) (or more generally using the effective sample size in the presence of autocorrelations). In contrast, the band defined here reflects the dispersion of the posterior distribution itself, i.e.\ the standard deviation, rather than the uncertainty on the mean estimator.}

    \item \textbf{Percentile Method:}  
    The percentile method is a fully Bayesian approach that constructs credible intervals directly from the empirical distribution of the replica ensemble at each value of \(x\) and \(Q^2\). It is widely employed in modern global PDF analyses, such as nNNPDF3.0~\cite{nNNPDF3.0:2022fyi}, and does not rely on assumptions of symmetry or normality of the posterior distribution.
    Let \(\{f_i(x,Q^2)\}_{i=1}^{N_{\rm rep}}\) denote the full set of replica predictions. For a desired confidence level \(\mathrm{CI}\) (e.g., \(68.3\%\)), the central value and bounds are defined through the corresponding percentiles of the distribution:
    \begin{align*}
    f_{\rm med}(x,Q^2) &= \text{50th percentile (median)}, \\
    f_{\rm low}(x,Q^2) &= \left(\frac{100-\mathrm{CI}}{2}\right)\text{th percentile}, \\
    f_{\rm high}(x,Q^2) &= \left(\frac{100+\mathrm{CI}}{2}\right)\text{th percentile}.
    \end{align*}
    For the canonical \(68.3\%\) interval, this corresponds approximately to
    \[
    \text{Band:}\quad [\, f_{\rm low}(x,Q^2),\; f_{\rm high}(x,Q^2) \,] \;\approx\;
    [\,15.85\%,\;84.15\%\,].
    \]
   This method corresponds directly to the construction of quantile-based Bayesian credible intervals and provides a robust, potentially asymmetric uncertainty band. It is particularly well suited for non-Gaussian posterior distributions, as it faithfully captures the local spread of the posterior at each point in phase space.

    \item \textbf{ Cumulative-$\boldsymbol{\chi^2}$ Method:}
    The cumulative-\(\chi^2\) method incorporates goodness-of-fit information to identify the most statistically consistent parameter configurations~\cite{Putze:2008ps}. The full set of replicas is first ordered in ascending order of their associated \(\chi^2\) values,
    \[
    \chi^2_1 \leq \chi^2_2 \leq \cdots \leq \chi^2_{N_{\rm rep}}.
    \]
    The best-fit replica is defined as
    \[
    f_{\rm best}(x,Q^2) = f_1(x,Q^2),
    \]
    corresponding to the minimum \(\chi^2\), i.e.\ the maximum-posterior (MAP) estimate under the assumption of Gaussian errors and weakly informative priors. The number of replicas retained is chosen as a fixed fraction of the sample (e.g., $68.3\%$),
    \[
    N_{68} = \left\lceil 0.683 \times N_{\rm rep} \right\rceil,
    \]
    thereby defining a subset of parameter configurations with the lowest \(\chi^2\) values.\footnote{Since the posterior distribution is proportional to $\exp(-\chi^2/2)$ in this setting, this selection effectively identifies the highest-posterior-density (HPD) samples.} The uncertainty band is then constructed as the pointwise envelope over this subset:
    \begin{gather*}
    f_{\min}(x,Q^2) = \min_{1 \le i \le N_{68}} f_i(x,Q^2), \\
    f_{\max}(x,Q^2) = \max_{1 \le i \le N_{68}} f_i(x,Q^2), \\
    \text{Band:}\quad
    [\, f_{\min}(x,Q^2),\; f_{\max}(x,Q^2) \,].
    \end{gather*}
    This procedure yields a conservative uncertainty envelope composed exclusively of replicas that provide a globally acceptable description of the data, thereby emphasizing the model’s predictive core.
\end{enumerate}
Although all three methods aim to characterize a \(68.3\%\) uncertainty region, they differ fundamentally in how this region is defined. The percentile method provides a local, pointwise characterization of the posterior spread at each value of \(x\) and \(Q^2\), corresponding directly to quantile-based Bayesian credible intervals. In contrast, the cumulative-\(\chi^2\) method selects a subset of parameter configurations based on global goodness-of-fit and constructs an uncertainty band from their envelope. As a result, the percentile method typically yields uncertainty bands that closely reflect the local structure of the posterior, while the \(\chi^2\)-based approach emphasizes global consistency with the data and generally produces more conservative uncertainty estimates. Together, these complementary methods provide a comprehensive characterization of the uncertainties associated with the extracted nuclear PDFs.
A detailed presentation of the resulting parameter estimates and nuclear PDFs, together with their associated uncertainties, is given in the next chapter.

\section{Overview of the Methodology}\label{sec:6.4}
The methodology presented in this work provides a systematic and statistically rigorous framework for performing a global analysis of nuclear parton distribution functions using Markov Chain Monte Carlo sampling. The approach is designed to thoroughly explore the posterior distribution of the nPDF parameters, to quantify uncertainties in a principled Bayesian manner, and to ensure robustness and internal consistency at all stages of the analysis. The main components of the methodology are summarized below.

\paragraph{nPDF Analysis Setup: Data Selection and Parametrization.}
The analysis begins with a careful selection of experimental datasets relevant for constraining nuclear modifications of parton distributions. Two complementary global fits are performed. The primary analysis, referred to as the \texttt{Pb-only} fit, includes only measurements involving lead nuclei ($A=208$). This dataset comprises inclusive charged-current neutrino DIS data from the CHORUS experiment, electroweak vector boson ($W^\pm$, $Z$) production in $p$Pb collisions measured by LHC, and heavy-flavor production in $p$Pb collisions from LHC. 
Furthermore, a \texttt{Multi-nuclei} analysis is also carried out, incorporating additional measurements on a variety of nuclear targets. These include fixed-target DIS data from EMC, BCDMS, SLAC, and NMC, as well as Drell--Yan production data from the E772 and E866 experiments. This broader analysis probes the consistency of nuclear modifications across different nuclei and provides further insight into the $A$-dependence of nPDFs.

The theoretical modeling of the nPDFs is implemented within the \textbf{nCTEQ} framework, using the CJ15 proton PDFs as the baseline reference. Nuclear modifications are parameterized at the initial scale $Q_0 = 1.3~\mathrm{GeV}$ using a flexible yet controlled functional form. In this work, the flavor and $x$-dependent nuclear effects are governed by a total of ten free nuclear-modification parameters, which are determined through the fit. The PDFs defined at the input scale are subsequently evolved to higher values of $Q^2$ by solving the DGLAP evolution equations at next-to-leading order (NLO), using \texttt{HOPPET} and \texttt{APFEL}.
To enable efficient evaluation of theoretical predictions during the fit, high-speed interpolation grids are constructed using \texttt{APPLgrid} and \texttt{APFEL++}. This setup allows for fast convolution of parton-level cross sections with nPDFs, thereby significantly reducing the computational cost associated with repeated likelihood evaluations in the MCMC sampling.

\paragraph{Posterior Construction and MCMC Sampling.}
The goal of the fit is to sample from the posterior distribution of nPDF parameters, conditioned on the experimental data and the corresponding theoretical predictions. The posterior probability density is constructed from a likelihood function derived from the global $\chi^2$ statistic, which consistently incorporates both uncorrelated and correlated experimental uncertainties. Under the assumption of Gaussian-distributed experimental errors and adopting flat priors for all fit parameters, the posterior distribution is proportional to $\exp(-\chi^2/2)$. Sampling of the posterior distribution is performed using the adaptive Metropolis–Hastings algorithm. To ensure robust exploration of the parameter space and to facilitate reliable convergence diagnostics, multiple independent Markov chains are run in parallel. Each chain is initialized using a distinct random seed and starting points drawn from the vicinity of the best-fit values obtained in a preliminary Hessian analysis. Following an initial non-adaptive phase, the proposal distribution is dynamically updated to reflect the evolving covariance structure of the sampled posterior. This adaptive strategy improves sampling efficiency, enhances mixing, and enables effective exploration of correlated directions in the high-dimensional parameter space.

\paragraph{Chain Convergence and Preprocessing.}
After the MCMC chains are generated, their convergence and quality are carefully assessed using a set of standard diagnostic tools. Trace plots, cumulative mean plots, and autocorrelation function analyses are employed to verify proper mixing and stability, thereby ensuring that the chains have reached their stationary distribution. Only chains exhibiting clear and consistent convergence behavior are retained for further analysis.
Once convergence is established, an initial thermalization (burn-in) period is removed from each chain. The remaining post-burn-in samples are subsequently used for autocorrelation analysis, thinning, chain merging, and the extraction of parameter estimates and uncertainties.

\paragraph{Thinning and Combining Approximately Independent Samples.}
To suppress autocorrelations in the post-burn-in chains, we apply a thinning procedure to each converged chain by retaining only every $\eta$-th sample. The thinning interval $\eta$ is determined using an iterative procedure, guided by estimates of the integrated autocorrelation time $\tau_{\text{int}}$ obtained with the $\Gamma$-method. The thinning is chosen such that the resulting thinned chains satisfy $\tau_{\text{int}} \simeq 0.5$ within statistical uncertainty, corresponding to approximately independent samples.
The thinned samples from all retained chains are then merged into a single combined ensemble. Since the samples are approximately uncorrelated and the chains are independently initialized, this merging procedure increases the effective statistical precision of the posterior representation without introducing bias.

\paragraph{Evaluation of nPDFs and Uncertainty Propagation.}
Each state of the combined chain corresponds to a unique point in the nPDF parameter space. For every sampled parameter vector, the full set of nuclear PDFs is evaluated as functions of $(x,Q^2)$ using the input parameterization at $Q_0$ and NLO DGLAP evolution equations. The resulting ensemble of nPDF replicas $\{f_i(x,Q^2)\}$ provides a Monte Carlo representation of the posterior distribution and enables the propagation of parameter uncertainties to physically relevant distributions and derived observables.
For practical applications, the replica ensemble is converted into the standard LHAPDF format~\cite{Buckley:2014ana}. Each replica is stored as an independent PDF member, preserving the statistical information contained in the sampled posterior. To summarize this ensemble in terms of central values and uncertainty bands, three complementary statistical constructions are employed:
\begin{itemize}
    \item Monte Carlo Standard Error: Assumes an approximately Gaussian posterior and defines a symmetric uncertainty interval based on the ensemble mean and standard deviation.
    \item Percentile Method: constructs pointwise credible intervals from the empirical quantiles of the replica distribution at each $(x,Q^2)$, without assuming symmetry or Gaussianity.
    \item Cumulative-$\chi^2$ Method: orders replicas according to their $\chi^2$ values and constructs a conservative envelope from the subset of best-fitting replicas.
\end{itemize}

\chapter{Results and Discussion}\label{Chap.7}
This chapter presents the main results of the Bayesian global QCD analysis of nuclear parton distribution functions (nPDFs) performed within the Markov Chain Monte Carlo (MCMC) framework developed in Chapter~\ref{chap.6}. The goal of this chapter is first to provide a detailed statistical characterization of the posterior distributions obtained from the MCMC sampling and second, to extract the corresponding nuclear PDFs together with their uncertainties and compare them to conventional Hessian-based results.
To balance statistical precision, computational feasibility, and phenomenological scope, two complementary analyses are performed. 
\begin{itemize}
    \item \texttt{Pb-only} Analysis: The primary fit focuses exclusively on the lead nucleus ($A=208$) and includes data from electroweak gauge boson production, heavy-flavor production in $p$Pb collisions, and neutrino deep inelastic scattering on lead. Restricting the dataset to a single heavy nucleus significantly reduces computational complexity in the evaluation of theoretical predictions, enabling the generation of a substantially larger MCMC ensemble. This high-statistics posterior sample allows for a detailed exploration of the nuclear PDF parameter space, including the investigation of parameter correlations, non-Gaussian structures, and potential multi-modal behavior. As such, the \texttt{Pb-only} analysis constitutes the central quantitative result of this thesis.

    \item \texttt{Multi-nuclei} Analysis: As a complementary study, the analysis is extended to include measurements from multiple nuclear targets. This \texttt{Multi-nuclei} fit probes the $A$-dependence of nuclear modifications and tests the stability and universality of the adopted parameterization across different mass numbers. Although computationally more demanding and therefore based on a smaller posterior ensemble, this study provides an essential robustness check of the Bayesian framework and places the \texttt{Pb-only} results in a broader phenomenological context.
\end{itemize}

The chapter is organized as follows. Section~\ref{sec:7.1} presents the \texttt{Pb-only} analysis in two distinct stages. We begin with a detailed study of a representative individual MCMC chain, focusing exclusively on statistical validation. Convergence diagnostics, thermalization behavior, autocorrelation properties, and mixing efficiency are examined to ensure that the Markov process has reached stationarity and provides a reliable sampling of the target posterior. Then, all properly thermalized and uncorrelated chains are combined to construct the full posterior ensemble. The combined sample is then used to analyze the posterior structure and its physical implications. Marginalized parameter distributions, parameter correlations, and possible secondary minima are investigated, and deviations from Gaussian behavior are quantified. From this ensemble, the lead nuclear PDFs and their associated credible uncertainty bands are extracted. These results are subsequently compared with those obtained using the traditional Hessian approach, allowing a direct assessment of differences in uncertainty estimation and sensitivity to non-Gaussian features.

Section~\ref{sec:7.2} begins with the analysis of the combined \texttt{Multi-nuclei} posterior ensemble, constructed from all converged and uncorrelated chains. The global posterior structure is examined, providing the statistical foundation for the physical interpretation. 
From this ensemble, the lead nuclear PDFs are extracted and presented with central values and Bayesian credible uncertainty bands obtained directly from the MCMC samples. The resulting uncertainties are then compared with those obtained using the traditional Hessian method. 
Sec.~\ref{sec:7.3} provides a systematic comparison between the \texttt{Multi-nuclei} and \texttt{Pb-only} analyses, considering both the MCMC-based uncertainty estimates and the corresponding Hessian results.
Finally, Sec.~\ref{sec:7.4} addresses the construction of a practical LHAPDF representation of the Bayesian result. Starting from the full MCMC posterior ensemble, two complementary compression strategies are applied and systematically compared in order to obtain a statistically controlled and computationally efficient set of representative replicas.

%%%%%%%%%%%%%%%%%%%%%%%%%%%%%%%%%%%%%%%%%%%%%%%%%%%%%%%%%%%%%
\section{Pb-only Analysis}\label{sec:7.1}
%%%%%%%%%%%%%%%%%%%%%%%%%%%%%%%%%%%%%%%%%%%%%%%%%%%%%%%%%%%%%
This section presents the results of the primary \texttt{Pb-only} fit. We begin with a statistical validation of the MCMC sampling using a representative single chain, followed by the construction and analysis of the combined posterior ensemble. From this ensemble, the lead nuclear PDFs are extracted and presented together with their associated uncertainty bands.
%%%%%%%%%%%%%%%%%%%%%%%%%%%%%%%
\subsection{Single Chain Analysis}\label{sec:7.1.1}
%%%%%%%%%%%%%%%%%%%%%%%%%%%%%%%
\begin{figure}
  \centering
  \includegraphics[angle=90,width=\paperheight,height=\paperwidth,keepaspectratio]
  {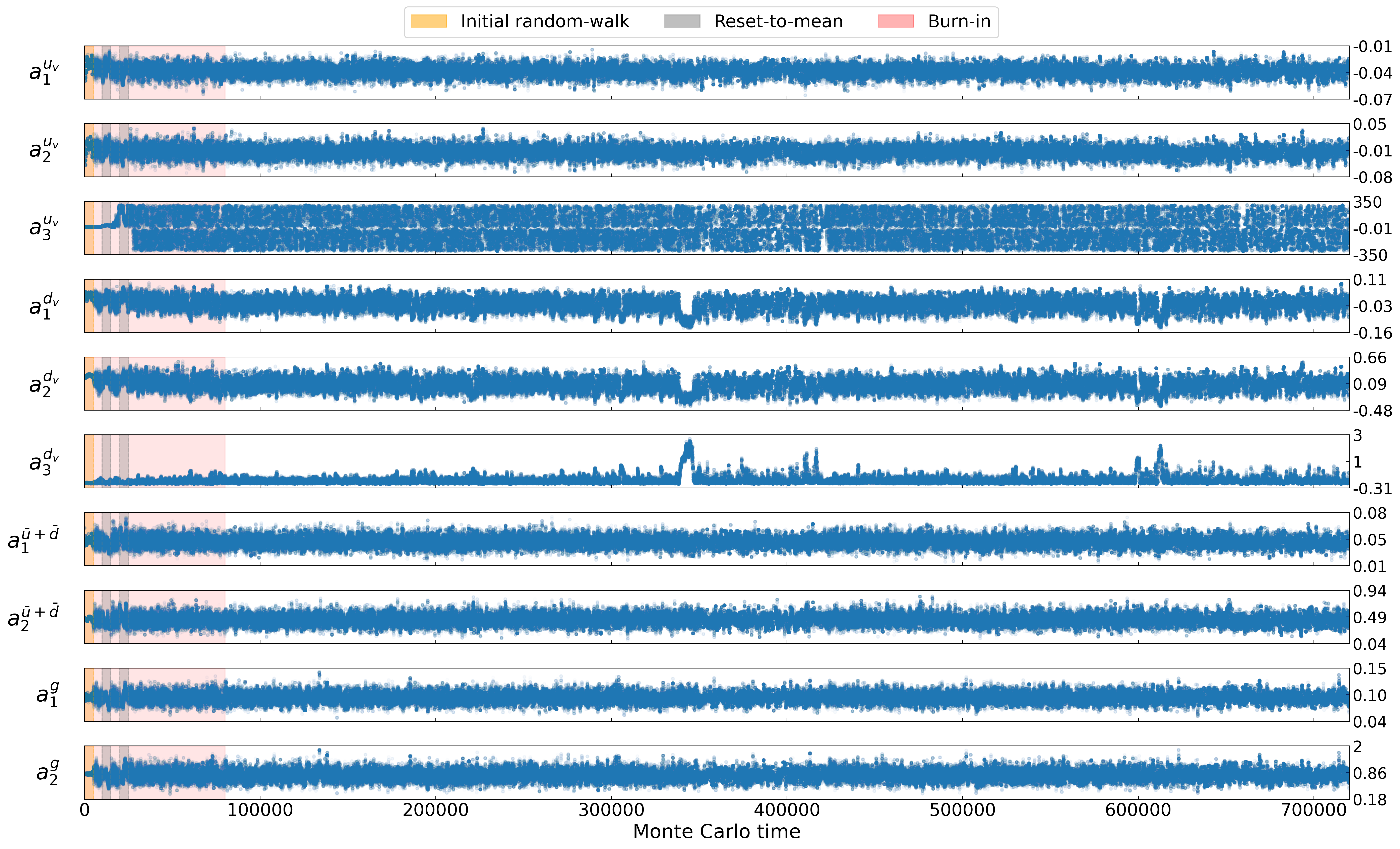}
  \caption[Trace plot for a representative 10-parameter \texttt{Pb-only} MCMC chain]{Trace plot of the parameter values as a function of Monte Carlo time for a representative single chain in the 10-parameter \texttt{Pb-only} MCMC analysis. The initial $5\,000$ non-adaptive steps used for covariance estimation are shaded in orange. Proposal covariance updates (RTM adaptations) are indicated in gray. The burn-in period of $80\,000$ steps is marked in red. The stationary fluctuations observed beyond the burn-in region indicate that the chain has reached equilibrium and is sampling the target posterior distribution.}
  \label{fig:7.1}
\end{figure}

Figure~\ref{fig:7.1} displays the full trace plot of a representative MCMC chain obtained in the \texttt{Pb-only} fit. The purpose of this subsection is strictly statistical: we assess convergence, thermalization, and mixing properties of the adaptive Metropolis–Hastings implementation in the ten-dimensional parameter space.
The \texttt{Pb-only} fit includes $1\,488$ experimental data points, including $W^\pm$ and $Z$ boson production, heavy-flavor production, and neutrino deep-inelastic scattering (DIS) on lead, as summarized in Tables~\ref{tab:WZdata}–\ref{tab:HQdata}. To ensure the validity of perturbative QCD, standard DIS kinematic cuts of $Q^2 > 4.0~\mathrm{GeV}^2$ and $W^2 > 12.25~\mathrm{GeV}^2$ are imposed on the neutrino DIS data. In addition, a transverse-momentum cut of $P_T > 3.0~\mathrm{GeV}$ is applied to the heavy-flavor production data to restrict the analysis to the perturbative regime.
Nuclear modifications are parameterized within the nCTEQ framework, following the functional forms defined in Eqs.~(\ref{eq:6.3})–(\ref{eq:6.6}), with the CJ15 proton PDFs used as the baseline input.
The representative chain was generated with the following configuration:
\begin{itemize}
    \item \textbf{Initial covariance estimation ($N_{\mathrm{pre}} = 5\,000$):}  
    A $5\,000$-step pre-run is performed to estimate the initial proposal covariance matrix. During this stage—and immediately after each RTM adaptation—the sampler operates in non-adaptive mode to stabilize the proposal distribution.
    \item \textbf{Proposal adaptation (RTM updates):}  
    The proposal covariance matrix is updated at $10\,000$ and $20\,000$ steps during burn-in, improving sampling efficiency and enabling exploration of correlated directions in parameter space.
    \item \textbf{Initial proposal scale ($\sigma_0 = 0.001$):}  
    A small initial scale ensures controlled exploration during the early sampling phase before adaptation stabilizes the covariance structure.
    \item \textbf{Mixture parameter ($\beta = 0.2$):}
    The mixing probability of the proposal in the AMH algorithm that ensures stable and ergodic sampling.
    \item \textbf{Prior constraint:}  
    A uniform prior is imposed on $a_{3}^{u_v}$ in the range $[-300,\,300]$ to prevent  the sampler from exploring unphysical regions of parameter space.
    \item \textbf{Initialization:}  
    The chain is started at the Hessian minimum to reduce initial transient behavior and accelerate convergence toward the high-probability region.
    \item \textbf{Total length:}  
    The chain consists of $720\,000$ proposed points (accepted and rejected), sampling the full ten-dimensional nuclear modification parameter space.
    \item \textbf{Burn-in removal ($N_{\mathrm{burn}} = 80\,000$):}  
    The first $80\,000$ steps (red shaded region in Fig.~\ref{fig:7.1}) are discarded to keep only equilibrium samples.
\end{itemize}

\begin{figure}[!h]
\centering
\includegraphics[width=\textwidth]{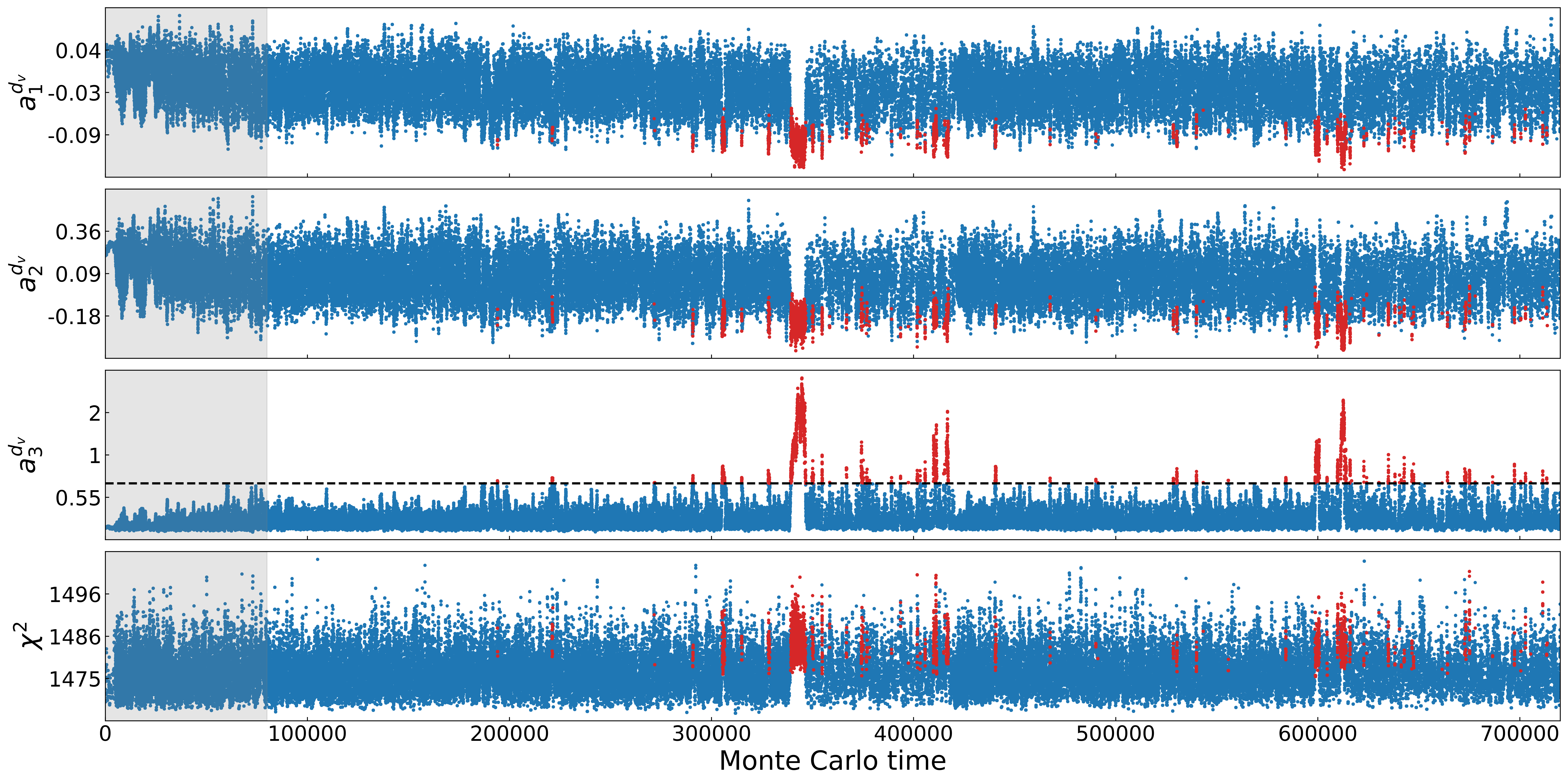}
\caption[Observed secondary minima in $a_3^{d_v}$ for a single MCMC chain]{To probe the non-trivial structure observed in the trace of $a_3^{d_v}$, a threshold $T= 0.8$ is applied to separate the two regions of parameter space. Samples with $a_3^{d_v} \le 0.8$ are shown in blue and those with $a_3^{d_v} > 0.8$ in red. The same color-coding is applied to the corresponding traces of $\chi^2$, $a_1^{d_v}$, and $a_2^{d_v}$, highlighting correlated shifts in the $d_v$ parameter values associated with transitions between these regions.}
\label{fig:7.2}
\end{figure}

After removal of the burn-in phase, the traces of most parameters exhibit clear stationary behavior, characterized by stable fluctuations around well-defined mean values and the absence of systematic long-term drift. The behavior observed in Fig.~\ref{fig:7.1} is consistent with equilibration of the Markov chain and indicates efficient sampling of the dominant posterior region throughout the full runtime. The valence parameters, however, display a more intricate structure. In particular, the parameter $a_3^{d_v}$ has occasional transitions between two distinct regions of parameter space. As illustrated in Fig.~\ref{fig:7.2}, where samples are separated using a threshold $T = 0.8$ and color-coded accordingly, the chain predominantly fluctuates around values near zero (blue region) but occasionally transitions to a region at larger values (red region). 
Applying the same color-coding to the traces of $a_1^{d_v}$, $a_2^{d_v}$, and $\chi^2$ reveals correlated shifts within the $d_v$ parameter subspace during these transitions. Importantly, the corresponding $\chi^2$ values are somewhat larger but remain within the typical range of the well-sampled region, indicating that these configurations are statistically comparable and not strongly disfavored by the data. A milder structural feature is also observed in the trace of $a_3^{u_v}$, suggesting a non-trivial posterior structure in the valence sector.
A definitive characterization of these features requires analysis of the merged and uncorrelated ensemble of all converged chains, which is presented in the following Sec. \ref{sec.7.1.2}.

\begin{figure}[h]
\centering
\includegraphics[width=\textwidth]{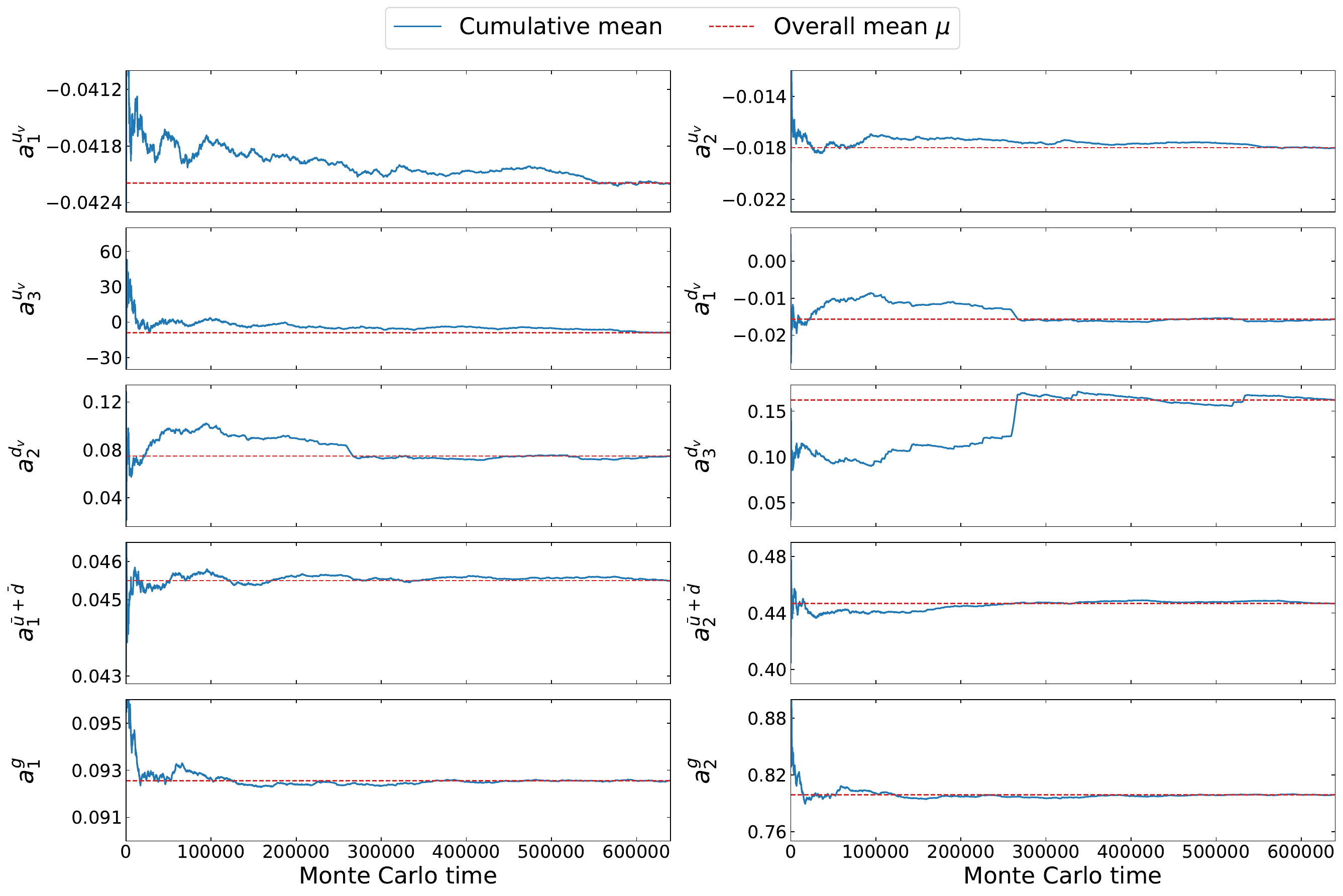}
\caption[Cumulative mean evolution of a representative chain]{Cumulative mean of all 10 fit parameters for the chain shown in Fig.~\ref{fig:7.1}, after removal of the burn-in phase. Each blue curve shows the running average of a parameter as a function of Monte Carlo time, while the red line indicates the mean value obtained from the converged chain.}
\label{fig:7.3}
\end{figure}

As a complementary convergence diagnostic, Fig.~\ref{fig:7.3} presents the cumulative mean (running average) for each parameter. The rapid stabilization of these averages indicates that the chain repeatedly samples the high-probability region and achieves numerical stability. After removal of the burn-in phase, the running averages approach constant values with fluctuations consistent with Monte Carlo uncertainty, demonstrating that the influence of the initial conditions has largely diminished. Nevertheless, stable running averages are a necessary but not sufficient condition for convergence, as they do not ensure that all relevant regions of the posterior have been adequately explored.

\begin{figure}[htb!]
  \centering
  % First Subplot
  %\subfloat[$\eta = 50$\label{fig:sub1}]{
    \includegraphics[width=0.49\textwidth]{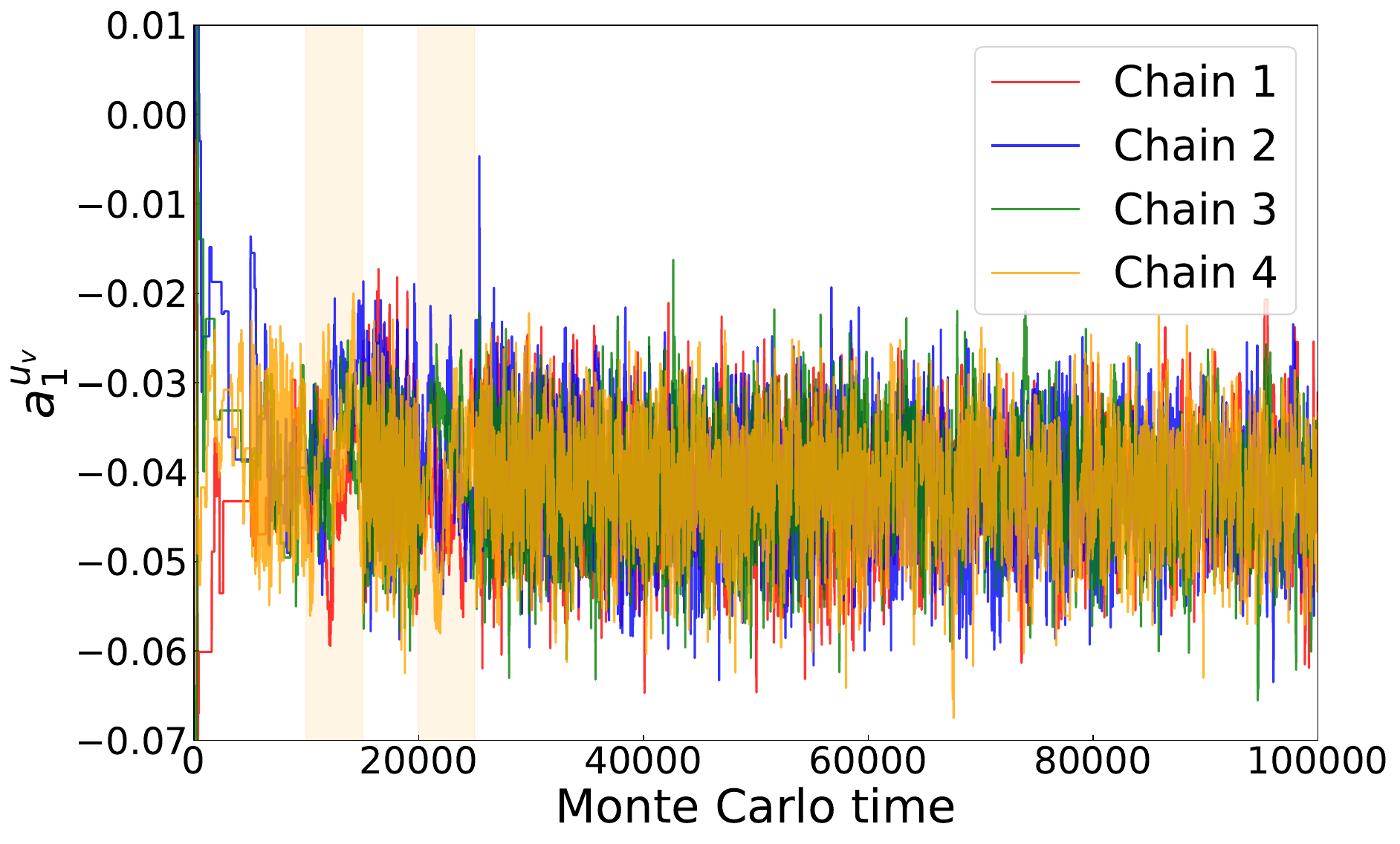}
  \hfill
  % Second Subplot
  %\subfloat[$\eta = 200$\label{fig:sub2}]{
    \includegraphics[width=0.49\textwidth]{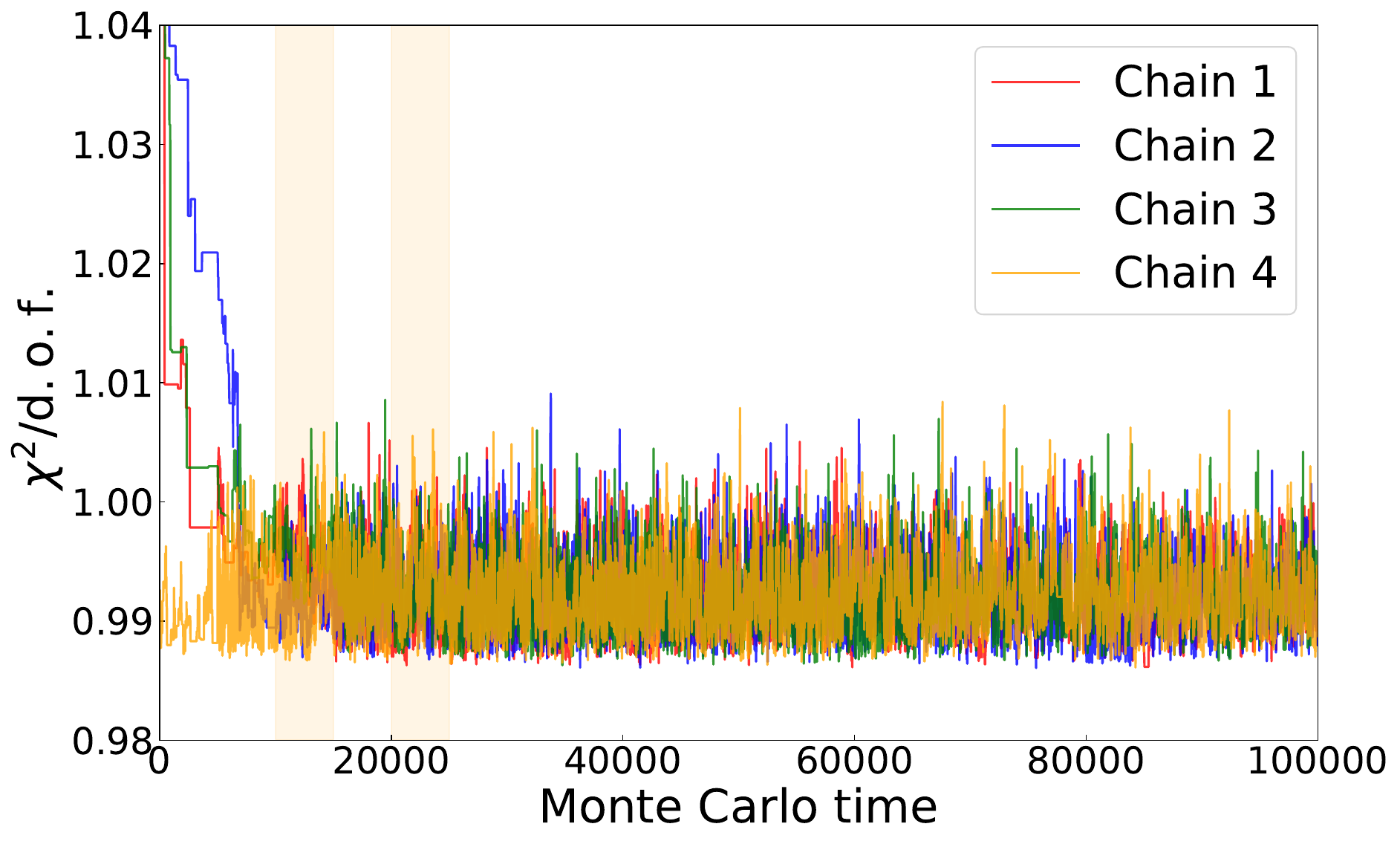}
  % Main Caption
  \caption[Cross-chain convergence test for the \texttt{Pb-only} analysis]{Cross-chain convergence test for the \texttt{Pb-only} analysis. Left: evolution of the parameter $a_1^{u_v}$ as a function of Monte Carlo time for five independent chains initialized from different random starting points near the Hessian minimum. Right: corresponding $\chi^2$ evolution. All chains converge to the same region, indicating consistent sampling of the common posterior.}
  \label{fig:7.4}
\end{figure}

Figure~\ref{fig:7.4} further assesses convergence by comparing five independent chains (including the representative chain discussed above), each initialized from different random starting points. The left panel shows the evolution of the parameter $a_1^{u_v}$, while the right panel displays the corresponding $\chi^2$ values. All chains converge toward the same stationary region of parameter space and exhibit consistent stabilization of $\chi^2$, indicating reproducibility of the sampling and independence from the choice of initial conditions. This behavior is not limited to $a_1^{u_v}$ but is observed consistently across other parameters and chains, providing strong evidence for robust convergence of the full parameter set. The absence of systematic offsets between chains further supports ergodic behavior within the explored region.
Taken together, these diagnostics demonstrate that the representative chain has reached equilibrium and yields statistically reliable samples from the posterior distribution of the nPDF parameters.

\begin{figure}[htb!]
\centering
\includegraphics[width=\textwidth]{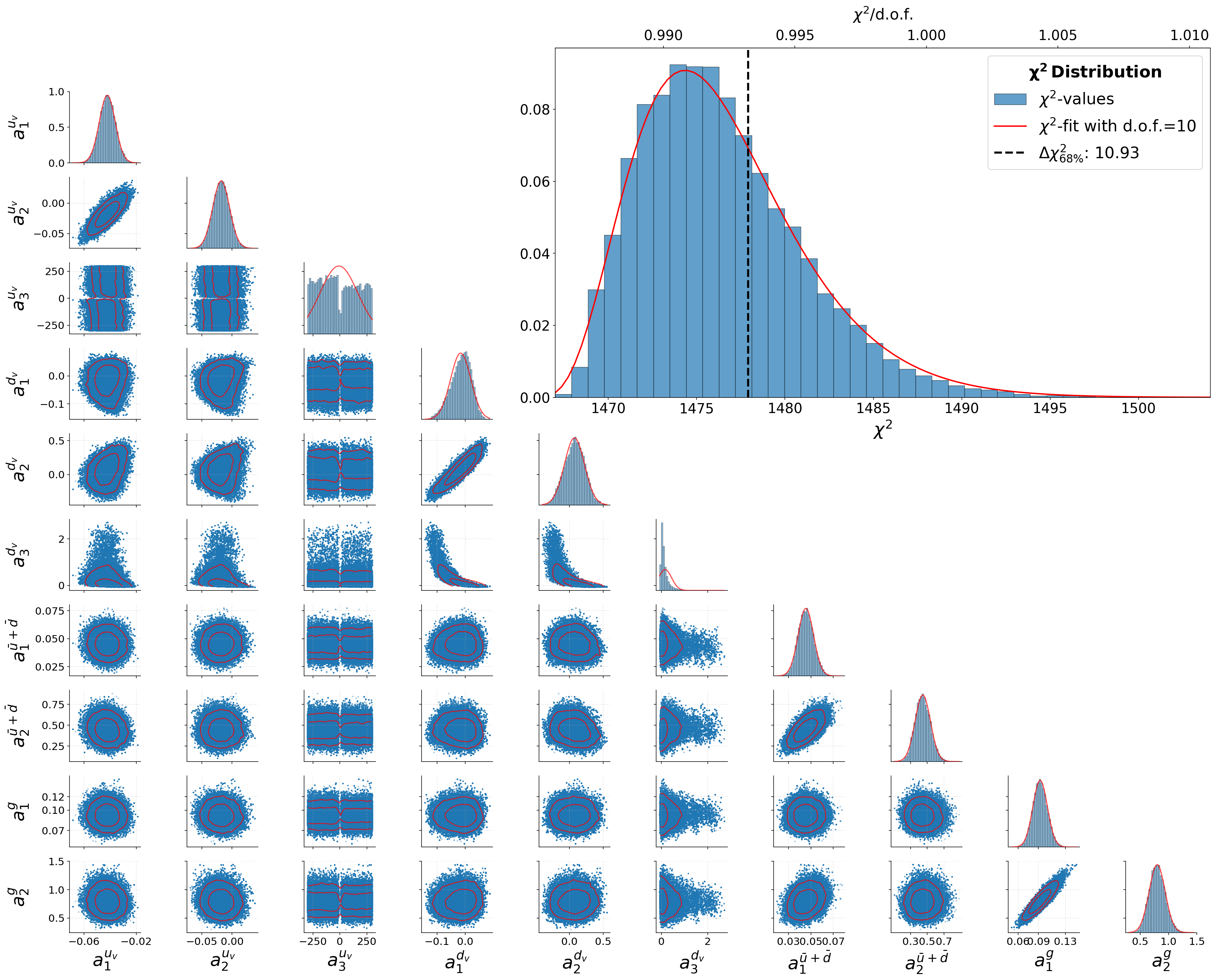}
\caption[Pairwise plot for a representative chain of \texttt{Pb-only} fit]{Pairwise plot for a representative \texttt{Pb-only} chain, showing two-dimensional projections of the posterior for all parameter combinations. The diagonal panels display one-dimensional marginal distributions, while the off-diagonal panels show pairwise correlations. Elliptical contours indicate the 68\% and 95\% credible regions. The top-right panel presents the $\chi^2$ distribution, shown in terms of both $\chi^2$ and $\chi^2/N_{\text{dof}}$, together with a fitted theoretical $\chi^2$ function (orange line).}
\label{fig:7.5}
\end{figure} 

In the next step, we investigate the structure of the posterior of the example chain (Fig. \ref{fig:7.1}) in more detail by studying marginal distributions and parameter correlations using a pairwise (corner) plot as shown in Fig.~\ref{fig:7.5}. The diagonal panels display the one-dimensional marginal posterior distributions, while the off-diagonal panels show the corresponding two-dimensional projections, revealing pairwise correlations among all parameter combinations. Elliptical contours indicate the 68\% ($1\sigma$) and 95\% ($2\sigma$) credible regions.\footnote{In pairwise plots, a tilted elliptical contour indicates correlation between parameters (positive if tilted up-right, negative if tilted down-right), while a circular or axis-aligned ellipse suggests weak or negligible correlation.}
Overall, correlations are moderate and largely confined to parameters within the same flavor combination, such as $a_1$ and $a_2$ for valence or gluon distributions, while cross-flavor correlations remain weak. This pattern indicates that the adaptive Metropolis–Hastings algorithm has successfully adjusted to the posterior covariance structure, enabling efficient exploration without strong linear dependencies. The projections involving $a_3^{d_v}$ show clear deviations from Gaussian behavior: in the two-dimensional pairwise plots, the samples do not form a single elliptical contour but instead populate extended and partially separated regions. This is consistent with the corresponding one-dimensional marginal distribution, which is asymmetric with a left-skewed distribution, rather than a symmetric Gaussian profile. Moreover, the parameter $a_3^{u_v}$ exhibits qualitatively different behavior. Its marginal distribution is distinctly non-Gaussian and remains approximately flat over a broad range of parameter values, indicating a lack of strong experimental constraints. Its two-dimensional projections reveal a split structure, with samples populating two separated regions. This behavior reflects an underconstrained direction in parameter space, where the likelihood provides only weak guidance and allows multiple statistically comparable configurations.\footnote{As discussed in Sec.~\ref{sec:6.2.2}, a uniform prior bound is imposed on the parameter $a_3^{u_v}$ to prevent runaway behavior into unphysical regions. The approximate flatness of the resulting marginal distribution is not induced by the prior, but rather reflects a lack of strong data constraints; within these bounds, the likelihood remains relatively uniform, leading to a flat posterior.} The top-right panel shows the distribution of goodness-of-fit values in terms of both the total $\chi^2$ and the reduced $\chi^2/N_{\text{d.o.f.}}$, where $N_{\text{d.o.f.}} \equiv \text{number of degrees of freedom}$. The sampled $\chi^2$ distribution closely follows the expected theoretical $\chi^2$ curve (orange line), providing an important validation of the statistical framework and confirming that the experimental uncertainties and covariance treatment are consistently incorporated. Overall, the corner plot analysis supports the convergence diagnostics while highlighting non-Gaussian and weakly constrained directions in the valence parameter space, which are further investigated using the combined chain.

\begin{figure}[!h]
  \centering
  \includegraphics[width=0.49\textwidth]{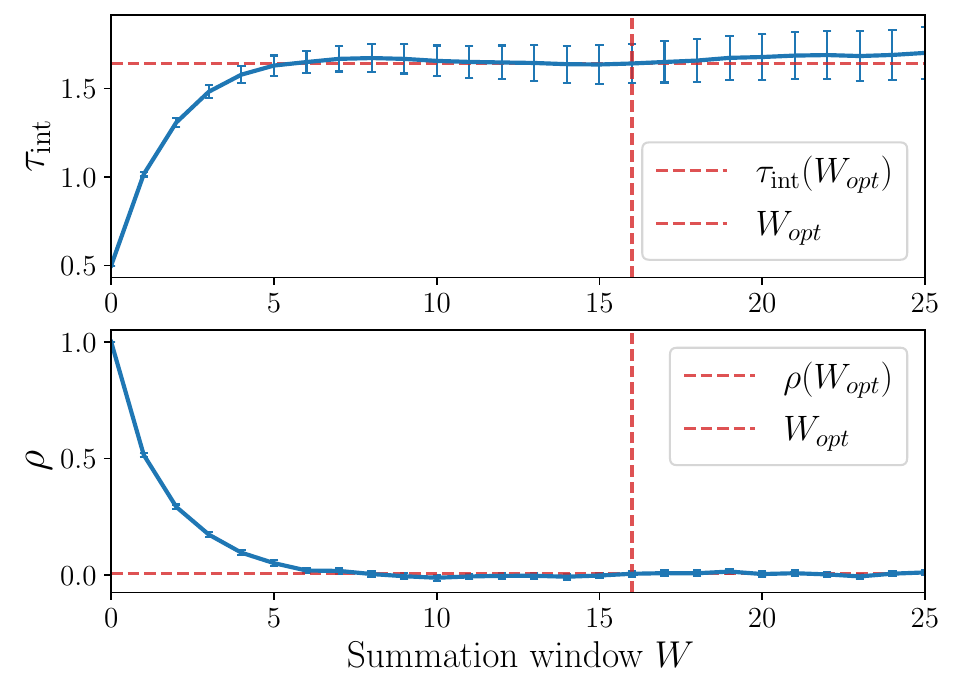}
  \includegraphics[width=0.49\textwidth]{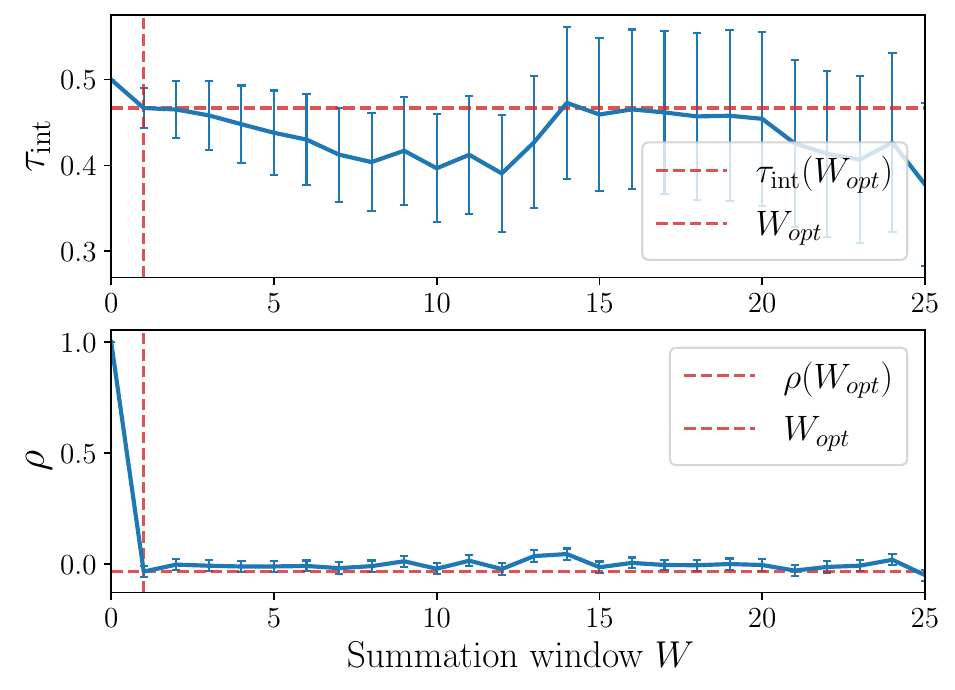}
  \caption[Autocorrelation analysis using the $\Gamma$-method]{Autocorrelation functions ($\rho$) and integrated autocorrelation times ($\tau_\text{int}$) as functions of the summation window $W$ for the \texttt{Pb-only} chain, shown for thinning factors $\eta = 50$ (left) and $\eta = 400$ (right). The dashed red vertical line indicates the optimal summation window, while the horizontal dashed red line marks the corresponding value of $\tau_\text{int}$ or $\rho$, as determined by the $\Gamma$-method.}
  \label{fig:7.6}
\end{figure}

A crucial aspect of the MCMC diagnostics is the analysis of autocorrelation within the chain. This determines the effective number of statistically independent samples available for uncertainty estimation. Since successive Markov states are inherently correlated, the raw sample size does not directly reflect the true statistical power of the ensemble. Instead, the integrated autocorrelation time $\tau_{\text{int}}$ provides a quantitative measure of how many steps are required to obtain one effectively independent draw from the posterior. Table~\ref{tab:7.1} summarizes the estimated values of $\tau_{\text{int}}$ for different thinning factors $\eta$, together with the corresponding number of samples retained after thinning.
\begin{table}[!h]
    \centering
    \begin{tabular}{lcc}
    \text{Thinning rate}($\eta$)  & $\tau_\text{int}$ & $N_\text{tot}$ \\
    \hline
    50 & 1.64 $\pm$ 0.11 & 12801\\
    100 & 1.09 $\pm$ 0.09 & 6401\\
    200 & 0.55 $\pm$ 0.04 & 3201\\
    300 & 0.53 $\pm$ 0.04 & 2134\\
    350 & 0.58 $\pm$ 0.05 & 1829\\
    \rowcolor{red!20}
    400 & 0.47 $\pm$ 0.02 & 1601\\
    450 & 0.46 $\pm$ 0.02 & 1423 \\
    500 & 0.50 $\pm$ 0.04 & 1281\\
    600 & 0.53 $\pm$ 0.06 &  1067\\
    \hline
    \end{tabular}
    \caption[The thinning rates along with the corresponding autocorrelation times for a representative chain]{The thinning rates $\eta$ along with the corresponding autocorrelation times $\tau_\text{int}$ and number of samples after thinning $N_\text{tot}$, confirms that a thinning rate of $\eta \simeq 400$ yields nearly uncorrelated samples suitable for independent replica generation.}
    \label{tab:7.1}
\end{table}
As expected, increasing $\eta$ reduces residual autocorrelation. For $\eta \simeq 400$, the values of $\tau_{\text{int}}$ approach a plateau close to the uncorrelated limit $\tau_{\text{int}} \approx 0.5$, indicating that further thinning yields only marginal improvement while unnecessarily reducing the total number of retained samples. We therefore adopt $\eta = 400$ as a practical compromise between minimizing correlations and preserving statistical precision, in line with the criterion discussed in Sec.~\ref{sec:6.3.4}.
This behavior is illustrated more explicitly in Fig.~\ref{fig:7.6}, which shows the autocorrelation function (ACF) and the corresponding $\tau_{\text{int}}$ as a function of the summation window $W$ for selected values of $\eta$.
In each panel, the vertical dashed red line marks the optimal summation window $W_{\text{opt}}$ determined using the $\Gamma$-method, while the horizontal dashed red line indicates the value of $\tau_{\text{int}}$ (upper panels) and $\rho(W_{\text{opt}})$ (lower panels) evaluated at that window.
In the left panel, corresponding to $\eta = 50$, the thinning rate is too small to fully eliminate correlations, as $\tau_{\text{int}}$ stabilizes at a value of approximately $\tau_{\text{int}} \sim 1.5$. Increasing the thinning factor to $\eta = 400$ leads to $\tau_{\text{int}} \approx 0.5$, consistent with a nearly uncorrelated sample.
For larger thinning factors, the ACF decays rapidly toward zero and $\tau_{\text{int}}$ stabilizes quickly, confirming that long-range correlations have been effectively suppressed. Overall, these diagnostics demonstrate that the retained ensemble consists of nearly uncorrelated samples and that the adopted thinning procedure provides a statistically reliable basis for subsequent posterior analysis and uncertainty propagation.
%%%%%%%%%%%%%%%%%%%%%%%%%%%%%%%%%%%%%%%%%%%%%%%%%%%%%%%%%
\subsection{Combined Posterior Sample}\label{sec:7.2}
\begin{figure}[!t]
\centering
\includegraphics[width=\textwidth]{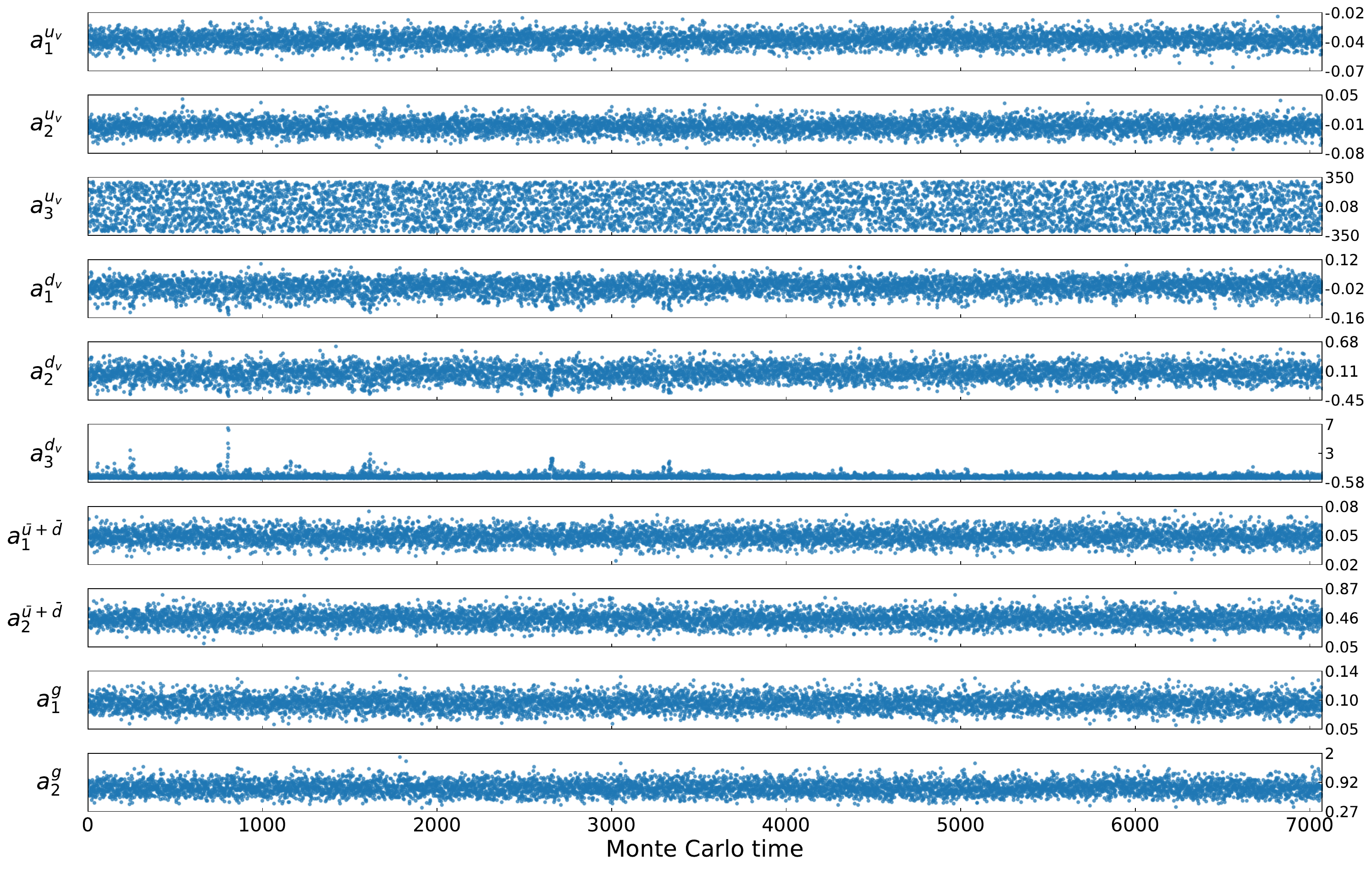}
\caption[Trace plot of the combined posterior samples for the \texttt{Pb-only} analysis]{Trace plot of the combined posterior samples obtained from the \texttt{Pb-only} MCMC analysis.}
\label{fig:7.7}
\end{figure}

%%%%%%%%%%%%%%%%%%%%%%%%%%%%%%%%%%%%%%%%%%%%%%%%%%%%%%%%%
To obtain a statistically robust representation of the posterior distribution, we combine multiple independent Markov chains. Several chains were generated, each initialized from a different random starting point in parameter space. After applying the convergence diagnostics described in Sec.~\ref{sec:7.1}, we retain 11 chains that exhibit consistent stationary behavior and satisfy all convergence criteria.\footnote{Chains that failed the convergence tests typically exhibited large $\chi^2$ values (well above the Hessian minimum) together with very low acceptance rates, resulting in long sequences of repeated states and inefficient exploration of parameter space.}
The final combined sample, which forms the basis of all subsequent analyses, is constructed in three steps:
\begin{enumerate}
    \item \textbf{Removal of thermalization phase:} For each retained chain, the initial burn-in (thermalization) phase is discarded. The length of the burn-in region varies slightly between chains, typically in the range of $60\,000$--$80\,000$ points. After this cut, the retained chains contain a total of $3.8\times 10^{6}$ samples (accepted and rejected) before thinning.
    
    \item \textbf{Autocorrelation reduction:} To suppress autocorrelations and obtain approximately independent samples, a thinning factor $\eta$ is applied to each chain. The optimal thinning rate varies between $\eta=400$ and $\eta=600$, chosen such that the integrated autocorrelation time satisfies $\tau_{\text{int}} \approx 0.5$ for all fitted parameters. For the final combined ensemble, the integrated autocorrelation time estimated is found to be $\tau_{\text{int}} = 0.49 \pm 0.01$, confirming that the final samples are effectively uncorrelated.

    \item \textbf{Combination of chains:} The resulting sets of nearly uncorrelated samples are merged into a single ensemble. The final combined dataset contains $N_{\text{tot}} = 7069$ effectively independent samples and serves as the definitive posterior sample used for the extraction of central values and uncertainty estimates of the nPDFs.
\end{enumerate}
After constructing the combined ensemble, we perform additional diagnostic checks to validate the quality and consistency of the aggregated posterior. Figure~\ref{fig:7.7} shows the trace plot obtained by concatenating the \emph{thinned} post-burn-in segments of the 11 converged chains. 
Across the full trace, all parameters fluctuate around stable mean values with no visible long-term drift, confirming that the retained chains are mutually consistent and that the merged sample provides a coherent representation of the posterior distribution.

\begin{figure}[!h]
\centering
\includegraphics[width=0.99\textwidth]{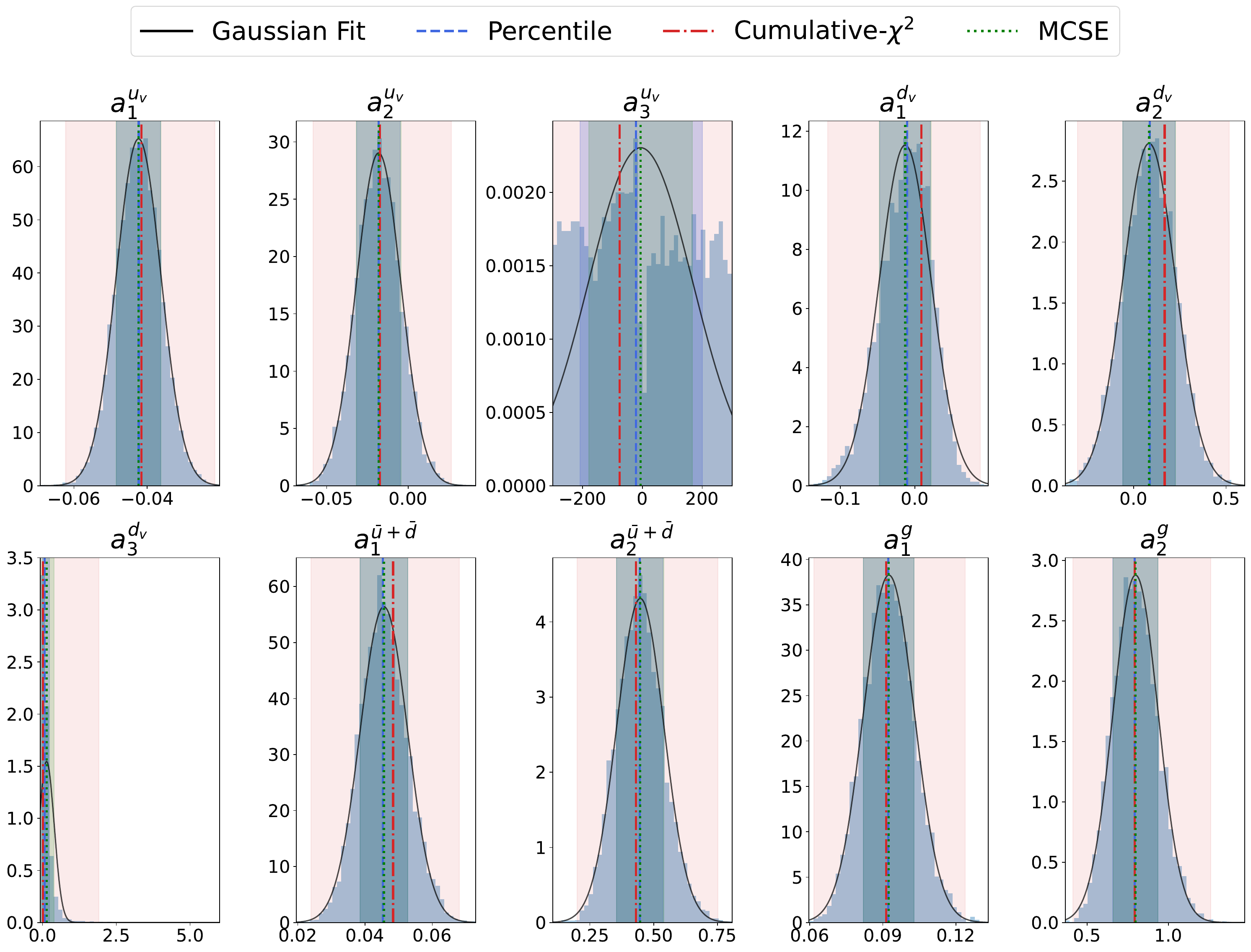}
\caption[Posterior histograms of the combined \texttt{Pb-only} chain]{Posterior histograms of the 10 fitted nPDF parameters from the combined \texttt{Pb-only} chain. The black curve shows a Gaussian fit to each distribution. Central values from three statistical methods are indicated: the blue dashed line (percentile method), the red dash-dotted line (cumulative-$\chi^2$ method), and the green dotted line (Monte Carlo standard error, MCSE). The shaded regions represent the corresponding 68\% credible intervals.}
\label{fig:7.8}
\end{figure}
Figure~\ref{fig:7.8} summarizes the statistical properties of the sampled parameters through the one-dimensional marginal posterior distributions for all ten fitted parameters. For reference, each histogram is overlaid with a normalized Gaussian fit (black curve). Vertical lines indicate central-value estimates obtained using three approaches: the Percentile Method (blue dashed), the Cumulative $\chi^2$ Method (red dotted), and the Monte Carlo Standard Error of the Mean (MCSE, green dash-dotted). The corresponding 68\% credible intervals (CIs) are shown as shaded bands for each method.
Several parameters, most notably $a_1^{u_v}$, $a_2^{u_v}$, $a_2^{\bar{u}+\bar{d}}$, $a_1^{g}$, and $a_2^{g}$, exhibit nearly symmetric, unimodal posteriors that are well approximated by Gaussian shapes. This is reflected in the close agreement between the histograms and the Gaussian overlays, as well as in the near coincidence of the three central-value estimators.
In contrast, the valence parameters display more complex behavior. The posterior distribution of $a_3^{u_v}$ is broad and clearly non-Gaussian, with substantial support across the prior range $[-300,\,300]$ and exhibits a split structure rather than a single dominant peak. As a result, the central-value estimates derived from the three methods show a noticeably larger spread. The posterior of $a_3^{d_v}$ deviates from Gaussianity in a distinct way, exhibiting a pronounced asymmetry with a long tail extending toward larger values. This extended tail reflects the presence of an elongated and structured region of the posterior, rather than a localized peak, and leads to a significant departure from the quadratic (Gaussian) approximation. This behavior is consistent with the intricate posterior structure observed in the trace diagnostics, where the chain explores distinct regions of parameter space that contribute non-negligibly to the overall distribution.

\begin{figure}[!b]
\centering
\includegraphics[width=\textwidth]{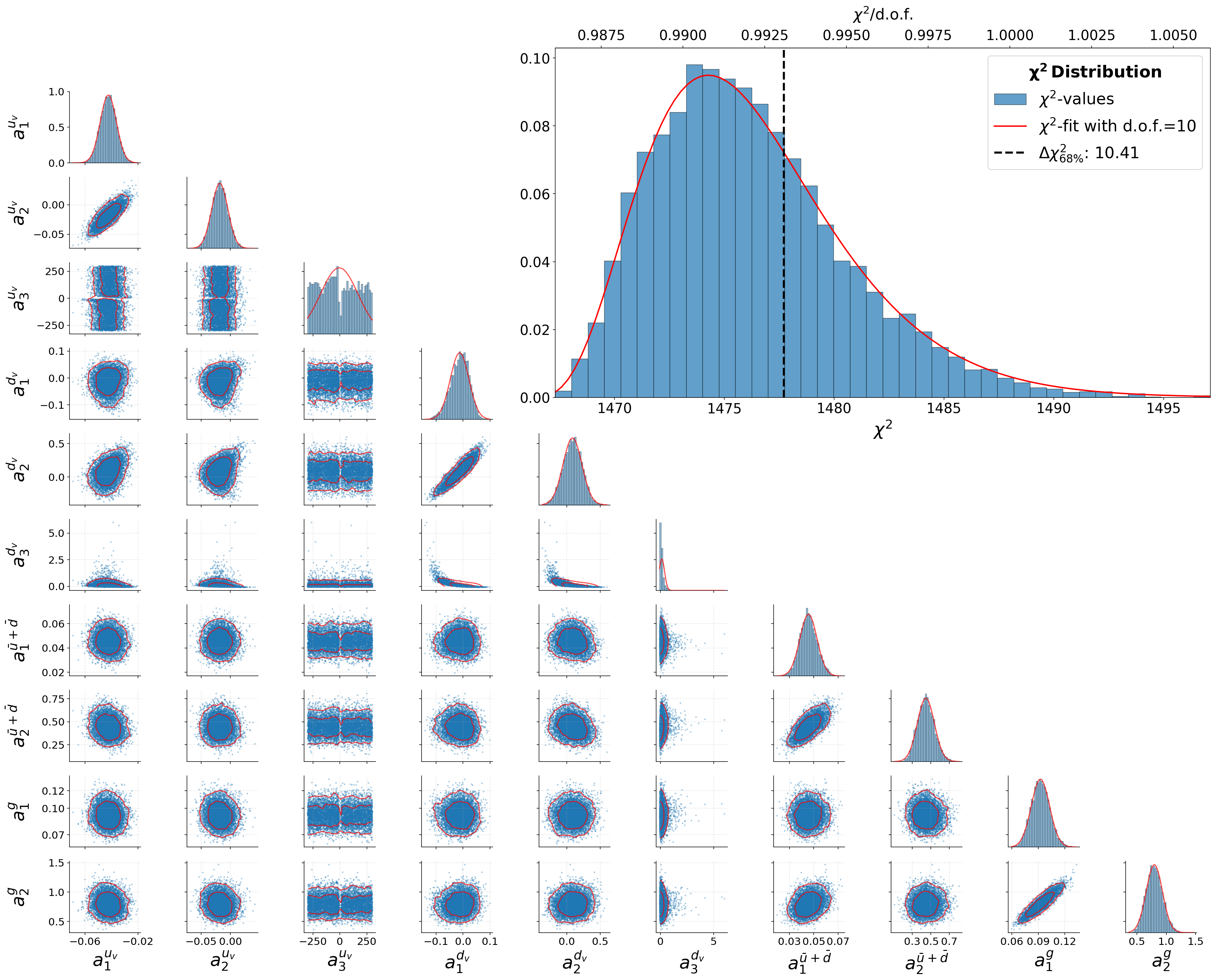}
\caption[Pairwise plot of the combined \texttt{Pb-only} posterior sample]{Pairwise (corner) plot of the combined \texttt{Pb-only} posterior sample. The diagonal panels show one-dimensional marginal distributions for each fitted parameter, while the off-diagonal panels display the corresponding two-dimensional projections and pairwise correlations. The 68\% and 95\% credible regions are indicated by elliptical contours. The top-right inset shows the $\chi^2$ distribution of the sampled ensemble, together with a fit to the expected theoretical $\chi^2$ form (red curve).}
\label{fig:7.9}
\end{figure}

Figure~\ref{fig:7.9} presents the pairwise plot obtained from the \texttt{Pb-only} combined posterior ensemble. The overall pattern of the one- and two-dimensional marginal distributions is consistent with that observed in the representative single-chain analysis discussed in Sec.~\ref{sec:7.1}, confirming that merging the converged chains preserves the posterior structure while improving statistical sampling. The combined ensemble therefore provides a more robust determination of parameter correlations and uncertainties.

The top-right panel shows the distribution of the $\chi^2$ values across the combined sample. The histogram exhibits a well-defined peak around $\chi^2 \approx 1475$ and is well reproduced by a theoretical $\chi^2$ distribution with 10 degrees of freedom (shown by the red curve), corresponding to the number of fitted parameters. The dashed vertical line marks the cumulative 68\% confidence region of the sampled distribution, which corresponds to $\Delta\chi^2_{68\%}=10.41$.\footnote{This value is consistent with the expectation from Wilks’ theorem \cite{Wilks:1938dza}, which predicts that $\Delta\chi^2$ follows a $\chi^2$ distribution with $k$ degrees of freedom, where $k$ is the number of fitted parameters.} In the subsequent analysis, this MCMC-derived tolerance is adopted as the $\Delta\chi^2$ criterion in the corresponding Hessian fit. Using the same tolerance definition in both frameworks ensures a consistent comparison between the Bayesian sampling of the full posterior distribution and the quadratic approximation employed in the Hessian method.

For the sea quark and gluon parameters ($a_1^{\bar{u}+\bar{d}}$, $a_2^{\bar{u}+\bar{d}}$, $a_1^{g}$, $a_2^{g}$), the dominant features are the strong positive correlations between the $a_1$ and $a_2$ parameters within each flavor, visible as narrow tilted elliptical contours. Aside from these expected correlations, the distributions remain compact and largely decoupled from the rest of the parameter space, indicating that these components are well constrained by the data. The valence parameters exhibit a more intricate correlation structure. The parameters $a_1^{u_v}$, $a_2^{u_v}$ and $a_1^{d_v}$, $a_2^{d_v}$ form well-defined correlated pairs, reflecting the constraints imposed by the valence sum rules. In contrast, the valence $a_3$ parameters display qualitatively different behavior. The projections involving $a_3^{d_v}$ show strongly non-elliptical structures, characterized by a dense concentration of samples at low values and a continuous, extended tail toward larger values. The resulting shapes are elongated and asymmetric, often exhibiting a wedge-like pattern in the two-dimensional projections rather than a compact elliptical contour. This behavior indicates a significant departure from the Gaussian approximation and reflects a highly non-trivial posterior geometry, where an extended region of parameter space remains statistically relevant. For $a_3^{u_v}$, the behavior is different: the distributions are more diffuse and organized into two clearly separated bands along the $a_3^{u_v}$ direction, with a noticeable absence of samples around $a_3^{u_v} \approx 0$, leading to an effectively excluded region between the two bands. 

Overall, the pairwise analysis highlights that while most parameters exhibit compact and approximately Gaussian correlation patterns, the valence $a_3$ parameters introduce non-trivial structure in the posterior geometry, reflecting weaker constraints and a more complex $\chi^2$ landscape. These features motivate a more detailed investigation of the parameter space, presented in the following section.
%%%%%%%%%%%%%%%%%
\subsubsection{Multi-Modal Structure of the Posterior}\label{sec.7.1.2}
%%%%%%%%%%%%%%%%%
The structure observed for the valence $a_3$ parameters in Fig.~\ref{fig:7.2} and Fig.~\ref{fig:7.7} indicates a non-trivial posterior structure. In this subsection, we examine this behavior in more detail and assess the statistical relevance of the secondary regions sampled by the Markov chains.
Inspection of the combined trace plots (Fig.~\ref{fig:7.7}) shows that the secondary region, previously hinted at in the representative single-chain diagnostics, is not sampled uniformly across all chains. Only a subset of the converged chains exhibits transitions into this region, while the remaining chains predominantly explore the vicinity of the primary mode.\footnote{In longer individual chains, transitions to the secondary region were observed more frequently. However, due to the stochastic nature of Markov chain sampling, it is difficult to determine whether this reflects genuine differences in sampling efficiency or simply statistical fluctuations.} This behavior suggests the presence of distinct regions in parameter space separated by suppressed posterior probability.

\begin{figure}[!h]
\centering
\subfloat[Secondary region in $a_3^{d_v}$]{
\label{fig:7.10a}
\includegraphics[width=\textwidth]{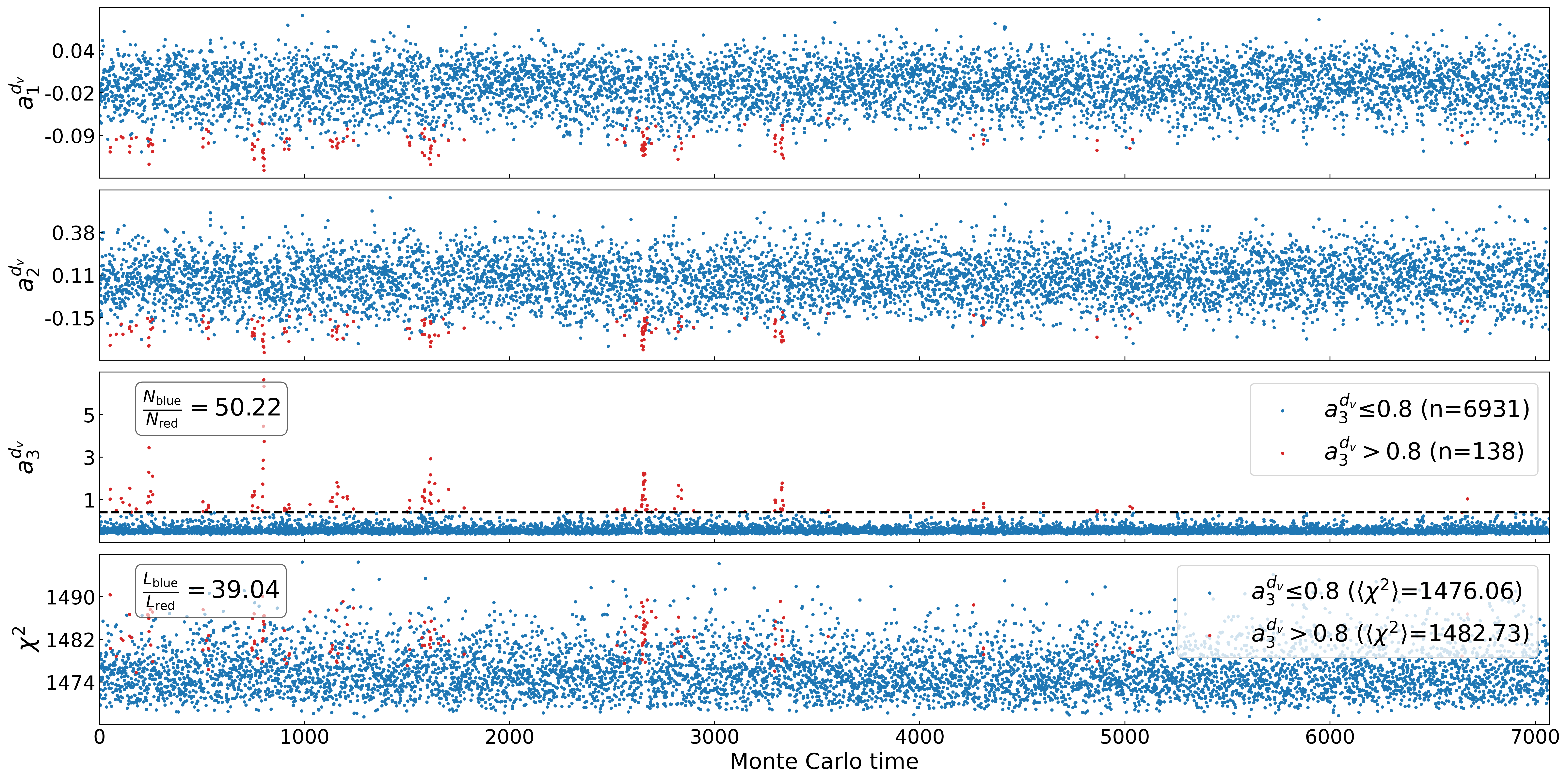}}\\
\subfloat[Secondary region in $a_3^{u_v}$]{
\label{fig:7.10b}
\includegraphics[width=\textwidth]{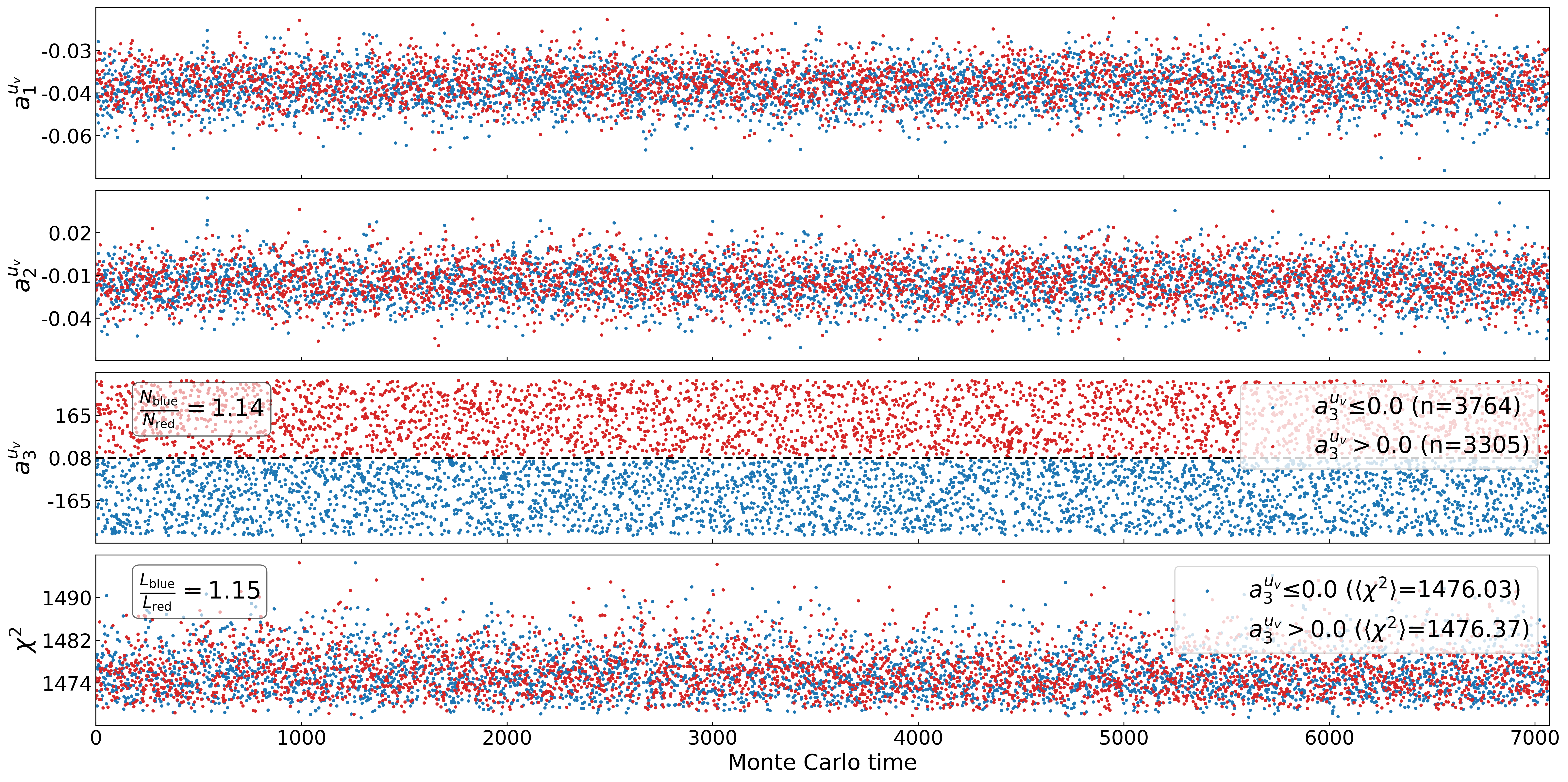}}
\caption[Minima test: trace plots for $a_3^{d_v}$ and $a_3^{u_v}$ in the \texttt{Pb-only} combined chain]{Identification of locally favored regions in the \texttt{Pb-only} combined chain using threshold partitions on the parameters $a_3^{d_v}$ (top) and $a_3^{u_v}$ (bottom). Samples with $a_3^{d_v} > 0.8$ ($a_3^{u_v} > 0.0$) are shown in red, while those with $a_3^{d_v} \le 0.8$ ($a_3^{u_v} \le 0.0$) are shown in blue. For each region, the mean $\chi^2$ and the corresponding likelihood ($L \propto \exp(-\chi^2/2)$) are computed.}
\label{fig:7.10} 
\end{figure}

To quantify this structure, Fig.~\ref{fig:7.10} partitions the combined sample according to threshold values of the parameters $a_3^{d_v}$ and $a_3^{u_v}$. In Fig.~\ref{fig:7.10a}, the samples are separated using the condition $a_3^{d_v} > T=0.8$ (shown in red) and $a_3^{d_v} \le T=0.8$ (shown in blue).\footnote{The threshold value $T=0.8$ is chosen to separate the visually distinct regions in the trace plots. Although the precise value of the cut is somewhat arbitrary, small variations of $T$ lead to qualitatively similar results.} A similar separation is applied in Fig.~\ref{fig:7.10b} for the parameter $a_3^{u_v}$ using the threshold $T = 0.0$, which reflects the gap observed around $a_3^{u_v} \approx 0$ in the sampled distributions. For each region, we compute the mean $\chi^2$ value and the corresponding likelihood weight, $L \propto \exp(-\chi^2/2)$, allowing for a direct comparison of the statistical preference of the dominant mode and the secondary region.
This procedure provides a systematic way to evaluate whether the secondary region corresponds to a genuinely competitive solution within the global posterior distribution, to assess whether it is properly sampled, and to determine its possible impact on the extracted nuclear PDF uncertainties. Compared to the representative single-chain example shown in Fig.~\ref{fig:7.2}, the apparent statistical relevance of the secondary region in the combined ensemble should be interpreted with caution. Due to the thinning procedure applied during post-processing, only a very small number of samples from the secondary parameter configuration survive in the final combined chain. Consequently, the population of this region in the thinned ensemble is small, resulting in larger statistical fluctuations.

Table~\ref{tab:second_min} summarizes the relative population, average $\chi^2$, and effective likelihood weight associated with the two regions as defined by the threshold partitions shown in Fig.~\ref{fig:7.10}. For the parameter $a_3^{d_v}$, the secondary region is populated by only a small fraction of the samples, corresponding to approximately $2\%$ of the total ensemble, with a ratio $N_{\mathrm{main}}/N_{\mathrm{sec}} \approx 50$, indicating a strong dominance of the primary region. This is accompanied by a noticeable increase in the mean $\chi^2$, $\Delta\langle\chi^2\rangle \approx 6.6$, which translates into a substantial suppression of the likelihood weight, $\langle L_{\mathrm{main}} \rangle/\langle L_{\mathrm{sec}} \rangle \approx 39$. The population and likelihood ratios are not significantly different from each other, indicating that the algorithm adequately explores this region, although it does not necessarily imply a strongly multi-modal structure.
\begin{table}[!h]
    \centering
    \caption[Summary of the main and secondary minima in the \texttt{Pb-only} posterior]{Summary of the main and secondary locally favored regions in the combined \texttt{Pb-only} posterior sample. The regions are defined by the threshold partitions shown in Fig.~\ref{fig:7.10}. The likelihood ratio is computed from $L \propto \exp(-\chi^2/2)$ by averaging $L$ over the samples in each region.}
    \label{tab:second_min}
    \renewcommand{\arraystretch}{1.2}
    \setlength{\tabcolsep}{10pt}
    \begin{tabular}{lcc}
        \hline
        & $a_3^{d_v}$ split ($T=0.8$) & $a_3^{u_v}$ split ($T=0.0$) \\
        \hline
        $N_{\mathrm{blue}}$ (main)          & 6931  & 3764  \\
        $N_{\mathrm{red}}$ (secondary)            & 138  & 3305 \\
        %$f_{\mathrm{sec}}$ [\%]             & 3.58  & 48.62 \\
        $N_{\mathrm{blue}}/N_{\mathrm{red}}$           & 50.22   & 1.14  \\
        $\langle\chi^2\rangle_{\mathrm{blue}}$  & 1476.06 & 1476.03 \\
        $\langle\chi^2\rangle_{\mathrm{red}}$   & 1482.73 & 1476.37 \\
        $\Delta\langle\chi^2\rangle$        & 6.6  & 0.34 \\
        $\langle L_{\mathrm{blue}} \rangle/\langle L_{\mathrm{red}} \rangle$ & 39.04 & 1.15  \\
        \hline
    \end{tabular}
\end{table}
In contrast, the structure observed for $a_3^{u_v}$ remains clearly visible in the combined sample. The two regions are populated with comparable frequency, with the secondary region accounting for approximately $47\%$ of the samples, as reflected by $N_{\mathrm{main}}/N_{\mathrm{sec}} \approx 1.14$. The corresponding difference in mean $\chi^2$ is very small, $\Delta\langle\chi^2\rangle \approx 0.34$, leading to nearly identical likelihood weights, with $\langle L_{\mathrm{main}} \rangle/\langle L_{\mathrm{sec}} \rangle \approx 1.15$. 
The close agreement between the population ratio $N_{\mathrm{main}}/N_{\mathrm{sec}}$ and the likelihood ratio $\langle L_{\mathrm{main}} \rangle/\langle L_{\mathrm{sec}} \rangle$ indicates that the two regions are sampled in proportion to their posterior probabilities. This consistency provides strong evidence that the MCMC algorithm adequately explores both regions and that the secondary region is properly represented in the ensemble. In this case, the posterior exhibits a genuinely multi-modal structure, with two statistically comparable regions contributing to the final ensemble.

\begin{figure}[!h]
\centering
\includegraphics[width=\textwidth]{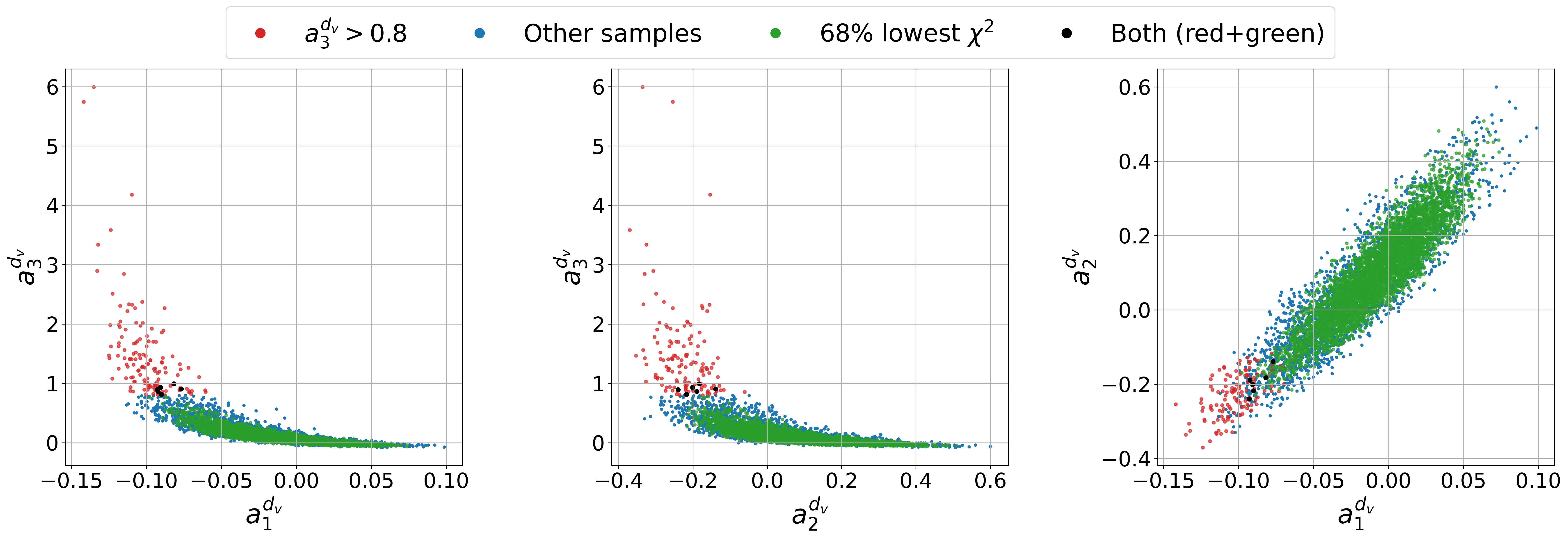}
\includegraphics[width=\textwidth]{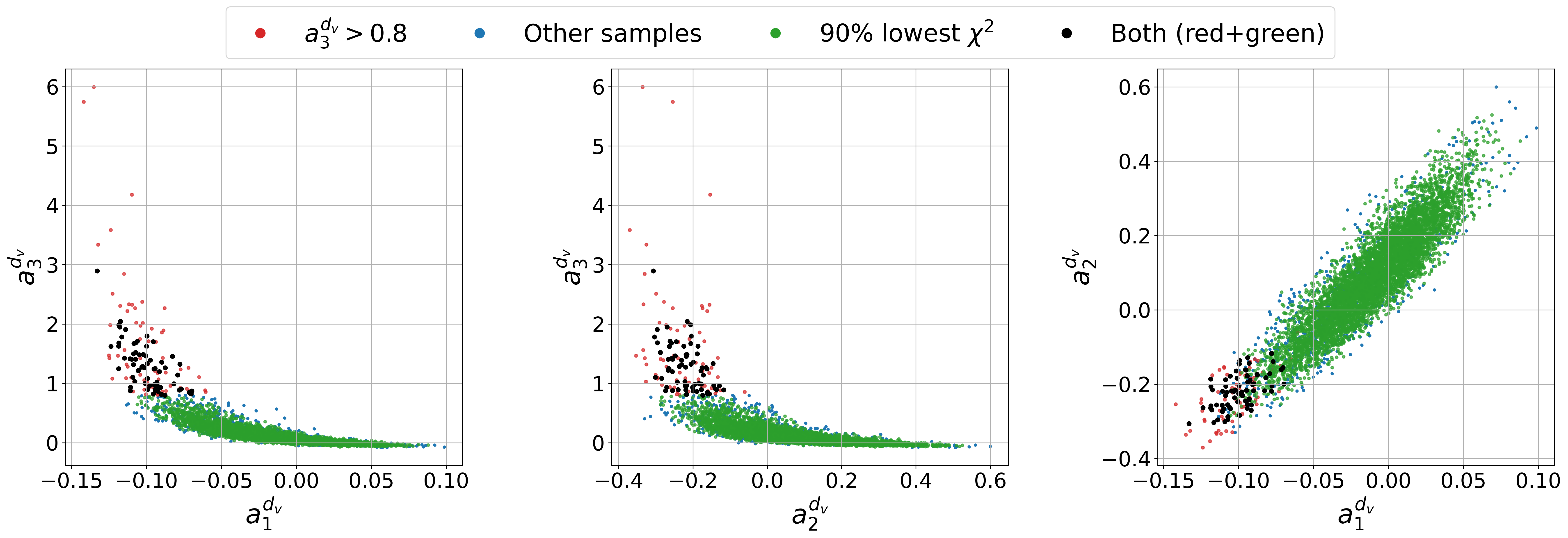}
\caption[Minima test: pairwise projections of the $d_v$ parameters in the \texttt{Pb-only} combined chain]{Pairwise scatter plots of the $d_v$ parameters at 68\% (top) and 90\% (bottom) confidence levels from \texttt{Pb-only} combined chain. Samples are color-coded to distinguish contributions from the secondary region: red for $a_3^{d_v} > 0.8$, green for points within the 68\% (90\%) $\chi^2$ confidence region, black for their overlap, and blue for the remaining samples.}
\label{fig:7.11}
\end{figure}

To further clarify the influence of the secondary region on the geometry of the posterior, Fig.~\ref{fig:7.11} shows the pairwise marginal distributions of the $d_v$ parameters obtained from the \texttt{Pb-only} combined ensemble. The color-coding distinguishes samples associated with the secondary configuration from those belonging to the dominant high-probability region. Points within the 68\% (upper panel) and 90\% (lower panel) confidence regions, defined using the cumulative-$\chi^2$ criterion, are shown in green. Samples satisfying $a_3^{d_v} > 0.8$ are highlighted in red, with their overlap with the corresponding confidence region indicated in black, while the remaining samples are shown in blue. 
This representation allows a direct assessment of the contribution of the secondary  region to the credible intervals. At the 68\% confidence level, only a small fraction of the threshold-selected samples overlaps with the highest-probability region, indicating that the secondary $d_v$ configuration has a negligible impact on the central posterior. When the confidence level is extended to 90\%, a larger fraction of these samples is included, reflecting a non-negligible but statistically suppressed contribution. Overall, the secondary $d_v$ region primarily affects the uncertainty estimate at higher confidence levels, consistent with its larger mean $\chi^2$ and reduced likelihood weight. It therefore does not represent a competitive global mode but rather a locally accessible region with suppressed posterior probability.

\begin{figure}[!h]
\centering
\includegraphics[width=\textwidth]{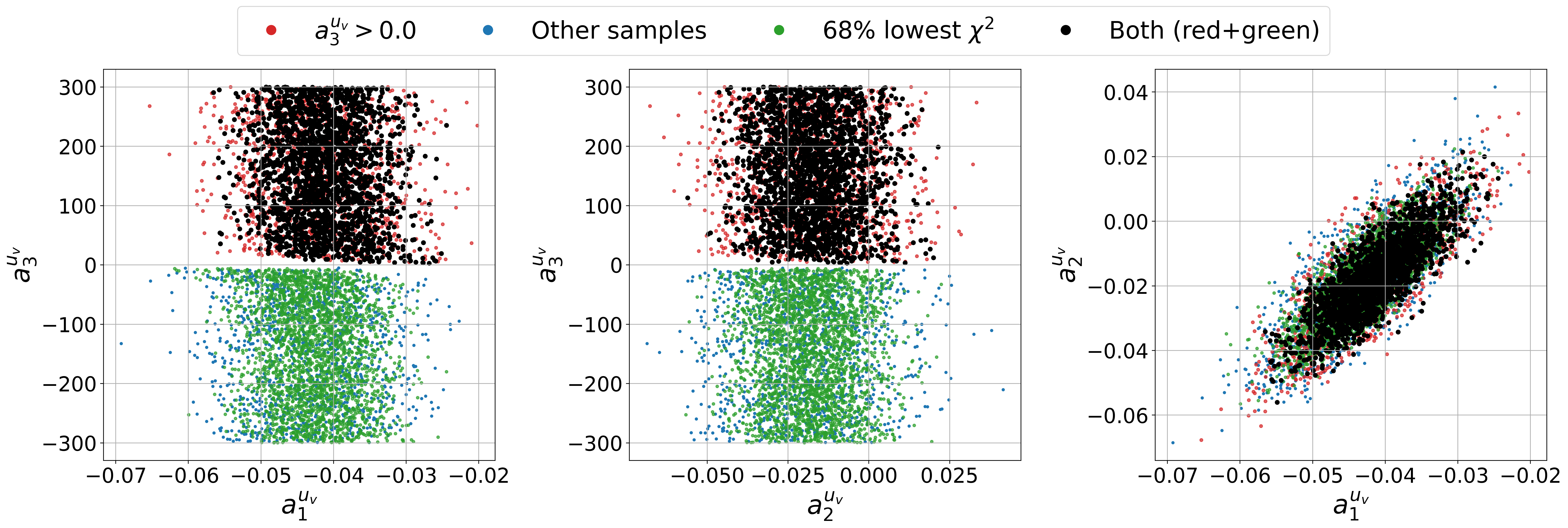}
\caption[Minima test: pairwise projections of the $u_v$ parameters in the \texttt{Pb-only} combined chain]{Same as Fig.~\ref{fig:7.11} but for $u_v$ parameters. Here 68\% CL is used.}
\label{fig:7.12}
\end{figure}

Figure~\ref{fig:7.12} presents the analogous pairwise distributions for the $u_v$ parameters at the 68\% confidence level. In this case, the posterior structure in $a_3^{u_v}$ exhibits a clear separation between positive and negative values, with very few samples around $a_3^{u_v} \simeq 0$. This pattern indicates the presence of two distinct regions of parameter space contributing to the posterior.
Unlike the $d_v$ case, however, this separation does not induce a corresponding structure in the $a_1^{u_v}$ or $a_2^{u_v}$ directions. Correlations with these parameters remain weak, indicating that the separation in $a_3^{u_v}$ is largely independent of variations in $a_1^{u_v}$ and $a_2^{u_v}$.
Taken together, these projections show that the multi-modal structure identified in the $a_3$ parameters manifests differently in the up- and down-valence distributions. While the secondary $d_v$ configuration corresponds to a statistically suppressed region of the posterior, the split in $a_3^{u_v}$ reflects two comparably populated regions of parameter space that both contribute to the credible uncertainty estimates.
It is important to emphasize that the multi-modal structure should not be interpreted as a deficiency of the analysis or an artifact of the sampling procedure. Instead, it reflects the intrinsic geometry of the posterior distribution in the full ten-dimensional parameter space, determined by the chosen parametrization and the constraining power of the available experimental data. The observed multi-modality in a small subset of parameters does not imply that the posterior is globally multi-modal. In well-constrained sectors of the fit, particularly the gluon and sea-quark parameters, the posterior remains close to Gaussian and is dominated by a single mode. The observed complexity is therefore confined to a restricted subspace associated with the valence $a_3$ parameters. Within this subspace, different regions can have varying statistical significance, ranging from suppressed configurations with larger $\chi^2$ to genuinely competitive solutions with comparable likelihood weights. The role of the MCMC algorithm is to faithfully sample this structure rather than to generate it. By exploring the full posterior, it naturally captures non-Gaussian features such as asymmetries and multiple locally preferred regions. These are intrinsic to the inference problem and must be consistently accounted for in the uncertainty estimation. The implications for the extracted nuclear PDFs are discussed in the following subsection.

%%%%%%%%%%%%%%%%%%%%%%%%%%%%%%%%%%%%%%%%%
\subsection{Nuclear PDFs and Uncertainty Estimates}\label{sec:7.1.3}
%%%%%%%%%%%%%%%%%%%%%%%%%%%%%%%%%%%%%%%%%
Using the final combined posterior ensemble from the primary \texttt{Pb-only} analysis, the nuclear PDFs for lead (Pb) are evaluated together with their corresponding uncertainty bands. As described in Sec.~\ref{sec:6.3.5}, three strategies are employed to estimate PDF uncertainties from the posterior samples: the percentile method, the cumulative-$\chi^2$ method, and the Monte Carlo standard error (MCSE). The percentile method provides a non-parametric, Bayesian estimate by constructing credible intervals directly from the empirical distribution, naturally capturing asymmetries. The cumulative-$\chi^2$ method selects the best-fitting replicas based on their $\chi^2$ values and builds a conservative envelope from this subset. The MCSE-based construction uses the ensemble mean and standard deviation, yielding a symmetric band that effectively assumes a Gaussian posterior.
The PDFs are evaluated for each posterior sample, and the uncertainty bands are constructed according to the chosen prescription. This procedure preserves the full probability structure of the parameter space, including correlations and non-Gaussian features identified in the previous subsection.

\begin{figure}[!h]
\centering
\includegraphics[width=0.49\textwidth]{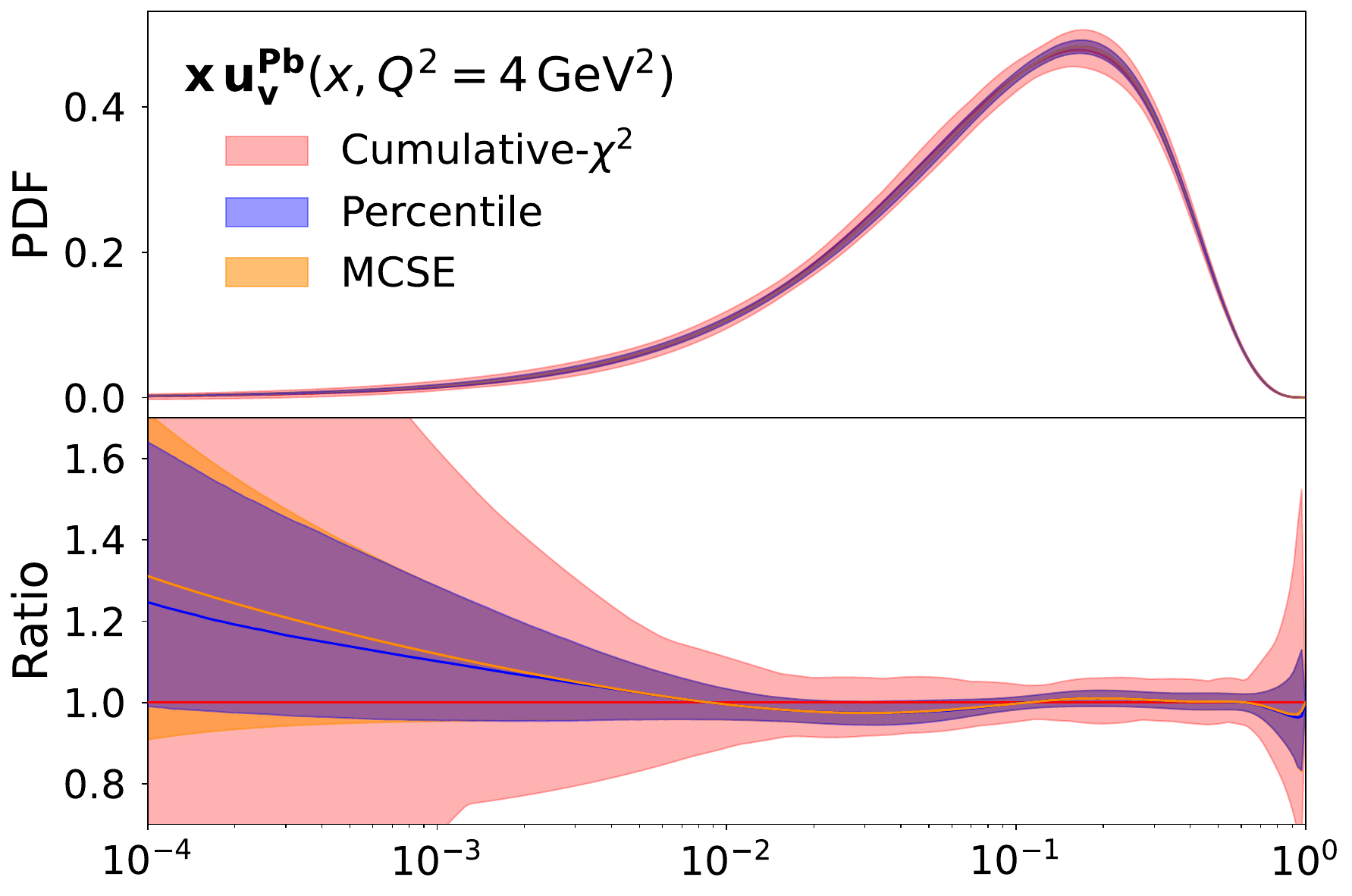}
\includegraphics[width=0.49\textwidth]{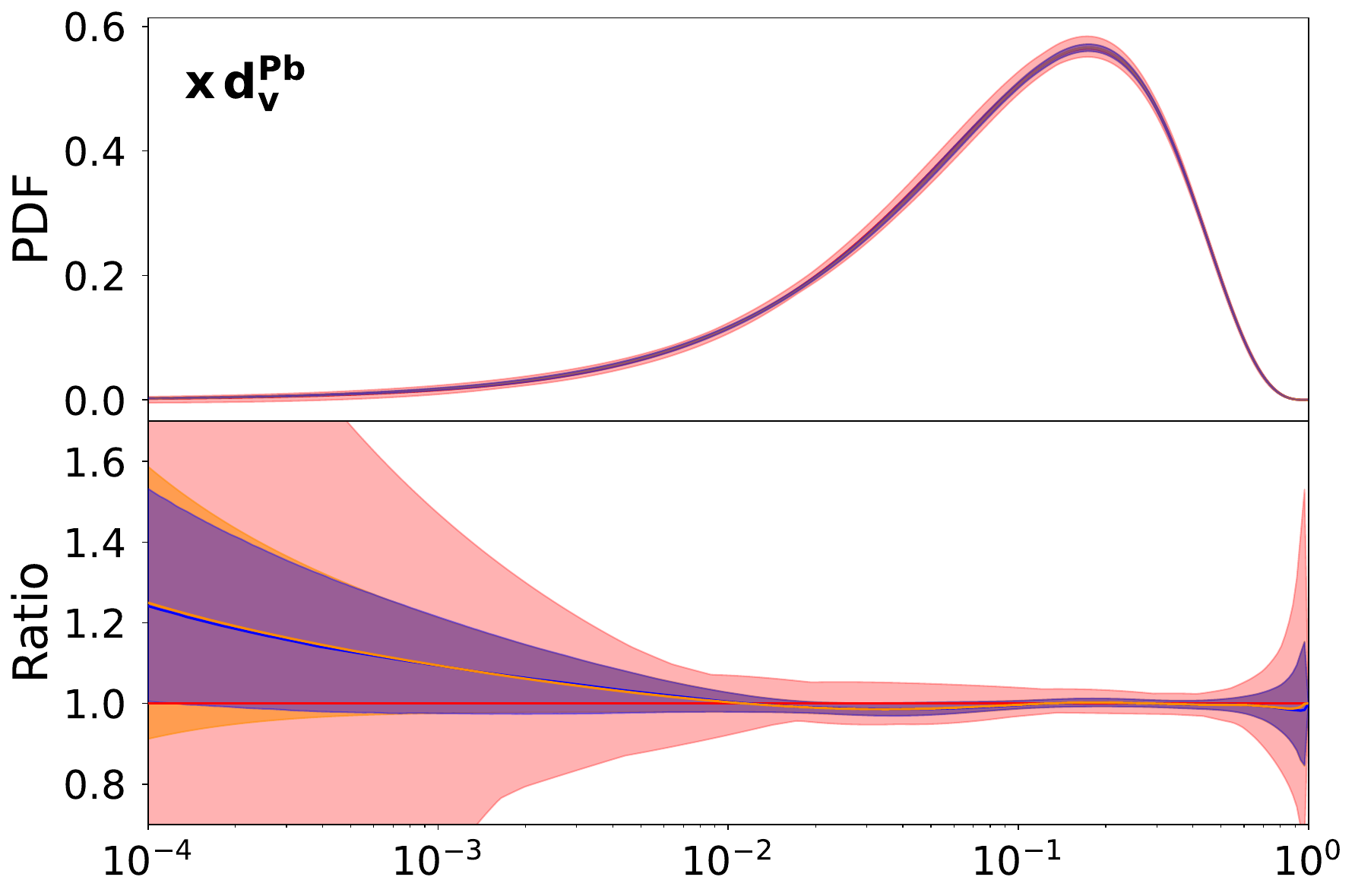}
\includegraphics[width=0.49\textwidth]{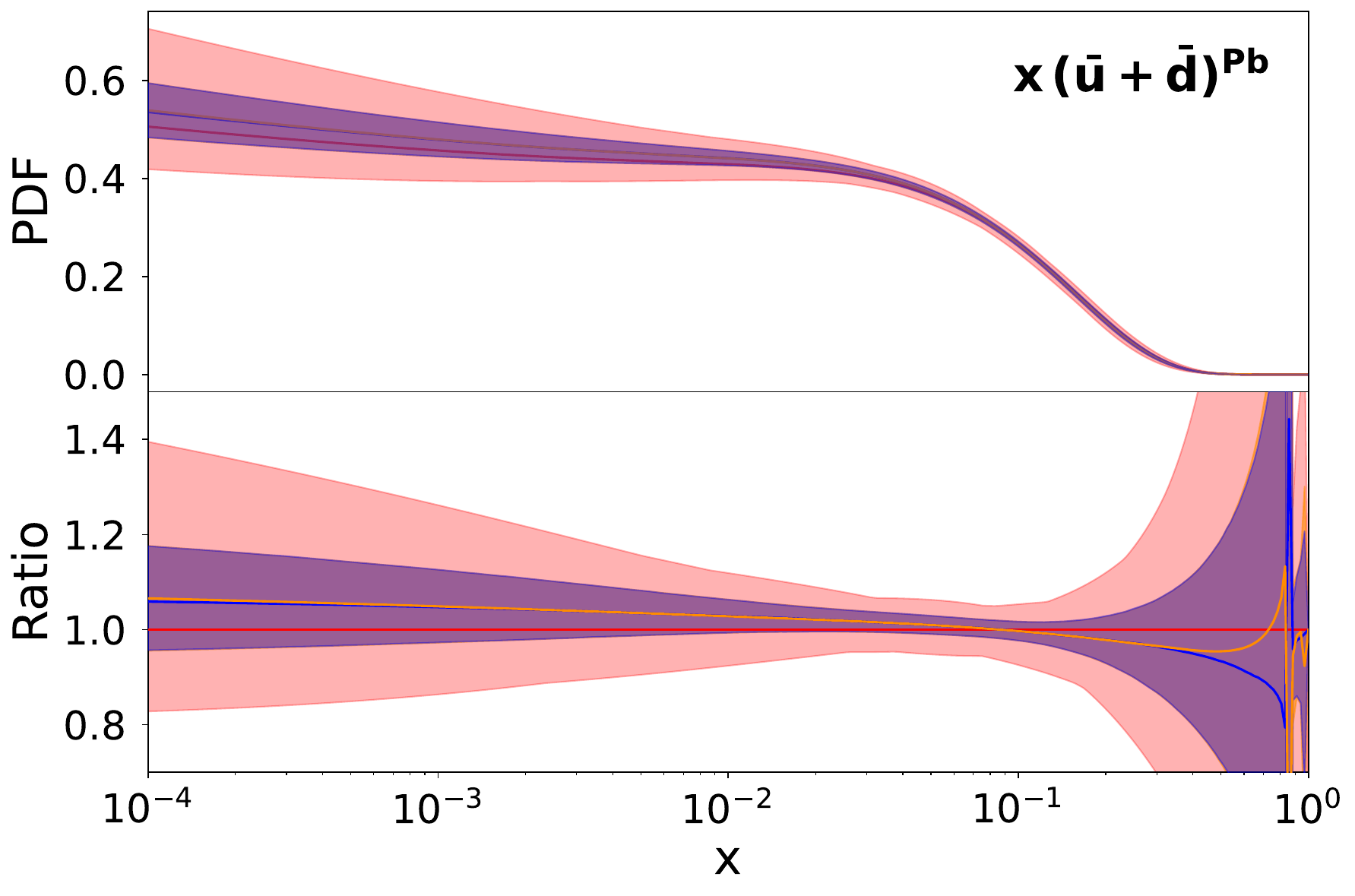}
\includegraphics[width=0.49\textwidth]{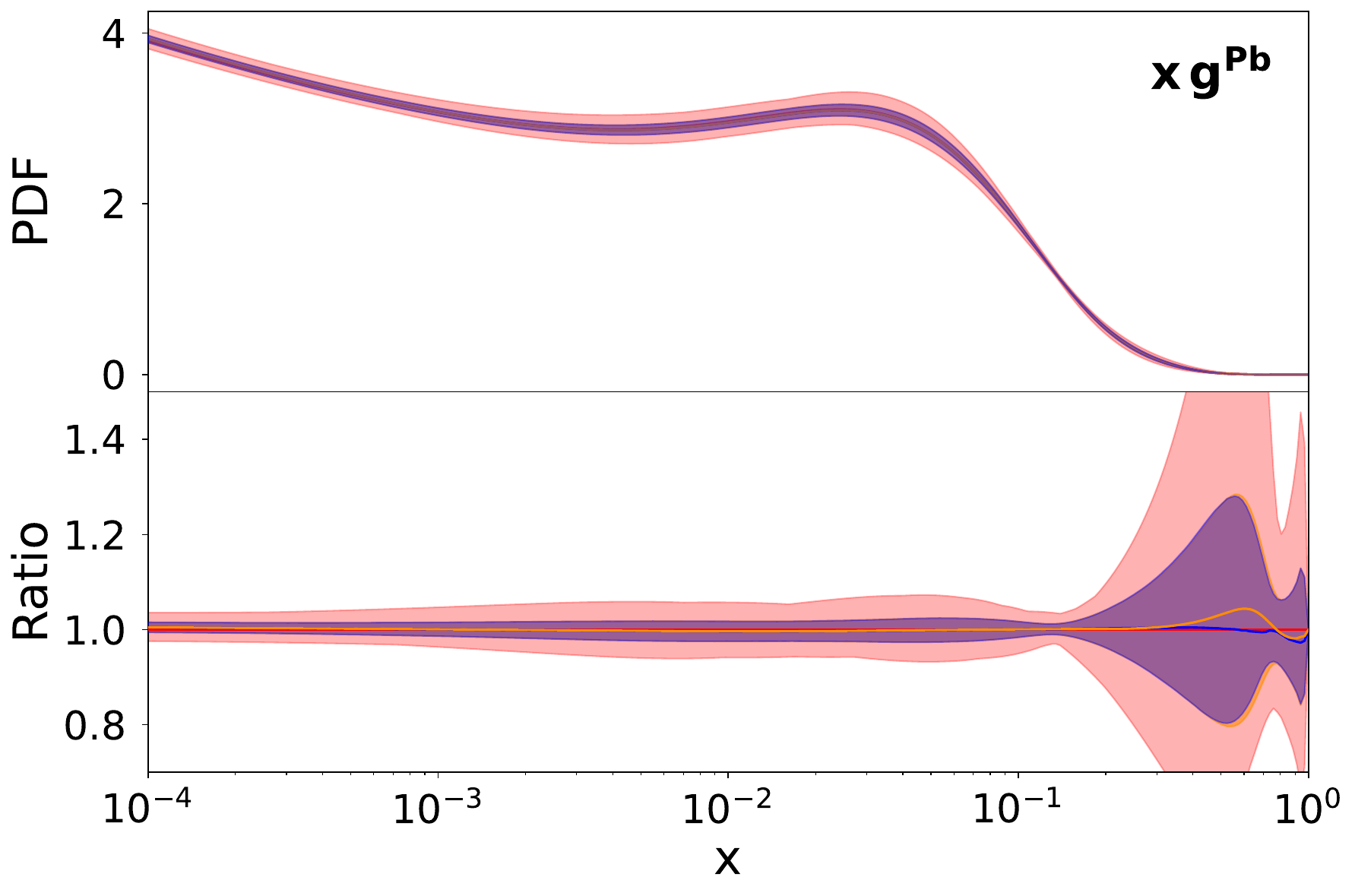}
\caption[\texttt{Pb-only} fit: MCMC estimator comparison for Pb PDFs]{\texttt{Pb-only} fit: Comparison of 68\% uncertainty bands for the Pb valence ($u_v$, $d_v$), light-sea ($\bar{u}+\bar{d}$), and gluon ($g$) distributions obtained using three MCMC-based estimators: cumulative-$\chi^2$ (red), Percentile (blue), and MCSE (orange). For each flavor, the upper panel shows the absolute distributions, while the lower panel displays ratios to the cumulative-$\chi^2$ best-fit prediction.}
\label{fig:7.13}
\end{figure}

We begin by comparing the three MCMC-based estimators. Figure~\ref{fig:7.13} shows the PDFs together with their ratios to the best-fit prediction (defined as the central value of the cumulative-$\chi^2$ method), along with the corresponding 68\% uncertainty bands for the $u_v$, $d_v$, $\bar{u}+\bar{d}$, and $g$ distributions. The results from the Percentile (blue), MCSE (orange), and cumulative-$\chi^2$ (red) methods are displayed. As seen in the ratio panels, the Percentile and MCSE central values are in very good agreement across most of the $x$ range. However, a clear separation appears in the $u_v$ distribution at low $x$ ($x \lesssim 10^{-3}$), where the Percentile estimate lies systematically below the MCSE. In contrast, the cumulative-$\chi^2$ central value shows a significantly larger deviation in this region, particularly for the valence distributions, where it lies noticeably below the other two methods as $x \to 10^{-4}$. This reflects the sensitivity of the $\chi^2$-based estimator to asymmetries in the posterior when data constraints are weak. The exception is the gluon distribution, which is comparatively better constrained and exhibits more gaussian behavior; here, all three central values remain largely consistent and overlap across the full $x$ range.

The differences are more pronounced in the uncertainty bands. While the cumulative-$\chi^2$ method consistently yields the widest (most conservative) uncertainties, the Percentile and MCSE bands overlap very closely across most of the $x$ range. A noticeable difference appears only at very small $x$ ($x \sim 10^{-4}$) in the valence sector, where the MCSE band becomes slightly wider than the Percentile band, with the orange envelope extending further than the blue. This indicates that the mean-based estimator is more sensitive to the tails of the posterior in this poorly constrained region, whereas the Percentile method is less affected by such excursions.
In the intermediate region ($10^{-2} \lesssim x \lesssim 10^{-1}$), all three methods show strong agreement: central values overlap and the uncertainty bands narrow significantly, reflecting the presence of robust experimental constraints. At large $x$ ($x \gtrsim 10^{-1}$), uncertainties increase again for all parton species, particularly for the sea quark and gluon distributions as $x \to 1$. This growth is driven by a lack of data and limited kinematic coverage in this region. The cumulative-$\chi^2$ band expands more rapidly than the others, while the Percentile and MCSE bands remain comparatively stable until very high $x$. Overall, these results demonstrate that while the three estimators are consistent in well-constrained regions, they differ substantially in extrapolation regions, where both central values and uncertainties become sensitive to the statistical definition of the estimator. For this reason, the cumulative-$\chi^2$ method is adopted as the preferred choice in this analysis.

\begin{figure}[!h]
\centering
\includegraphics[width=0.49\textwidth]{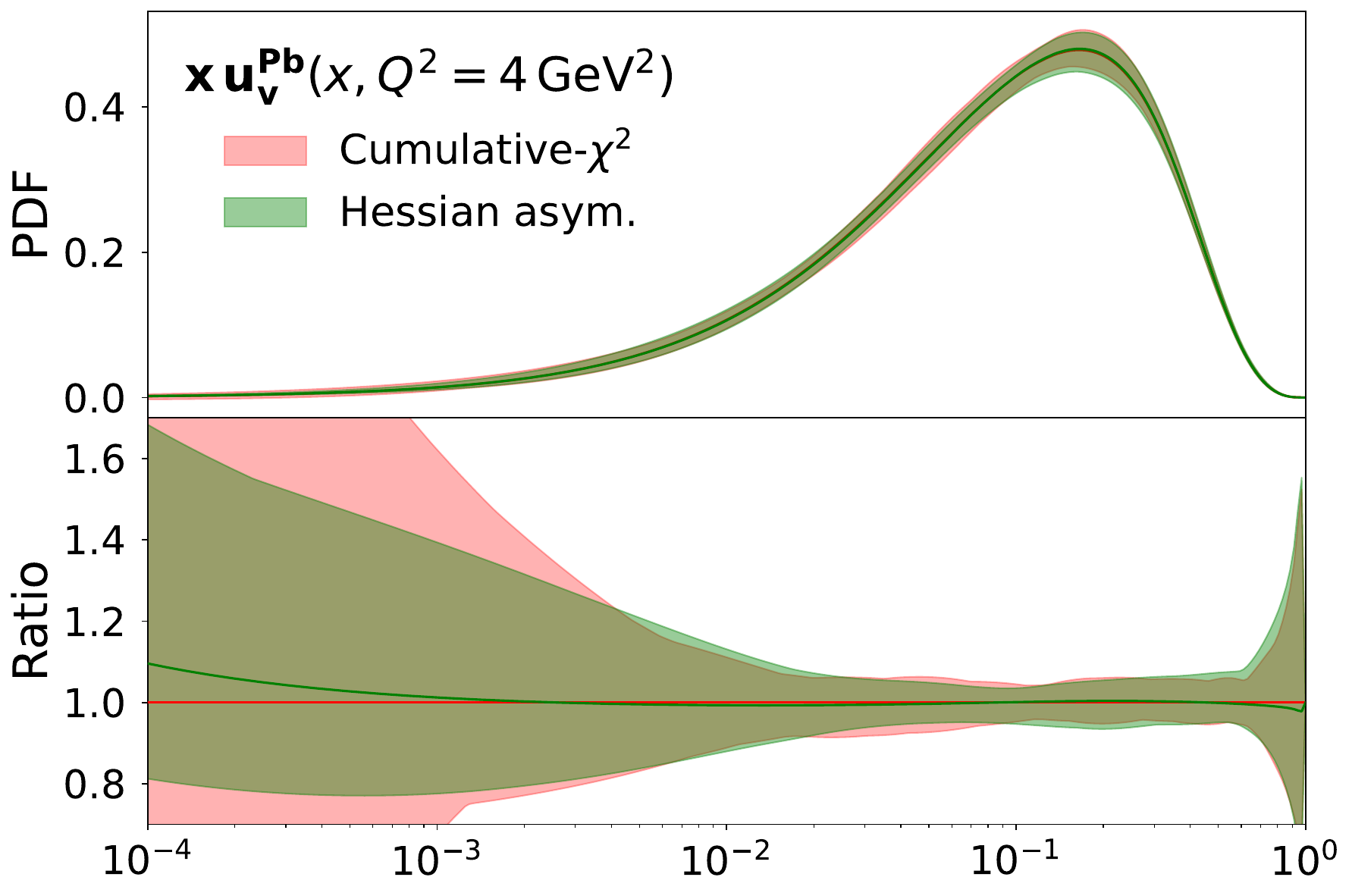}
\includegraphics[width=0.49\textwidth]{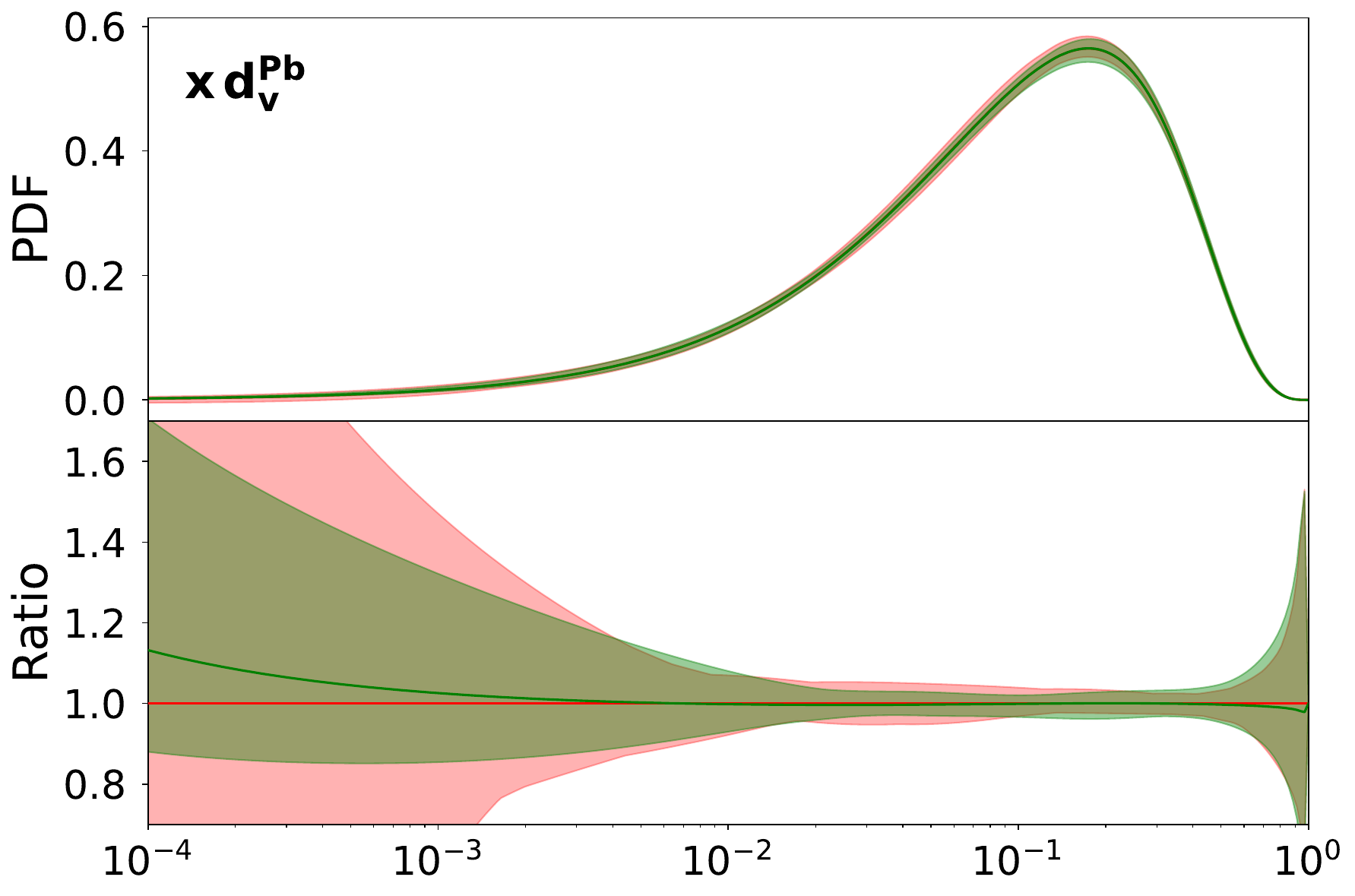}
\includegraphics[width=0.49\textwidth]{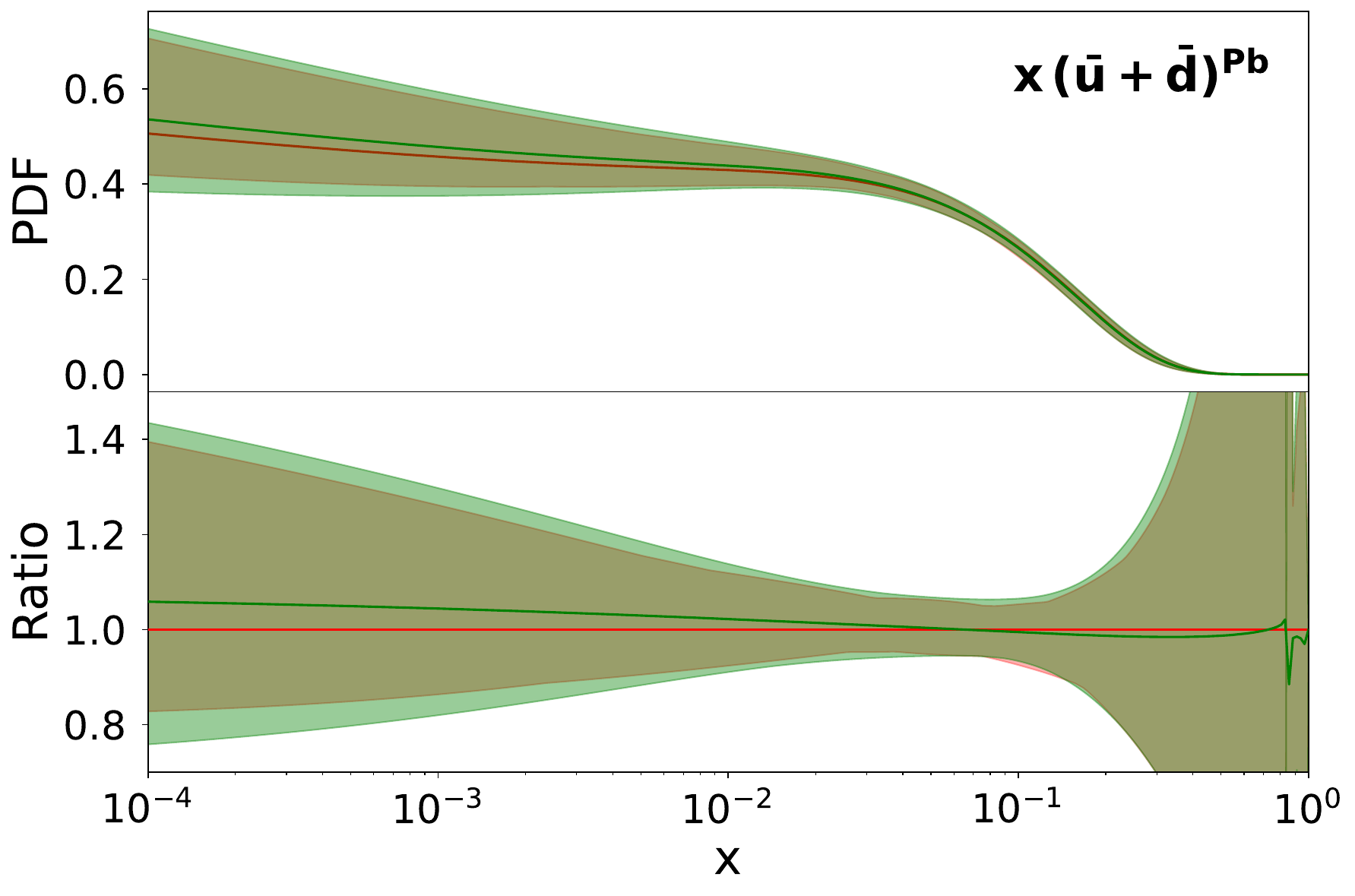}
\includegraphics[width=0.49\textwidth]{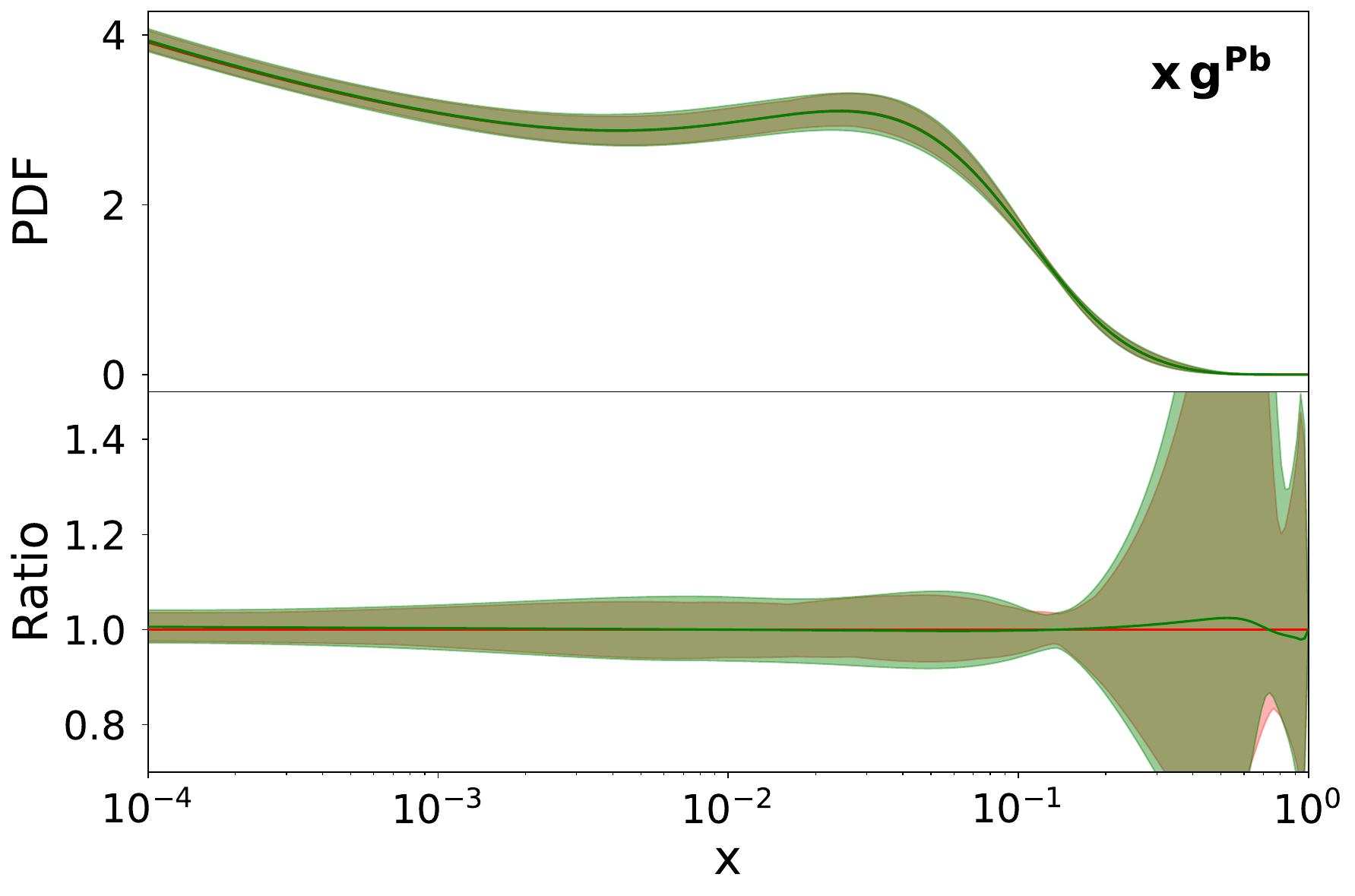}
\caption[\texttt{Pb-only} fit: MCMC vs Hessian uncertainty comparison]{\texttt{Pb-only} fit: Comparison of 68\% uncertainty bands for the Pb valence ($u_v$, $d_v$), light-sea ($\bar{u}+\bar{d}$), and gluon ($g$) distributions obtained using the cumulative-$\chi^2$ MCMC approach (red) and the asymmetric Hessian method (green). For each flavor, the upper panel shows the absolute distributions, while the lower panel displays ratios to the cumulative-$\chi^2$ best-fit prediction.}
\label{fig:7.14}
\end{figure}

We now turn to Figure~\ref{fig:7.14}, which compares the cumulative-$\chi^2$ approach with the traditional Hessian method. To ensure a consistent comparison between the two frameworks, the tolerance used in the Hessian approach is derived from the MCMC-based $\Delta\chi^2$ criterion illustrated in Fig.~\ref{fig:7.9}. This comparison provides a direct assessment of how the two fundamentally different uncertainty propagation frameworks (sampling-based versus quadratic approximation) behave across the kinematic range. At the level of central values, the agreement between the two methods is remarkably good across all parton species over most of the $x$ range. For both valence ($u_v$, $d_v$) and sea quark ($\bar{u}+\bar{d}$) distributions, the predictions are nearly identical, with only small deviations appearing at the edges of the kinematic range, particularly at very low $x$ ($10^{-4} \lesssim x \lesssim 10^{-3}$). The gluon distribution shows excellent consistency across the full $x$ range. This overall agreement indicates that the position of the global minimum is robust and consistently identified by both approaches.

The main differences arise in the uncertainty bands. In the intermediate region ($10^{-2} \lesssim x \lesssim 10^{-1}$), the two methods produce very similar uncertainty estimates, with significant overlap of the bands for all parton species. This reflects the fact that, in this region, the likelihood is strongly constrained and approximately Gaussian, making the Hessian approximation reliable.
However, in the poorly constrained regions, both at low $x$ and high $x$, systematic differences emerge. At low $x$, particularly in the valence sector, the cumulative-$\chi^2$ band becomes noticeably broader and more asymmetric than the Hessian band, indicating that the sampling-based approach captures a wider range of acceptable solutions beyond the quadratic approximation. The Hessian uncertainties, in contrast, remain more compact, reflecting the limitations of the quadratic approximation in regions where the parameter space is less constrained and potentially non-Gaussian. At large $x$ ($x \to 1$), the uncertainty bands grow for all parton species in a broadly similar manner. Moreover, two features can be observed in this figure: an almost perfect agreement between the two methods for the gluon distribution, and a slightly larger uncertainty band from the Hessian approach for the sea quark ($\bar{u}+\bar{d}$) distribution.

Overall, the comparison shows that while both methods provide consistent central values and similar uncertainties in well-constrained regions, the cumulative-$\chi^2$ approach yields systematically larger, asymmetric, and more flexible uncertainty estimates in poorly constrained regions. This reflects the ability of the MCMC-based framework to capture non-Gaussian features of the posterior distribution, providing a more complete representation of the underlying uncertainties, particularly where deviations from Gaussian behavior become important.

\paragraph{Impact of Secondary Minima on the Extracted nPDFs}
While the comparison above clarifies the impact of different statistical prescriptions on the extracted uncertainties, it does not yet address a more intricate question: how does the multi-modal structure identified in the valence parameters manifest itself at the level of the physical PDFs? To investigate this, we isolate the subset of replicas associated with the secondary locally favored regions and examine the corresponding PDF shapes. This allows us to determine whether the secondary parameter configurations correspond to genuinely distinct physical solutions or instead reflect parametrization degeneracies that yield similar distributions.

\begin{figure}[]
\centering
\subfloat{
\includegraphics[width=0.49\textwidth]{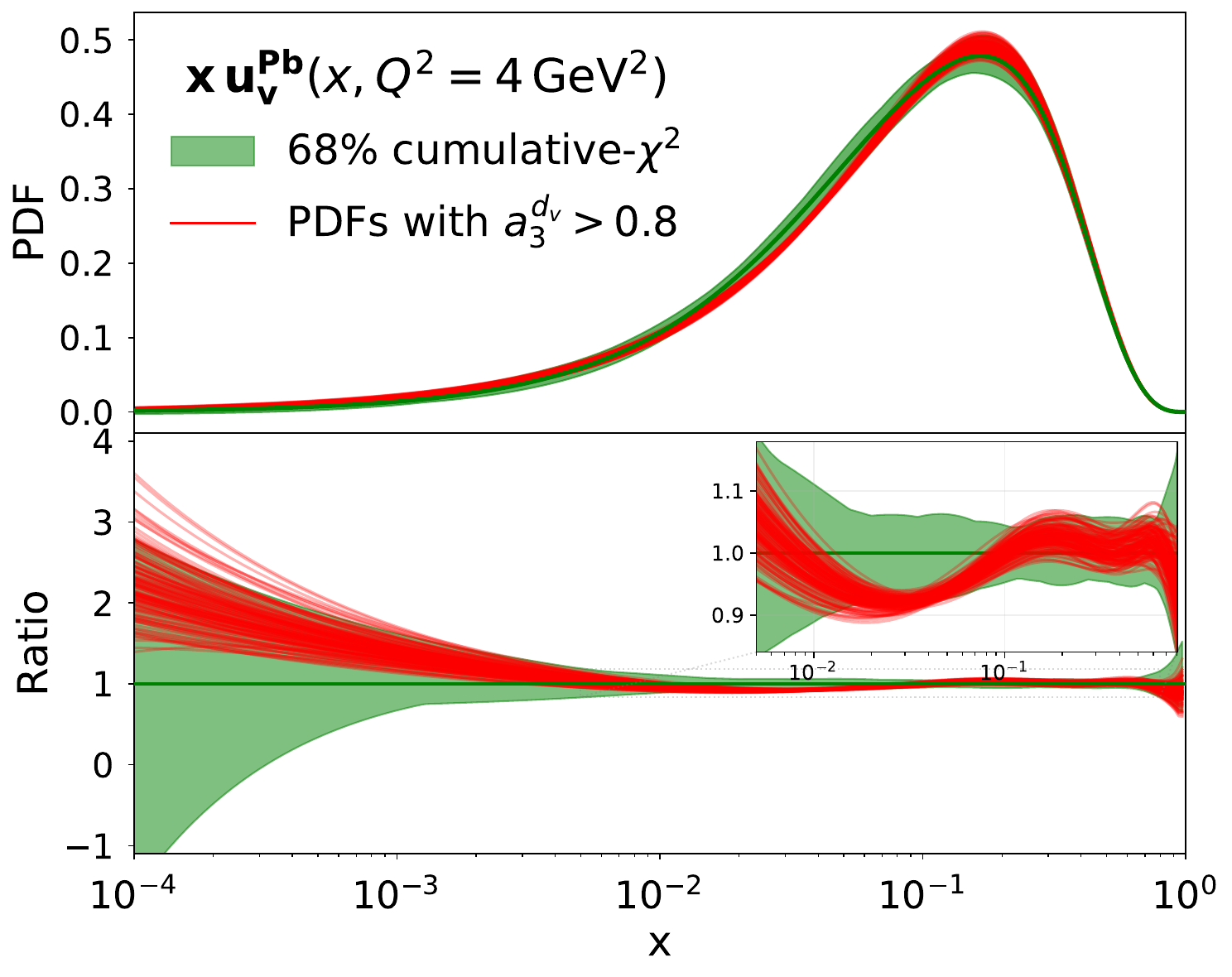}
\includegraphics[width=0.49\textwidth]{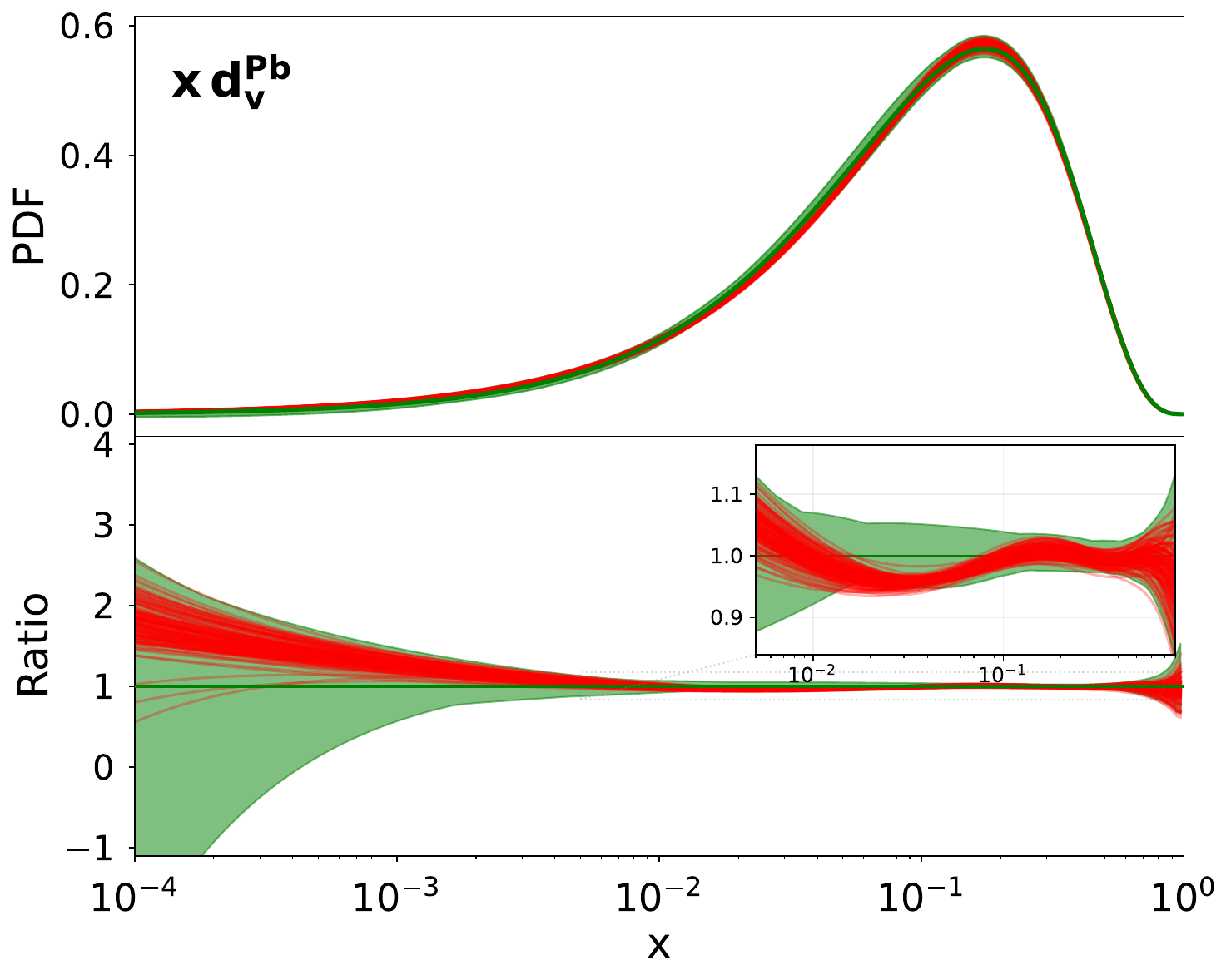}}\\
\addtocounter{subfigure}{-1}
\subfloat[Full Pb PDFs]{
\label{fig:7.15a}
\includegraphics[width=0.49\textwidth]{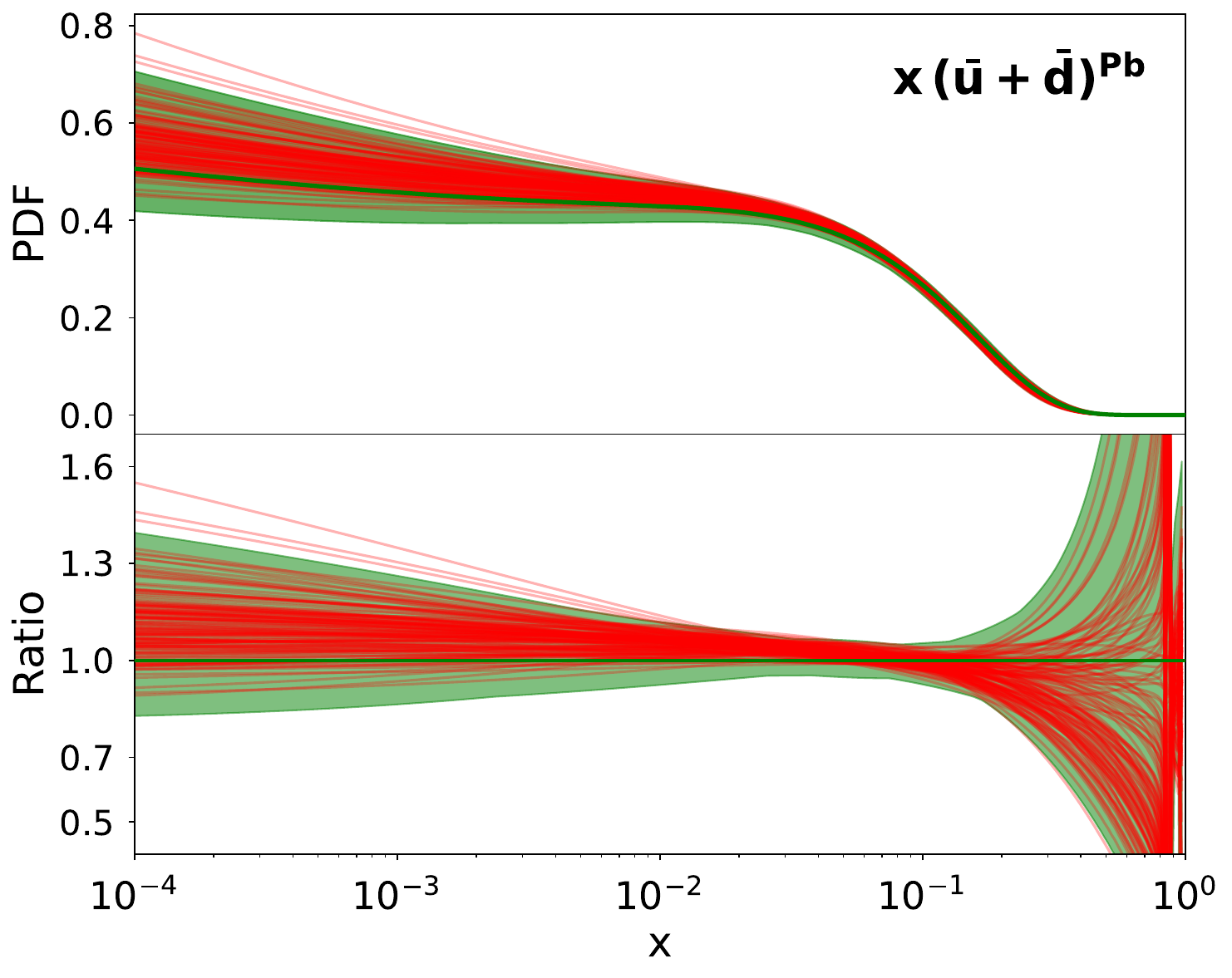}
\includegraphics[width=0.49\textwidth]{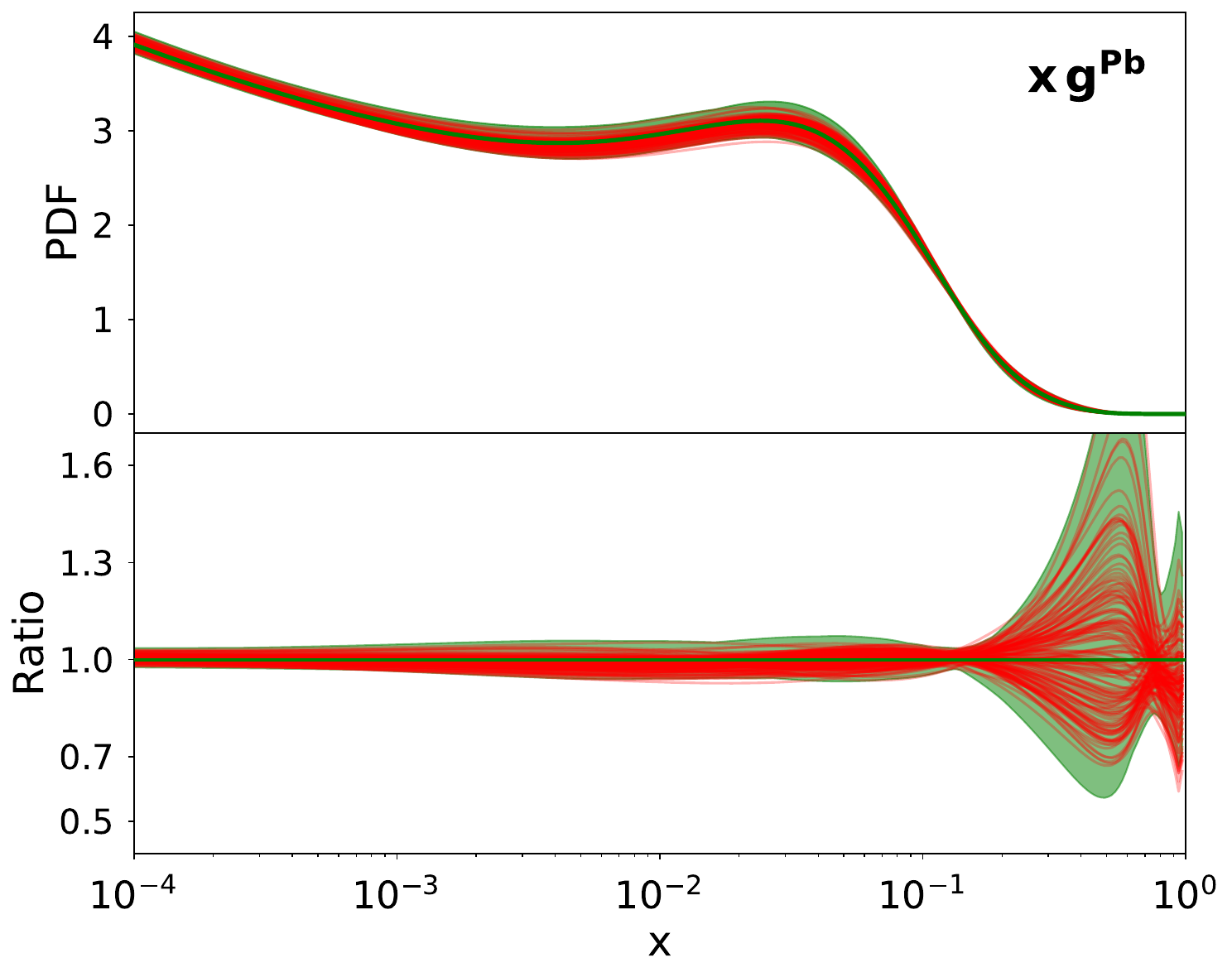}}\\
\subfloat[Bound-proton Pb PDFs]{
\label{fig:7.15b}
\includegraphics[width=0.49\textwidth]{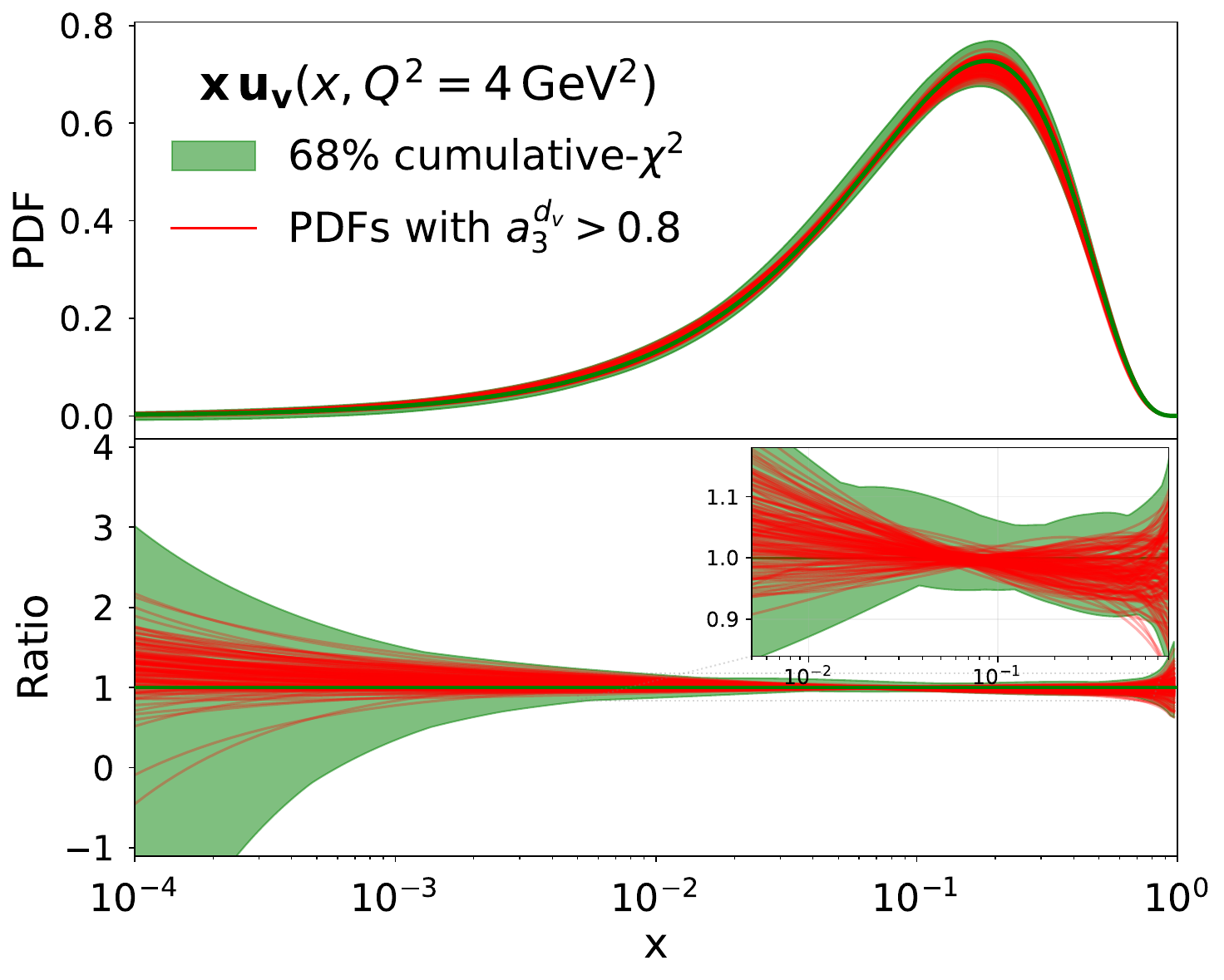}
\includegraphics[width=0.49\textwidth]{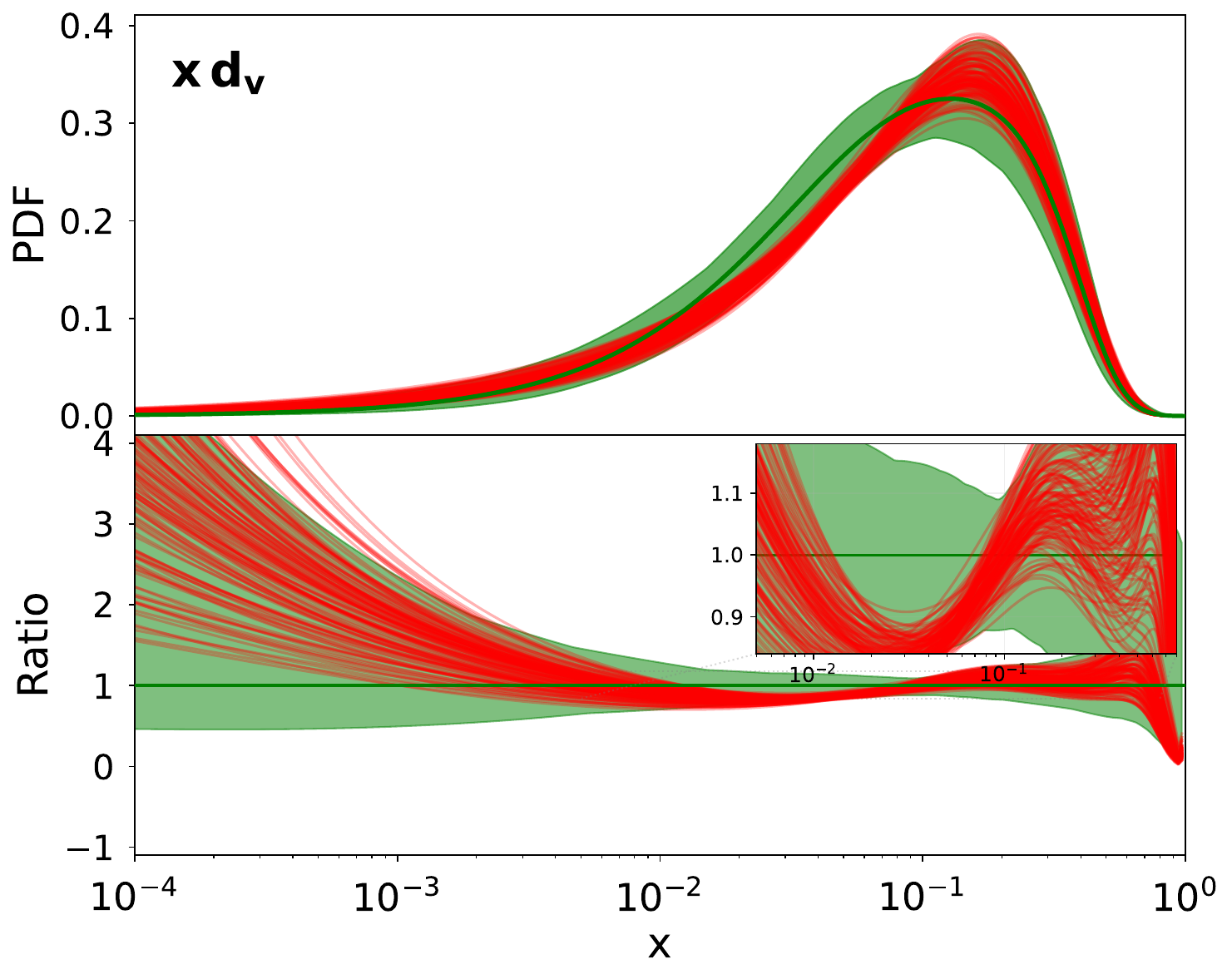}}
\caption[Secondary minimum nPDFs (full \& bound) for $a_3^{d_v}$]{Pb PDFs with 68\% CL cumulative-$\chi^2$ uncertainty bands (green), overlaid with individual PDF replicas (red) corresponding to the secondary $d_v$ configuration defined by $a_3^{d_v} > 0.8$. The selected replicas match the subset highlighted in Fig.~\ref{fig:7.10a}. In all panels we represent both the absolute PDF together with the corresponding ratio to the cumulative-$\chi^2$ central value. Panel (a) displays the full Pb PDFs, and Panel (b) shows the corresponding bound-proton distributions, defined in Eqs.~(\ref{eq:6.2}) and (\ref{eq:6.3}), respectively.}
\label{fig:7.15}
\end{figure}

Figure~\ref{fig:7.15} displays the 68\%\,CL uncertainty bands obtained with the cumulative-$\chi^2$ method (green), together with individual PDF replicas corresponding to the secondary $d_v$ configuration ($a_3^{d_v}>0.8$), shown in red. 

All distributions are presented as ratios relative to the cumulative-$\chi^2$ central value. Figure~\ref{fig:7.15a} shows the full nuclear Pb PDFs defined in Eq. (\ref{eq:6.2}), while Figure~\ref{fig:7.15b} displays the corresponding bound-proton distributions defined in Eq.~(\ref{eq:6.3}). 
The comparison demonstrates that the secondary configuration does not correspond to the same physical solution as the dominant minimum. Instead, it leads to a clearly distinguishable modification of the valence distributions. In particular, the secondary $d_v$ configuration enhances the distribution at small $x$, while producing a suppression in the region around $x \sim 10^{-2}$. Although these variations remain largely within the 68\% uncertainty band, they contribute significantly to both the asymmetry and the overall broadening of the uncertainty envelope.

In contrast, the gluon distribution remains essentially unaffected by the secondary $d_v$ configuration. This behavior is consistent with the absence of significant correlations between the $d_v$ parameters associated with the secondary region and the gluon parameters, as observed in the pairwise posterior projections. The gluon PDF therefore provides an internal consistency check: the secondary solution modifies only those PDF components that are statistically correlated with the $d_v$ parameters. Although the full nuclear PDFs represent the primary quantities of interest in a nuclear fit, the bound-proton PDFs provide a more direct view of how the parameter-level multi-modality propagates into PDF space. This is most clearly seen in the valence distributions. In the bound-proton representation (Fig.~\ref{fig:7.15b}), the secondary configuration induces a pronounced modification of the $d_v$ shape, occupying a distinctly different region of PDF space compared to the dominant solution. By contrast, the bound-proton $u_v$ distribution exhibits only minor variations under the same selection, reflecting the more limited sensitivity of the up-valence component to $a_3^{d_v}$. In the full nuclear Pb PDFs shown in Fig.~\ref{fig:7.15a}, the averaging between up- and down-quark contributions partially smooths this effect, making the difference less pronounced but visible in both valence distributions.

\begin{figure}[]
\centering
\includegraphics[width=0.49\textwidth]{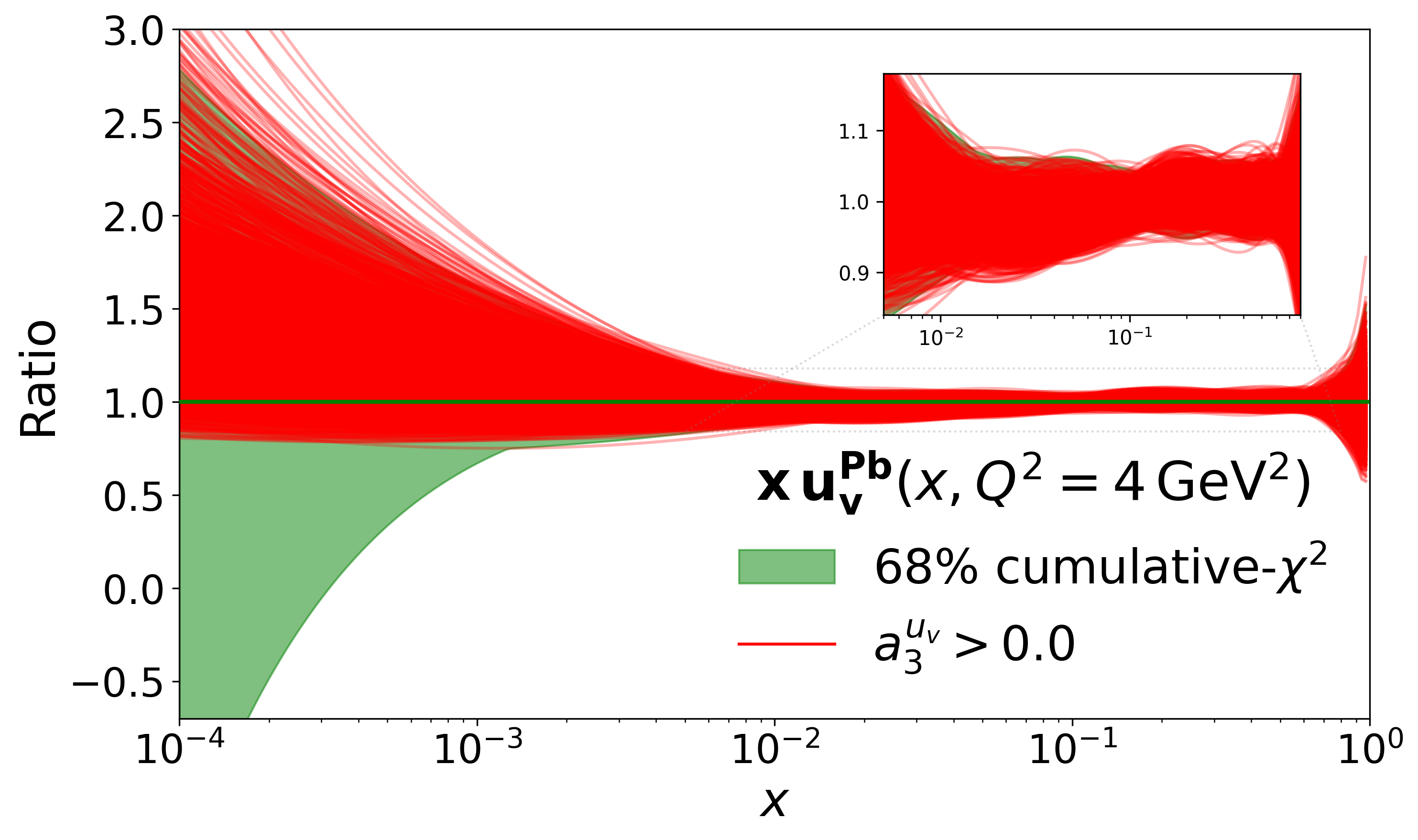}
\includegraphics[width=0.49\textwidth]{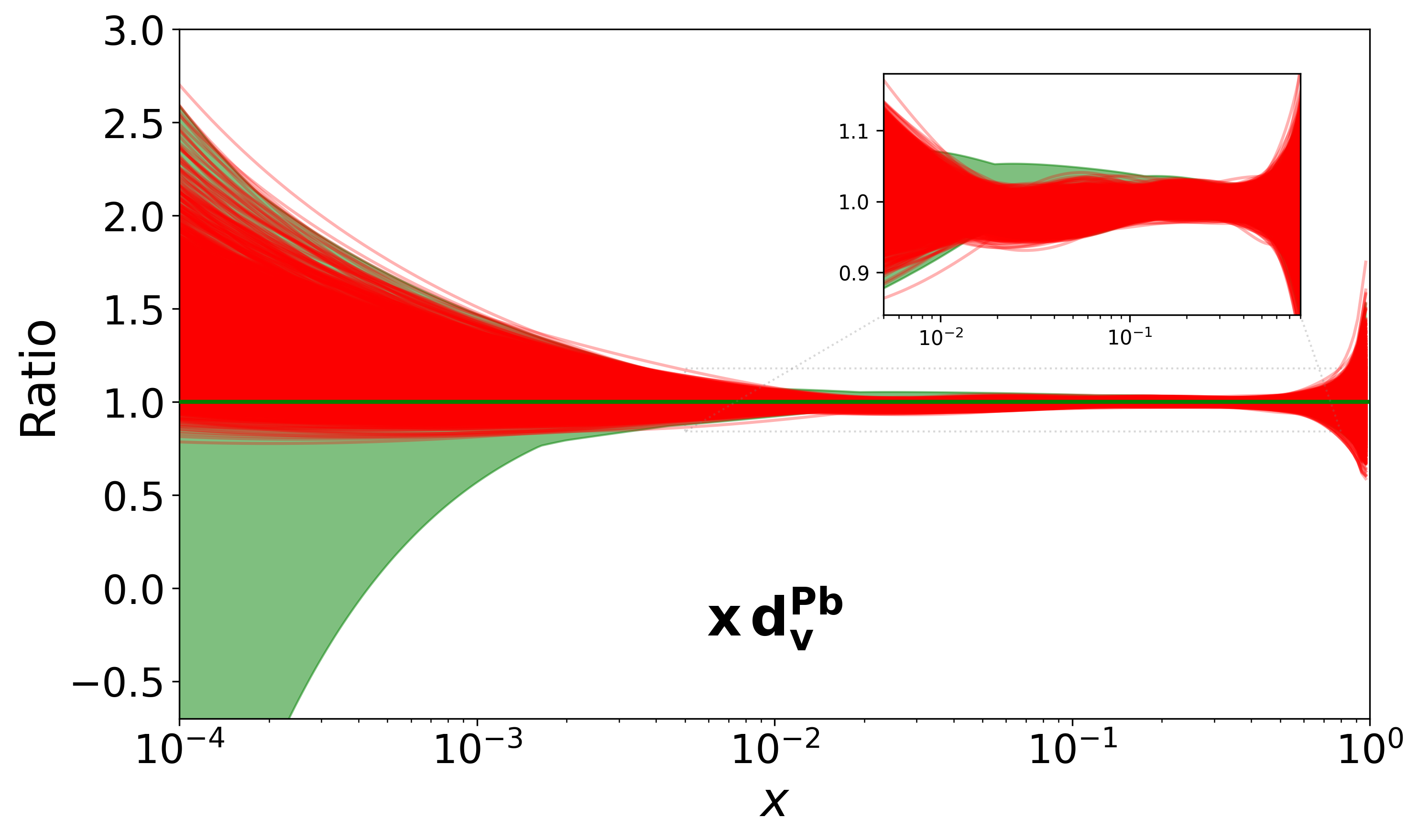}
\includegraphics[width=0.49\textwidth]{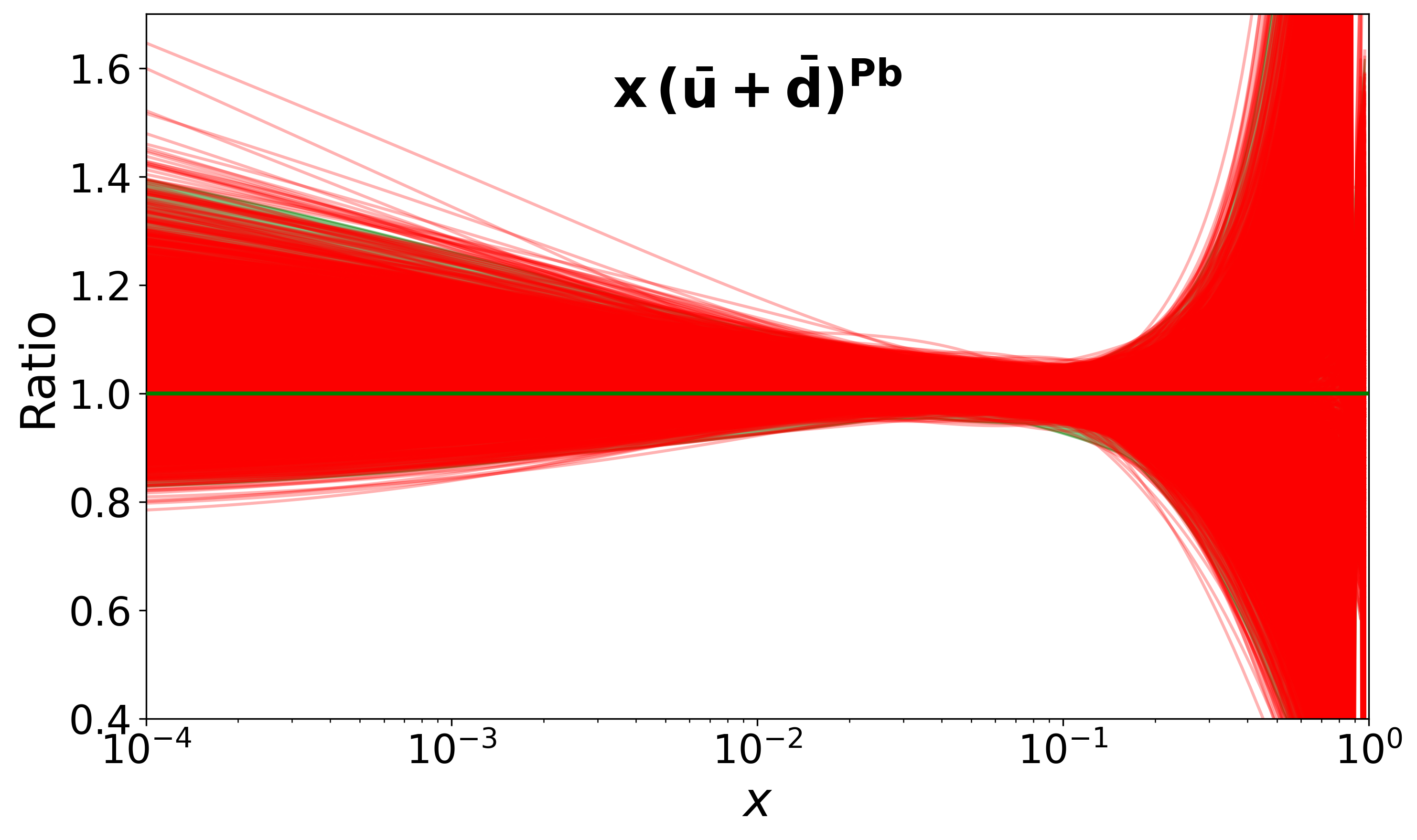}
\includegraphics[width=0.49\textwidth]{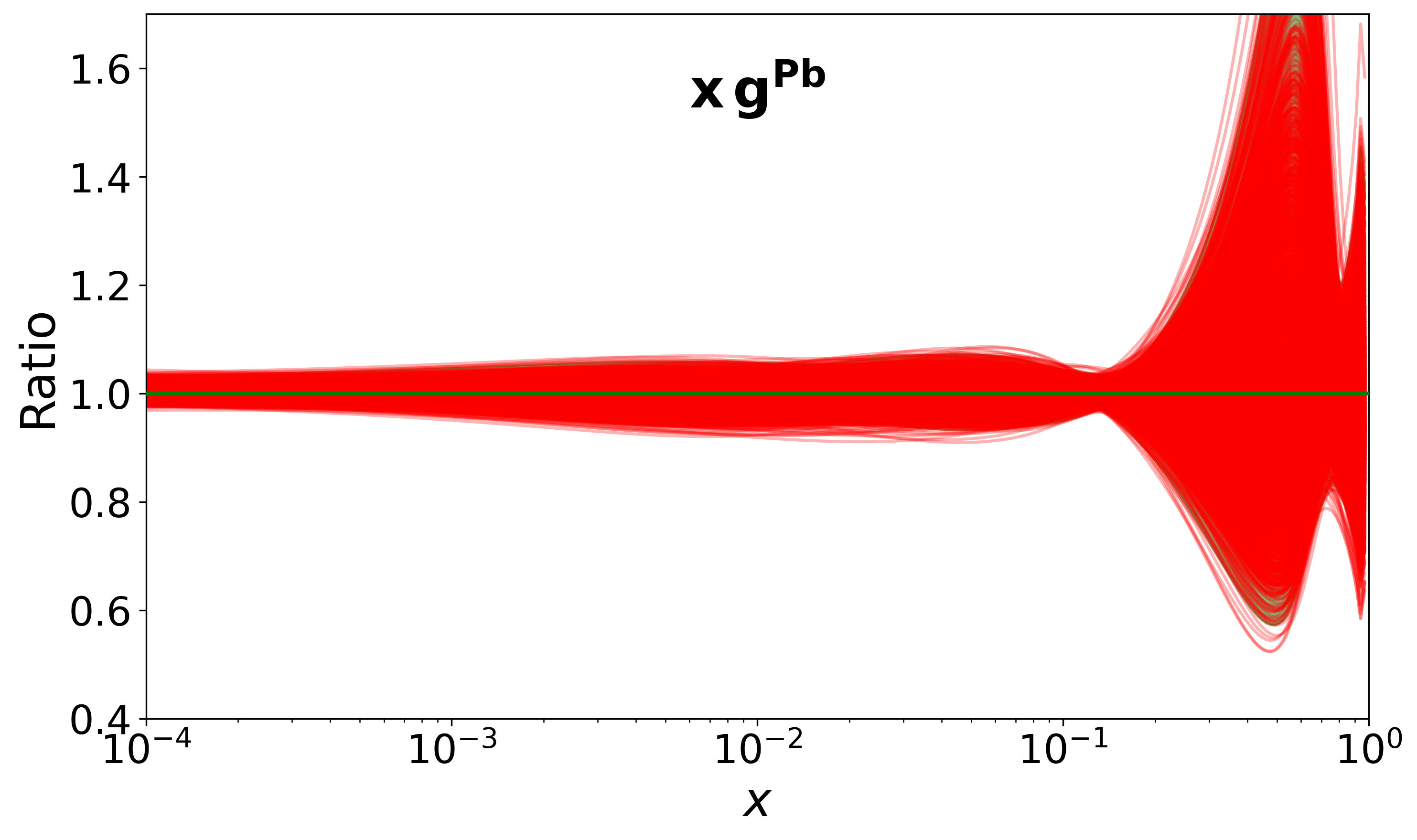}
\caption[Secondary minimum nPDFs for $a_3^{u_v}$]{Same as Fig.~\ref{fig:7.15}, but focused on the secondary minimum associated with the $a_3^{u_v}$ parameter. The red replicas correspond to the samples with $a_3^{u_v}>0$, highlighted in Fig.~\ref{fig:7.10b}. Shown here are only ratios compared to the minimal $\chi^2$ replica}
\label{fig:7.16}
\end{figure}

An analogous analysis for the secondary configuration in $a_3^{u_v}$ is presented in Fig.~\ref{fig:7.16}. In this case, the red curves correspond to samples satisfying $a_3^{u_v}>0$. The separation between the two posterior regions is now more symmetric, consistent with the comparable statistical weight of the two modes identified earlier (see Figure \ref{fig:7.10b}). The resulting PDF replicas form two branches that are clearly distinguishable. This can be seen at low $x$ where the two regions in $a_3^{u_v}$ parameter lead to very different distributions. Hence the multi-modal posterior structure translates directly into distinct realizations of the valence distribution.

\begin{figure}[]
\centering
\includegraphics[width=0.49\textwidth]{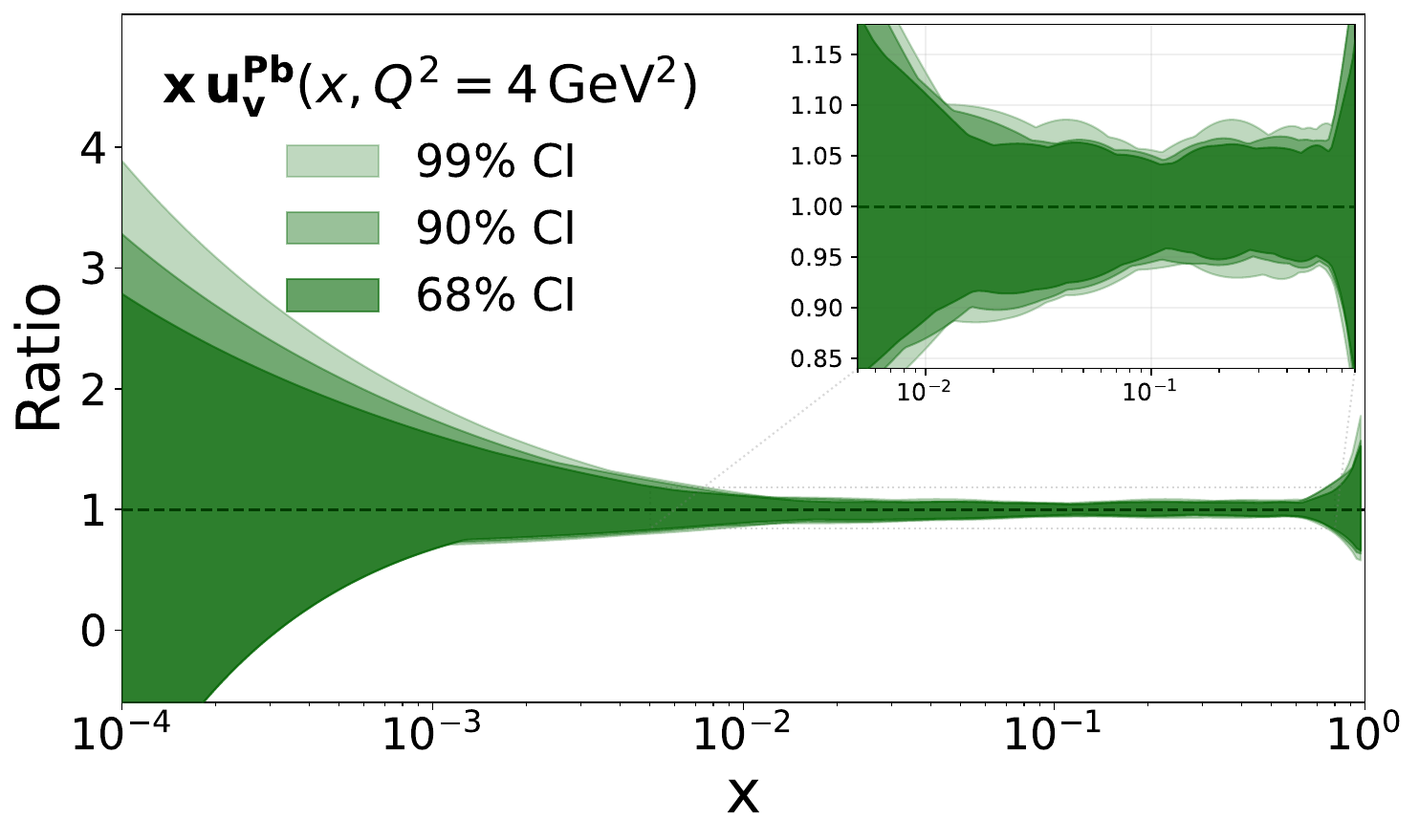}
\includegraphics[width=0.49\textwidth]{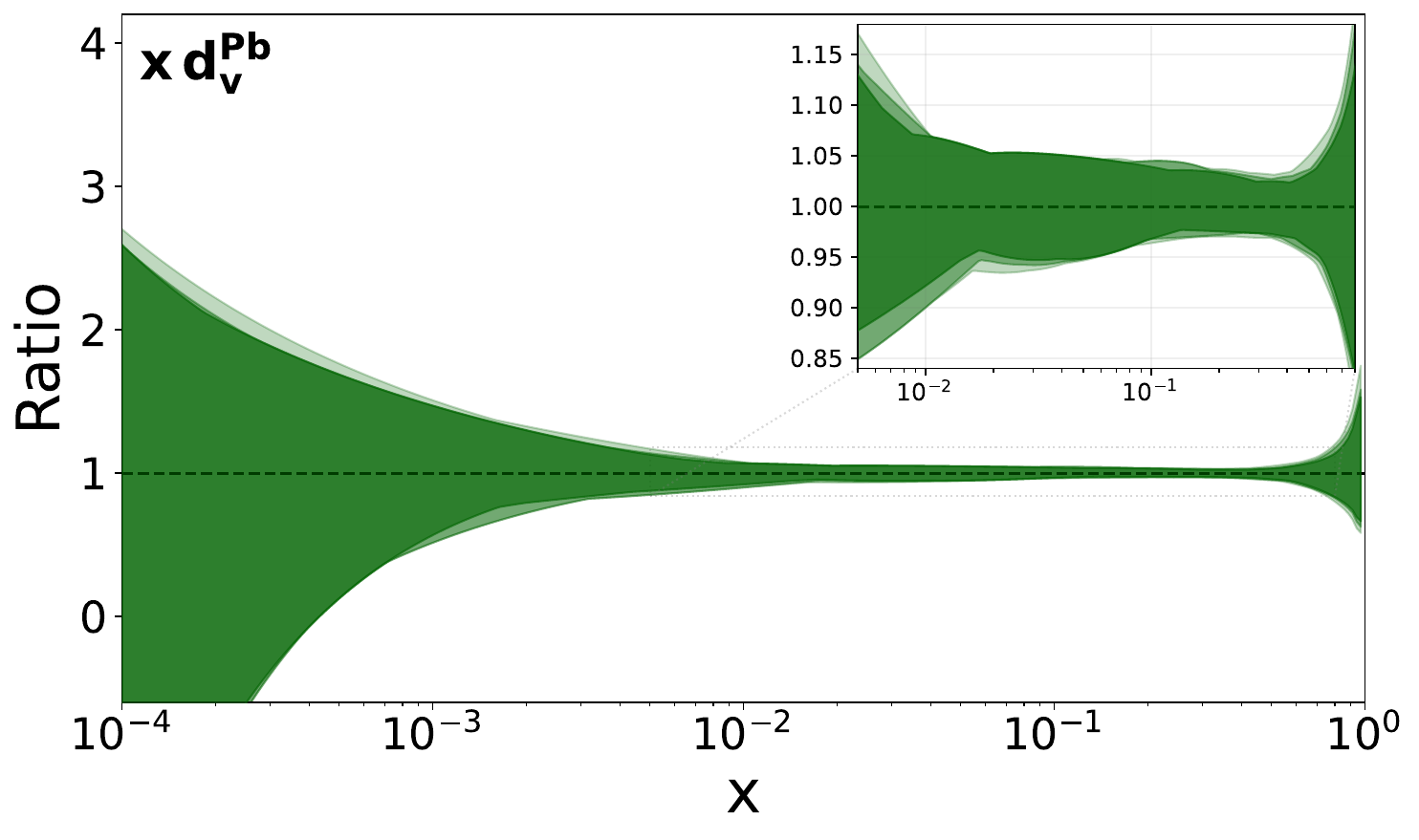}
\includegraphics[width=0.49\textwidth]{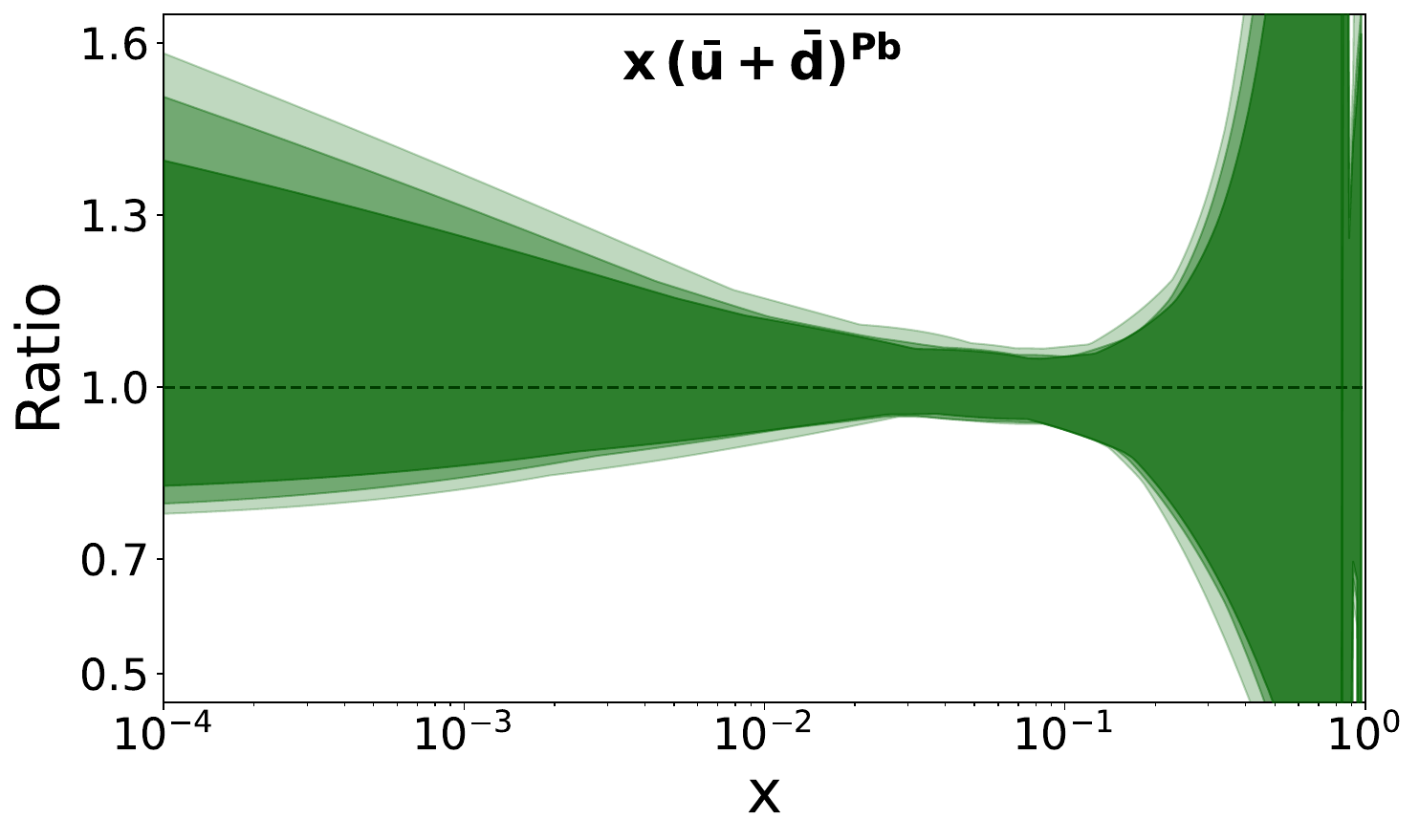}
\includegraphics[width=0.49\textwidth]{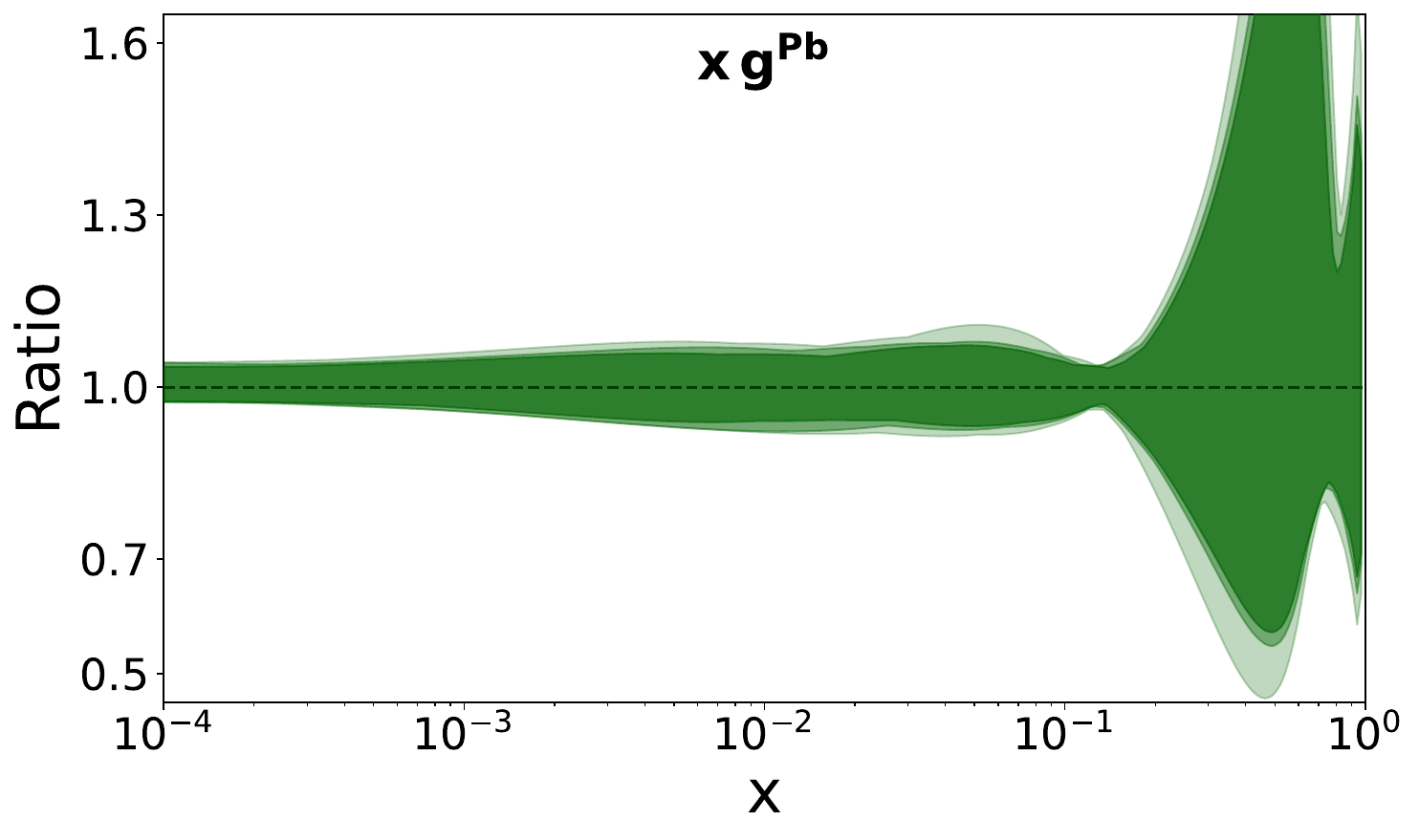}
\caption[Comparison of Pb PDF uncertainties at different cumulative-$\chi^2$ CLs]{Comparison of Pb PDF uncertainty estimates at \{68\%, 90\%, 99\%\} confidence levels, obtained using the cumulative-$\chi^2$ method. The panels show the ratios of each flavor distribution to its corresponding central value.}
\label{fig:7.17} 
\end{figure}

Figure~\ref{fig:7.17} further illustrates this effect by comparing the cumulative-$\chi^2$ uncertainty bands for the full Pb PDFs at different confidence levels (68\%, 90\%, and 99\%). As the confidence level increases, the influence of the secondary region in the $a_3$ parameters becomes progressively more visible, particularly in the valence distributions. This behavior reflects the fact that the multi-modal structure primarily affects the tails of the posterior: while its impact at 68\% CL remains moderate, it becomes increasingly pronounced at higher credibility levels, where the global goodness-of-fit constraints of the cumulative-$\chi^2$ method include these secondary minima.
A characteristic feature of this effect is the emergence of asymmetric uncertainty bands, most notably in the $u_v$ and $\bar{u}+\bar{d}$ distributions, where the upper and lower bounds expand differently as the confidence level increases. This asymmetry directly reflects the non-Gaussian and multi-modal structure of the posterior. In contrast, a symmetric uncertainty band, such as that obtained in the Hessian approach, would not capture this behavior, even at higher confidence levels, as it relies on a quadratic approximation around a single minimum.

Taken together, these results demonstrate that the observed multi-modality is not merely a parametrization degeneracy producing equivalent PDF shapes but instead corresponds to genuinely distinct physical solutions coexisting within the posterior. The MCMC framework captures these competing configurations and propagates their statistical weight directly into the final uncertainty bands.
The \texttt{Pb-only} analysis therefore shows that Bayesian posterior sampling provides a faithful mapping of the nuclear PDF parameter space, revealing structures, most notably the multi-modality associated with $a_3$ valence parameters, that remain inaccessible in a purely quadratic approximation. In addition, it naturally captures non-Gaussian and non-symmetric features of the posterior distribution. While the gluon and light-sea distributions are well described by an approximately Gaussian posterior, the valence PDFs exhibit pronounced non-Gaussian behavior. Such features cannot be consistently accommodated within the standard Hessian framework without introducing an enlarged tolerance parameter, which would artificially inflate uncertainties across other parton flavors and $x$ regions.

%%%%%%%%%%%%%%%%%%%%%%%%%%%%%%%%%%%%%%%%%%%%%%%%%%%%%%%%%%%%%%%%%%%%
\section{Multi-nuclei Analysis}\label{sec:7.2}
%%%%%%%%%%%%%%%%%%%%%%%%%%%%%%%%%%%%%%%%%%%%%%%%%%%%%%%%%%%%%%%%%%%%%
In this section, we extend the previous study by incorporating data from a range of nuclei and performing a global fit using the standard $A$-dependent parametrization described in Chapter~\ref{chap.6}. In contrast to the \texttt{Pb-only} analysis, which isolates lead data in order to probe the intrinsic structure of the parameter space, the \texttt{Multi-nuclei} setup introduces additional experimental constraints that correlate different nuclear systems through the imposed $A$-dependence of Eq. (\ref{eq:6.6}). As a result, information from lighter nuclei contributes indirectly to the determination of the lead PDFs, potentially modifying both the geometry of the posterior distribution and the resulting uncertainty estimates.

The MCMC setup for the \texttt{Multi-nuclei} fit follows that of the \texttt{Pb-only} analysis, with one notable modification in the proposal distribution. In particular, the scale of the fixed covariance matrix $C_0 = \mathrm{diag}(\sigma_0 \times \mathbf{1})$ was reduced. While the \texttt{Pb-only} fit employed $\sigma_0 = 10^{-3}$, the \texttt{Multi-nuclei} analysis required a significantly smaller value, $\sigma_0 = 10^{-5}$. This adjustment was necessary to obtain a sufficiently high acceptance rate during sampling. For larger proposal widths, candidate points were systematically rejected, indicating that the posterior distribution becomes substantially more constrained once correlations between different nuclear systems are introduced through the $A$-dependent parametrization. In contrast to the \texttt{Pb-only} case, the sampler does not explore unphysical regions of parameter space, and therefore no additional priors are required to stabilize the sampling.

Applying the same convergence criteria as before, seven chains were retained for the final analysis. After removing the burn-in phase, the combined chains contain approximately $N_{\mathrm{tot}} \simeq 873\,000$ samples. Optimal thinning reduces this number to $N_{\mathrm{tot}} = 1515$ statistically independent points, which form the final posterior ensemble used in the following analysis.
Given the reduced statistics relative to the \texttt{Pb-only} fit, the discussion focuses on the merged posterior and its global correlation structure, rather than on individual chain behavior, before turning to the extracted PDFs and their associated uncertainties.

%%%%%%%%%%%%%%%%%%%%%%%%%%%%%
\subsection{Posterior of the Combined Chain}\label{sec:7.2.1}
%%%%%%%%%%%%%%%%%%%%%%%%%%%%%
\begin{figure}[h]
\centering
\includegraphics[width=\textwidth]{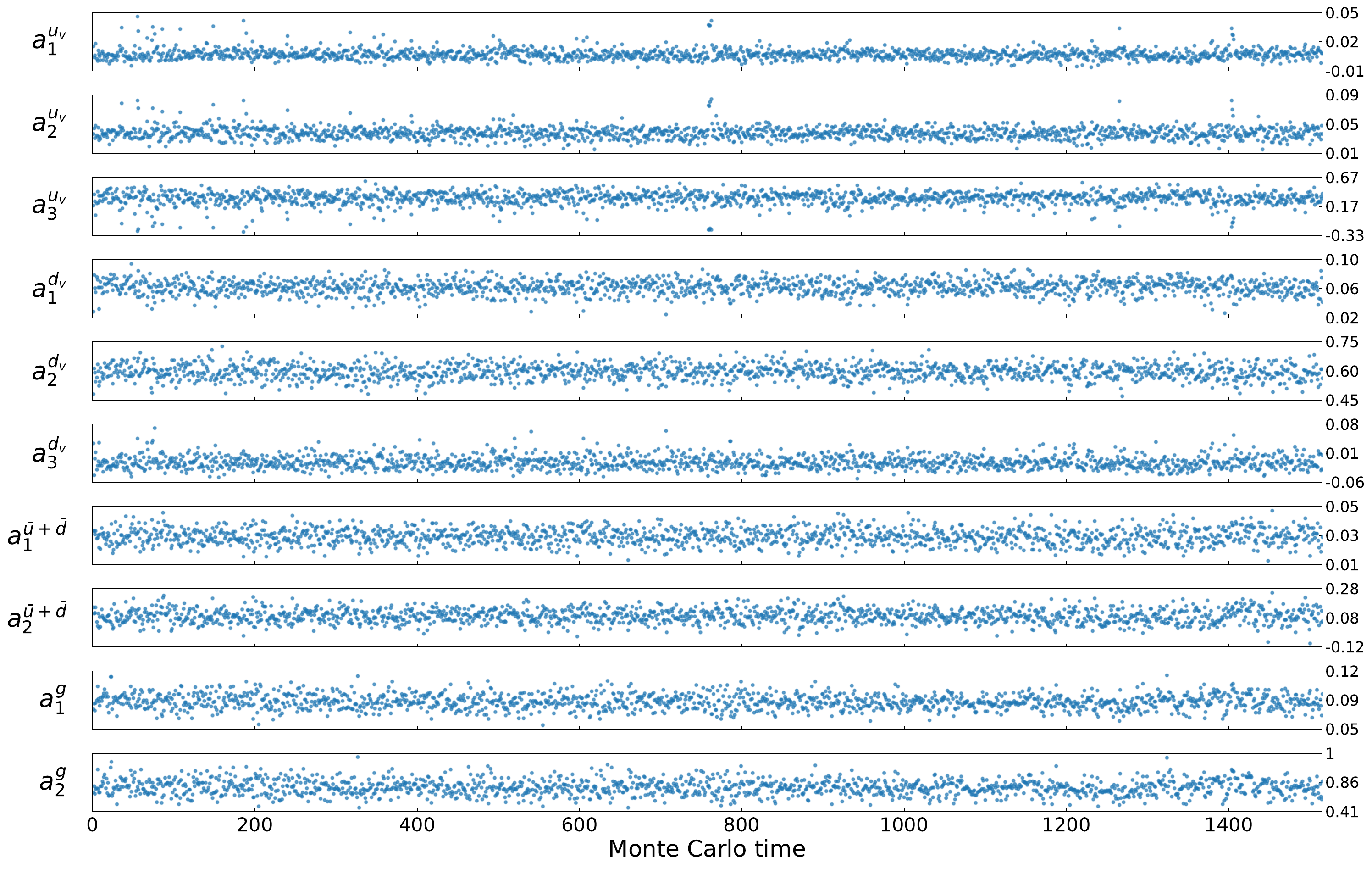}
\caption[Combined \texttt{Multi-nuclei} posterior samples]{Trace plot for the combined \texttt{Multi-nuclei} posterior samples. }
\label{fig:7.18}
\end{figure}

We begin by examining the global sampling behavior of the combined posterior ensemble obtained in the \texttt{Multi-nuclei} fit. Figure~\ref{fig:7.18} shows the trace plots for the concatenated samples from the seven converged chains after burn-in removal and thinning. Across the full ensemble, all parameters fluctuate around stable mean values without visible long-term drift, indicating that the chains are consistently sampling the same stationary distribution.
Compared to the \texttt{Pb-only} analysis, the fluctuations are more localized, reflecting the additional constraints introduced by the inclusion of data from multiple nuclei. The acceptance rate remains stable throughout the sampling process, with no evidence of systematic exploration of unphysical regions of parameter space. This behavior confirms that the smaller proposal width enables controlled and stable exploration of the posterior distribution.

\begin{figure}[h]
\centering
\includegraphics[width=\textwidth]{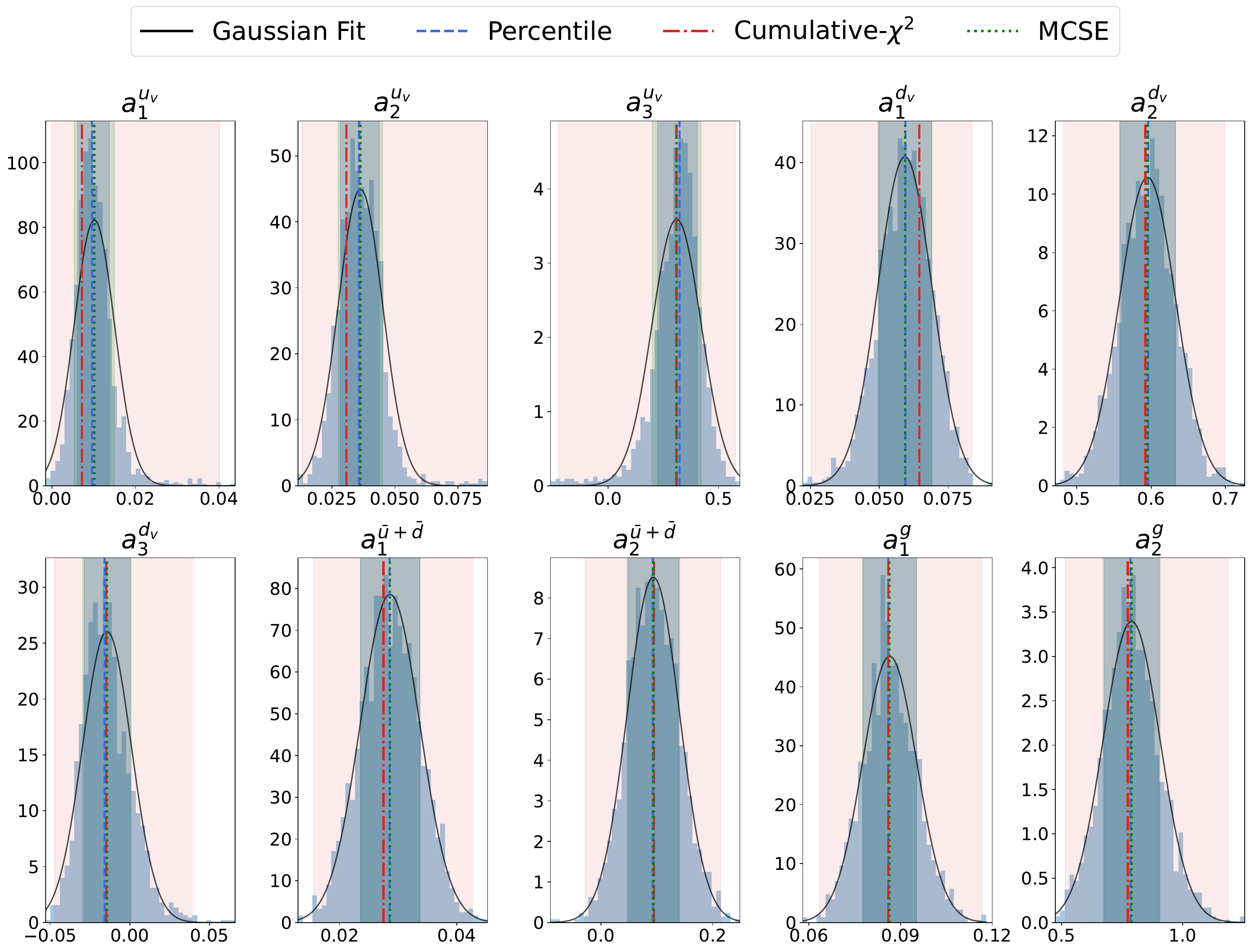}
\caption[Histogram plot for the combined \texttt{Multi-nuclei} posterior samples.]{Histogram of marginal distributions for individual parameters for the combined \texttt{Multi-nuclei} posterior samples.}
\label{fig:7.19}
\end{figure}

Before examining correlations in detail, we first consider the one-dimensional marginal distributions of the fitted parameters shown in Fig.~\ref{fig:7.19}. The histograms are generally symmetric and sharply peaked, indicating well-localized posterior distributions. In most cases, Gaussian fits provide a good approximation with some exceptions in the valence sector, consistent with the stronger constraints in the \texttt{Multi-nuclei} analysis. The central values obtained using the Monte Carlo standard error (green), percentile (blue), and cumulative-$\chi^2$ (red) prescriptions are closely aligned, with only minor shifts reflecting residual asymmetries, most notably in $a_1^{u_v}$, $a_2^{u_v}$, and $a_1^{d_v}$.
Despite this overall Gaussian appearance, some non-Gaussian features persist, most prominently in the form of extended tails in the $a_1^{u_v}$, $a_2^{u_v}$, and $a_3^{u_v}$ distributions. In particular, the $a_1^{u_v}$ and $a_2^{u_v}$ parameters exhibit a visible positive skew, with the distributions extending further toward larger values. This leads to a slight rightward shift of the percentile-based median relative to the cumulative-$\chi^2$ peak. The $a_3$ valence parameters, while significantly more stable than in the \texttt{Pb-only} analysis due to the additional constraints from neutral current DIS and Drell–Yan data, still display mild tail structures.
Compared to the \texttt{Pb-only} case, the distributions are significantly narrower and no clear multi-modal structure is visible in the one-dimensional projections, particularly for the $a_3$ valence parameters. However, the presence of extended tails alongside otherwise sharply localized peaks suggests a persistent structural asymmetry in the posterior. These features provide evidence of the suppressed multi-modal structure observed in the \texttt{Pb-only} fit, where the additional cross-nuclear constraints reduce secondary modes into subleading, non-Gaussian tails without fully eliminating the underlying tensions in parameter space.

\begin{figure}[h]
\centering
\includegraphics[width=\textwidth]{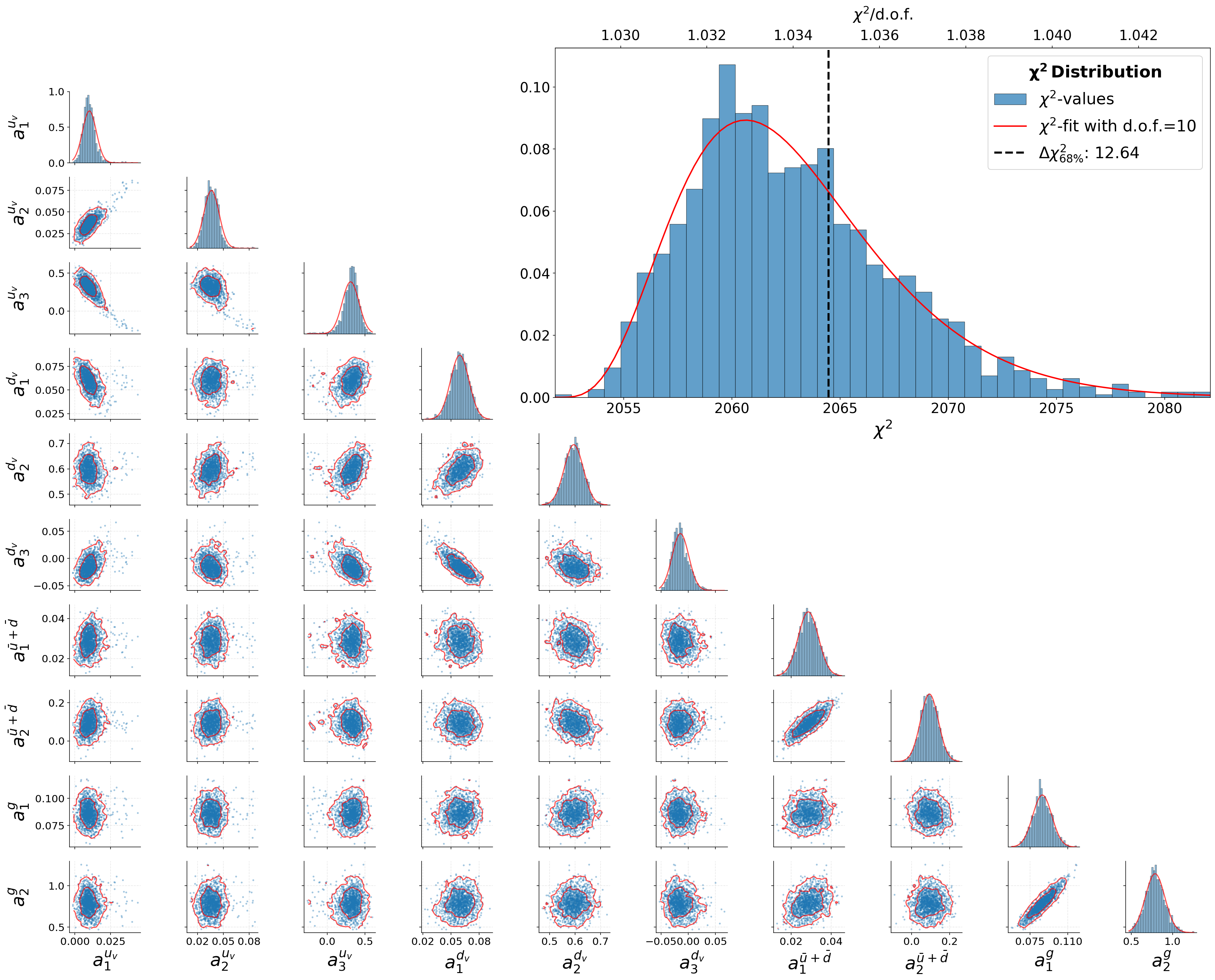}
\caption[Pairwise plot for Combined \texttt{Multi-nuclei} fit]{Pairwise plot for the combined \texttt{Multi-nuclei} posterior samples. }
\label{fig:7.20}
\end{figure}

We now turn to the global structure of the posterior ensemble. Figure~\ref{fig:7.20} illustrates this through the $\chi^2$ distribution and the pairwise marginal distributions of the fitted parameters. The samples exhibit tighter clustering than in the \texttt{Pb-only} analysis, especially for the valence parameters, reflecting the increased constraints from the \texttt{Multi-nuclei} dataset. A prominent feature of the pairwise distributions is the structure of correlations within the valence parameters. In particular, a strong negative correlation is observed between $a_1^{u_v}$ and $a_3^{u_v}$, visible as a narrow, tilted ellipse. Similar, though less pronounced, correlations are also present among other valence parameters. At the same time, non-trivial deviations from purely Gaussian behavior remain visible. In the $(a_1^{u_v}, a_2^{u_v})$ and $(a_2^{u_v}, a_3^{u_v})$ projections, a slight broadening and extension of the density away from the main cluster can be observed. These features indicate a residual secondary structure in the posterior, consistent with the non-Gaussian tails seen in the one-dimensional distributions. Compared to the \texttt{Pb-only} analysis, however, these structures no longer form clearly separated regions but instead appear as diffuse extensions of the dominant mode, reflecting a strongly suppressed multi-modal behavior. This residual structure is investigated further in the following analysis.

In contrast, the sea-quark and gluon parameters ($a_1^{\bar{u}+\bar{d}}$, $a_2^{\bar{u}+\bar{d}}$, $a_1^{g}$, $a_2^{g}$) exhibit more regular and compact structures. The dominant features are the strong positive correlations between the $a_1$ and $a_2$ parameters within each flavor, visible as narrow, tilted elliptical contours. Aside from these expected correlations, the distributions remain well localized and largely decoupled from the detailed fluctuations of the valence parameters and are independently constrained by the data.
The top-right panel of Fig.~\ref{fig:7.20} shows the distribution of $\chi^2$ values obtained from the sampling. The distribution is well described by a standard $\chi^2$ function with $N_{\text{d.o.f.}}= 10$, indicating good statistical consistency between the model and the data. Nevertheless, we observe larger fluctuations compared to the \texttt{Pb-only} results in Fig.~\ref{fig:7.9}. This is expected due to the smaller sample size. The agreement between the sampled histogram and the fitted curve suggests that the experimental uncertainties are appropriately estimated and that the fit does not exhibit significant over- or under-fitting. Furthermore, the distribution is sharply peaked around $\chi^2 \approx 2060$, with only a narrow range of variation corresponding to $\chi^2/N_{\text{d.o.f.}} \approx 1.033$–$1.040$. 

\begin{figure}[!h]
\centering
\includegraphics[width=\textwidth]{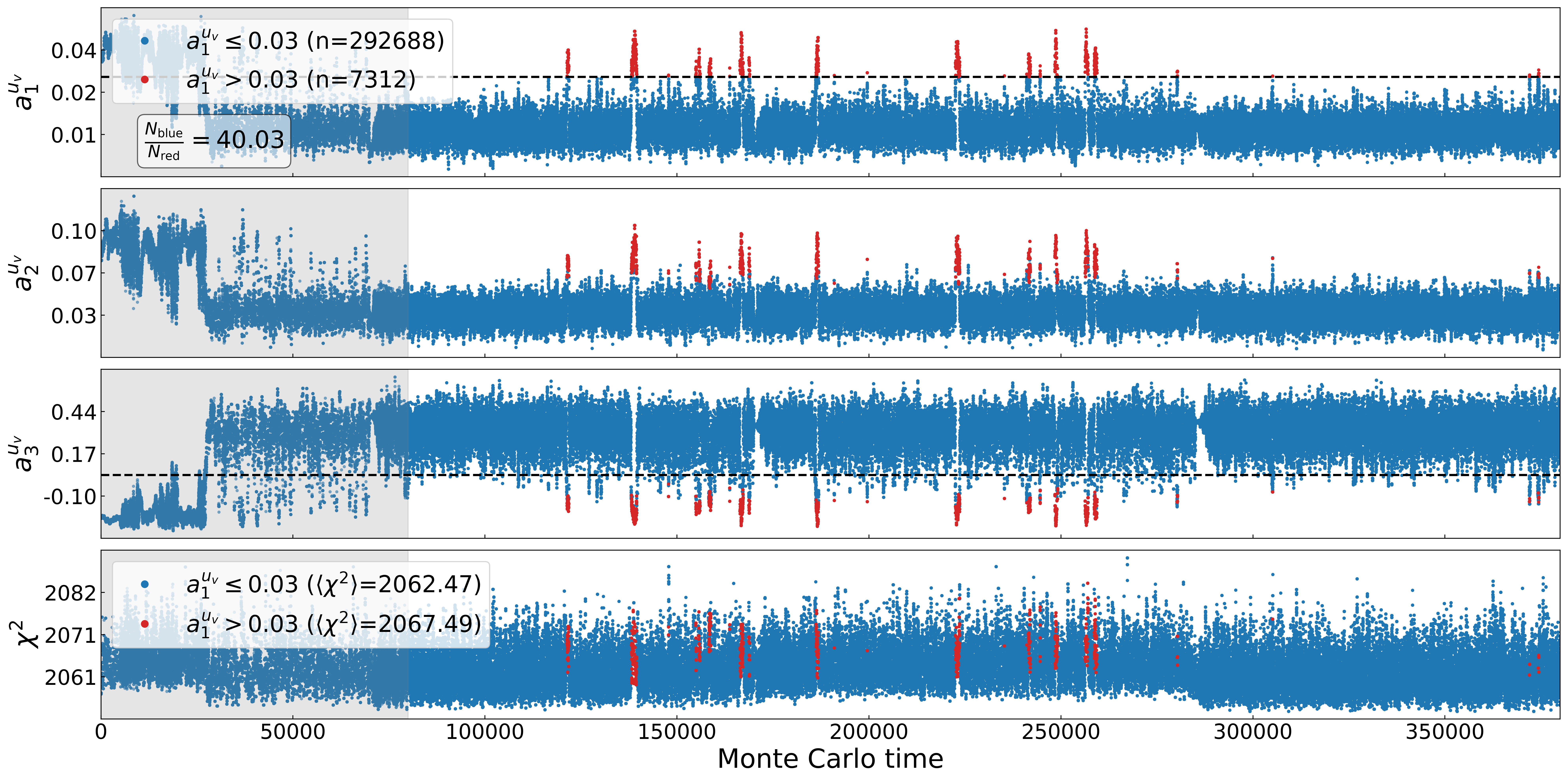}
\includegraphics[width=\textwidth]{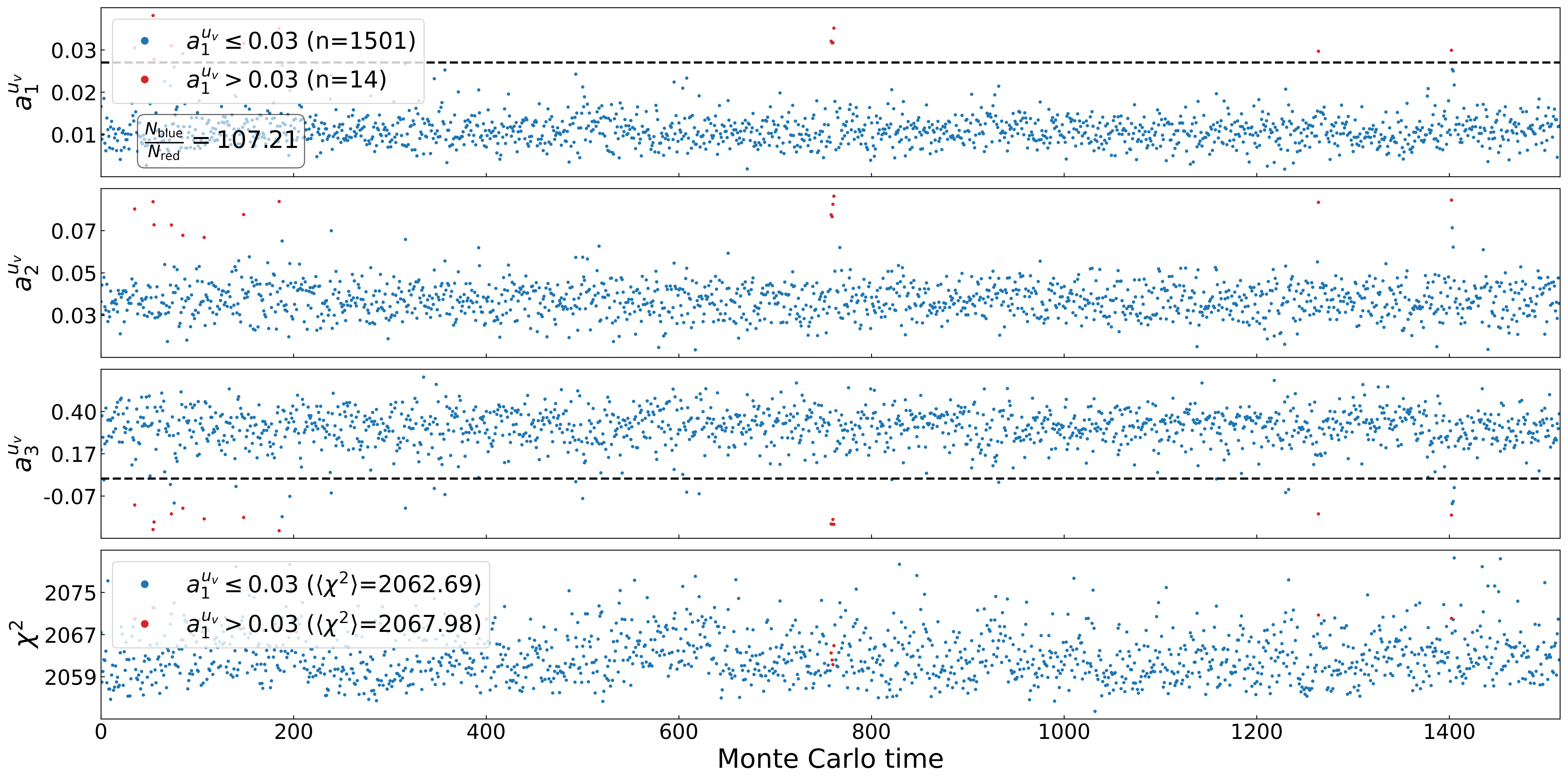}
\caption[Trace plot of $u_v$ parameters in the \texttt{Multi-nuclei} fit, indicating potential remnants of multi-modal structure]{Trace plot of $u_v$ parameters in the \texttt{Multi-nuclei} fit, indicating potential remnants of multi-modal structure. Samples are separated using the threshold $T = 0.03$, with $a_1^{u_v} < T$ shown in blue and $a_1^{u_v} > T$ shown in red. Top: representative single chain. Bottom: combined posterior sample.}
\label{fig:7.21} 
\end{figure}

To further assess whether the multi-modal structure observed in the \texttt{Pb-only} analysis persists in the \texttt{Multi-nuclei} fit, we perform a simplified minima test focusing on the $a_1^{u_v}$ parameter. Figure~\ref{fig:7.21} shows the trace behavior for a representative chain (top) and for the combined posterior sample (bottom), where samples are separated using a threshold $T = 0.03$, with points satisfying $a_1^{u_v} > T$ shown in red and those with $a_1^{u_v} < T$ shown in blue.
In contrast to the \texttt{Pb-only} case, the \texttt{Multi-nuclei} fit exhibits a strongly suppressed secondary population. While occasional samples beyond the threshold are observed, they are sparse and do not form a statistically significant secondary mode. This is quantitatively reflected in the strong dominance of one subset over the other: out of 1515 total samples, 1501 belong to the primary region ($a_1^{u_v} < T$), while only 14 (less than 1\%) fall into the secondary region ($a_1^{u_v} > T$). This significant difference indicates that the posterior is effectively governed by a single region in parameter space. The corresponding $\chi^2$ values for the two subsets are very similar, with only a modest difference between the primary and secondary regions. This separation is comparable to that observed in the \texttt{Pb-only} analysis ($\Delta\chi^2 \sim 5$–$6$), indicating that the relative quality of the two regions remains similar. However, in the present case the effect is less pronounced due to the additional experimental constraints, which reduce the relative impact of secondary structures in the posterior. Moreover, the smaller sample size makes this behavior more difficult to resolve statistically.
These observations are consistent with the mild non-Gaussian features observed in both the marginal and pairwise distributions. In particular, the residual tails and asymmetric distortions can be interpreted as suppressed remnants of the multi-modal structure identified in the \texttt{Pb-only} analysis. The inclusion of \texttt{Multi-nuclei} data therefore does not fully eliminate the residual structure in parameter space but instead compresses them into subleading, non-Gaussian features within a single dominant posterior region.

Despite the absence of clearly separated modes, the posterior is not strictly Gaussian. These residual asymmetries and tail structures represent deviations from the quadratic approximation underlying the Hessian method and therefore cannot be consistently captured within that framework. In particular, a Hessian approach would approximate the posterior as a single symmetric Gaussian, thereby missing the asymmetric tails and subtle distortions induced by the suppressed multi-modality. This highlights the advantage of the MCMC framework, which is able to faithfully represent both Gaussian and non-Gaussian features of the posterior without imposing restrictive assumptions on its shape.

%%%%%%%%%%%%%%%%%%%%%%%%%%%%
\subsection{Nuclear PDFs and Uncertainties}
\begin{figure}[!b]
\centering
\includegraphics[width=0.49\textwidth]{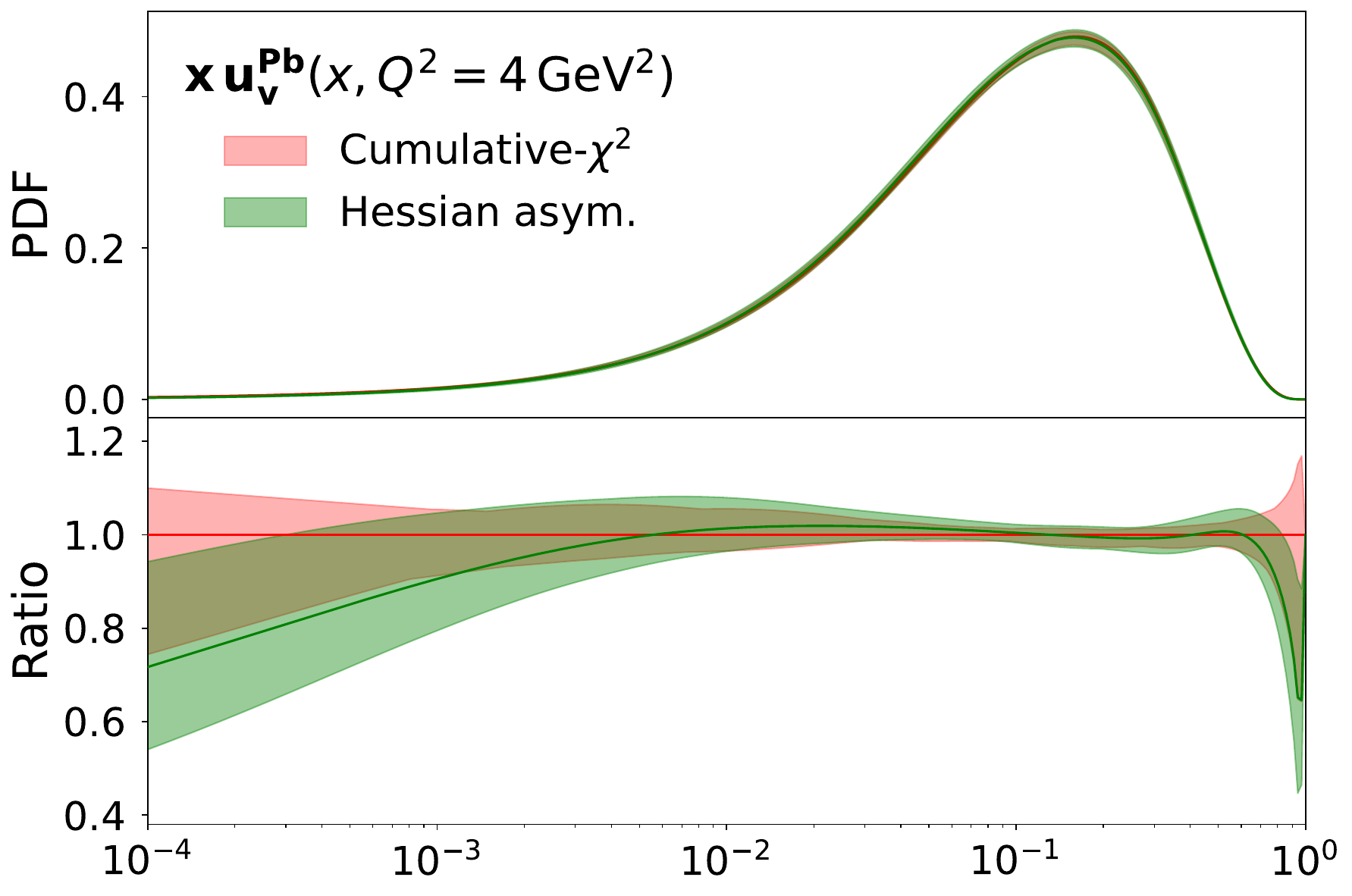}
\includegraphics[width=0.49\textwidth]{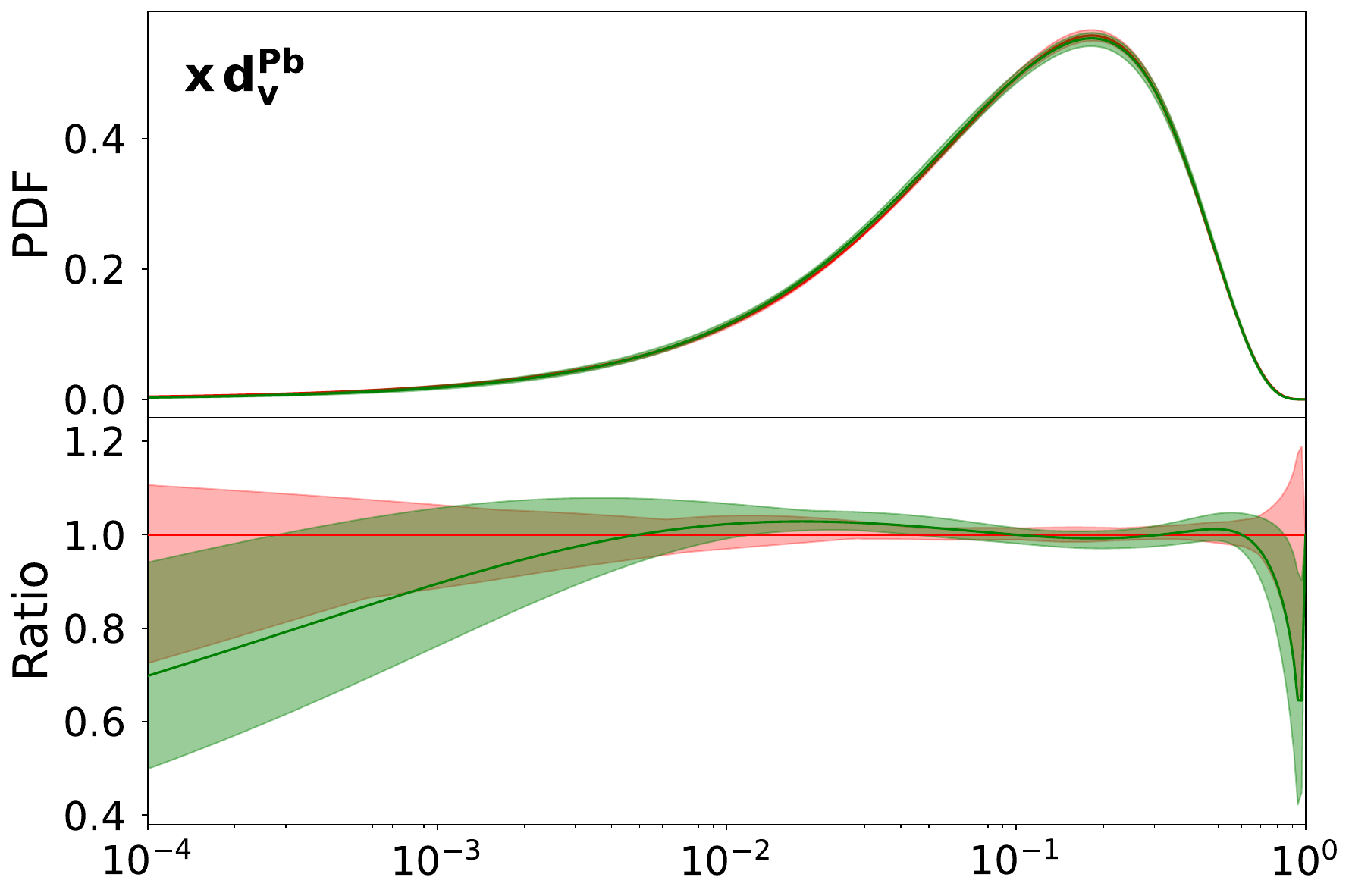}
\includegraphics[width=0.49\textwidth]{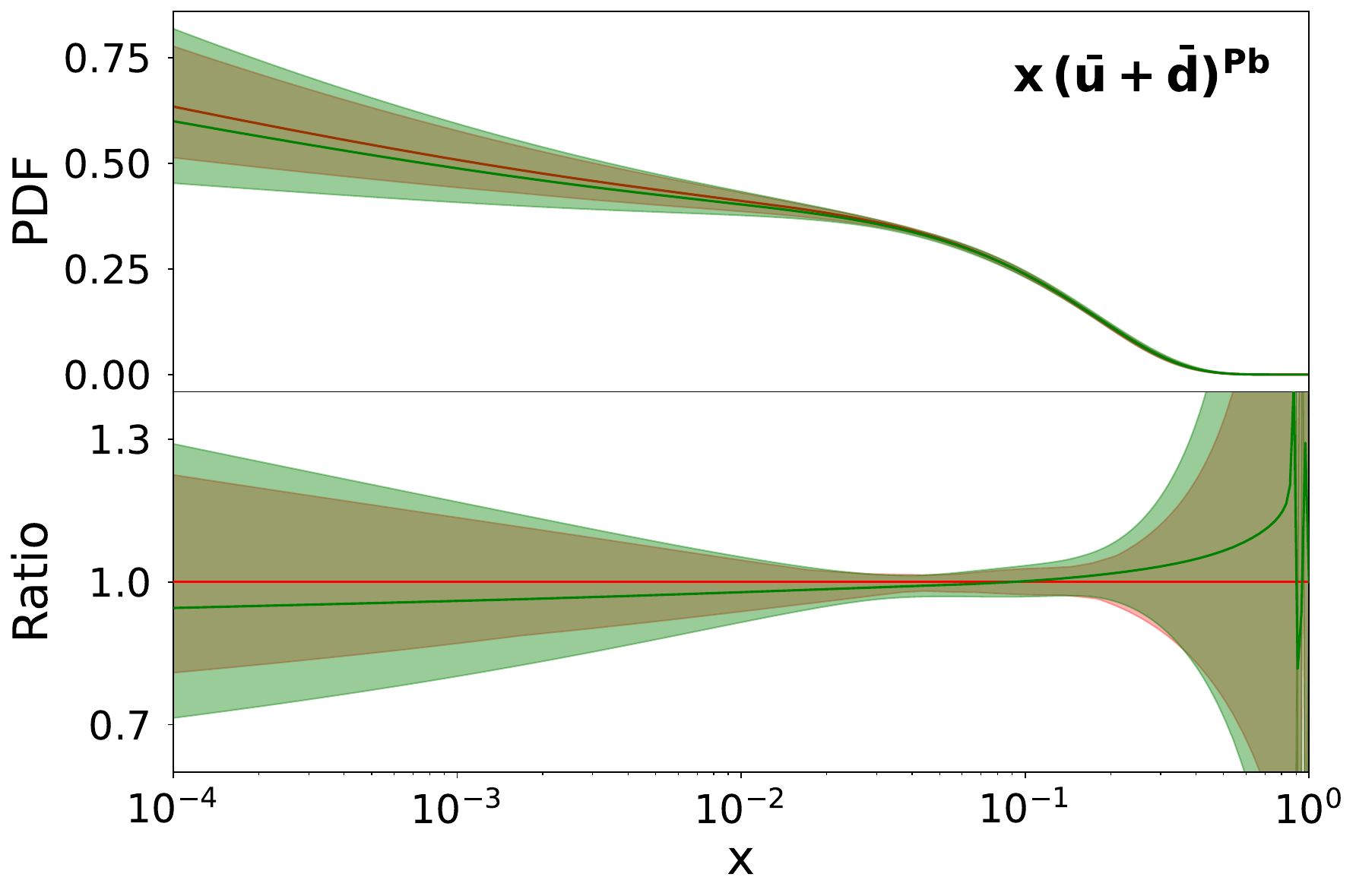}
\includegraphics[width=0.49\textwidth]{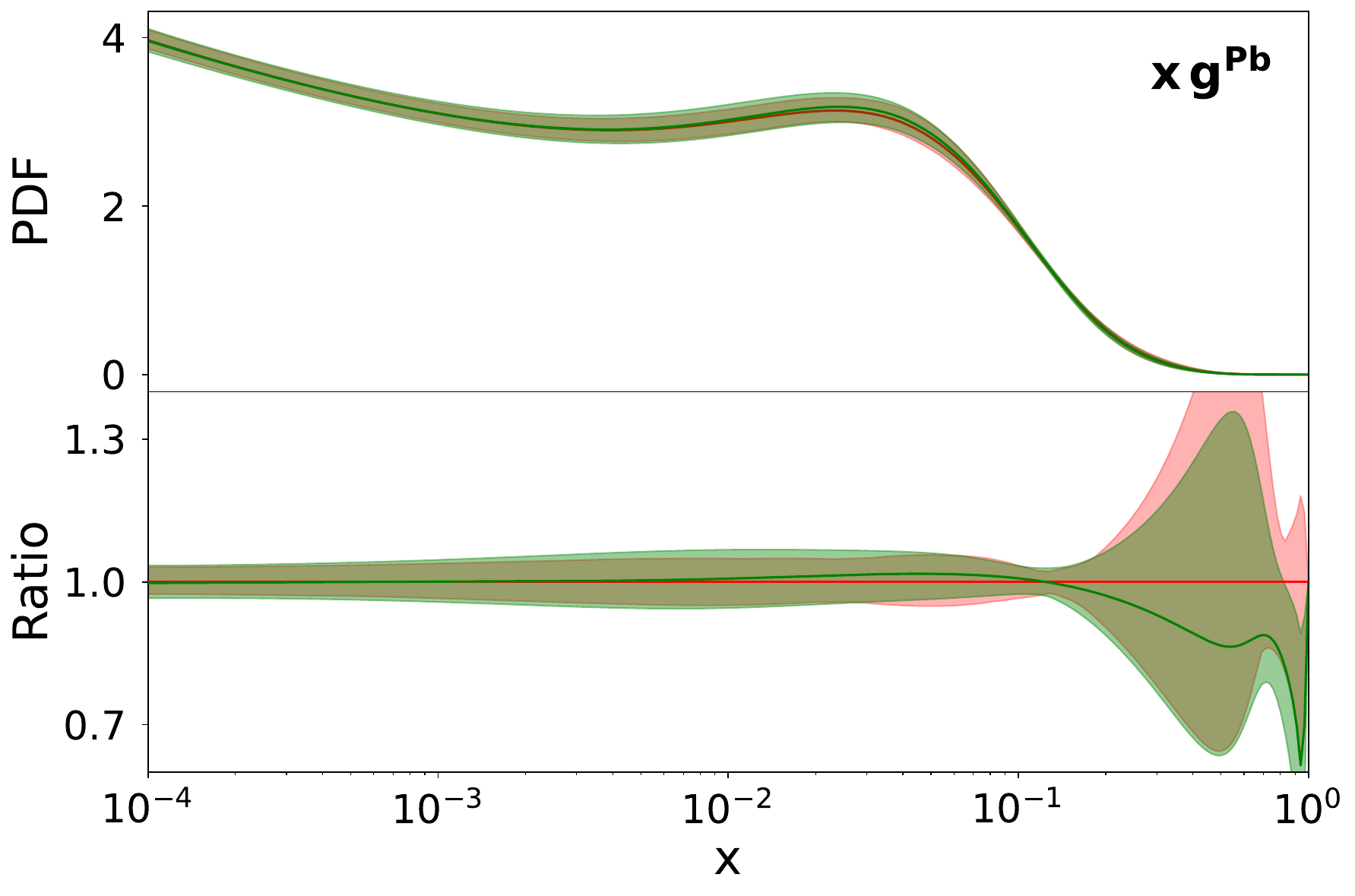}
\caption[\texttt{Multi-nuclei} uncertainties comparison for the Pb PDFs]{\texttt{Multi-nuclei} fit: Comparison of 68\% uncertainty bands for the Pb valence ($u_v$, $d_v$), light-sea ($\bar{u}+\bar{d}$), and gluon ($g$) distributions. For each flavor, the upper panel shows the absolute distribution, while the lower panel shows uncertainty bands normalized to the best-fit (minimum-$\chi^2$) prediction.}
\label{fig:7.22} 
\end{figure}

%%%%%%%%%%%%%%%%%%%%%%%%%%%%
Using the combined posterior ensemble of $N_{\mathrm{tot}} = 1515$ statistically independent samples, we evaluate the nuclear PDFs within the \texttt{Multi-nuclei} framework. Figure~\ref{fig:7.22} presents the extracted lead PDFs with their 68\% uncertainty bands obtained using the cumulative-$\chi^2$ (red) and asymmetric Hessian (green) approaches, where the Hessian tolerance is fixed by the MCMC-based $\Delta\chi^2$ criterion shown in Fig.~\ref{fig:7.20}. For each flavor, the upper panels show the absolute PDFs, while the lower panels display the corresponding uncertainty bands normalized to the cumulative-$\chi^2$ central value, defined by the minimum-$\chi^2$ sample. As in the \texttt{Pb-only} case, the PDFs are evaluated for each posterior replica, ensuring that parameter correlations and residual non-Gaussian features are consistently propagated to the observable level.

For the valence distributions, the differences between the uncertainty estimates are reduced compared to the \texttt{Pb-only} analysis but remain clearly visible in specific kinematic regions. In the intermediate-to-high-$x$ region ($10^{-2} \lesssim x \lesssim 0.7$), all three methods are in excellent agreement, with nearly identical central values and uncertainty bands. This reflects the strong experimental constraints from DIS and Drell–Yan data, which effectively localize the posterior and suppress non-Gaussian effects. In contrast, significant differences emerge at small $x$ ($x \lesssim 10^{-3}$). In this region, the Hessian uncertainties are considerably broader and exhibit a downward shift relative to the MCMC-based estimates (about a 15–20\% downward shift in central value), while the cumulative-$\chi^2$ and percentile bands remain tighter and more stable. This behavior indicates that the Hessian method overestimates the uncertainty by enforcing a symmetric quadratic approximation of the likelihood. In reality, the posterior exhibits asymmetric tail structures, as seen in Sec.~\ref{sec:7.2.1}, which effectively act as a soft boundary that limits the allowed variation of the PDFs. As a result, the MCMC-based methods capture a more realistic, constrained uncertainty, while the Hessian approach fails to account for these non-linear effects. At large $x$ ($x \gtrsim 0.6$), the uncertainty bands display a characteristic dip followed by a widening, with the Hessian results again remaining more conservative than the MCMC-based estimates. Nevertheless, they do not cover the full extent of the cumulative-$\chi^2$ band in this region.

For the light-sea combination $\bar{u}+\bar{d}$, a similar pattern is observed. In the intermediate-$x$ region, all three methods yield consistent and nearly symmetric uncertainty bands. In contrast, at small $x$ the Hessian uncertainties are slightly broader than the MCMC-based estimates, while at large $x$ the different prescriptions show increasing deviations. These differences again reflect the presence of non-Gaussian features and parameter correlations observed in the posterior.

The gluon distribution shows good agreement between the different uncertainty methods in the intermediate- and low- $x$ region, where the constraints from the data are strongest and the posterior is close to Gaussian. However, noticeable differences emerge toward the kinematic boundaries. At large $x$ the uncertainty bands begin to diverge, with the cumulative-$\chi^2$ estimate exhibiting a stronger upward widening and the Hessian result showing a characteristic downward trend. These features indicate that residual parameter correlations and non-linear effects persist in the poorly constrained regions.

Overall, the inclusion of \texttt{Multi-nuclei} data leads to a more regular and constrained posterior, reducing the impact of the strong multi-modality observed in the \texttt{Pb-only} analysis. As a result, the different uncertainty prescriptions show improved agreement, particularly in the intermediate-$x$ region where the experimental constraints are strongest. However, the remaining discrepancies at small and large $x$, especially in the valence PDFs, demonstrate that the posterior is not fully Gaussian. These deviations originate from residual asymmetries and tail structures, which are consistently propagated by the MCMC framework but cannot be captured within the quadratic approximation underlying the Hessian method. This highlights the importance of Bayesian sampling for a faithful and comprehensive determination of nPDF uncertainties, even in the presence of strong cross-nuclear constraints.

%%%%%%%%%%%%%%%%%%%%%%%%%%%%%%%%%%%%%%%%%%%%%%%%%%%%%%%
\section{Comparison Between Pb-only and Multi-Nuclei Fits}\label{sec:7.3}
%%%%%%%%%%%%%%%%%%%%%%%%%%%%%%%%%%%%%%%%%%%%%%%%%%%%%%%%
We now perform a direct comparison between the \texttt{Pb-only} and \texttt{Multi-nuclei} analyses in order to quantify the impact of the imposed $A$-dependence and the addition of cross-nuclear data on both the fit quality and the extracted PDFs.

A first quantitative comparison is provided in Table~\ref{tab:chi2_comparison}, which summarizes the minimum $\chi^2$ values obtained from the Hessian and MCMC fits in the two setups. In both cases, the agreement between the two methods is very good, indicating that consistent minima are identified. In the \texttt{Pb-only} analysis, the Hessian fit yields a marginally lower $\chi^2$ value than the MCMC result; however, this difference is numerically negligible and should not be overinterpreted. The reduced $\chi^2$ values increase slightly in the \texttt{Multi-nuclei} analysis, from approximately $0.99$ in the \texttt{Pb-only} case to about $1.03$, reflecting the larger and more diverse dataset included in the global fit. Nevertheless, the values remain close to unity for both methods, confirming the overall consistency and good quality of the fit. In this setup, the MCMC result yields a slightly lower $\chi^2$ than the Hessian fit, although the difference remains small.

\begin{table}[!h]
\centering
\caption[$\chi^2_{\min}$ values obtained from the Hessian and MCMC fits for the \texttt{Pb-only} and \texttt{Multi-nuclei} analyses]{Comparison of the minimum $\chi^2$ values obtained from the Hessian and MCMC fits for the \texttt{Pb-only} and \texttt{Multi-nuclei} analyses. Values are shown both as total $\chi^2_{\min}$ and as reduced $\chi^2$ normalized by the number of degrees of freedom $N_{\mathrm{d.o.f.}}$.}
\setlength{\tabcolsep}{6pt}
\renewcommand{\arraystretch}{1.25}
\begin{tabular}{lcc}
\hline
\textbf{$\chi^2_{\min}$} & \textbf{\texttt{Pb-only}} & \textbf{\texttt{Multi-nuclei}} \\
\hline
Hessian (per $N_{\mathrm{d.o.f.}}$) 
  & 1466.57\; (0.985) 
  & 2056.10\; (1.035) \\
MCMC (per $N_{\mathrm{d.o.f.}}$) 
  & 1467.30\; (0.992) 
  & 2051.82\; (1.028) \\
\hline
\end{tabular}
\label{tab:chi2_comparison}
\end{table}

A detailed comparison of the best-fit parameters is given in Table~\ref{tab:param_comparison}, where both the fit-level coefficients $a_j^i$ and the corresponding nuclear-modified parameters $c_j^i(A)$, evaluated for lead nuclei ($A=208$), are reported.\footnote{The coefficients $c_j^i$ represent the nuclear-modified parameters evaluated at the fixed mass number $A$, obtained from the fit-level parameters $a_j^i$ via the relation $c_j^i = p_j^i + a_j^i \ln(A)$, where $p_j^i$ are the baseline coefficients defined in Table~\ref{tab:fit-parameters}. The extracted PDFs are evaluated using these $c_j^i$ coefficients.} The latter incorporates the logarithmic term $c_j^i = p_j^i + a_j^i \ln(A=208)$, with $\ln(208) \simeq 5.34$, implying that variations in $a_j^i$ are systematically amplified by a factor of $\mathcal{O}(5)$ when translated into $c_j^i$. As a result, the $c_j^i$ coefficients provide a more direct and sensitive probe of the physical differences between the \texttt{Pb-only} and \texttt{Multi-nuclei} fits.

\begin{table}[!h]
\centering
\caption[Comparison of fit parameters for the \texttt{Pb-only} and \texttt{Multi-nuclei} analyses]{Comparison of the fit-level parameters $a_j^i$ and the corresponding nuclear-modified coefficients $c_j^i(A=208)$ for the \texttt{Pb-only} and \texttt{Multi-nuclei} analyses. The coefficients $c_j^i$ are computed using Eq.~(\ref{eq:6.6}) with the baseline parameters $p_j^i$ given in Table~\ref{tab:fit-parameters}.}
\renewcommand{\arraystretch}{1.2}
\begin{tabular}{lcc}
\toprule
\textbf{Parameter} & \textbf{\texttt{Pb-only}} & \textbf{\texttt{Multi-nuclei}} \\
\midrule

\multicolumn{3}{c}{$a_j^i$ parameters} \\
\midrule
$a_1^{u_v}$ & -0.04152 & 0.00717 \\
$a_2^{u_v}$ & -0.01727 & 0.03068 \\
$a_3^{u_v}$ & -75.973  & 0.30950 \\
\addlinespace
$a_1^{d_v}$ & 0.00897 & 0.06449 \\
$a_2^{d_v}$ & 0.16770 & 0.59227 \\
$a_3^{d_v}$ & 0.01083 & -0.01486 \\
\addlinespace
$a_1^{\bar{u}+\bar{d}}$ & 0.04837 & 0.02748 \\
$a_2^{\bar{u}+\bar{d}}$ & 0.43032 & 0.09484 \\
\addlinespace
$a_1^{g}$ & 0.09135 & 0.08593 \\
$a_2^{g}$ & 0.79180 & 0.77323 \\

\midrule
\multicolumn{3}{c}{$c_j^i = p_j^i + a_j^i \ln(A=208)$} \\
\midrule
$c_1^{u_v}$ & 0.39376 & 0.65364 \\
$c_2^{u_v}$ & 3.45112 & 3.70706 \\
$c_3^{u_v}$ & -405.50878 & 1.65197 \\
\addlinespace
$c_1^{d_v}$ & 1.20738 & 1.50372 \\
$c_2^{d_v}$ & 7.44651 & 9.71266 \\
$c_3^{d_v}$ & -3.44519 & -3.58232 \\
\addlinespace
$c_1^{\bar{u}+\bar{d}}$ & 0.05043 & -0.06107 \\
$c_2^{\bar{u}+\bar{d}}$ & 10.62545 & 8.83481 \\
\addlinespace
$c_1^{g}$ & 1.09065 & 1.06172 \\
$c_2^{g}$ & 10.70746 & 10.60834 \\

\bottomrule
\end{tabular}
\vspace{2mm}
\label{tab:param_comparison}
\end{table}

This amplification is particularly evident in the up-valence parameters. The parameter $a_3^{u_v}$ undergoes a dramatic shift from $-75.97$ in the \texttt{Pb-only} fit to $0.31$ in the \texttt{Multi-nuclei} analysis. When propagated to the physical level, this results in an extreme change in $c_3^{u_v}$, from $-405.5$ to $1.65$, corresponding to a variation of more than two orders of magnitude. This behavior confirms that the $a_3$ parameter in the \texttt{Pb-only} fit is very poorly constrained and driven to unphysical values,\footnote{In the \texttt{Pb-only} analysis, a flat prior is imposed on this parameter, which allows it to explore a wide range of values in the absence of sufficient experimental constraints.} while the inclusion of \texttt{Multi-nuclei} data introduces constraints that restore the coefficient to a natural scale. In contrast, the behavior of the down-valence parameter $a_3^{d_v}$ is much more stable. At the fit level, $a_3^{d_v}$ changes significantly, including a sign reversal, from $0.0108$ to $-0.0149$. However, after including the free proton component, the corresponding coefficient $c_3^{d_v}$ exhibits only a modest shift from $-3.45$ to $-3.58$, i.e., at the few-percent level. This indicates that, unlike the up-valence case, the up-valence $a_3$ parameter is less strongly affected by the inclusion of \texttt{Multi-nuclei} data, although some residual structure persists. In particular, the mild shifts in $c_3^{d_v}$ suggest that the $d_v$ sector is comparatively more stable, while still exhibiting non-Gaussian features consistent with the posterior behavior observed earlier.

The $a_1$ and $a_2$ parameters exhibit more moderate but still significant variations. For example, $a_2^{d_v}$ increases from $0.168$ to $0.592$, leading to a corresponding shift in $c_2^{d_v}$ from $7.45$ to $9.71$, i.e. an increase of approximately $30\%$. Similarly, in the sea-quark parameters, $a_2^{\bar{u}+\bar{d}}$ decreases from $0.430$ to $0.095$, which translates into a reduction of $c_2^{\bar{u}+\bar{d}}$ from $10.63$ to $8.83$, corresponding to a $\sim 17\%$ change. In addition, $c_1^{\bar{u}+\bar{d}}$ exhibits a pronounced shift, changing sign from $0.0504$ to $-0.061$. The gluon sector, by contrast, remains comparatively stable. The parameter $a_2^g$ changes only slightly, from $0.792$ to $0.773$, resulting in a variation of $c_2^g$ from $10.71$ to $10.61$, i.e. below the $1\%$ level. This stability reflects the strong experimental constraints on the gluon distribution, particularly from heavy-flavor production data in proton–lead collisions, and explains why the gluon parameters are less sensitive to the extension of the dataset.

Overall, this comparison demonstrates that while some flavors, most notably the gluon, are robust against the inclusion of additional nuclear data, others, particularly $C_3^{u_v}$ and $c_1^{\bar{u}+\bar{d}}$ parameters, are highly sensitive. The logarithmic amplification factor $\ln(208)$ plays a crucial role in exposing these differences, emphasizing that the physical impact of nuclear effects is more appropriately assessed at the level of the coefficients $c_j^q$, rather than solely through the fitted parameters $a_j^q$.

\begin{figure}[!h]
\centering
\includegraphics[width=\textwidth]{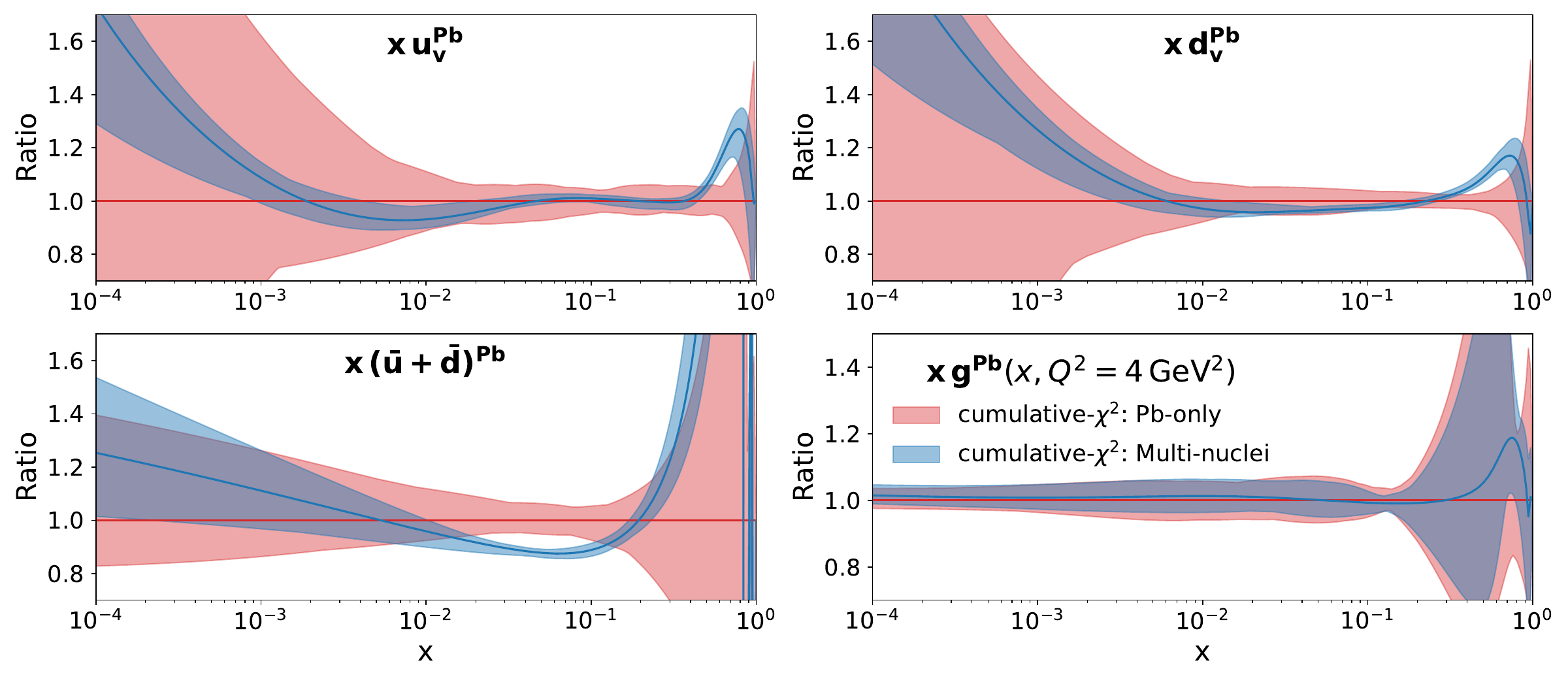}
\caption[PDFs Comparison of \texttt{Pb-only} and \texttt{Multi-nuclei} MCMC analyses]{Comparison of PDFs from the \texttt{Pb-only} and \texttt{Multi-nuclei} MCMC analyses plotted as a ratio to the best-fit PDF from the \texttt{Pb-only} analysis.}
\label{fig:7.23} 
\end{figure}

We now compare the results at the PDF level. Figure~\ref{fig:7.23} shows the ratios of the central values and 68\% uncertainty bands, obtained using the cumulative-$\chi^2$ method,\footnote{The MCMC sample size in the \texttt{Multi-nuclei} analysis (1515 samples) is smaller than in the \texttt{Pb-only} case (7069 samples). While this may introduce slightly larger statistical fluctuations in the extracted uncertainty bands, it does not affect the stability of the central values or the qualitative conclusions of the comparison.} in the two setups, normalized to the \texttt{Pb-only} central result. In this representation, deviations from unity directly reflect the impact of the \texttt{Multi-nuclei} constraints. 

Overall, the central values of the two fits are broadly consistent within uncertainties across most of the $x$-range, indicating that the inclusion of additional nuclear data does not lead to drastic modifications of the extracted PDFs. Nevertheless, systematic and flavor-dependent differences are clearly visible. In the valence sector, both $u_v$ and $d_v$ exhibit an enhancement at small $x$ ($x < 10^{-2}$) in the \texttt{Multi-nuclei} fit, followed by a suppression in the intermediate-to-large
$x$ regions relative to the \texttt{Pb-only} baseline. This effect is particularly pronounced in the up-valence distribution and reflects a redistribution of momentum strength across $x$ induced by the tighter constraints of the global dataset. The light-sea combination ($\bar{u}+\bar{d}$) shows a more pronounced suppression in the intermediate region ($10^{-2}<x < 10^{-1}$) in the \texttt{Multi-nuclei} fit. In particular, the central value develops a distinct dip around $x \simeq 0.1$, which is significantly more pronounced than in the \texttt{Pb-only} fit. Moreover, in this region the change in shape of the \texttt{Multi-nuclei} PDFs extends beyond the uncertainty bands of the \texttt{Pb-only} PDFs, indicating that the constraints are strongly driven by non-lead data. By contrast, the gluon distribution remains largely stable in its central value across the small-to-intermediate $x$ range. Toward very large $x$, both fits show increasing spread, specifically for light-sea and valence PDFs, reflecting the extrapolation beyond the region directly constrained by data.

The most striking feature of the comparison, however, is the behavior of the uncertainty bands. Across all flavors, the \texttt{Multi-nuclei} uncertainties are significantly reduced relative to the \texttt{Pb-only} case. In the valence PDFs, the \texttt{Pb-only} fit exhibits a strong growth of uncertainties at small $x$ ($x \lesssim 10^{-2}$), reflecting the lack of direct constraints on lead in this region. The \texttt{Multi-nuclei} analysis substantially reduces this spread, pulling the distribution toward a more localized range of values. For the light-sea combination $\bar{u}+\bar{d}$, a similar reduction of uncertainties is observed, while at small $x$ the reduction in uncertainties is less pronounced than in the valence sector. Furthermore, the \texttt{Multi-nuclei} fit exhibits a more pronounced and well-constrained structure in the intermediate-$x$ region around $x \sim 0.1$, where the \texttt{Pb-only} uncertainties remain comparatively broad. 

The gluon distribution shows a qualitatively different behavior. Both the central values and the uncertainty bands remain nearly unchanged between the two fits across most of the $x$-range, with only minor differences appearing at very large $x$. This indicates that the gluon PDF is already well constrained in the \texttt{Pb-only} analysis, and the data from other nuclei do not introduce additional information.

These features reflect the different origins of experimental constraints. The reduction of uncertainties in the valence and, to a lesser extent, sea quark distributions arises from the additional information provided by neutral-current DIS and fixed-target Drell–Yan data on lighter nuclei. Through the imposed $A$-dependence, these data propagate constraints to the lead PDFs. In contrast, the gluon distribution is primarily constrained by data taken directly on lead, in particular heavy-flavor production measurements at the LHC, and is therefore less sensitive to the inclusion of lighter nuclei, remaining essentially unchanged. Beyond the reduction in uncertainties, the observed modifications in the central values follow a systematic pattern. 

The enhancement at low $x$ and suppression at intermediate $x$ resemble a reversal of the typical nuclear modification pattern. This suggests that the constraints propagated from lighter nuclei effectively weaken the magnitude of nuclear effects in lead. From a methodological perspective, this highlights that the imposed $A$-dependence introduces correlations between different nuclear systems, which can have a nontrivial impact on the extracted PDFs. The differences observed here can therefore be interpreted as an estimate of the systematic effects associated with this modeling assumption.

\begin{figure}[!h]
\centering
\includegraphics[width=\textwidth]{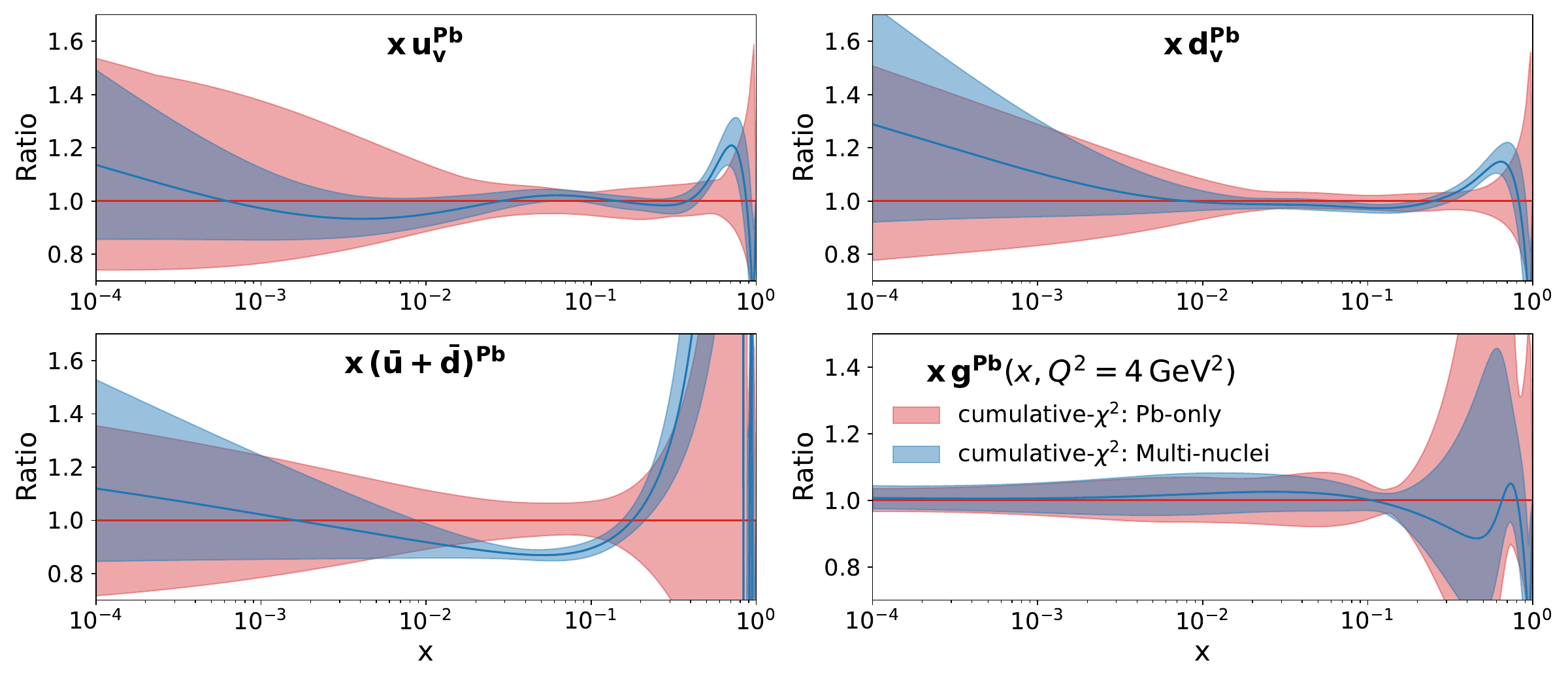}
\caption[PDFs Comparison of \texttt{Pb-only} and \texttt{Multi-nuclei} Hessian analyses]{Comparison of PDFs from the \texttt{Pb-only} and \texttt{Multi-nuclei} Hessian fits plotted as a ratio to the central PDF from the \texttt{Pb-only} Hessian fit.}
\label{fig:7.24} 
\end{figure}

A complementary comparison is provided in Fig.~\ref{fig:7.24}, where the PDFs obtained from the \texttt{Pb-only} and \texttt{Multi-nuclei} Hessian fits are shown as ratios to the \texttt{Pb-only} central result. This allows for a direct assessment of how the addition of \texttt{Multi-nuclei} data modifies both central values and uncertainties within the quadratic approximation.

Comparing the central values of the two fits in this Hessian-based analysis shows that the qualitative behavior is consistent with that observed in the MCMC analysis displayed in Fig.~\ref{fig:7.23}. The valence and sea-quark distributions exhibit similar patterns, with an enhancement at small $x$ and a suppression in the intermediate-$x$ region, indicating that both methods capture the same underlying shifts induced by the additional constraints, although the magnitude of the enhancement is more moderate in the Hessian approach. Moreover, the light-sea quark distribution exhibits a similarly pronounced suppression, or ``dip'' in the region around $x \simeq 0.1$.

The behavior of the uncertainty bands, however, differs more substantially. While the \texttt{Multi-nuclei} Hessian fit leads to a reduction of uncertainties in the valence sector, particularly at small $x$, this effect is noticeably weaker than in the MCMC analysis. In regions where the \texttt{Pb-only} fit is poorly constrained, the Hessian uncertainties for \texttt{Multi-nuclei} analysis remain comparatively large and show only limited reduction, suggesting that the quadratic approximation is less effective in incorporating the additional constraints.
This difference is particularly evident in the gluon distribution. Although the central values remain largely unchanged between the two fits, reflecting the dominant role of lead data, the uncertainty bands exhibit a stronger dependence on the fit setup in the Hessian framework, especially at intermediate and large $x$ where constraints are weaker. In contrast, the MCMC results show a more uniform and stable behavior across the two analyses, indicating a more consistent propagation of constraints.

Overall, these observations indicate that, while both methods reproduce the same qualitative trends in the central values, their treatment of uncertainties differs significantly in regions with limited experimental input. The MCMC approach, by sampling the full posterior distribution, provides a more robust and stable estimate of uncertainties, including the impact of residual non-Gaussian features. In contrast, the Hessian method, based on a local quadratic approximation, is more sensitive to the local structure of the parameter space and may fail to fully capture the effect of additional constraints, particularly in regions affected by non-linearities or weak data coverage.

%%%%%%%%%%%%%%%%%%%%%%%%%%%%%%%%%%%%%%%%%%%%%%%%%%%%%%%
\section{Posterior Compression}\label{sec:7.4}
A key practical goal of this work is to provide a compact yet statistically representative nPDF ensemble suitable for public release in the form of LHAPDF grids \cite{Buckley:2014ana}. In the following, we focus on the posterior obtained from the \texttt{Pb-only} analysis, for which the combined MCMC ensemble contains $N_{\mathrm{tot}} = 7069$ approximately independent samples. While this ensemble provides a faithful representation of the posterior distribution, publishing the full set would be impractical for phenomenological applications. We therefore investigate additional reduction strategies in which a smaller subset of replicas is selected while preserving the essential statistical properties of the original posterior.
To quantify the information loss induced by further reduction, we construct reduced ensembles using different thinning factors and compare their resulting nPDF central values and uncertainty bands to those obtained from the full sample. The objective is to assess how reducing the number of replicas affects the statistical properties of the posterior, particularly in regions where the distributions are non-Gaussian or weakly constrained by data.

\subsection*{Standard Thinning}
Standard thinning consists of selecting replicas at uniform intervals from the posterior ensemble, thereby reducing the total number of samples without introducing an explicit selection bias. In this work, this procedure was already employed at an earlier stage to reduce the autocorrelation within individual Markov chains and to obtain a set of approximately independent samples for the combined posterior ensemble (see Sec.~\ref{sec:6.3.4}). In the additional thinning considered here, the replica corresponding to the minimum $\chi^2$ is always retained, while the remaining samples are uniformly subsampled to achieve the desired ensemble size. Provided that the original chains are sufficiently decorrelated, such uniform thinning preserves the underlying probability distribution in expectation. However, any further reduction in the number of replicas decreases the effective sample size, which directly impacts the precision of uncertainty estimates, particularly in regions where the posterior exhibits large variance or non-Gaussian structure.

\begin{figure}[!h]
\centering
\includegraphics[width=0.49\textwidth]{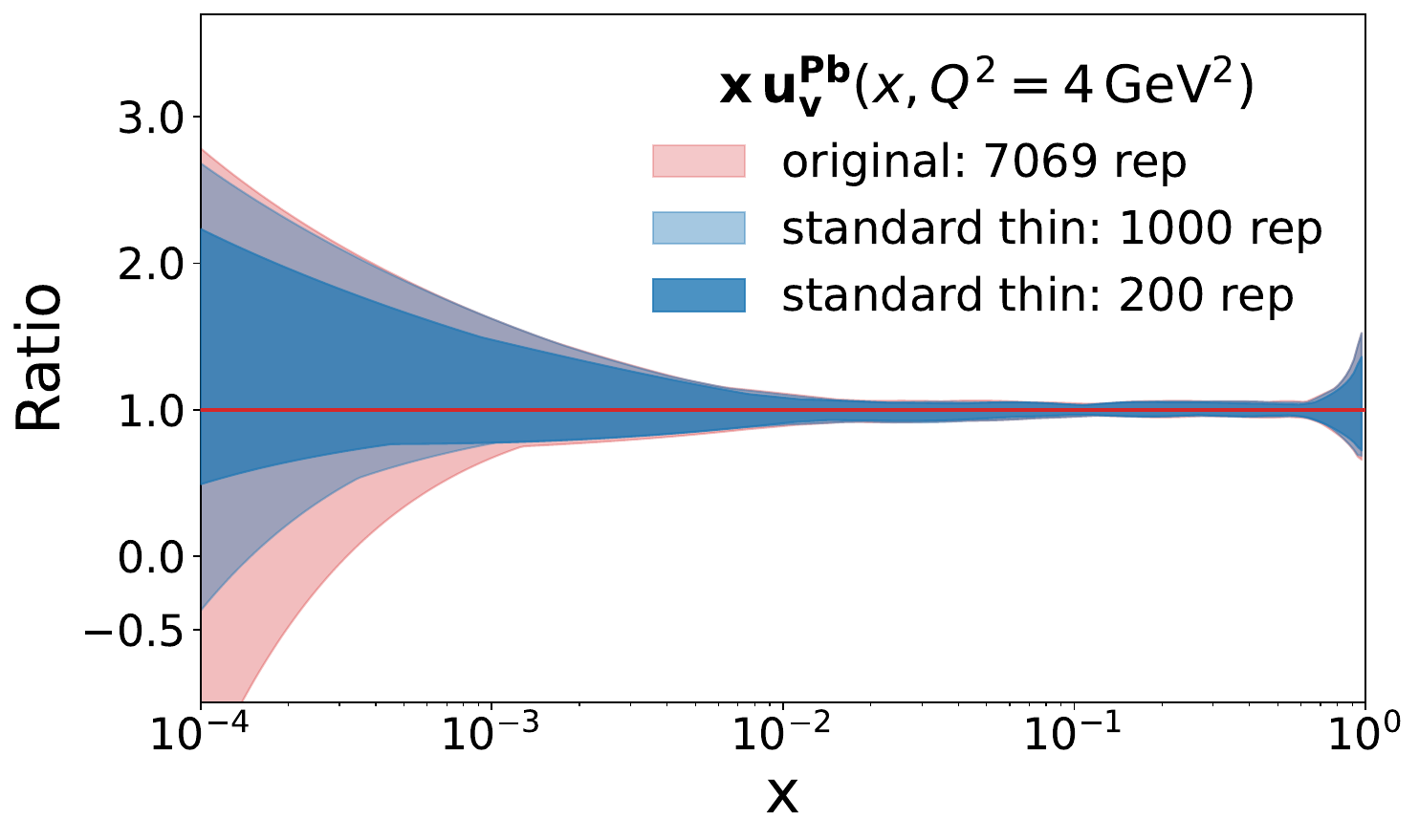}
\includegraphics[width=0.49\textwidth]{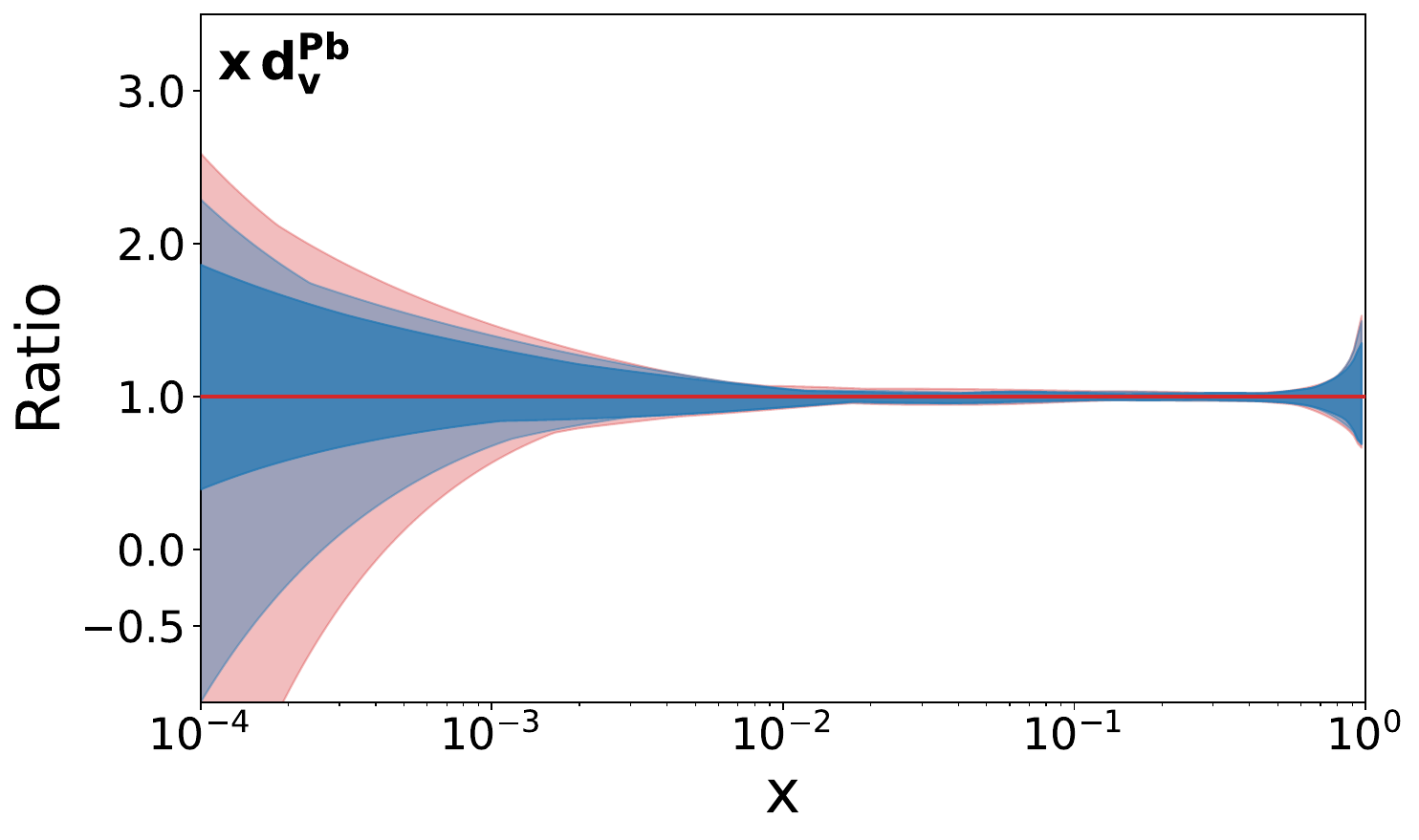}
\includegraphics[width=0.49\textwidth]{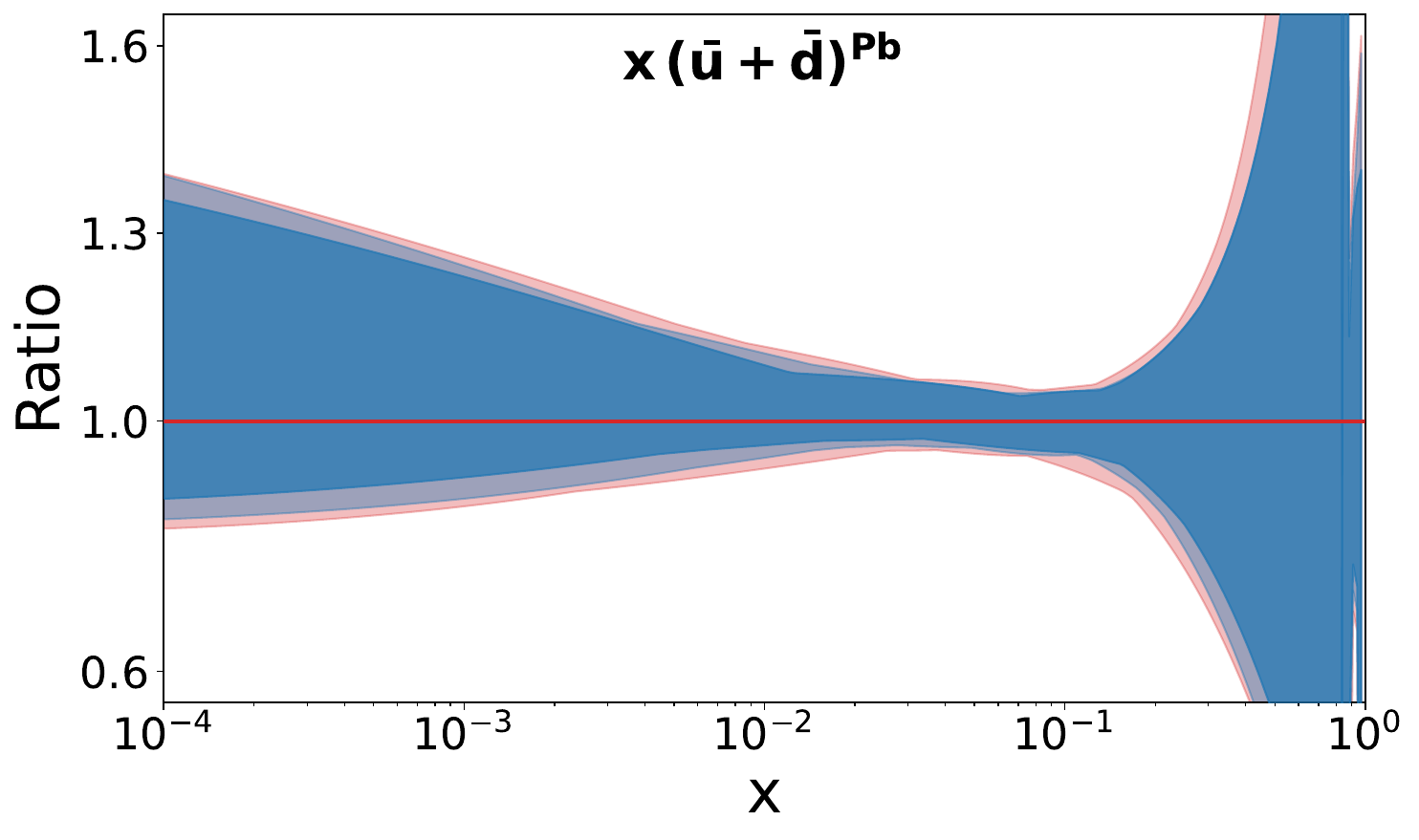}
\includegraphics[width=0.49\textwidth]{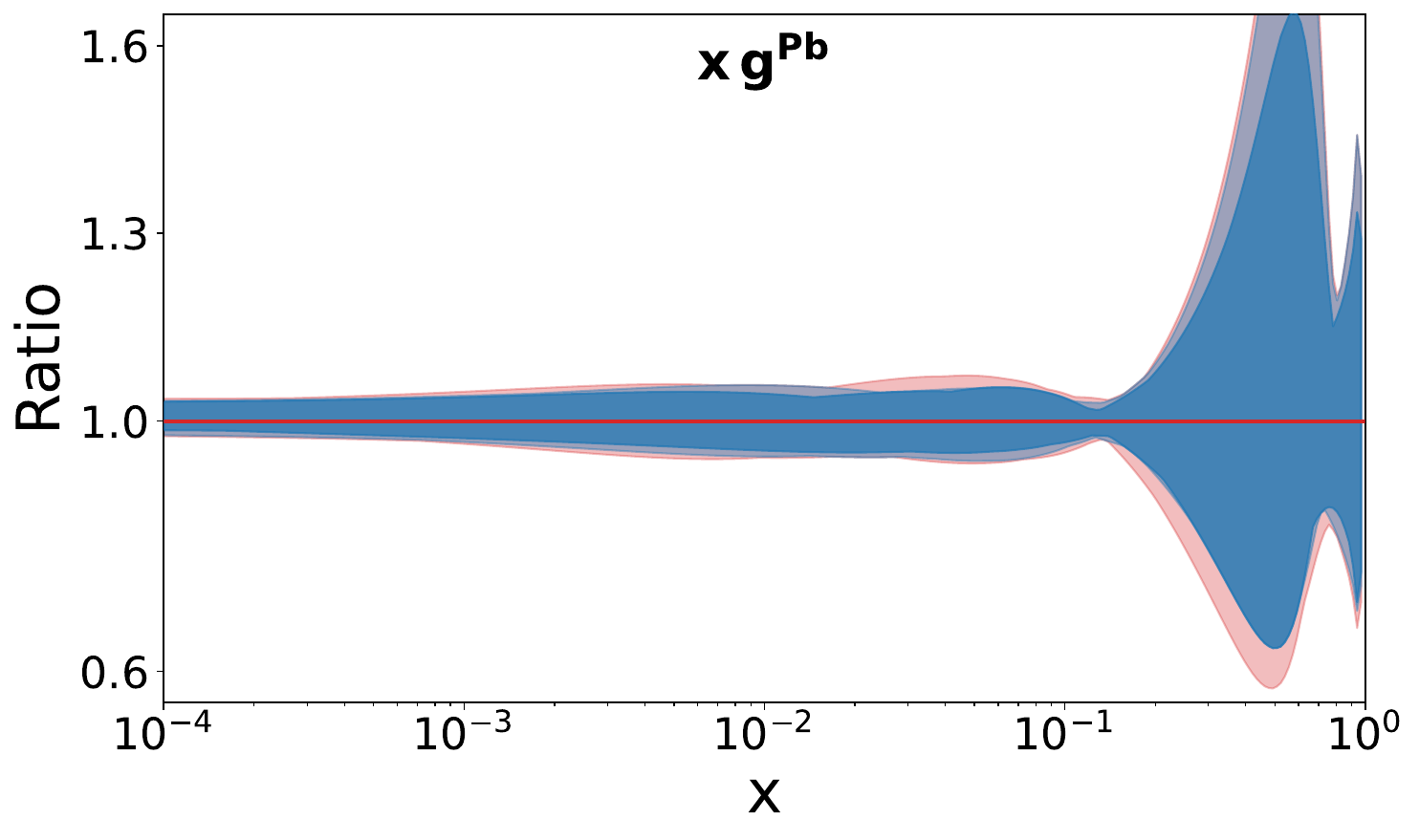}
\caption[Comparison of Pb PDFs uncertainties at different thinning rates]{Standard thinning: Comparison of Pb PDFs uncertainty estimates at different thinning rates, derived using the cumulative-$\chi^2$ method.}
\label{fig:7.25}
\end{figure}

Figure~\ref{fig:7.25} shows the results obtained using the cumulative-$\chi^2$ method for the original ensemble of 7069 replicas and its thinned subsets of 1000 and 200 replicas. The central values remain generally stable under thinning, as the minimum-$\chi^2$ replica is explicitly retained and therefore not affected by the subsampling procedure. Any residual variations arise only from the reduced sampling of the surrounding posterior. The degree of stability is, nevertheless, flavor dependent. For $u_v$, $d_v$, the uncertainty bands exhibit a clear and systematic dependence on the number of retained replicas. In the intermediate-to-large $x$ region ($ 10^{-2}<x<0.5$), where the fit is strongly constrained by data, the uncertainty bands for both the 1000- and 200-replica ensembles closely follow those of the full sample. In these regions, the posterior is approximately Gaussian and sufficiently narrow that even a reduced ensemble provides an adequate statistical representation.
At small $x$ ($x \lesssim 10^{-3}$), the impact of thinning becomes more pronounced. While the 1000-replica ensemble still reproduces the overall shape and width of the uncertainty bands reasonably well, the 200-replica case shows visible distortions. These effects reflect insufficient sampling of the tails of the posterior, where fluctuations are larger and the distributions deviate from Gaussian behavior, typically leading to enhanced asymmetries.

In contrast, the light-sea and gluon distributions exhibit a higher degree of stability under thinning. When comparing the uncertainty bands across different replica counts (7069, 1000, and 200), a strong level of statistical convergence is observed. The bands largely overlap across the full $x$-range, indicating that the global features of the underlying probability distribution are well captured even at reduced sample sizes. This behavior suggests that the corresponding posteriors are relatively smooth and do not exhibit pronounced localized structures that would require a larger ensemble to resolve.

Overall, these results demonstrate that while standard thinning preserves the central values of the PDFs by construction,  it can substantially reduce the accuracy of the uncertainty estimates when the number of replicas becomes too small. In particular, regions characterized by limited experimental constraints and non-Gaussian posterior structure, most notably the valence PDFs at small and large $x$, require a sufficiently large ensemble to ensure stable and reliable uncertainty quantification.

\subsection*{Compression via Genetic Algorithm}
To achieve a more efficient reduction of the posterior ensemble, we employ the \textit{Compressed Monte Carlo} (CMC) approach introduced in Ref.~\cite{Carrazza:2015hva}. In this method, a subset of replicas is selected such that it optimally reproduces the statistical properties of the full ensemble, including central values, uncertainties, correlations, and higher moments. Unlike standard thinning, which samples replicas uniformly, the CMC approach performs an optimized selection based on the global structure of the posterior.

The quality of the compressed set is quantified using an error function (ERF), which measures the deviation between the reduced and original ensembles across multiple statistical descriptors. These include the central value (CV), standard deviation (STD), skewness (SKE), correlation matrices (Corr), kurtosis (KUR), and distribution-level comparisons via a Kolmogorov–Smirnov distance (KS). For example, the contribution from the central values is defined as
\begin{equation}
\mathrm{ERF}_{\text{CV}} = \frac{1}{N_{\text{CV}}} 
\sum_{i = -n_f}^{n_f} \sum_{j = 1}^{N_x} 
\left( 
\frac{ f_i^{\text{CV}}(x_j, Q_0) - g_i^{\text{CV}}(x_j, Q_0)} 
{g_i^{\text{CV}}(x_j, Q_0)} 
\right)^2,
\end{equation}
where $f_i^{\text{CV}}$ and $g_i^{\text{CV}}$ denote the central values of the compressed and original ensembles, respectively. Analogous contributions are included for the variance, higher moments, and correlations, leading to the total error function
\begin{equation}
\mathrm{ERF}_{\text{total}} = \mathrm{ERF}_{\text{CV}} + \mathrm{ERF}_{\text{STD}} + \mathrm{ERF}_{\text{SKE}} + \mathrm{ERF}_{\text{KUR}} + \mathrm{ERF}_{\text{KS}} + \mathrm{ERF}_{\text{Corr}}.
\end{equation}
The optimal subset of replicas is identified using a genetic algorithm, which iteratively searches for the combination that minimizes the ERF. Starting from random candidate subsets, the algorithm applies mutation and selection steps to progressively improve the agreement with the full ensemble until convergence is reached.

\begin{figure}[!h]
\centering
\includegraphics[width=0.49\textwidth]{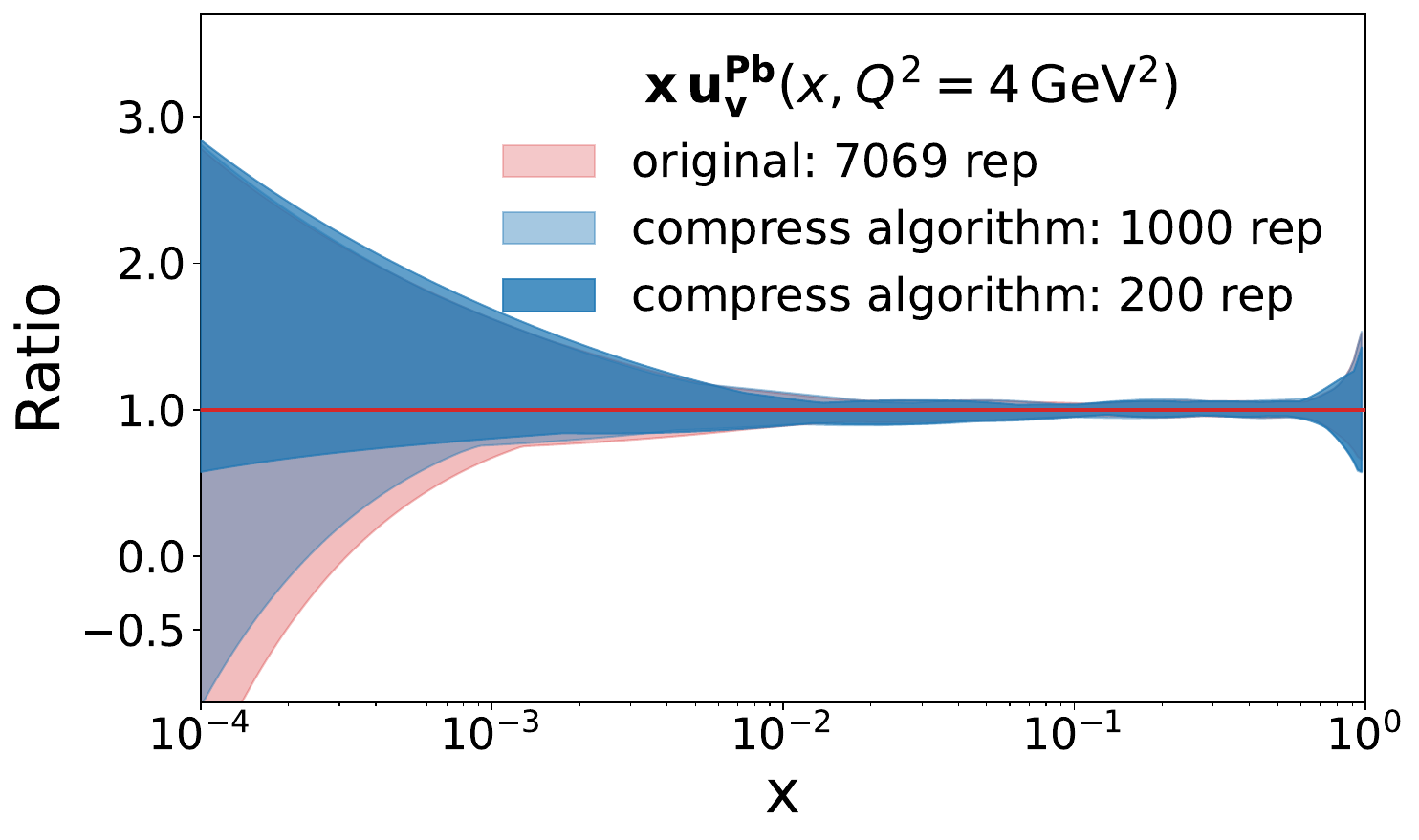}
\includegraphics[width=0.49\textwidth]{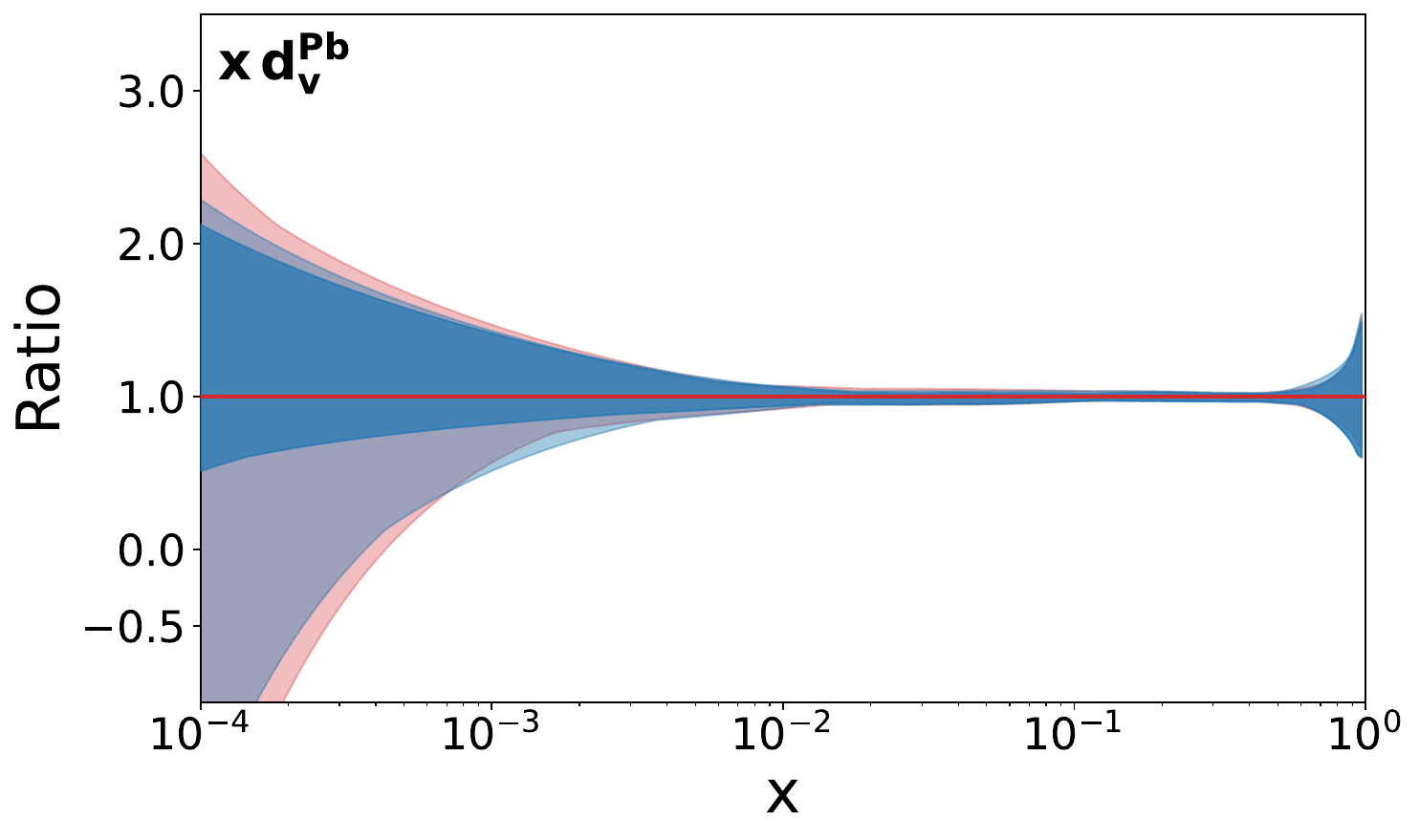}
\includegraphics[width=0.49\textwidth]{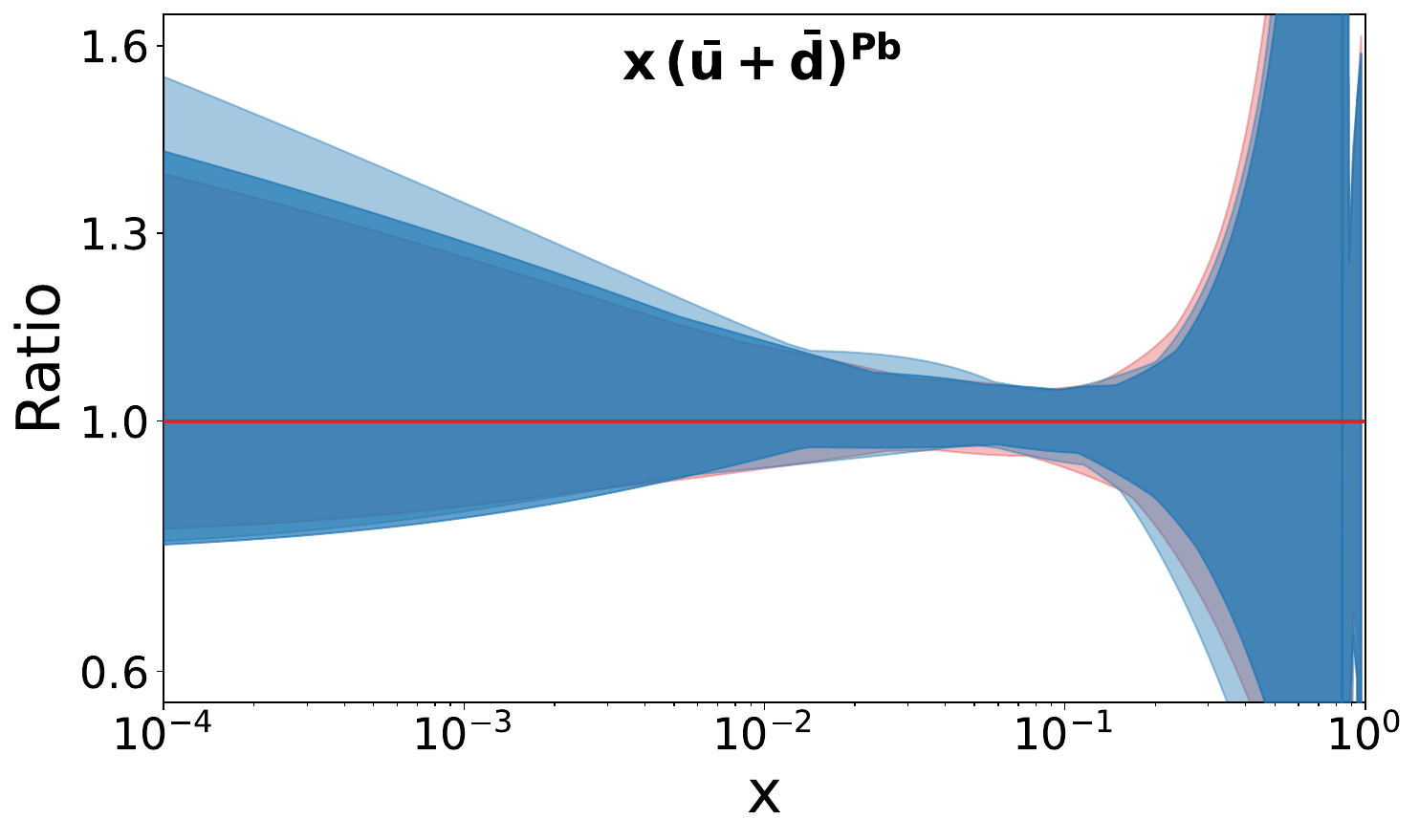}
\includegraphics[width=0.49\textwidth]{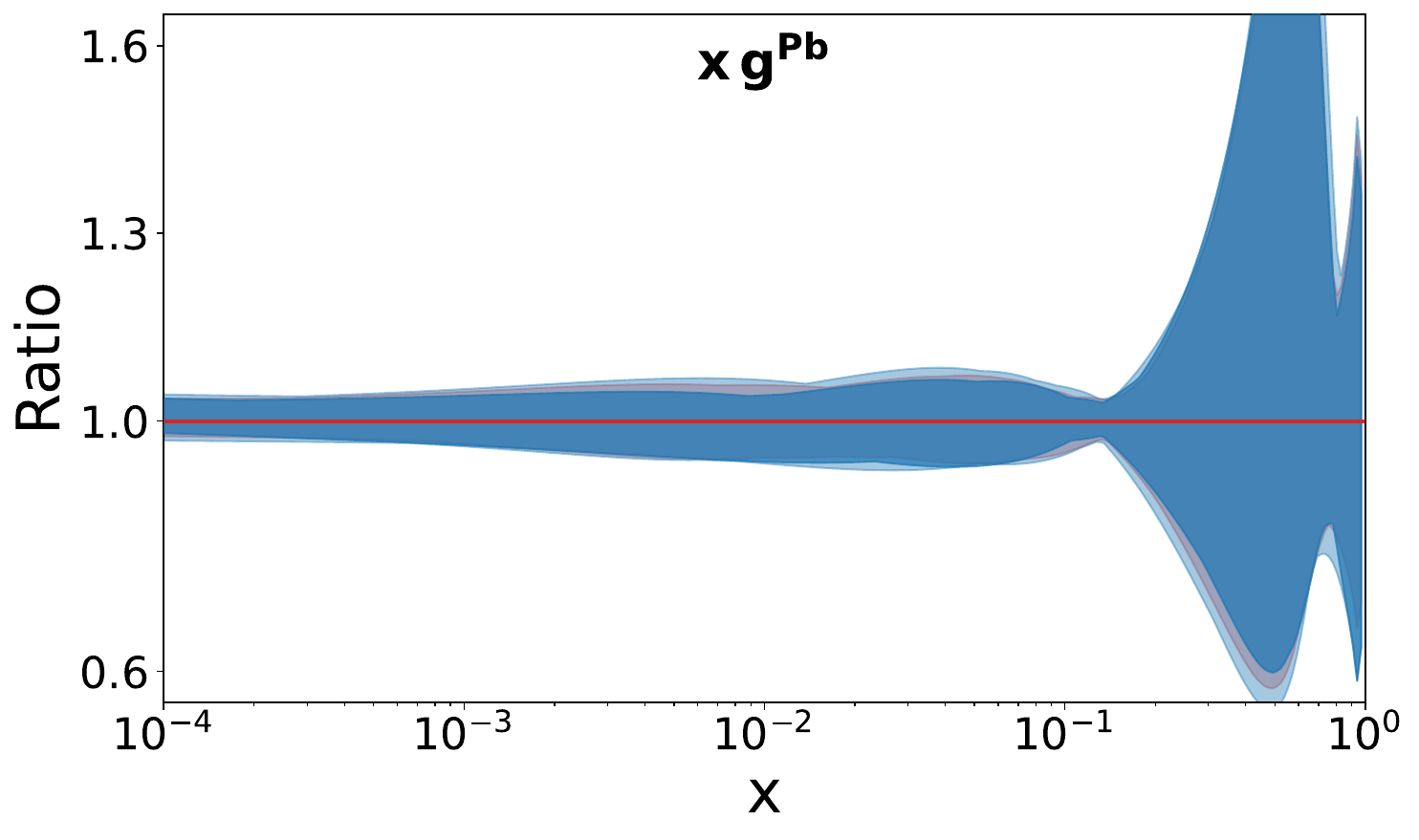}
\caption[Comparison of Pb PDFs uncertainties at different thinning rate]{Compression algorithm: Comparison of Pb PDFs uncertainty estimates at different thinning rates, derived using the cumulative-$\chi^2$ method.}
\label{fig:7.26}
\end{figure}

The performance of the compression procedure is illustrated in Fig.~\ref{fig:7.26}, where compressed ensembles of 1000 and 200 replicas are compared to the original sample. As in the standard thinning approach, the replica corresponding to the minimum $\chi^2$ is explicitly retained, ensuring that the central value remains fixed by construction. For moderately reduced ensembles (1000 replicas), the compression method provides a clear improvement over standard thinning, yielding a more accurate reproduction of uncertainty bands. In particular, the algorithm efficiently preserves correlations and higher-order features of the posterior, including the asymmetric and non-Gaussian structures observed in the valence sector. However, for more aggressive reductions (200 replicas), the performance of the compression method degrades and becomes less reliable than standard thinning. In this regime, the optimization procedure struggles to adequately represent the full complexity of the posterior using a very limited number of replicas.
As a result, the compressed set may introduce distortions in the uncertainty bands, whereas standard thinning, despite its simplicity, provides a more stable representation by maintaining a more uniform sampling of the original distribution. This suggests that 200 replicas are not sufficient to retain the information required to accurately describe more complex distributions, regardless of the reduction method employed.

Overall, the compression approach enables a substantial reduction in the number of replicas while preserving the statistical accuracy of the original posterior, provided that the target ensemble size remains sufficiently large. Compared to standard thinning, the CMC method is more effective at retaining non-Gaussian features and correlations when moderate compression is applied. However, for very small ensemble sizes, its performance can degrade, highlighting a trade-off between compression efficiency and statistical robustness. These results indicate that compression is most beneficial in the regime of moderate reduction, while simple thinning may remain preferable for extremely compact replica sets.
\chapter{Conclusion}\label{chap.8}
In this thesis, we have developed and applied a fully Bayesian framework for the determination of nuclear parton distribution functions using Markov Chain Monte Carlo (MCMC) methods. The primary objective has been to investigate the statistical structure of nuclear PDF fits beyond the limitations of conventional approaches, with particular emphasis on the geometry of the posterior distribution and its implications for uncertainty estimation of PDFs.

The \texttt{Pb-only} analysis, based on a high-statistics posterior ensemble, revealed a highly nontrivial structure of the parameter space. In particular, the posterior distributions exhibit pronounced non-Gaussian features, including asymmetries, strong parameter correlations, and the presence of multiple local minima in the $a_3$ valence parameter. These effects lead to broadened and asymmetric uncertainty bands, which cannot be consistently described within the standard Hessian framework. This demonstrates that the commonly assumed quadratic approximation to the $\chi^2$ landscape is insufficient to capture the full complexity of nuclear PDF fits.

The extension to a \texttt{Multi-nuclei} analysis shows that the inclusion of data from different nuclear targets provides essential constraints on the parameter space. The resulting cross-nuclear constraints significantly reduce flat directions and poorly constrained combinations of parameters and suppress the pronounced multi-modal behavior observed in the \texttt{Pb-only} case. While the posterior distribution remains non-Gaussian, its structure becomes more regular and closer to a Gaussian-like regime. Consequently, the extracted PDFs exhibit smoother and more symmetric uncertainty bands, and the agreement between different uncertainty estimation methods improves, particularly in regions that are well constrained by the data.

A direct comparison between the two analyses indicates that the main features of the lead PDFs are already determined in the \texttt{Pb-only} fit, while the \texttt{Multi-nuclei} framework enhances the overall precision and stability of the extraction. In particular, it provides stronger constraints in poorly determined regions, such as the valence sector and the kinematic boundaries at small and large momentum fractions. Moreover, the increased regularity of the posterior improves the validity of the Hessian approximation, which becomes progressively more reliable when global nuclear constraints are incorporated.

From a methodological perspective, this work demonstrates that a Bayesian MCMC approach provides a fundamentally more complete description of the statistical properties of nuclear PDF fits. By directly sampling the posterior distribution, this framework naturally captures non-Gaussian features, complex correlations, and multi-modal structures that remain inaccessible in conventional analyses. The performed \texttt{Pb-only} analysis is the first single-nucleus extraction of nuclear PDFs. Together with the \texttt{Multi-nuclei} analysis, it provides a first-ever estimate of systematic effects (uncertainties) that are introduced to nPDF analysis by the routinely used $A$-dependence tying together data for different nuclear targets.

Despite these advances, several limitations remain. The computational cost of high-dimensional MCMC sampling restricts both the number of fit parameters that can be efficiently explored, in contrast to the Hessian approach, and the overall statistical precision achievable in more complex analyses. In addition, due to these computational constraints, not all available nuclear datasets could be included in the present study, particularly in the \texttt{Multi-nuclei} framework.

Future developments may focus on improving the computational efficiency of the framework, for example through parallelization of the $\chi^2$ evaluation across individual datasets as well as through the implementation of more advanced sampling algorithms. Such improvements would enable the inclusion of a larger number of fit parameters and a more comprehensive set of experimental data. The Bayesian approach developed in this work also provides a natural foundation for studying the impact of future measurements, for example at the Electron-Ion Collider, and for constructing statistically robust PDF ensembles for phenomenological applications. It also provides a means of determining and validating the $\Delta\chi^2$ tolerance used in the Hessian method, which is not uniquely defined and is an important limitation of this method.

In summary, this thesis establishes a consistent Bayesian framework for nuclear PDF determination and demonstrates its advantages in revealing the underlying structure of the parameter space. The combination of MCMC-based inference with comprehensive experimental constraints provides a more reliable and physically transparent characterization of nuclear PDFs and their uncertainties, representing a significant step toward precision QCD studies in nuclear environments.

% ********************************** Back Matter *******************************
% Backmatter should be commented out, if you are using appendices after References
%\backmatter

% ********************************** Bibliography ******************************
\begin{spacing}{0.9}

% To use the conventional natbib style referencing
% Bibliography style previews: http://nodonn.tipido.net/bibstyle.php
% Reference styles: http://sites.stat.psu.edu/~surajit/present/bib.htm
%\bibliographystyle{utphys}
%\bibliographystyle{apalike}
%\bibliographystyle{unsrtnat}
%\bibliographystyle{unsrt} % Use for unsorted references  
%\bibliographystyle{plainnat} % use this to have URLs listed in References
\cleardoublepage
% If you would like to use BibLaTeX for your references, pass `custombib' as
% an option in the document class. The location of 'reference.bib' should be
% specified in the preamble.tex file in the custombib section.
% Comment out the lines related to natbib above and uncomment the following line.
\printbibliography[heading=bibintoc, title={References}]

\end{spacing}

% ********************************** Appendices ********************************

%\begin{appendices} % Using appendices environment for more functunality

%\include{Appendix1/appendix1}
%\include{Appendix2/appendix2}

%\end{appendices}

% *************************************** Index ********************************
\printthesisindex % If index is present

\end{document}